\documentstyle[aps,prd,eqsecnum,epsfig]{revtex}

\begin{document}
\title{Dispersion approach to quark-binding effects\\ 
in weak decays of heavy mesons}
\author{Dmitri Melikhov}
\address{Institut f\"ur Theoretische Physik der Universit\"at Heidelberg, \\
Philosophenweg 16, D-69120, Heidelberg, Germany\thanks{On leave from: 
{\it Nuclear Physics Institute, Moscow State University, Moscow, Russia}}}
\maketitle
\thispagestyle{empty}
\vspace{2cm}
\begin{abstract}
The dispersion approach based on the constituent quark picture and its
applications to weak decays of heavy mesons are reviewed. 
Meson interaction amplitudes are represented within this approach 
as relativistic spectral integrals over the mass variables 
in terms of the meson wave functions and spectral densities of the 
corresponding Feynman diagrams. Various applications of this approach 
are discussed: 

\vspace{.1cm}
Relativistic spectral representations for meson elastic and transition form 
factors at spacelike momentum transfers are constructed. 
Form factors at $q^2>0$ are obtained by the analytical continuation. 
As a result of this procedure, form factors 
are given in the full $q^2$ range of the weak decay in terms of the wave 
functions of the participating mesons. 

\vspace{.1cm}
The $1/m_Q$ expansion of the obtained spectral representations 
for the form factors for the particular limits of the heavy-to-heavy and 
heavy-to-light transitions are analysed.
Their full consistency with 
the constraints provided by QCD for these limits is demonstrated.  

\vspace{.1cm}
Predictions for form factors for $B_{(s)}$ and $D_{(s)}$ decays to 
light mesons are given.  

\vspace{.1cm}
The $B\to\gamma\ell\nu$ decay and the weak annihilation in rare radiative 
decays are considered. Nonfactorizable corrections to the $B^0-\bar B^0$ 
mixing are calculated.  

\vspace{.1cm}
Inclusive weak $B$ decays are analysed and the differential distributions 
are obtained in terms of the $B$ meson wave function. 

\vspace{.4cm}
\end{abstract}
\newpage
\tableofcontents
\newpage
\section{Introduction}
\label{sec:introduction}
Weak decays of hadrons provide an important source of information on
the parameters of the standard model, the structure
of weak currents, and internal structure of hadrons. 
Therefore for many years weak decays have been in the focus 
of experimental and theoretical investigations. 

In the last decade, the main emphasis has been laid on weak $B$ decays which  
allow to measure the unknown parametrs of the Cabibbo-Kobayashi-Maskawa (CKM) matrix 
describing the mixing of heavy $c$, $b$ and $t$ quarks, and the CP-violation:  
semileptonic and nonleptonic $B$ decays induced 
by the charged-current quark transition $b\to u, c$ provide a direct 
access to the CKM matrix elements $V_{ub}$ and $V_{cb}$ and the weak 
CP-violating phase. 

Rare semileptonic $B$ decays induced by the 
flavour-changing neutral current transitions $b\to s$ and $b\to d$ measure  
$V_{ts}$ and represent an important test of the standard model and of the physics beyond it.  
Rare decays are forbidden at the tree level in the standard model and occur through 
loop diagrams. Thus they provide the possibility to probe at the 
relatively low energies of $B$ decays 
the structure of the electroweak sector at large mass scales from contributions of virtual particles 
in the loop diagrams. 
Interesting information about the structure of the theory is contained in the 
Wilson coefficents in the effective Hamiltonian which describes the $b\to s,d$ transition 
at low energies. These Wilson coefficients take different values in different theories with 
testable consequences in rare $B$ decays. 

However, along with the interesting parameters of the standard model, the decay rates and 
differential distributions measurable in the decay process contain also quantites 
related to the presence of hadrons in the decay process. 
Therefore the extraction of the interesting standard model parameters
from the decay experiments requires reliable information on hadron structure and hadronic 
amplitudes of the weak quark currents.  

Theoretical description of hadronic amplitudes of the quark currents is one of the key problems of particle 
physics as such amplitudes provide a bridge between QCD formulated in the language of quarks and gluons and
observable phenomena which deal with hadrons. The difficulty in such calculations lie in the fact that hadron
formation occurs at large distances where perturbative QCD methods are not applicable and therefore 
a nonperturbative consideration is necessary.

The presence of heavy quarks with the masses much bigger than the confinement scale of QCD provide 
important constraints upon the long distance effects due to a new spin-flavour symmetry which emerges 
in this limit \cite{shifmanvoloshin,iwhh}. This symmetry leads to a heavy quark effective 
theory (HQET) 
\cite{hqet} for QCD with heavy quarks. HQET restricts the structure of the expansion 
of the hadronic transition amplitudes in inverse powers of the heavy quark mass. 

In inclusive $B$ decays the combination of the heavy-quark and the operator product expansions allow one 
to connect the decay rate of the quark bound in a heavy hadron to the decay rate of the free quark. 
An important consequence of the operator product expansion is the appearance of the 
corrections to the free quark decay rate only at the second power of the 
$1/m_Q$ \cite{cgg,bsuv93}. Therefore, these corrections are numerically small. 
Providing quite reliable predictions for the integrated rates, the theoretical approach based on the OPE turns out to be less efficient for the
description of the differential distributions. In this case a more precise information on the 
details of the $b$ quark motion inside the $B$ meson is necessary. 

For the exclusive $b\to c$ decays, HQET provides constraints on the structure of the expansion of 
the meson transition form factors in the inverse powers of the $b$ and $c$ quark masses 
and leads to the appearance of universal process-independent form factors at each order 
of the $1/m_Q$ expansion \cite{luke,neubert}. Moreover, heavy quark symmetry provides the absolute value of the 
leading order universal form factor - the Isgur-Wise function at zero recoil (maximum momentum transfer) 
thus allowing for a relible description of the $b\to c$ decay in this kinematical region.  
HQET however cannot calculate the  universal form factors as functions of the momentum transfer. 
Moreover, the $1/m_c$ corrections for the 
$b\to c$ transiton form factors turn out to be not small because the $c$ quark 
is not sufficiently heavy. 

For the exclusive $b\to u,s$ transitions, when the final quark is light, 
relations between the form factors describing meson transition induced by
different currents emerge in the region near zero recoil \cite{iwhl}. 
In the opposite region of large recoils, where the final quark is fast, one can construct 
another effective theory, so called large energy effective theory (LEET) which allows 
the double expansion of the transition 
form factors in inverse powers of $m_b$ and the energy $E$ of the light quark produced 
in the weak decay \cite{leet}. LEET predicts the appearance of several universal form factors at the leading order 
of the $1/E$ and $1/m_b$ expansion, but does not calculate these form factors, and also does not constrain 
the structure of higher order corrections. 

Thus, heavy-quark symmetry provides important constraints for the form factors but does not allow to 
calculate them in full detail. This task requires a detailed treatment of the nonperturbative effects. 

Theoretical approaches for calculating transition form factors are
quark models \cite{wsb,koerner,isgw,isgw2,jaus,jauswyler,dubin,beyer,faustov,orsay,aley,sim95,simula,gns,m1,mplb1,mplb2,m2}, 
QCD sum rules \cite{sr1,bbd,bhr,ball,sr4,sr7,colangelo,lcsr,lcsr1,braun,aliev,likhoded}, 
and lattice QCD \cite{lat1,lat2,lat3,lat4,lat5,lattice,ape,ukqcd,latg}.  
Approaches combining different methods are also extensively used  
\cite{lr,burdmandonoghue,stech,lat,anal1,anal2,lellouch,soares}.  

In spite of the considerable progress made in the recent years 
theoretical uncertainties are still uncomfortably large, around 10-15\%. 
The main difficulty in obtaining the full picture of the form factors for various decays and for
all $q^2$ lies in the fact that methods more directly related to QCD, 
such as lattice QCD and QCD sum rules
have only a limited range of applicability, whereas the results from quark models are
sensitive to the details of the model and the parameters used. 

QCD sum rules are suitable for describing the low $q^2$
region of the form factors. Sum rule calculations require various inputs such 
as the condensates or the distribution amplitudes of the light mesons produced 
in the heavy meson decay. The higher $q^2$ region is hard to get and
higher order calculations are not likely to give real progress because of the
appearance of many new parameters. The accuracy of the method
cannot be arbitrarily improved because of the necessity to isolate the
contribution of the states of interest from others. 

Lattice QCD gives good results for the high $q^2$ region. But because of the
many numerical extrapolations involved this method does not provide for
a full picture of the form factors and for the relations between the various
decay channels.

Quark models do provide such relations connecting different 
processes through the meson soft wave functions, and give 
the form factors in the full $q^2$ range. 
However, quark models are not closely related to the QCD
Lagrangian (or at least this relationship is not well understood yet) and
therefore have input parameters which are not directly measurable and may
not be of fundamental significance.

Since quark models are not directly deduced from QCD it is important to match 
the results for the transition form factors obtained within quark models  
to rigorous QCD predictions for these form factors in the specific limit of 
heavy quark masses. 

The application of various versions of the constituent quark picture 
to decay processes has a long history.  
First models were based on a relativistic \cite{wsb} or 
nonrelativistic \cite{koerner,isgw} considerations. They could not fully 
incorporate the quark dynamics and spin structure and therefore could not satisfy 
rigorous relations for the transition 
form factors based of the spin-flavour symmetry in the heavy quark limit of QCD. 
 
A self-consistent relativistic treatment of the quark spins can be performed 
within the light-front quark model \cite{lcqm,lcqm1}. The model allows for a calculation 
of the reference-frame dependent partonic contribution to the form factor. However, the
non-partonic contribution cannot in general be calculated. At spacelike momentum transfers 
the nonpartonic part can be killed by an appropriate choice of the reference frame. Therefore, 
the partonic contribution calculated in this specific reference frame gives the full form factor. 
At timelike momentum transfers the nonpartonic contribution does not vanish for any accessible
choice of the reference frame, and the knowledge of the frame-dependent partonic contribution is
not sufficient to determine the form factor. 

A relativistic dispersion approach to the decay processes based on the constituent 
quark picture overcomes the above difficulties. It has been formulated in 
\cite{m1,mplb1,mplb2,m2} and applied to the study of the long-distance effects in 
many pocesses involving the $B$ meson: such as 
exclusive semileptonic \cite{m1,mplb1,mplb2,m2,mb1,mb2,ms} and rare 
\cite{mnsplb1,mns} $B$ decays, the $B\to\gamma l\nu$ form factor and the weak annihilation 
in the rare radiative $B\to \rho\gamma$ decay \cite{bmns,ms1}, nonfactorizable corrections 
for the $B^0-\bar B^0$ mixing \cite{mn}, calculation of the differential distributions in 
inclusive $B\to X_cl\nu$ decays \cite{msimula}. 

The approach is based on a consistent treatment of the two-particle singularities of the 
Feynman diagrams describing meson interactions. Amplitudes of these processes are given by the
relativistic spectral representations over the mass variables in terms of the wave functions
of the participating mesons and spectral densities of the corresponding Feynman diagrams. 
In particular, meson transition form factors both at spacelike and timelike momentum transfers
are given by the double spectral representations. 

We discuss in this paper applications of the dispersion approach to various processes 
involving heavy mesons, laying the main emphasis on the calculation of the form factors 
for heavy meson decays. 

Let us highlight the main features of our dispersion formulation of the constituent quark
model:

\vskip0.5cm
\centerline{\it 1. The physical picture}
\vskip0.5cm
\noindent 
The constituent quark picture is based on the following phenomena expected from QCD: 
\begin{itemize}
\item
the chiral symmetry breaking in the low-energy region which provides for 
the masses of the constituent quarks;
\item
a strong peaking of the nonperturbative meson wave functions
in terms of the quark momenta with a width of the order of
the confinement scale;  
\item
a $q\bar q$ composition of mesons in terms of constituent quarks. 
\end{itemize}
As is well known, the $q\bar q$ component of the meson wave function in terms of  
current quarks dominates the exclusive form factors in the deep inelastic region, 
i.e. for large spacelike momentum transfers. A successesfull description of the 
mass spectrum of mesons as $q\bar q$ states in terms of the constituent quarks
obtained in \cite{gi} shows that the $q\bar q$ approximation works well also 
in the soft region, where one however has to take into account the transition 
of the current quarks to constituent quarks. The $q\bar q$ meson component 
in terms of the constituent quarks leads to a good description of the elastic 
form factors at small and intermediate momentum transfers \cite{amn1,amn2}. 
Therefore, one can expect the two-particle approximation to be quite reliable for 
the description of the range of momentum transfers relevant for 
heavy meson decays.  

An important shortcoming of previous quark model predictions was a strong
dependence of the results on the special form of the quark model
and on the parameter values. We demonstrate that once 
\begin{itemize}
\item[(a)]
a proper relativistic formalism is used for the description of the transition 
form factors and 
\item[(b)]
the numerical parameters of the model are chosen properly (we discuss
criteria for such a proper choice below), 
\end{itemize}
the quark model yields results
in full agreement with the existing experimental data and in accord with the
predictions of more fundamental theoretical approaches. 
In addition, our approach allows to predict many other form factors which
have not yet been measured.

\vskip0.5cm
\centerline{\it 2. The formalism}
\vskip0.5cm

\noindent For the description of the transition form factors in their full
$q^2$-range and for various initial and final mesons, a fully relativistic
treatment is necessary. The dispersion formulation of the quark
model provides for such a relativisitc treatment and guarantees the correct 
spectral and analytical properties of the obtained form factors. 

The form factors are given by the double spectral representations 
over the variables $s_1$ and $s_2$, the squares of the invariant masses of the initial 
and final $Q\bar q$ pairs, respectively. 
The integrations in $s_1$ and $s_2$ run along the two-particle cuts in the complex $s_1$ and
$s_2$ planes. 
The spectral functions of these spectral representations involve the 
wave functions of the participating mesons and the double discontinuities of the 
corresponding triangle Feynman diagrams. 

We start our analysis of the transition form factors from the spacelike region $q^2<0$, 
where the double discontinuities can be calculated by means of the Landau-Cutkosky rules. 

The form factors in the decay region $q^2>0$ are obtained by the analytical
continuation in $q^2$. A specific feature of the timelike decay region of $q^2$,  
is the 
appearance of the anomalous cuts in the complex $s_1$ and $s_2$ planes and, respectively, of 
the
anomalous contributions to the form factors. Let us point out that the anomalous contribution 
as well as the normal one is completely determined by the wave functions of the participating 
mesons in the physical region. The anomalous contribution is small for small positive $q^2$ 
but becomes increasingly important as $q^2$ rises.     

The form factors obtained by this procedure obey all rigorous constraints from QCD 
on the structure of the long-distance corrections in the heavy quark limit:  
they develop the correct heavy-quark expansion at
leading and next-to-leading orders in $1/m_Q$ in accordance with QCD 
in the case when both quarks participating in the weak transition are heavy, i.e. 
have masses much larger than the confinement scale of QCD.

For the heavy-to-light transition, i. e. when only the initial quark is treated as heavy, 
the transition form factors of the dispersion approach 
satisfy the relations between the form factors
of vector, axial-vector, and tensor currents valid at small recoil. 
In the limit of the heavy-to-light transitions at small $q^2$ the form
factors obey the lowest order $1/m_Q$ and $1/E$ relations of the Large
Energy Effective Theory. 

We want to emphasise once more that form factors for meson decays  
in the physical decay region $q^2>0$ are directly calculated through the 
meson wave functions. 

\vskip0.5cm
\centerline{\it 3. Parameters of the model}
\vskip0.5cm

\noindent 
In previous applications of quark models the transition form factors
turned out to be sensitive to the numerical parameters, such as the
quark masses and the slopes of the meson soft wave functions.

A possible way to control quark masses and the meson soft wave functions
is to use the lattice results for the $B\to \rho$
form factors at large $q^2$ as 'experimental' inputs. 
The $b$ and $u$ constituent quark masses
and slope parameters of the $B$, $\pi$, and $\rho$ wave functions assuming for 
them a simple Gaussian form are obtained through this procedure \cite{mb1}. 
The Gaussian wave functions of the charm and strange mesons and the effective masses 
$m_c$ and  $m_s$ are fixed by fitting the measured rates
for the decays $D\to (K, K^*)l\nu$ \cite{ms}.  

With these few inputs, numerous predictions for the form 
factors for the $D_{(s)}$ and $B_{(s)}$ decays into light mesons 
are obtained \cite{ms} which nicely agree with the experimental results 
at places where data are available. The calculated transition form factors 
are also found to be in good agreement with the results from lattice QCD 
and from sum rules in their regions of validity. 

Thus, in spite of the rather different masses and properties of mesons
involved in weak transitions, all existing data on the form factors can be 
understood in the quark picture, i.e. all 
form factors can be described by the few degrees of
freedom of constituent quarks. Details of the soft wave functions are not 
crucial; only the spatial extention  of these wave functions of order of 
the confinement scale is important. In other words, only the meson radii are essential 
for the decay processes. 

\vspace{.5cm}
The paper is organised as follows: 

\vspace{.5cm}

In Chapter II we present details of the composite system description using spectral
representations over mass variables. The start with the amplitude of the constituent 
interaction in the low-energy region and its analytical properties. 
We then the consider the interaction of the bound state 
with the external electromagnetic field and construct the gauge and relativistic invariant
amplitude of the bound state interaction. The elastic electromagnetic form factor is 
discussed and the relativistic wave function is introduced. The normalization condition 
for the wave function is obtained which corresponds to the electric charge conservation. 

Properties of pseudoscalar mesons are studied. A single spectral representation 
for the weak leptonic decay constant and double spectral representations for 
the elastic electromagnetic and the weak transition form factors in terms of the wave 
function are obtained at $q^2<0$. For comparison with the light-cone quark model results 
we rewrite our explicitly invariant spectral representations in terms of the light-cone 
variables and demonstarte the equivalence of the form factors obtained within our dispersion 
approach and the light-cone quark model for $q^2<0$. 

The weak transition form factor at $q^2>0$ is obtained by the 
analytical continuation which discussed in full detail. 

We then consider the case when one of the quarks in the pseudoscalar meson is heavy, 
$m_Q\to\infty$ and analyse the size of the $1/m_Q$ effects in the
form factors for quark masses in the range of the $b$ and $c$ quarks. 

\vspace{.5cm}

Chapter III contains a detailed discussion of the weak transitions of the pseudoscalar mesons
to pseudoscalar and vector mesons. The double spectral representation for both cases are 
constructed starting from $q^2<0$ and going to $q^2>0$ by the analytical continuation. 
A procedure of fixing the subtraction prescription in the double spectral representations 
is discussed. 

We then perform the $1/m_Q$ expansion of the dispersion form factors for the 
heavy-to-heavy transition to next-to-leading order accuracy in $1/m_Q$.  
For the heavy-to-light transition the expansion within the leading order in $1/m_Q$ 
is developed. 
Full consistency with the structure of the heavy-quark expansion in QCD 
for both of these cases to the orders considered is verified. 

Spectral representation for the Isgur-Wise function and subleading universal form factors 
is obtained in terms of the wave function of an infinitely heavy meson. Numerical estimates
for the universal form factors are given.   

\vspace{.5cm}

In Chapter IV the choice of the numerical paremeters of the model is discussed and form
factors for many $D(D_s)$ and $B(B_s)$ to light mesons are calculated. 
Convenient parametrizations for the calculated form factors are given. Strong coupling
constants of heavy mesons are estimated by analysing the residues of the form factors 
in the poles located beyond the physical region of meson decay. 
The strong coupling constant obtaibed by this procedure are found in agreement with the 
results of the direct calculation within our dispersion approach. 

A detailed comparison with the experimental data and results from other approaches
is presented. In all cases a good agreement with the available experimental data on form 
factors and strong coupling constants is observed.  

\vspace{.5cm}

In Chapter V the $B\to\gamma l\nu$ form factors and the weak annihilation in rare 
$B\to\rho\gamma$ decay is analysed within the factorization approximation. 
A detailed analysis of the contact terms in the weak annihilation amplitude is presented. 
A new contribution missed in the previous analyses of the weak annihilation is reported. 
Parameter-free numerical estimates of the relevant form factors are given. 

\vspace{.5cm}

Chapter VI contains the analysis of the nonfactorizable effects in the $B^0-\bar B^0$ mixing 
due to soft gluon exchanges. Assuming the dominance of the local gluon condensate, the
correction to factorization can be expressed in terms of the specific $B$ meson transition
form factors at zero momentum transfers. It is shown that the correction is strictly 
negative independent of the values of the form factors. The form factors are calculated within the
dispersion approach and numerical estimates for them are obtained. 

\vspace{.5cm}

In Chapter VII the application of the dispersion approach to inclusive $B\to X_c l\nu$ decays 
is presented. Spectral representation for the integrated semileptonic rate in terms of the 
$B$ meson wave function is constructed. The subtraction prescription is discussed and the
absence of the $1/m_Q$ correction in the ratio of the  bound to free quark decay rates is 
verified. Differential distributions are calculated in terms of the $B$ meson wave function. 

\vspace{.5cm}

Conclusion summarises the main results.

\newpage
\section{\label{sec:i}Spectral representation for bound state transition form factors}

This Chapter presents a formalism for the relativistic description of hadron form factors 
within the constituent quark picture \cite{m1,amn1,amn2,akmsyaf,akms,ammp,ammpyaf,myaf}.

Our approach is based on representing the amplitudes of hadron interactions in the form of the 
dispersion integrals over the hadron mass in terms of the hadron soft wave function.   
This procedure corresponds to a consistent relativistic treatment of the 
leading two-particle singularities of the scattering amplitude 
and the bound state form factors at spacelike momentum transfer $q$.  
The form factors at the timelike momentum transfers corresponding to the decay process 
$M_{ini}\to M_{final}$ are obtained by performing the analytical continuation in the 
variable $q^2$ from its negative to positive values. 
As a result, the weak decay form factors in the kinematical region $q^2\le (M_{ini}-M_{final})^2$ 
can be expressed through the bound state wave function. 

The application of spectral representations to the description of composite systems has
a long history \cite{chew,mandelstam,blank1,blank2,burton}. Usually, 
spectral representations in $q^2$ are considered, $q$ being the momentum transfer. 
In this case anomalous singularities in $q^2$ appear in the explicit form as separate 
contributions to spectral representations. 

We present here an approach to the bound state description based on spectral 
representations in mass variables. Within this approach anomalous contributions turn out 
to be included in the usual dispersion integrals for the form factors at spacelike momentum 
transfers, relevant for the scattering problems. Form factors at timelike momentum transfers,  
corresponding to the decay processes, are obtained by performing the analytic continuation 
from the spacelike region. In this case, anomalous cuts give separate contributions to 
spectral representations for the form factors. 

An important advantage of spectral representations in mass variables is the possibility to 
introduce in a consistent way a relativistic-invariant function which describes the 
motion of the constituents inside the bound state and which can be interpreted as the 
bound state wave function. We show how this function emerges when keeping in a 
relativistic and gauge-invariant way only two-particle 
singularities of the Feynman graphs. 

In Section \ref{i.1} we give some details of describing relativistic bound state 
using spectral representations. 

In Section \ref{i.2} we present all technical details of the
description of a pseudoscalar meson within the dispersion
approach (leptonic weak decay, two-photon decay, elastic
electromagnetic form factor) and demonstrate the equivalence of the 
dispersion approach and the light-cone constituent quark model \cite{lcqm}. 

In Section \ref{i.3} we study form factors describing weak transitins between pseudoscalar
mesons starting with the region of spacelike momentum transfers. 
As the next step, we perform the analytic continuation to the region $q^2>0$ and show how 
the anomalous contribution to the form factor in this region emerges. The origin of this 
anomalous contribution is connected with the non-Landau-type singularities of the Feynman 
triangle diagrams.

In Section \ref{i.4} we analyse electroweak properties of pseudoscalar mesons 
using a simplified parameterization of the meson wave function based on the heavy quark symmetry. 
This allows us to study the dependence of the axial-vector decay constant $f_P$ and the
heavy-meson elastic form factor on the heavy quark mass $m_Q$. In particular, we study 
the transition to the heavy quark limit $m_Q\to\infty$ and discuss the size of the 
subleading $1/m_Q$-corrections for the heavy quark mass in the region of 
$b$ and $c$ quark masses. We calculate weak decay form factors at timelike momentum transfers 
and compare our results with other theoretical analyses and the experimental data. 

\subsection{\label{i.1} Bound state description within dispersion relations}
In this section we present some aspects of the dispersion approach to the 
relativistic description of the bound states. 
For the sake of argument we consider the case of two spinless constituents with the 
masses $m_1$ and $m_2$ interacting via
exchanges of a meson with the mass $\mu$. We start with the scattering
amplitude of the real constituents 
\begin{eqnarray}
A(s,t)=\langle k'_1,k'_2|S|k_1,k_2\rangle, 
\quad &&s=(k_1+k_2)^2,\; t=(k_1-k'_1)^2,\nonumber\\
&&k_1^2=k_1'^2=m_1^2,\;k_2^2=k_2'^2=m_2^2.  
\end{eqnarray}
The amplitude as a function of $s$ has the
threshold singularities in the complex $s$-plane connected with
elastic rescatterings of the constituents and production of new mesons at
\begin{equation}
s= (m_1+m_2)^2,\;(m_1+m_2+\mu)^2,\;(m_1+m_2+2\mu)^2\ldots
\end{equation}
We assume that an $S$-wave bound state with the mass $M<m_1+m_2$ exists, then
the partial wave amplitude $A_0(s)$ has a pole at $s=M^2$.
The amplitude $A(s,t)$ has also $t$-channel singularities with thresholds at 
$t=(n\mu)^2;\;\; n=1,2,3\ldots$ connected with meson exchanges.
If one needs to construct the amplitude in the low-energy region $s\geq
(m_1+m_2)^2$
the dispersion $N/D$ representation turns out to be convenient.
Consider the $S$-wave partial amplitude
\begin{equation}
A_0(s)=\int\limits^1_{-1} dz\, A(s,t(s,z)),
\end{equation}
where
$t(z)=-(1-z)\lambda(s,m_1^2,m_2^2)/2s$, $z=\cos\theta$ in the c.m.s.
The $A_0(s)$ as a function of complex $s$ has the right-hand
singularities related to $s$-channel singularities of $A(s,t)$.
In addition, it has left-hand singularities located at
$s=(m_1+m_2)^2-(n\mu)^2;\;\; n=1,2,3\ldots$. They come from $t$-channel
singularities
of $A(s,t)$.
The unitarity condition in the region $s\approx (m_1+m_2)^2$ reads
\begin{equation}
{\rm Im}\, A_0(s)= \rho(s)\;|A_0(s)|^2,\qquad
\rho(s) =\frac{\lambda(s,m_1^2,m_2^2)}{16\pi s}
\end{equation}
with
$\rho(s)$ the two-particle phase space.
The $N/D$ method represents the partial amplitude as $A_0(s)=N(s)/D(s)$, where
the function $N$ has only left-hand singularities and $D$ has only right-hand
ones. The unitarity condition yields
\begin{eqnarray}
D(s) &=& 1 - \int\limits^\infty_{(m_1+m_2)^2}\frac{d\tilde{s}}{\pi}\,
\frac{\rho(\tilde{s})N(\tilde{s})}{\tilde{s}- s}\; 
\nonumber\\
&\equiv& \; 1-B(s).
\end{eqnarray}
Assuming the function $N$ to be positive we introduce
$G(s)=\sqrt{N(s)}$.
\begin{figure}
\begin{center}
\mbox{   \epsfig{file=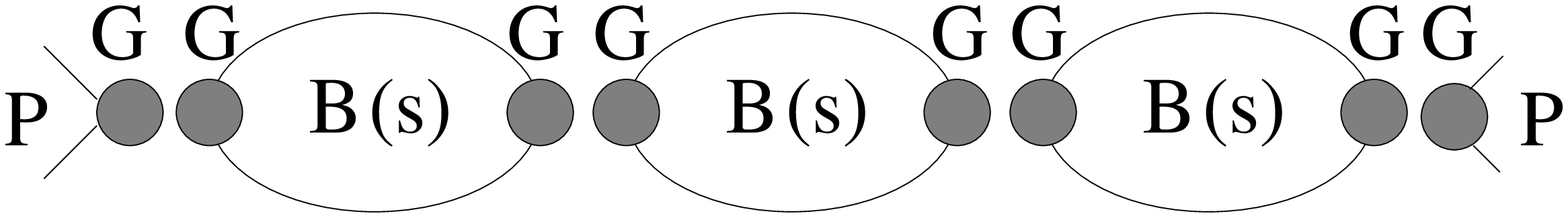,width=10cm}    }
\caption{ One of the terms in the expansion of $A_0(s)$
\label{fig:a1}}
\end{center}
\end{figure}
Then the partial amplitude takes the form
\begin{eqnarray}
A_0(s)&=&G(s)\left[1+B(s)+B^2(s)+B^3(s)+\ldots\right] G(s)
\nonumber\\
&=&\frac{G(s)G(s)}{1-B(s)}.
\end{eqnarray}
This expression can be interpreted as a series of loop diagrams of
Fig.\ref{fig:a1}
with the basic loop diagram
\begin{equation}
B(s)=\int\limits^\infty_{(m_1+m_2)^2}\frac{d\tilde{s}}{\pi}\,
\frac{\rho(\tilde{s})\;
G^2(\tilde{s})}{\tilde{s}- s}.
\end{equation}
The bound state with the mass $M$ corresponds to a pole both in the total
and partial amplitudes at $s=M^2$ so $B(M^2)=1$.
Near the pole one has for the total amplitude
\begin{eqnarray}
\label{amplitude-bs}
A&=&\langle k'_1,k'_2|p\rangle \frac1{M^2-p^2}\langle p|k_1,k_2 \rangle+{\rm regular\; terms} \nonumber \\
&\equiv&\chi^*_p(k'_1,k'_2)\frac1{M^2-s}\chi_p(k_1,k_2)+\ldots
\end{eqnarray}
where $\chi_p(k_1,k_2)$ is the amputated Bethe-Salpeter amplitude
of the bound state.
The dispersion amplitude near the pole reads
\begin{eqnarray}
\label{amplitude-nd}
A&=&N/D+{\rm regular\; terms\; related\; to\; other\; partial\; waves}
\nonumber\\
&=&\frac{G^2(M^2)}{(M^2-s)B'(M^2)}+\ldots
\nonumber\\
&\equiv&\frac{G_v^2(M^2)}{M^2-s}+\ldots
\end{eqnarray}
where $G_v$ is a vertex of the bound state transition to the constituents and 
\begin{equation}
\label{bprime}
B'(M^2)=\int\limits^\infty_{(m_1+m_2)^2}\frac{d\tilde{s}}{\pi}\,
\frac{\rho(\tilde{s})\;
G^2(\tilde{s})}{(\tilde{s}- M^2)^2}.
\end{equation}
The singular terms in Eqs (\ref{amplitude-bs}) and (\ref{amplitude-nd}) 
correspond to each other and hence
\begin{equation}
\label{bsa}
\chi_p(k_1,k_2)\to G_v(s)\equiv \frac{G(s)}{\sqrt{B'(M^2)}}.  
\end{equation}
Underline that among right-hand singularities only the two-particle cut
is taken into account in the constructed dispersion amplitude. 

\begin{center}
\begin{figure}[h]
\mbox{\epsfig{file=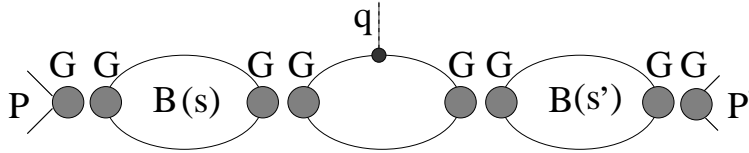,width=10cm}}
\caption{ One of the terms in the series for $T_\mu$.
\label{fig:a2}}
\end{figure}
\end{center}

Let us turn to the interaction of the two-constituent system with an external
electromagnetic field. The amplitude of this process
$T_\mu=\langle k'_1,k'_2|J_\mu|k_1,k_2\rangle$ in the case of a bound 
state takes the form
\begin{eqnarray}
\label{tmu}
T_\mu&=& \langle k'_1,k'_2|p'\rangle 
\frac1{p'^2-M^2}
\langle p'|J_\mu|p\rangle 
\frac1{p^2-M^2}
\langle p|k_1,k_2\rangle+\ldots   
\nonumber \\
&=&\chi^*_p(k'_1,k'_2)
\frac1{P'^2-M^2}
(p+p')_\mu F(q^2)
\frac{1}{p^2-M^2}
\chi_p(k_1,k_2)+\ldots
\end{eqnarray}
where the bound state form factor is defined according to the relation 
\begin{equation}
\langle p'|J_\mu|p\rangle=(p'+p)_\mu F(q^2). 
\end{equation}
The dispersion amplitude
$T_\mu$ with only two-particle singularities in the $p^2$- and
$p'^2$-channels taken into account is given \cite{akms} by the series of graphs
in Fig.\ref{fig:a2}.
These graphs are obtained from the dispersion scattering
amplitude series by inserting a photon line into constituent lines.
The amplitude has the form 
\begin{eqnarray}
T_\mu(p',p,q)&=&2p_\mu(q)T(s',s,q^2)+\frac{q_\mu}{q^2}C,
\\
\nonumber
p^2=s,\;p'^2&=&s',\;q=p'-p,\;p_\mu(q)=(p-\frac{qp}{q^2}\,q)_\mu
\end{eqnarray}
The dispersion method allows one to determine the transverse part 
$T(s,s',q^2)$ of the amplitude. 
Summation of the series of dispersion graphs in Fig.\ref{fig:a2} gives
\begin{equation}
T(s',s,q^2)=\frac{G(s)}{1-B(s)}\Gamma(s',s,q^2)\frac{G(s')}{1-B(s')}.
\end{equation}
Here
\begin{eqnarray}
\Gamma(s',s,q^2)=\int\frac{d\tilde{s}\,G(\tilde{s})}{\pi(\tilde{s}-s)}
\frac{d\tilde{s}'\,G(\tilde{s}')}{\pi(\tilde{s}'-s')}\Delta(\tilde{s}',\tilde{s},q^2),
\nonumber
\end{eqnarray}
and $\Delta(\tilde{s}',\tilde{s},q^2)$
is the double spectral density of the three-point Feynman graph with a
pointlike vertex of the constituent interaction.

The longitudinal part $C$ is determined by the Ward identity
\begin{equation}
C=\frac{G(s)}{1-B(s)}\left\{ {B(s')-B(s)} \right\} \frac{G(s')}{1-B(s')}.
\end{equation}
In the region $s\simeq s'\simeq M^2$, $T_\mu$ develops 
both $s$ and $s'$ poles, so we write 
\begin{equation}
\label{ff}
T_\mu(p',p,q)=\frac{G_v(M^2)}{M^2-s}(p'+p)_\mu F(q^2)
\frac{G_v(M^2)}{M^2-s'}+{\rm less\;singular\;terms}
\end{equation}
where
\begin{equation}
\label{ffv}
F(q^2)=\int\limits_{(m_1+m_2)^2}^\infty
\frac{d\tilde s\, G_v (\tilde s)}{\pi(\tilde s -M^2)}
\frac{d\tilde s'\, G_v (\tilde s')}{\pi(\tilde s'-M^2)}\Delta(\tilde s',\tilde s,q^2).
\end{equation}
is the bound-state form factor  (see (\ref{bsa}) and (\ref{tmu})).
Thus, the quantity $\langle p'|J_\mu|p \rangle$ corresponds to the three-point 
dispersion graph with the vertices $G_v$.
One can derive the following important relation 
\begin{eqnarray}
\label{deltai}
\Delta(\tilde s',\tilde s,0)=\pi\delta(\tilde s'-\tilde s)\rho(\tilde s). 
\end{eqnarray}
This is a consequence of the Ward identity which relates the
three-point graph at zero
momentum transfer to the loop graph. Making use of the relations (\ref{deltai}) 
and (\ref{bprime}), one obtains the normalization of the form factor at zero momentum
\begin{eqnarray}
F(0)=1,  
\end{eqnarray}
which is just the charge conservation condition. 
The expression (\ref{ffv}) gives the form factor in
terms of the $N$-function of the constituent scattering amplitude and double
spectral density of the Feynman graph. 

If the constituent is a nonpoint particle,
the expression (\ref{ffv}) should be multiplied by form factor of an on-shell
constituent.


In general, for constructing the spectral representation 
of the amplitude describing the bound state interaction within a  
two-particle approximation, the following prescription is valid: 
the spectral density of the amplitude is just the spectral density of the corresponding 
Feynman graph multiplied by $G_v$. The amplitude obtained through this procedure 
takes into account in a consitent relativistic-invariant way only two-particle 
singularities of the corresponding amplitude.

\subsection{\label{i.2}Quark structure of pseudoscalar mesons}
The pseudoscalar meson $P$ with the mass $M$ is considered to be a bound state
of the constituent quark with the mass $m_1$ and the antiquark with the mass
$m_2$. Therefore in order to derive the meson interaction amplitude for e.g. 
leptonic decay $P\to l\nu$, two-photon decay of the neutral pseudoscalar meson 
$P^0\to \gamma\gamma$, and the elastic electromagnetic interaction 
$\langle P(p')|J_\mu|P(p)\rangle$, we start with the corresponding amplitudes 
of the constituent quark interactions $\langle l\nu|Q\bar Q\rangle$, 
$\langle \gamma\gamma|Q\bar Q\rangle$, and $\langle Q\bar
Q|J_\mu|Q\bar Q\rangle$ and single out poles corresponding
to the pseudoscalar meson. The amplitude of the $Q\bar Q$ interaction 
turns out to be the basic quantity for describing
the bound state properties. Near the pole corresponding to the bound state with the
quantum numbers $J^P=0^-$, the amplitude is dominated by the $S$-wave partial 
amplitude. The latter can be expressed in the two-particle approximation through
the dispersion loop graph $B_{ps}$ with the vertex
\begin{equation}
\label{vert}
\frac{\bar Q^a(k_1,m_1) i\gamma_5 Q^a(-k_2,m_2)}{\sqrt{N_c}}G(s)
\end{equation}
with
$a$ a color index, $N_c=3$ the number of quark colors,
$k_1^2=m_1^2$, $k_2^2=m_2^2$,
and $(k_1+k_2)^2=s\ne M^2$.
For on-shell constituents, the expression (\ref{vert}) is the only independent
spinorial structure.

\begin{figure}
\begin{center}
\mbox{\epsfig{file=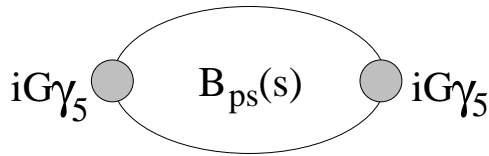,height=2.cm}  }
\end{center}
\caption{Meson dispersion loop graph $B_{ps}(P^2)$.\label{fig:1}}
\end{figure}

The dispersion loop graph Fig. \ref{fig:1}, which is connected
with the meson vertex normalization, reads
\begin{equation}
B_{ps}(p^2)=\int\limits^\infty_{(m_1+m_2)^2}\frac{ds\;G^2(s)}{\pi(s-p^2)}\rho_{ps}(s),
\quad B_{ps}(M^2)=1, 
\end{equation}
with $\rho_{ps}(s)$ the spectral density of the Feynman loop graph
\begin{eqnarray}
\rho_{ps}(s,m_1,m_2)&=&\frac1{8\pi^2}\int dk_1
dk_2\delta(k^2_1-m_1^2)\delta(k^2_2-m_2^2)
\delta(\tilde p-k_1-k_2)
\nonumber\\
&&\times (-1)\;Sp\left({ (\hat k_1+m_1)i\gamma_5(m_2-\hat k_2)i\gamma_5 }\right)
\nonumber\\
&=&\frac{\lambda^{1/2}(s,m_1^2,m_2^2)}{8\pi s}(s-(m_1-m_2)^2)\;\theta(s-(m_1+m_2)^2),
\end{eqnarray}
where $s=\tilde p^2$ and 
\begin{eqnarray}
\nonumber
\lambda(s,m_1^2,m_2^2)\equiv(s+m^2_1-m^2_2)^2-4sm_1^2.
\end{eqnarray}
Taking into account constituent-quark rescatterings leads to the
renormalization of $G$ (Section \ref{i.1}) and the soft 
constituent-quark structure of the pion is given by the vertex 
\begin{equation}
\label{vertex}
\frac{\bar Q^a(k_1,m_1) i\gamma_5 Q^a(-k_2,m_2)}{\sqrt{N_c}}G_v(s)
\end{equation}
where $G_v(s)=G(s)/B'(M^2)$, such that 
\begin{equation}
\label{vertnorm}
\int \frac{G^2_v(s)\rho_{ps}(s,m_1,m_2)ds}{\pi(s-M^2)^2}=1
\end{equation}
Once the soft vertex (\ref{vertex}) is fixed,
we can proceed with calculating meson interaction amplitudes.

\subsubsection{Leptonic decay constant}
Let us consider the decay $P\to l\nu$. The corresponding amplitude reads 
\begin{equation}
\langle 0|A_\mu(0)|P(p)\rangle =ip_\mu\;f_P. 
\end{equation}
In this expression $A_\mu=\bar q^a(0) \gamma_\mu\gamma_5 q^a(0)$ 
is the axial-vector current where summation over quark colours is implied 
and $f_P$ is the meson axial-vector decay constant. 

We must first take into account that the pion structure is described in terms of the 
constituent quarks whereas the current $A_\mu(0)=\bar q\gamma_\mu\gamma_5 q$ is given 
in terms of the current quarks. So let us consider the quantity
\begin{eqnarray}
\langle 0|A_\mu(0)|Q\bar Q\rangle.
\end{eqnarray} 
The bare matrix element which emerges before we take into account the soft 
interctions of the constituent quarks has the structure
\begin{eqnarray}
A_\mu^0&=&\langle 0|A_\mu(0)|Q(k_1)\bar Q(k_2)\rangle_{bare}
\nonumber\\
&=&\bar Q(-k_2)\left[{
\gamma_\mu\gamma_5g^0_A(p^2)+p_\mu\gamma_5h^0_+(p^2)+(k_1-k_2)_\mu
\gamma_5h^0_-(p^2)}\right]Q(k_1)
\end{eqnarray}
If current quarks were identical to constituent ones we would have had
\begin{equation}
\label{barevalues}
g_A^0(p^2) \equiv 1, \quad h_+^0(p^2) \equiv 0, \quad h_-^0(p^2) \equiv 0.
\end{equation}
It is reasonable to assume that the form factors
$g^0$ and $h^0_{\pm}$ are not far from these values \cite{weinberg,jaffe}. 

\begin{figure}
\begin{center}
\mbox{\epsfig{file=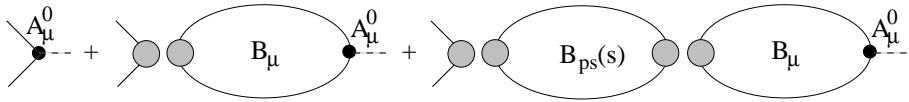,width=12.cm}}
\end{center}
\caption{The series of dispersion graphs for $\langle Q\bar Q|A_\mu(0)|0\rangle$.
\label{fig:2}}
\end{figure}

Soft rescatterings of the constituent quarks lead to the series of the
dispersion graphs of Fig. \ref{fig:2}.
The loop diagram $B_{ps}$ is already known, so let us discuss $B_\mu(p)$ 
The latter contains the pseudoscalar vertex and the bare matrix element $A_\mu^0$. 
The spectral density of the corresponding Feynman graph reads ($\tilde p^2=s$) 
\begin{eqnarray}
\label{fpdensity}
&&-\frac{\sqrt{N_c}}{8\pi^2}\int dk_1
dk_2\delta(k^2_1-m_1^2)\delta(k^2_2-m_2^2) \delta(\tilde p-k_1-k_2)\;
\nonumber\\
&&\times Sp\left({
\left\{\gamma_\mu\gamma_5 g_A^0(s)+P_\mu \gamma_5h_+^0(s)+(k_1-k_2)_\mu
\gamma_5h_-^0(s)\right\}
(\hat k_1+m_1)\;i\gamma_5(m_2-\hat k_2)}\right). 
\end{eqnarray}
The trace is equal to
\begin{eqnarray}
\label{fptrace}
-4i(k_{1\mu}m_2+k_{2\mu}m_1)g^0_A(s)&+&4i(k_1+k_2)_\mu(k_1k_2+m_1m_2)h^0_+(s)\nonumber\\
&+&4i(k_1-k_2)_\mu(k_1k_2+m_1m_2)h^0_-(s)
\end{eqnarray}
The expression for the loop graph $B_\mu$ takes the form
\begin{eqnarray}
\label{bmu}
B_\mu(p)&=&4i p_\mu\sqrt{N_c}
\int\limits_{(m_1+m_2)^2}^\infty
\frac{ds\;G(s)}{\pi(s-p^2)}\frac{\lambda^{1/2}(s,m_1^2,m_2^2)}{16\pi s}
\frac{s-(m_1-m_2)^2}{2s}
\nonumber\\
&&\times[(m_1+m_2)g_A^0(s)-s\;h^0_+(s)-(m_1^2-m_2^2)h^0_-(s)].
\end{eqnarray}
The amplitude with the quark rescatterings taken into account has the same
spinorial structure as the bare amplitude 
\begin{eqnarray}
\label{qqbartomunu}
&&\langle 0|A_\mu(0)|Q(k_1)\bar Q(k_2)\rangle 
=\bar Q(-k_2)\left[{
\gamma_\mu\gamma_5g_A(p^2)
+p_\mu\gamma_5h_+(p^2)+(k_1-k_2)_\mu\gamma_5h_-(p^2)}\right]Q(k_1)
\end{eqnarray}
with
\begin{eqnarray}
\nonumber
g_A(p^2)&=& g_A^0(p^2)
\\
\nonumber
h_-(p^2)&=& h_-^0(p^2)
\\
\nonumber
h_+(p^2)&=& h_+^0(p^2)
-\frac{G(p^2)}{1-B_{ps}(p^2)}\int\frac{ds\; G(s)}{\pi(s-p^2)}
\frac{\lambda^{1/2}(s,m_1^2,m_2^2)}{16\pi s}\frac{s-(m_1-m_2)^2}{2s}
\\
\nonumber
&&
\qquad\qquad\qquad
\times 4[(m_1+m_2)g_A^0(s)-s\;h^0_+(s)-(m_1^2-m_2^2)h^0_-(s)]
\end{eqnarray}
The form factor $h_+$ develops the pole at $p^2=M^2$ since $B_{ps}(M^2)=1$.
Near $p^2=M^2$ the pole dominates the amplitude
\begin{equation}
\label{Ptomunu}
\langle 0|A_\mu|\bar QQ\rangle=\langle 0|A_\mu|P(p)\rangle
\frac1{M^2-p^2}\langle P(p)|Q\bar Q\rangle+{\rm regular\; terms}.
\end{equation}
Comparing the pole terms in (\ref{qqbartomunu}) and (\ref{Ptomunu}) and making use of 
the relation
\begin{eqnarray}
\langle P(p)|Q\bar Q\rangle\to \frac{\bar Q i\gamma_5Q}{\sqrt{N_c}}G_v
\end{eqnarray}
we find that 
\begin{equation}
\langle 0|A_\mu|P(p)\rangle=ip_\mu f_P
\end{equation}
with
\begin{eqnarray}
f_P&=&4\sqrt{N_c}\int\frac{ds\;G_v(s)}{\pi(s-M^2)}\frac{\lambda^{1/2}(s,m_1^2,m_2^2)}{16\pi s}
\frac{s-(m_1-m_2)^2}{2s}
\nonumber\\
&&\times[(m_1+m_2)g_A^0(s)-s\;h^0_+(s)-(m_1^2-m_2^2)h^0_-(s)]. 
\end{eqnarray}
Assuming that in reality $g_A^0$ and $h^0$ are not far from the
limit (\ref{barevalues}), we come to the relation
\begin{equation}
\label{decconst}
f_P=\sqrt{N_c}(m_1+m_2)g_A^0(M^2)\int\limits_{(m_1+m_2)^2}^\infty
\frac{ds\;G_v(s)}{\pi(s-M^2)}\frac{\lambda^{1/2}(s,m_1^2,m_2^2)}{8\pi s}
\frac{(s-(m_1-m_2)^2)}{s}.
\end{equation}

\subsubsection{Two-photon decay of a neutral pseudoscalar meson}
Let us consider the two-photon decay of the neutral pseudoscalar meson $P_0$. 
Its constituent quark structure is described by the vertex
\begin{equation}
\frac{\bar Qi\gamma_5 Q}{\sqrt{N_c}}G_v. 
\end{equation}
The rate of the decay $P_0(p)\to2\gamma$ can be written as
\begin{equation}
\Gamma=\frac\pi4\alpha^2M^3g^2_{P\gamma\gamma},\quad
g_{P\gamma\gamma}=G_{P\gamma\gamma}(M^2,0,0),
\end{equation}
where the form factor $G_{P\gamma\gamma}$ is connected with the amplitude
\begin{equation}
\langle 0|J^{em}_{\alpha_2}(q_2)J^{em}_{\alpha_3}(q_3)|P(p)\rangle 
=2\epsilon_{\alpha_2\alpha_3\beta_2\beta_3}
q_2^{\beta_2}q_3^{\beta_3}G_{P\gamma\gamma}(M^2,q^2_2,q^2_3).
\end{equation}
The electromagnetic current $J^{em}_\mu(0)=\bar q(0)\gamma_\mu q(0)$
is defined through current quarks, whereas the meson structure is described
in terms of the constituent quarks.
So, for calculating the meson amplitude
the constituent quark amplitude of the electromagnetic current is necessary.
The latter is assumed to have the following structure
\begin{equation}
\label{constcurrent}
\langle Q(k')|\bar q(0)\gamma_\mu q(0)|Q(k)\rangle =\bar Q(k')\gamma_\mu Q(k) f_c(q^2),\quad
q=k'-k.
\end{equation}
The constituent charge form factor $f_c(q^2)$ is normalized such that
$f_c(0)=e_c$,
the constituent charge. The anomalous magnetic moment of the constituent quark
is neglected in the expression (\ref{constcurrent}),
but it can be included into consideration straightforwardly.

Hereafter we skip the intermediate steps and consider only the residue of the
constituent interaction amplitude which determines the bound-state amplitude.  
From now on we shall use the following notations: 

\noindent $\tilde p$ is the off-shell momentum $\tilde p^2=s$
which is used for the calculation of the imaginary parts of the Feynman diagrams,  

\noindent $p$ is the on-shell bound state momentum, $p^2=M^2$.  

\begin{figure}
\begin{center}
\mbox{\epsfig{file=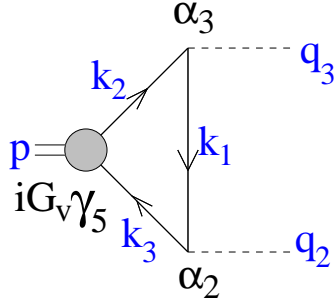,height=4.cm}}
\end{center}
\caption{The graph describing the decay $P^0\to\gamma\gamma$.
\label{fig:p2g}}
\end{figure}

The single dispersion representation for the form factor $G_{P\gamma\gamma}$ reads
\begin{equation}
\label{pg}
G_{P\gamma\gamma}(M^2,q_2^2,q_3^2)=f_c(q^2_2)f_c(q_3^2)
\int\frac{ds\,G_v(s)}{\pi(s-M^2)}
\Delta_{P\gamma\gamma}(s,q_2^2,q_3^2). 
\end{equation}
Here $\Delta_{P\gamma\gamma}$ is the spectral of the triangle 
Feynman graph of Fig.\ref{fig:p2g} and can be obtained from the following relation
\begin{eqnarray}
\label{pden}
&&-\epsilon_{\alpha_2\alpha_3\beta_2\beta_3}q_2^{\beta_2}q_3^{\beta_3}
\Delta_{P\gamma\gamma}(s,q_2^2,q_3^2)
\nonumber\\
&&\qquad=-\frac{\sqrt{N_c}}{8\pi^2}\int dk_1
dk_2 dk_3 \delta(\tilde p-k_2-k_3)\delta(k_2-k_1-q_3)
\frac{\delta(k^2_2-m^2)\delta(k^3_2-m^2)}{m^2-k_1^2}
\nonumber\\
&&\qquad\times Sp\left({
i\gamma_5(m-\hat k_3)\gamma_{\alpha_2}(m+\hat k_1)\gamma_{\alpha_3}(m+\hat k_2)}\right), 
\end{eqnarray}
where $\tilde p^2=s$. 
The trace reads
\begin{equation}
Sp\left({
i\gamma_5(m-\hat k_3)\gamma_{\alpha_2}(m+\hat k_1)\gamma_{\alpha_3}(m+\hat k_2)
}\right)=4m\epsilon_{\alpha_2\alpha_3\beta_2\beta_3}q_2^{\beta_2}q_3^{\beta_3},
\end{equation}
and we find
\begin{equation}
\label{pgden}
\Delta_{P\gamma\gamma}(s,q_2^2,q_3^2)=\frac{m\sqrt{N_c}}{4\pi}
\frac{\theta(s-4m^2)}{\lambda^{1/2}(s,q_2^2,q_3^2)}
\log
\left({
\frac{s-q_2^2-q_3^2+\lambda^{1/2}(s,q^2_2,q_3^2)\sqrt{1-4m^2/s}}
{s-q^2_2-q_3^2-\lambda^{1/2}(s,q_2^2,q_3^2)\sqrt{1-4m^2/s}}}\right)
\end{equation}
Substituting (\ref{pgden}) into (\ref{pg}), one obtains 
$G_{P\gamma\gamma}$ for off-shell photons. For real photons one finds
\begin{equation}
g_{P\gamma\gamma}=\frac{m\sqrt{N_c}}{4\pi}e_c^2\int\limits_{4m^2}^{\infty}
\frac{dsG_v(s)}{\pi(s-M^2)}
\frac1s
\log\left({
\frac{1+\sqrt{1-4m^2/s}}{1-\sqrt{1-4m^2/s}}
}\right)
\end{equation}

\subsubsection{Elastic electromagnetic form factor}
The elastic electromagnetic form factor of a pseudoscalar meson with the mass $M$ 
is given by the following matrix element
\begin{equation}
\langle P(p')|J^{em}_\mu(0)|P(p)\rangle=(p'+p)_\mu\;F^{el}(q^2),  
\end{equation}
where $p^2=p'^2=M^2$, $q=p-p'$ and $q^2<0$. 
The form factor $F^{el}(q^2)$ describes the amplitude of the photon emission from the 
bound state which depends on the three independent Lorentz invariants.  
It is convenient to choose such invariants as the squares of the three external momenta 
$p^2$, $p'^2$, and $q^2$. Therefore the form factor $F^{el}(q^2)$ is in fact the function of
the three variables $F^{el}(p^2, p'^2, q^2)$, but considered for fixed values of the two
invariants $p^2=M^2$, $p'^2=M^2$. 
 
\begin{figure}
\begin{center}
\mbox{\epsfig{file=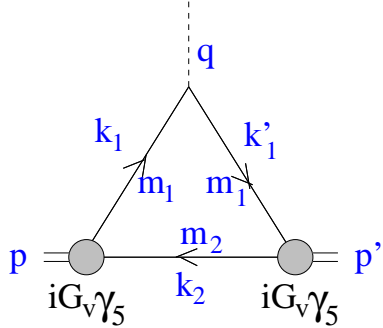,width=5.cm}}
\end{center}
\caption{Triangle diagram describing $H(q^2,m_1^2,m_2^2)$. 
\label{fig:4}}
\end{figure}

Assuming again the structure of the constitent quark matrix element of the 
the electromagnetic as given by Eq. (\ref{constcurrent}), the meson elastic charge 
form factor can be written in the form
\begin{equation}
F^{el}(q^2)=f_1(q^2) H(q^2,m_1^2,m_2^2)+f_2(q^2) H(q^2,m_2^2,m_1^2)
\end{equation} 
in terms of the form factors $H$. The quantity $H(q^2,m_1^2,m_2^2)$ describes
the subprocess when
the constituent $1$ interacts with the photon, while the constituent $2$
remains spectator (Fig.\ref{fig:4}). 
Clearly, the form factor $H(q^2,m_1^2,m_2^2)$ also implicitly depends on the 
invariant variables $p^2$ and $p'^2$, such that 
$H(q^2,m_1^2,m_2^2)=H(q^2,p^2=M^2,p'^2=M^2,m_1^2,m_2^2)$. 
As well known \cite{burton} the form factor $H(q^2,p^2,p'^2,m_1^2,m_2^2)$  
is an analytic function of the external mass variables $p^2$ and $p'^2$ and 
at $q^2<0$ satisifes the double spectral representation 
\begin{equation}
\label{ffh}
H(q^2,m_1^2,m_2^2)=
\int\frac{ds\;G_v(s)}{\pi(s-M^2)}
\frac{ds' G_v(s')}{\pi(s'-M^2)}\Delta_V(s',s,q^2|m_1,m_1,m_2).
\end{equation}
Here $\Delta_V$ is the double spectral density of $H$ over the variables $P^2$ and $P'^2$. 
According to the Landau-Cutkosky rules, $\Delta_V$ can be calculated through 
the following procedure: one should place all internal particles on their mass shell, 
$k_1^2=m_1^2$, $k_1'^2=m_1^2$, $k_2^2=m_2^2$
but go off the mass shell for the variables $p^2$ and $p'^2$, i.e. 
instead of the on-shell external momenta $p$ and $p'$ consider off-shell 
external momenta $\tilde p$ and $\tilde p'$, such that 
$\tilde p^2=s$, $\tilde p'^2=s'$, and $(\tilde p'-\tilde p)^2=q^2$, but 
$\tilde p'-\tilde p=\tilde q\ne q$.  
The double spectral density $\Delta$ is then determined from the following relation 
\begin{eqnarray}
\label{deltapi}
&&\frac1{8\pi}
\int dk_1 dk'_1 dk_2 \delta(k^2_1-m_1^2)\delta(k'^2_1-m_1^2)\delta(k^2_2-m_2^2)
\delta(\tilde p-k_1-k_2)\delta(\tilde p'-k'_1-k_2)
\nonumber\\
&&\times (-1)\,Sp\left({ (\hat k'_1+m_1)\gamma_\mu(\hat k_1+m_1)
i\gamma_5(m_2-\hat k_2)i\gamma_5 }\right)=2\tilde p_\mu(q)
\Delta_V(s',s,q^2|m_1,m_1,m_2)
\end{eqnarray}
with
\begin{eqnarray}
\nonumber
\tilde p_\mu(q)=(\tilde p-\frac{\tilde q\tilde p}{\tilde q^2}\,\tilde q)_\mu.
\end{eqnarray}
The trace reads
\begin{eqnarray}
\label{trace1}
&&\frac14 Sp\left({ (\hat k'_1+m_1)\gamma_\mu(\hat k_1+m_1)
\gamma_5(m_2-\hat k_2)\gamma_5 }\right)
\nonumber\\
&&=
2k'_{1\mu} (s-(m_1-m_2)^2)+2k_{1\mu} (s'-(m_1-m_2)^2)+2k_{2\mu}q^2
\end{eqnarray}
Multiplying both sides of (\ref{deltapi}) by $P_\mu$ and using (\ref{trace1})
one obtains for $q^2<0$
\begin{eqnarray}
&&\Delta_V(s',s,q^2|m_1,m_1,m_2)
\nonumber\\
&&\qquad=\frac{-q^2}{4\lambda^{3/2}(s',s,q^2)}\left({s's+(s'+s-q^2)m_2(m_1-m_2)-(m_1+m_2)(m_1-m_2)^3}\right)
\nonumber\\
&&\qquad\times\theta(s-(m_1+m_2)^2)\theta(s'-(m_1+m_2)^2)
\nonumber\\
&&\qquad\times\theta\left({
-q^2(s'+s-q^2+2(m_1^2-m_2^2))^2
+\lambda(s',s,q^2)(q^2-4m_1^2)
}\right)
\end{eqnarray}
with $\lambda(s',s,q^2)=(s'+s-q^2)^2-4s's$.

At $q^2=0$ one finds
\begin{equation}
\Delta_V(s',s,q^2=0|m_1,m_1,m_2)=\pi\rho_{ps}(s,m_1,m_2)\;\delta(s'-s),
\end{equation}
and
\begin{eqnarray}
\label{ffpiat0}
F^{el}(0)&=&(e_1+e_2)\int\limits_{(m_1+m_2)^2}^\infty\frac{ds\;
G_v^2(s)}{\pi(s-M^2)^2}
\rho_{ps}(s,m_1,m_2)
\nonumber\\
&=&e_1+e_2. 
\end{eqnarray}
This relation is the direct consequence of the Ward identity and corresponds to the 
electric charge conservation. 

\subsubsection{Dispersion approach in terms of the light-cone variable} 
For some applications and for comparison with the light-cone technique, 
it is convenient to rewrite our explicitly relativistic-invariant spectral 
representations in terms of the light-cone variables
\begin{equation}
\label{lcvariables}
k_-=\frac{1}{\sqrt{2}}(k_0-k_z);\quad
k_+=\frac{1}{\sqrt{2}}(k_0+k_z);\quad
k^2=2k_+k_-k^2_\perp. 
\end{equation}
Most easily this can be done by introducing the light-cone variables directly 
into the integral representation for the form factor spectral density (\ref{deltapi}). 
The variables (\ref{lcvariables}) should be connected with some specific reference
frame, which can be specified by fixing components of the physical momenta $p$ and $q$.
It is convenient to choose the reference frame in which
\begin{eqnarray}
\label{refframe}
p_\perp=0,\quad q_+=0,\quad q^2_\perp=-q^2. 
\end{eqnarray}
Notice that this choice is only possible in the region $q^2<0$.

The choice of the reference frame in the form (\ref{refframe}) 
allows us to choose components of the momenta in a convenient way. 
For the physical on-shell momenta\footnote{We use the notation $a_\mu=(a_+,a_-,a_\perp)$.} 
\begin{eqnarray}
\label{physicalvectors}
p &=&(p_+,\frac{M^2}{2p_+},0) \nonumber\\
p'&=&(p_+,\frac{M_2^2+q_\perp^2}{2p_+},-q_\perp)\nonumber\\
q&=&(0,\frac{-q_\perp^2}{2p_+}, q_\perp);
\end{eqnarray}
For the dispersion off-shell momenta:  
\begin{eqnarray}
\label{tildevectors}
\tilde p&=&(p_+,\frac{s}{2p_+},0)\nonumber\\
\tilde p'&=&(p_+,\frac{s'+q_\perp^2}{2p_+},-q_\perp)\nonumber\\
\tilde q&=&(0,\frac{s-s'-q_\perp^2}{2p_+}, q_\perp). 
\end{eqnarray}
A specific and good feature of this choice is that the $(+)$ and $(\perp)$ components of the 
on-shell vectors and the corresponding off-shell vectors are equal to each other, such that   
all off-shell effects in the description of a bound state
are completely shifted to the $(-)$ component of the momenta. 

Now, let us consider (\ref{deltapi}) and set $\mu=+$ in both sides of this equation.  
Using the result for the trace (\ref{trace1}) for $\mu=+$, and performing the 
$k_-$ integration, we obtain 
\begin{eqnarray}
\label{specden}
\Delta_V(s',s,q^2|m_1,m_1,m_2)=
\frac1{16\pi}\int \frac{dx d^2k_\perp}{x(1-x)}
\delta\left({s-\frac{m_1^2}{1-x}-\frac{m_2^2}{x}-\frac{k^2_\perp}{x(1-x)}}\right)
\nonumber\\
\times\delta\left({s'-\frac{m_1^2}{1-x}-\frac{m_2^2}{x}-\frac{(k_\perp+xq_\perp)^2}{x(1-x)}}\right)
(s'+s-2(m_1-m_2)^2-\frac{x}{1-x}q^2)
\end{eqnarray}
Here $x=k_{2+}/p_+$ and $k_\perp=k_{2\perp}$.

Substituting (\ref{specden}) into (\ref{ffh}) and performing $s$ and $s'$
integrations, one derives
\begin{equation}
\label{ffpilc}
H(q^2_\perp,m_1,m_2)=\int dx d^2k_\perp
\psi(x,k_\perp)\psi(x,k_\perp+xq_\perp)\beta(x,k_\perp,q_\perp)
\end{equation}
where the radial light-cone wave function of a pseudoscalar meson is introduced
\begin{eqnarray}
\label{lcwf}
\psi(x,k_\perp)&=&\frac{G_v(s)\sqrt{s-(m_1-m_2)^2}}{\pi^{3/2}\sqrt{8}(s-M^2)\sqrt{x(1-x)}},
\nonumber\\
&&\qquad s=\frac{m_1^2}{1-x}+\frac{m_2^2}{x}+\frac{k^2_\perp}{x(1-x)}
\end{eqnarray}
and 
\begin{eqnarray}
\beta=\frac{s-(m_1-m_2)^2+k_\perp q_\perp/(1-x)}
{\sqrt{s-(m_1-m_2)^2}\sqrt{s'-(m_1-m_2)^2}},
\qquad \beta(q_\perp=0)=1
\end{eqnarray}
The function $\beta$ accounts for the contribution of spins.
It is different from unity  at $q_\perp\ne 0$ because both the spin-nonflip and
spin-flip amplitudes of the interacting quark contribute.
The Eq.(\ref{ffpiat0}) is the normalization condition for the soft radial wave
function
\begin{equation}
\int dx\;d^2k_\perp\;|\psi(x,k_\perp)|^2=1.
\end{equation}

In terms of this wave function, the pseudoscalar meson axial-vector decay
constant $f_P$
is represented as
\begin{equation}
\label{fpilc}
f_P=g_A\frac{\sqrt{N_c}}{\sqrt2 \pi^{3/2}}
\int{dx\;d^2k_\perp}\;\psi(x,k_\perp)\frac{m_2(1-x)+m_1x}{\sqrt{s-(m_1-m_2)^2}}
\end{equation}
This expression can be easily deduced by introducing the light-cone variables
into the dispersion
representation (\ref{fpdensity}), making use of (\ref{fptrace}) and examining
the $\mu=+$ component
of the axial current.

Similarly, introducing the light-cone variables into (\ref{pden}) yields the
following expression
for $g_{P\gamma\gamma}$
\begin{equation}
\label{pi2glc}
g_{P\gamma\gamma}=\frac{m\sqrt{N_c}}{\sqrt{2}\pi^{3/2}}\int\frac{dxd^2k_\perp}{\sqrt{x(1-x)}}
\psi(x,k_\perp)\frac x{(m^2+k^2_\perp)\sqrt{s}}
\end{equation}

The same expressions for the speudoscalar meson elastic form factor, 
leptonic decay constant, and the two-photon decay constant as
(\ref{ffpilc})-(\ref{pi2glc}) were derived within the light-cone quark model in
refs \cite{jaus,gns}, with our $G_v(s)$ just equal to $h_0(P)$ 
of ref.\cite{jaus}. As we see later, the light-cone quark model and the dispersion 
approach also lead to the same expression for the form factor $F_+$ describing weak 
transitions between pseudoscalar mesons at $q^2<0$. This is not at all strange as both of 
these approaches are based on the assumption of dominance of the $q\bar q$ components 
in the description of the meson properties.  

However, the dispersion approach has several advantages compared to the light-cone quark 
model. In particular, the light-cone quark model faces at least two difficulties 
in applications to inelastic transitions: 
The first problem emerges already in the spacelike region of the momentum transfers 
in considering transitions induced by the current of higher spin. It is related to the 
proper choice of the current components to be used for the extraction of the form factors 
(so-called 'good' and 'bad' components and the problem to satisfy the 
angular condition, see \cite{gns} for discussion and references). 

The second problem emerges for the description of hadron transition processes at $q^2>0$. 
The light-cone quark model can consistently take into account the spectator contribution, 
whereas contributions of the so-called Z-graphs (also called non-partonic, or pair-creation 
subprocesses) are in general beyond the scope of the light-cone treatment. 
The Z-graph can be calculated only for some exceptional simple forms of the light-cone 
wave functions. Unfortunately, in the light-cone treatment such contribution is inevitably 
present for processes in the region of timelike momentum transfers where 
it cannot be suppressed by an appropriate choice of the reference frame. 

As we shall see, both of these problems find their solution in the dispersion approach. 

\subsection{\label{i.3}Form factors of meson transitions}
In this section we examine the electroweak transitions of pseudoscalar mesons.
First, we derive the transition form factors at
$q^2<0$ in the form of the relativistic dispersion representations.  
Second, these dispersion representations allow us to perform the analytic
continuation in $q^2$ and derive the form factors of semileptonic decays of 
pseudoscalar mesons at $q^2>0$. 

\subsubsection{The pseudoscalar meson transition form factor at $q^2<0$}
\begin{figure}
\begin{center}
\mbox{\epsfig{file=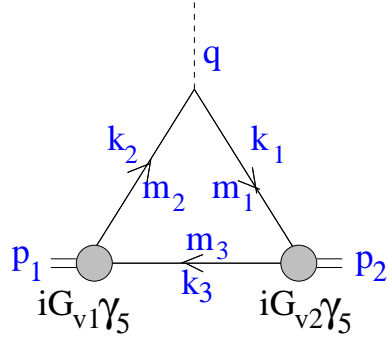,width=5.cm}}
\end{center}
\caption{The dispersion graph for the decay $\langle P_2|V^{aa}_\mu|P_1\rangle$.
\label{fig:trans12}}
\end{figure}

The amplitude of the weak transition of pseudoscalar mesons $M_1\to M_2$
(Fig.\ref{fig:trans12})
is determined by the two form factors $F_+$ and $F_-$
\begin{eqnarray}
\label{3transamp}
\langle p_{2},M_2|V_\mu|p_{1},M_1\rangle &=&(p_{1}+p_{2})_\mu F_+(q^2)+(p_{1}-p_{2})_\mu F_-(q^2) 
\nonumber\\
\langle p_{2},M_2|A_\mu|p_{1},M_1\rangle&=&0, 
\nonumber\\
p_{1}^2=M_1^2,\; && p_{2}^2=M_2^2,\qquad p_{1}-p_{2}=q.  
\end{eqnarray}
The weak currents are defined in terms of the current quarks
\begin{eqnarray}
\label{3currents}
V_\mu &=& \bar q_1^a(0) \gamma_\mu q_2^a(0), 
\nonumber\\
\quad A_\mu &=& \bar q_1^a(0)\gamma_\mu \gamma_5 q_2^a(0).
\end{eqnarray}
The structure of the mesons is described in terms of the constituent quarks
by the vertices
\begin{eqnarray}
\label{3vert1}
&&M_1:\qquad\frac{\bar Q_2(k_2)i\gamma_5 Q_3(-k_3)}{\sqrt{N_c}}G_{v1}(\tilde p_1^2),
\nonumber\\
&&M_2:\qquad\frac{\bar Q_1(k_1)i\gamma_5 Q_3(-k_3)}{\sqrt{N_c}}G_{v2}(\tilde p_2^2)
\end{eqnarray}
and $G_{v1,2}$ are normalized according to Eq. (\ref{vertnorm}).

For calculating the transition amplitude (\ref{3transamp})
we again need the constituent quark matrix element of the weak current which is
taken in the form
\begin{equation}
\label{3constamp}
\langle Q_1(k_1)|\bar q_1(0) \gamma_\mu q_2(0)|Q_2(k_2)\rangle=\bar Q_1(k_1)\gamma_\mu
Q_2(k_2)f_{21}(q^2)
\end{equation}
As in the case of the elastic form factor, the transition form factors $F_{\pm}$ 
can be treated 
as the functions of the three invariants $p_1^2$, $p_2^2$, and $q^2$, taken at 
$p_1^2=M_1^2$ and 
$p_2^2=M_2^2$. The form factors are analytic functions of the external mass variables 
$p_1^2$ and 
$p_2^2$ and can be represented by the following double spectral representations
\begin{equation}
\label{3ffs}
F_{\pm}(q^2)=f_{21}(q^2)\int\frac{ds_1G_{v1}(s_1)}{\pi(s_1-M_1^2)}\frac{ds_2G_{v2}(s_2)}{\pi(s_2-M_2^2)}
\Delta_\pm(s_1,s_2,q^2|m_1,m_2,m_3)
\end{equation}
Here $\Delta_\pm$ are the double spectral densities of the triangle Feynman graph 
corresponding to Fig.\ref{fig:trans12} in the $p_1^2$ and $p_2^2$-channels
To calculate $\Delta_{\pm}$ at $q^2<0$, we put internal particles 
on their mass shell, but consider external off-shell momenta 
$\tilde p_1=\tilde p_2+\tilde q$, $\tilde p_1^2=s_1$, $\tilde p_2^2=s_2$, 
$\tilde q^2=q^2$. 
Then the double spectral densities can be obtained from the following equation
\begin{eqnarray}
\label{3delta}
&&\frac1{8\pi}\int dk_1
dk_2 dk_3 \delta(k^2_1-m_1^2)\delta(k^2_2-m_2^2)\delta(k^3_2-m_3^2)
\delta(\tilde p_1-k_2-k_3)\delta(\tilde p_2-k_3-k_1)\nonumber\\
&&\qquad\times (-1)\,Sp\left({ (\hat k_1+m_1)\gamma_\mu(\hat k_2+m_2)
i\gamma_5(m_3-\hat k_3)i\gamma_5 }\right)\nonumber\\
&&=(\tilde p_1+\tilde p_2)_\mu\Delta_+(s_1,s_2,q^2|m_1,m_2,m_3)
+(\tilde p_1-\tilde p_2)_\mu \Delta_-(s_1,s_2,q^2|m_1,m_2,m_3)
\end{eqnarray}
The trace has the following form
\begin{eqnarray}
\label{3trace}
&&-Sp\left({ (\hat k_1+m_1)\gamma_\mu(\hat k_2+m_2)
i\gamma_5(m_3-\hat k_3)i\gamma_5 }\right)\nonumber\\
&&\qquad=2k_{1\mu} (s_1-(m_2-m_3)^2)+2k_{2\mu}
(s_2-(m_3-m_1)^2)+2k_{3\mu}(q^2-(m_1-m_2)^2).
\end{eqnarray}
Explicit calculations give 
\begin{eqnarray}
\Delta_\pm(s_1,s_2,q^2|m_1,m_2,m_3)=\frac{B_\pm(s_1,s_2,q^2)}{\lambda(s_1,s_2,q^2)}
\Delta(s_1,s_2,q^2|m_1,m_2,m_3)
\end{eqnarray}
where $B_{\pm}$ are the following polynomials 
\begin{eqnarray}
B_+(s_1,s_2,q^2)&=&b_+(s_1,s_2,q^2)
\left\{a(s_1,m_2,m_3)+a(s_2,m_3,m_1)-a(q^2,m_1,m_2)\right\}
\nonumber\\
&&+a(q^2,m_1,m_2)\lambda(s_1,s_2,q^2),
\nonumber\\
B_-(s_1,s_2,q^2)&=&b_-(s_1,s_2,q^2)
\left\{a(s_1,m_2,m_3)+a(s_2,m_3,m_1)-a(q^2,m_1,m_2)\right\},
\nonumber\\
&&+\left(a(s_2,m_3,m_1)-a(s_1,m_2,m_3)\right)\lambda(s_1,s_2,q^2)
\nonumber\\
b_+(s_1,s_2,q^2)&=&-q^2(s_1+s_2-q^2+m_1^2+m_2^2-2m_3^2)-(m_1^2-m_2^2)(s_1-s_2),
\nonumber\\
b_-(s_1,s_2,q^2)&=&(m_1^2-m_2^2)(2s_1+2s_2-q^2)
-(s_1-s_2)(s_1+s_2-q^2+m_1^2+m_2^2-2m_3^2), 
\nonumber\\
a(s,\mu_1,\mu_2)&=&s-(\mu_1-\mu_2)^2. 
\end{eqnarray}
In the above relations we have introduced $\Delta$, the double spectral density in 
$p_1^2$ and $p_2^2$-channels of the
Feynman triangle graph $\Gamma(p_1^2,p_2^2,q^2)$ with the scalar constituents 
defined according to the relation 
\begin{eqnarray}
\label{3scalar}
\Gamma(p_1^2,p_2^2,q^2)&=&\frac{1}{(2\pi)^4\,i}\int
\frac{dk_1 dk_2 dk_3 \delta(P_1-k_2-k_3)\delta(P_2-k_3-k_1)}
{(m_1^2-k_1^2-i0)(m_2^2-k_2^2-i0)(m_3^2-k_3^2-i0)}\nonumber\\
&=&\int\frac{ds_1}{\pi(s_1-p_1^2)}\frac{ds_2}{\pi(s_2-p_2^2)}
\Delta(s_1,s_2,q^2|m_1,m_2,m_3).
\end{eqnarray}
At $q^2<0$, $\Delta$ is given by the following integral 
\begin{eqnarray}
\Delta(s_1,s_2,q^2|m_1,m_2,m_3)&=&\frac1{8\pi}
\int dk_1 dk_2 dk_3 \delta(\tilde p_1-k_2-k_3)\delta(\tilde p_2-k_3-k_1)\nonumber\\
&&\times\delta(k^2_1-m_1^2)\delta(k^2_2-m_2^2)\delta(k^3_2-m_3^2). 
\end{eqnarray}
Introducing the light-cone variables we derive a useful expression 
\begin{eqnarray}
\label{3scalarlc}
\Delta(s_1,s_2,q^2|m_1,m_2,m_3)&=&
\frac1{16\pi}\int \frac{dx_1 dx_2 dx_3 }{x_1 x_2 x_3}d^2k_{3\perp}\delta(x_1-x_2)\delta(1-x_1-x_3)
\nonumber\\
&&\times\delta\left({
s_1-\frac{m_2^2}{x_2}-\frac{m_3^2}{x_3}-\frac{k^2_{3\perp}}{x_2x_3}
}\right)
\delta\left({
s_2-\frac{m_1^2}{x_1}-\frac{m_3^2}{x_3}-\frac{(k_{3\perp}+x_3q_{\perp})^2}{x_2x_3}
}\right)
\end{eqnarray}
Hereafter we denote $x_i=k_{i+}/p_{+}, p_{1+}=p_{2+}=p_{+}, 
-q_\perp^2=q^2, x\equiv x_3, k_\perp \equiv k_{3\perp}$. 
For $q^2<0$ one obtains  
\begin{eqnarray}
\label{thetafunction}
\Delta(s_1,s_2,q^2|m_1,m_2,m_3)=\frac{\theta\left({
b_+^2(s_1,s_2,q^2)-
\lambda(s_1,s_2,q^2)\lambda(q^2,m_1^2,m_2^2)}\right)}
{16\lambda^{1/2}(s_1,s_2,q^2)}. 
\end{eqnarray}
The solution of this $\theta$-function gives the following allowed intervals 
for the integration variables $s_1$ and $s_2$ 
\begin{eqnarray}
\label{limits}
(m_1+m_3)^2&<&s_2,
\nonumber\\
s_1^-(s_2,q^2)&<&s_1<s_1^+(s_2,q^2);
\end{eqnarray}
where  
\begin{eqnarray}
\label{s1pm}
s_1^\pm(s_2,q^2)&=&
\frac{s_2(m_1^2+m_2^2-q^2)+q^2(m_1^2+m_3^2)-(m_1^2-m_2^2)(m_1^2-m_3^2)}{2m_1^2}
\nonumber\\
&&\pm\frac{\lambda^{1/2}(s_2,m_3^2,m_1^2)\lambda^{1/2}(q^2,m_1^2,m_2^2)}{2m_1^2}
\end{eqnarray}
The final dispersion representation for the form factors at $q^2<0$ takes the form
\begin{equation}
\label{fftrans}
F_{\pm}(q^2)=f_{21}(q^2)\int\limits^\infty_{(m_1+m_3)^2}\frac{ds_2G_{v2}(s_2)}{\pi(s_2-M_2^2)}
\int\limits^{s_1^+(s_2,q^2)}_{s_1^-(s_2,q^2)}\frac{ds_1G_{v1}(s_1)}{\pi(s_1-M_1^2)}
\frac{B_\pm(s_1,s_2,q^2)}{\lambda^{3/2}(s_1,s_2,q^2)}
\end{equation}
This representation will be the starting point for the consideration of the
meson decays in the next section.

\newpage
\vspace{.4cm}
\underline{\it The light-cone representation}
\vspace{.4cm}

For comparison with the light-cone quark model results it is convenient to represent 
the spectral representation (\ref{fftrans}) as the integral over the light-cone variables. 
Following the same lines for the elastic form factor, we turn back to the
equation (\ref{3delta}) and again make use of the light-cone variables
(\ref{lcvariables}), choosing the reference frame $q_+=0, p_{1\perp}=0$
(see \cite{amn1,ammp,myaf} for details).  
Setting $\mu=+$ and making use of the first line of (\ref{3trace}) gives 
for $\Delta_+$
\begin{eqnarray}
\label{3deltalc}
\Delta_+(s_1,s_2,q^2|m_1,m_2,m_3)&=&
\frac{1}{16\pi}\int \frac{dx_1 dx_2 dx_3 }{x_1 x_2 x_3}d^2k_{3\perp}\delta(x_1-x_2)\delta(1-x_1-x_3)
\nonumber\\
&&\times\delta\left({
s_1-\frac{m_2^2}{x_2}-\frac{m_3^2}{x_3}-\frac{k^2_{3\perp}}{x_2x_3}
}\right)
\delta\left({
s_2-\frac{m_1^2}{x_1}-\frac{m_3^2}{x_3}-\frac{(k_{3\perp}+x_3q_{\perp})^2}{x_2x_3}}\right)
\nonumber\\
&&\times
\left\{
x_1(s_1-(m_2-m_3)^2)+
x_2(s_2-(m_1-m_3)^2)
+x_3(-q_\perp^2-(m_1-m_2)^2)
\right\}
\end{eqnarray}
Substituting (\ref{3deltalc}) into (\ref{3ffs}) yields the following expression
for the form factor $F_+$
\begin{eqnarray}
F_{+}(q_\perp^2)&=&f_{21}(q_\perp^2)
\int\frac{dxd^2k_\perp}{16\pi^3 x(1-x)}
\frac{ G_{v1}(s_1)}{\pi(s_1-M_1^2)}
\frac{ G_{v2}(s_2)}{\pi(s_2-M_2^2)}
\nonumber\\
&\times&\left({
s_1+s_2-(m_1-m_3)^2-(m_2-m_3)^2+\frac{x}{1-x}(-q_\perp^2-(m_1-m_2)^2)
}\right)
\end{eqnarray}
Introducing the radial light-cone wave funcion according to (\ref{lcwf}) leads
to the familiar light-cone expression (cf.\cite{jaus})
\begin{eqnarray}
\label{fpluslc}
F_{+}(q_\perp^2)&=&f_{21}(q_\perp^2)\int dx d^2k_\perp
\psi_1(x,k_\perp)\psi_2(x,k_\perp+xq_\perp)\beta_+(x,k_\perp,q_\perp),
\end{eqnarray}
with $\beta_+$ given by the expression 
\begin{eqnarray}
\beta_+&=&\frac{s_1+s_2-(m_1-m_3)^2-(m_2-m_3)^2+\frac{x}{1-x}(-q_\perp^2-(m_1-m_2)^2)}
{2\sqrt{s_1-(m_2-m_3)^2}\sqrt{s_2-(m_3-m_1)^2}}
\nonumber\\
&=&\frac{(m_1x+m_3(1-x))(m_2x+m_3(1-x))+k_\perp(k_\perp+xq_\perp)}
{x(1-x)\sqrt{s_1-(m_2-m_3)^2}\sqrt{s_2-(m_3-m_1)^2}}\nonumber
\end{eqnarray}
The same expression for $F_+(q_\perp^2)$ was obtained in \cite{jaus} 
within the light-cone quark model. The $(\perp)$ component in Eq. (\ref{3delta}) 
allows to determine the form factor $F_-(q_\perp^2)$. The $(-)$ component of  
Eq. (\ref{3delta}) cannot be used for isolating the form factor because of the 
difference between the $(-)$ components of the on-shell and the off-shell vectors 
as given by Eqs. (\ref{physicalvectors}) and (\ref{tildevectors}). 

\subsubsection{Transition form factors at $q^2>0$}
For the description of decay processes the form factors in the region
$0<q^2<(M_1-M_2)^2$
are necessary. For deriving the form factors at $q^2>0$ the dispersion representation
(\ref{fftrans})
turns out to be a convenient starting point. In general, this representation has 
the following form
\begin{equation}
\label{4ffs}
F(q^2)=f_{21}(q^2)\int\frac{ds_1G_{v1}(s_1)}{\pi(s_1-M_1^2)}\frac{ds_2G_{v2}(s_2)}{\pi(s_2-M_2^2)}
\frac{B(s_1,s_2,q^2)}{\lambda(s_1,s_2,q^2)}\Delta(s_1,s_2,q^2|m_1,m_2,m_3)
\end{equation}
where $\Delta$ is the double spectral density of the Feynman graph $\Gamma$
with scalar constituents Eq. (\ref{3scalar}). This double dispersion representation 
defines the analytic function of $q^2$ both at negative and positive values. 
It is important to point out that the functions $G_v(s)$ have no
singularities in the right-hand side of the complex $s$-plane \cite{akms},
and $B$ and $\lambda$ are polynomials. 
So the details of the dispersion integration at $q^2>0$ are determined by the
behavior of the quantity $\Delta$.

A detailed discussion of the double spectral representation and the anomalous
singularities can be found in \cite{burton} in connection with the deuteron elastic 
form factor. Anomalous singularities in decay processes were considered in \cite{sr1} 
for the case $m_1=m_3=0$. We perform the analysis for arbitrary nonzero masses.  

Let us start with the single dispersion represenation for $\Gamma$ in the 
variable $p_2^2$. A standard calculation yields
\begin{eqnarray}
\label{4scalar}
\Gamma(p_1^2,p_2^2,q^2)=\int\limits_{(m_1+m_3)^2}^{\infty}
\frac{ds_2}{\pi(s_2-p_2^2)}\sigma_2(p_1^2,s_2,q^2),
\end{eqnarray}
where
\begin{eqnarray}
\label{sigma2}
\sigma_2(p_1^2,s_2,q^2)&=&\sigma_2^+(p_1^2,s_2,q^2)-\sigma_2^-(p_1^2,s_2,q^2),
\nonumber\\
\sigma_2^\pm(s_1,s_2,q^2)&=&\frac1{16\pi\lambda^{1/2}(s_1,s_2,q^2)}
\nonumber\\
&\times&\log\left({
-s_2(s_1+q^2-s_2+m_1^2+m_3^2-2m_2^2)-(s_1-q^2)(m_1^2-m_3^2)}
\pm\lambda^{1/2}(s_2,m_1^2,m_3^2)\lambda^{1/2}(s_1,s_2,q^2)\right).\nonumber
\end{eqnarray}
Hereafter we assume $m_2>m_1$.
The single dispersion representation reproduces the exact value of the 
Feynman integral (\ref{3scalar}).

\begin{figure}
\begin{center}
\mbox{\epsfig{file=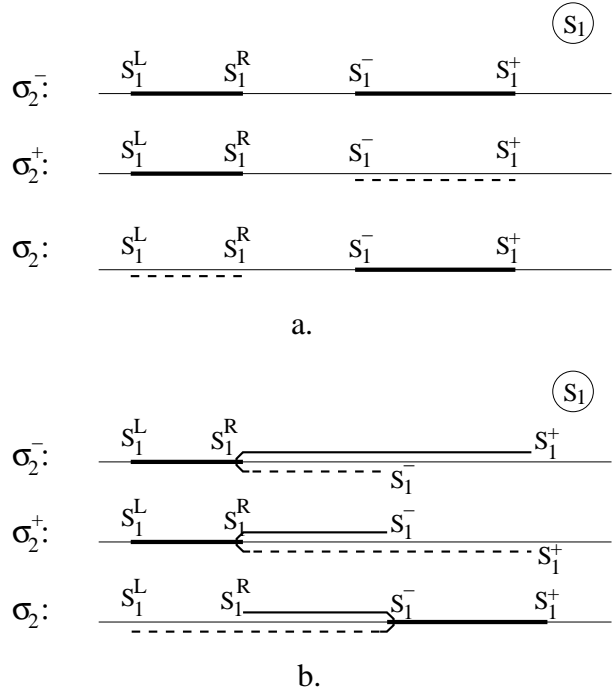,width=8cm}}
\end{center}
\caption{\label{fig:anal}
The location of the singularities of $\sigma_2$ in the complex $s_1$
plane as a function of $s_2$ for $q^2>0$: 
a). $s_2<s_2^0$; b). $s_2>s_2^0$. Solid lines cuts located on the physical sheet, 
dashed lines - on the second sheet. }
\end{figure}

Let us consider the function $\sigma_2(p_1^2,s_2,q^2)$ as the analytic
function of the variable $s_1=p_1^2$ at fixed $s_2$ and $q^2>0$.
As $s_2<s_2^0$ such that
\begin{equation}
\label{s20}
\sqrt{s_2^0}=-\frac{q^2+m_1^2-m_2^2}{2\sqrt{q^2}}+
\sqrt{
\left({
\frac{q^2+m_1^2-m_2^2}{2\sqrt{q^2}}
}\right)^2+(m_3^2-m_1^2)
},\quad q^2<(m_2-m_1)^2,
\end{equation}
both of the functions $\sigma_2^+$ and $\sigma_2^-$ have square-root branch points
on the physical
sheet at
$s_1^L=(\sqrt{s_2}-\sqrt{q^2})^2$ and $s_1^R=(\sqrt{s_2}+\sqrt{q^2})^2$,
connected by the cut, see Fig.\ref{fig:anal}. 
The function $\sigma_2^-$ has in addition a logarithmic cut from $s_1^-$ to 
$s_1^+$ on the physical sheet. Here $s_1^\pm$ given by (\ref{limits}) are the 
zeros of the argument of the logarithm in $\sigma_2^-$. 
The function $\sigma_2^+$ also has a 
logarithmic cut from $s_1^-$ to $s_1^+$ but on the second
unphysical sheet of the Riemann surface of the square-root 
(dashed line in Fig.\ref{fig:anal}a). 

The square-root cuts cancel in $\sigma_2=\sigma_2^+ -\sigma_2^-$, and the
logarithmic cut is the only singularity of $\sigma_2$ on the physical sheet.
Notice however that $\sigma_2$ has the two cuts from $s_1^L$ to $s_1^R$
and from $s_1^-$ to $s_1^+$ of the second sheet.  

The situation changes at $s_2=s_2^0$ which is determined by the condition
$s_1^R(s_2^0)=s_1^-(s_2^0)$. For this value of $s_2$ the logarithmic and square-root 
branch points coincide. As $s_2$ increases, $s_2>s_2^0$,  the branch point $s_1^-$ of 
$\sigma_2^+$ moves up through the square-root cut onto the physical sheet. At the same 
time the branch point $s_1^-$ of $\sigma_2^-$ moves onto the second sheet 
(Fig.\ref{fig:anal}b).
Hence, the function $\sigma_2^+$ acquires the logarithmic
cut from $s_1^R$ to $s_1^-$ on the physical sheet, and $\sigma_2^-$ still has the 
logarithmic cut which now lasts from $s_1^R$ to $s_1^+$. 
Both of the functions have also square-root cuts from $s_1^L$ to
$s_+^R$. In the difference $\sigma_2=\sigma_2^{+}-\sigma_2^{-}$ the square-root cuts 
cancel each other, but the logarithmic cuts add. 
The resulting expression for the double spectral density takes the form
\begin{eqnarray}
\label{4deltas}
\Delta(s_1,s_2,q^2|m_1,m_2,m_3)&=&
\frac{\theta(s_2-(m_1+m_3)^2)\theta(s_1^-<s_1<s_1^+)}{16\lambda^{1/2}(s_1,s_2,q^2)}
\nonumber\\
&&+\frac{2\theta(q^2)\theta(s_2-s_2^0)\theta(s_1^R<s_1<s_1^-)}{16\lambda^{1/2}(s_1,s_2,q^2)}.
\end{eqnarray}
One can check the double dispersion representation (\ref{3scalar}) with the
spectral density
$\Delta$ given by (\ref{4deltas}) to reproduce correctly the Feynman
expression.
The first term in (\ref{4deltas}) relates to the Landau-type contribution
emerging when all
intermediate particles go on mass shell, while the second term describes the
{\it anomalous} contribution.

\begin{center}
\begin{figure}[h]
\begin{tabular}{lr}
\mbox{\epsfig{file=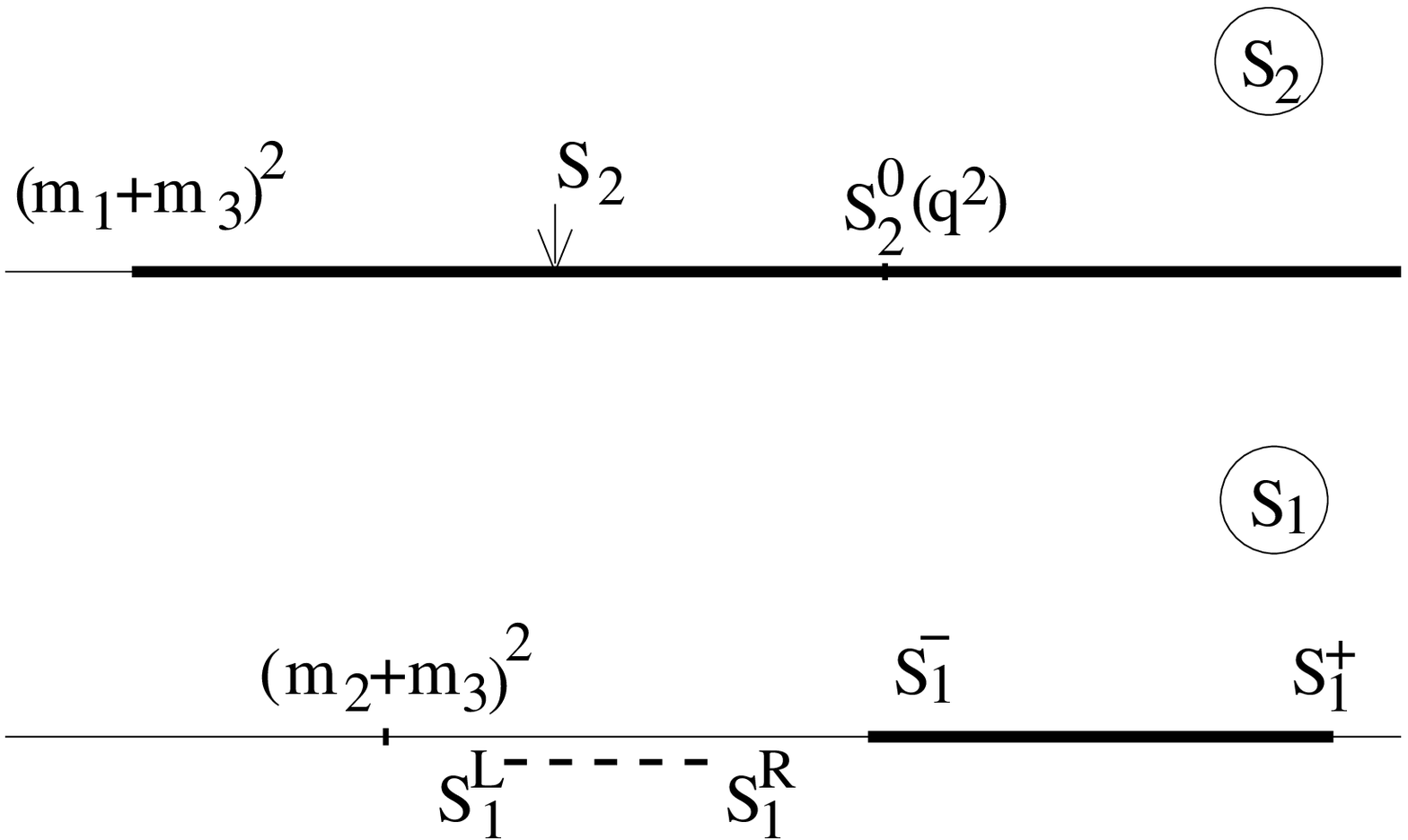,width=7.cm}}
\qquad
& 
\qquad
\mbox{\epsfig{file=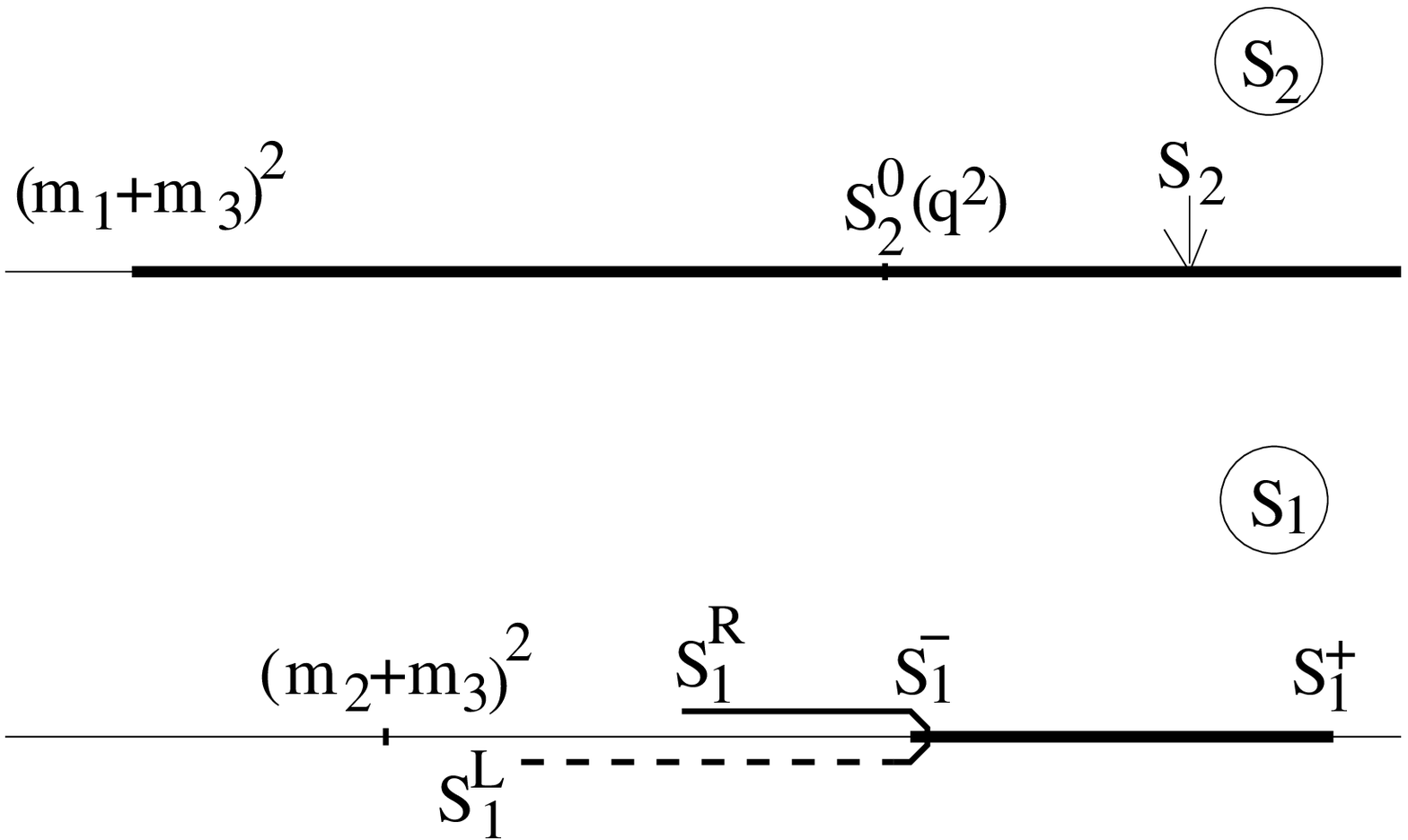,width=7.cm}}
\end{tabular}
\vspace{.5cm}
\caption{\label{fig:locations2s1}
The integration region for $m_2>m_1$ and $q^2>0$ for the order of integration 
$\int ds_2\int ds_1$: a. $s_2<s_2^0$ b. $s_2>s_2^0$. 
Solid lines - cuts on the physical sheet, dashed lines - cuts on the second sheet.}
\vspace{.5cm}
\end{figure}
\end{center}

The expression (\ref{4deltas}) for $\Delta$ is derived for $m_2>m_1$ implying the 'external' 
$s_2$ integration, and the 'internal' $s_1$-integration. The location of the integration 
region for this case is shown in Fig. \ref{fig:locations2s1}.  
In addition to the quantity $\Delta$, the spectral density of the
representation (\ref{4ffs}) involves the factor $1/\lambda(s_1,s_2,q^2)$ 
which is singular at $s_1^R$, the lower integration limit in the anomalous term: because 
\begin{eqnarray}
\nonumber
\lambda(s_1,s_2,q^2)=(s-s_1^L)(s-s_1^R).
\end{eqnarray}
As it has been discussed in \cite{sr1}, in this case an accurate
application of the Cauchy theorem yields the subtracion term in the non-Landau
contribution.
Representing $\sigma_2$ as a contour integral, we must take into account the
nonvanishing
contribution of the small circle around the point $s_1^R$. Underline once more
that the presence of the factor $G_{v1}(s_1)$ does not change the argumentation
as the function $G_v(s)$ has no singularities at $s_1>(m_2+m_3)^2$.
The final properly regularized representation for the form factors at
$0<q^2<(m_2-m_1)^2$
takes the form (omitting the constituent transition form factor $f_{21}(q^2)$)
\begin{eqnarray}
\label{final1}
F(q^2)&=&
\int\limits_{(m_1+m_3)^2}^\infty\frac{ds_2G_{v2}(s_2)}{\pi(s_2-M_2^2)}
\int\limits_{s_1^-}^{s_1^+}
\frac{ds_1G_{v1}(s_1)}{\pi(s_1-M_1^2)}
\frac{B(s_1,s_2,q^2)}{16\lambda(s_1,s_2,q^2)}
\nonumber\\
&+&
2\theta(q^2)\int\limits_{s_2^0}^\infty\frac{ds_2G_{v2}(s_2)}{\pi(s_2-M_2^2)}
\int\limits_{s_1^R}^{s_1^-}
\frac{ds_1}{16\pi(s_1-s_1^R)^{3/2}}
\left[{
\frac{G_{v1}(s_1)B(s_1,s_2,q^2)}{(s_1-s_1^L)^{3/2}(s_1-M_1^2)}-
\frac{G_{v1}(s_1^R)B(s_1^R,s_2,q^2)}{(s_1^R-s_1^L)^{3/2}(s_1^R-M_1^2)}
}\right]
\end{eqnarray}
It should be pointed out, that although the representations (\ref{4ffs}) and
(\ref{final1}) were derived for the case of pseudoscalar mesons, transition form 
factors of any hadrons have the similar structure. A particular choise of the initial 
and final hadrons yields a specific form of the function $B$. In the next 
chapter we shall calculate the double spectral densities for the form factors 
describing the pseudoscalar to vector meson transitions. 

\vspace{.4cm}
\underline{\it Bound state vertex and bound state wave function}
\vspace{.4cm}

It is convenient to introduce the bound state wave function $\varphi(s)$ 
related to the bound state vertex as follows
\begin{equation}
\varphi(s)=G_v(s)/(s-M^2). 
\end{equation}
In the nonrelativistic limit one finds $\varphi(s)\simeq\Psi_{NR}(\vec k)$, 
where the variable $\vec k$ is connected with $s$
\begin{eqnarray}
\nonumber
s=\sqrt{m_Q^2+\vec k^2}+\sqrt{m_q^2+\vec k^2}.
\end{eqnarray} 
In the nonrelativistic limit the elastic electromagnetic form factor of a 
pseudoscalar meson takes a well-known form 
\begin{equation}
F(\vec q^2)=\int \Psi_{NR}(\vec k)\Psi_{NR}(\vec k+\vec q)d\vec k, \qquad \vec q^2=-q^2. 
\end{equation}
The relationship between the bound state wave function $\varphi(s)$ and the light-cone wave function
is given by the relation (\ref{lcwf}). 

It should be pointed out that the analytic properties of $\varphi(s)$ in the region 
of $s$ near the two-particle threshold are different for a truly bound state 
with a negative binding energy (like a deuteron as a two-nucleon system) 
and a bound state in a confined potential with a positive binding energy 
(like a meson as a $q\bar q$ bound state): in the deuteron case $G_v(s)$ is a smooth regular 
function near the threshold and the pole at $s=M^2$ is located 
only slightly below the two-particle 
threshold at $s=4M_N^2$. Therefore $\varphi(s)$ is strongly peaked near the threshold and it is
more convenient to analyse the deuteron form factors in terms of the vertex functions $G_v(s)$. 

For the confined potential the situation is different: the pole at $s=M^2$ would have appeared 
in the physical region as the pole in $\varphi(s)$ at $s=M^2>(m_Q+m_q)^2$. However this does not
happen. As well known from e.g. the behaviour of the bound-state wave function in the harmonic 
oscillator potential, the function $\varphi(s)$ is a smooth exponential function of $s$ above 
$s=(m_Q+m_q)^2$. This means that the would-be pole in $\varphi(s)$ at $s=M^2$ is completely 
washed out by the interaction. Therefore, for the analysis of the meson transitions $\varphi(s)$ 
turns out to be more appropriate than $G_v(s)$. 
  
\subsection{\label{i.4}A simple model for pseudoscalar mesons}

We are now in a position to apply the developed formalism to
the analysis of the properties of pseudoscalar mesons and to the
direct calculation of the decay form factors. To this end
we must specify the parameters of the model, i.e. 
constituent quark masses and wave functions of pseudoscalar mesons. 

We consider in this section a simple choice of quark model parameters for 
pseudoscalar mesons $\pi$, $K$, $D$, and $B$ which gives a relalistic 
description of this sector and allows us to illustrate main features of the
weak form factors and study the transition to the heavy-quark limit. 
We leave a detailed discussion of criteria for fixing parameters of
the model, including also vector mesons, and analysis of the corresponding 
transition form factors for later sections. 
 
For a pseudoscalar meson built up of quarks with the masses $m_Q$ and $m_q$ 
the wave function $\varphi(s)=G_v(s)/(s-M^2))$ can be written in the form 
\begin{equation}
\label{4vertex}
\varphi(s)=\frac{\pi}{\sqrt{2}}\frac{\sqrt{s^2-(m_Q^2-m_q^2)^2}}{\sqrt{s-(m_Q-m_q)^2}}
\frac{1}{s^{3/4}}w(k),\qquad
k=\frac{\lambda^{1/2}(s,m_Q^2,m_q^2)}{2\sqrt{s}}. 
\end{equation}
The normalization condition (\ref{vertnorm}) for $G_v$ is equivalent to the 
following normalization condition for $w$
\begin{equation}
\label{normaw}
\int w^2(k)k^2dk=1.
\end{equation}
The function $w$ is the ground-state $S$-wave radial wave function of a
pseudoscalar meson for which we choose in this section a simple exponential form
\begin{equation}
\label{5exppar}
w(k)=\exp\left({-4\alpha {k^2}/{\mu^2_P}}\right)
\end{equation}
where $\mu_P=m_Qm_q/(m_Q+m_q)$ is the reduced mass.
The parameterization of the wave function in the form (\ref{5exppar}) is inspired 
by the nonrelativistic quantum
mechanics and is convenient for the analysis of the dependence of the observables 
on $m_Q$, in particular for analysing the case $m_Q\to\infty$. 

\begin{center}
\begin{figure}[h]
\begin{tabular}{cc}
\mbox{\epsfig{file=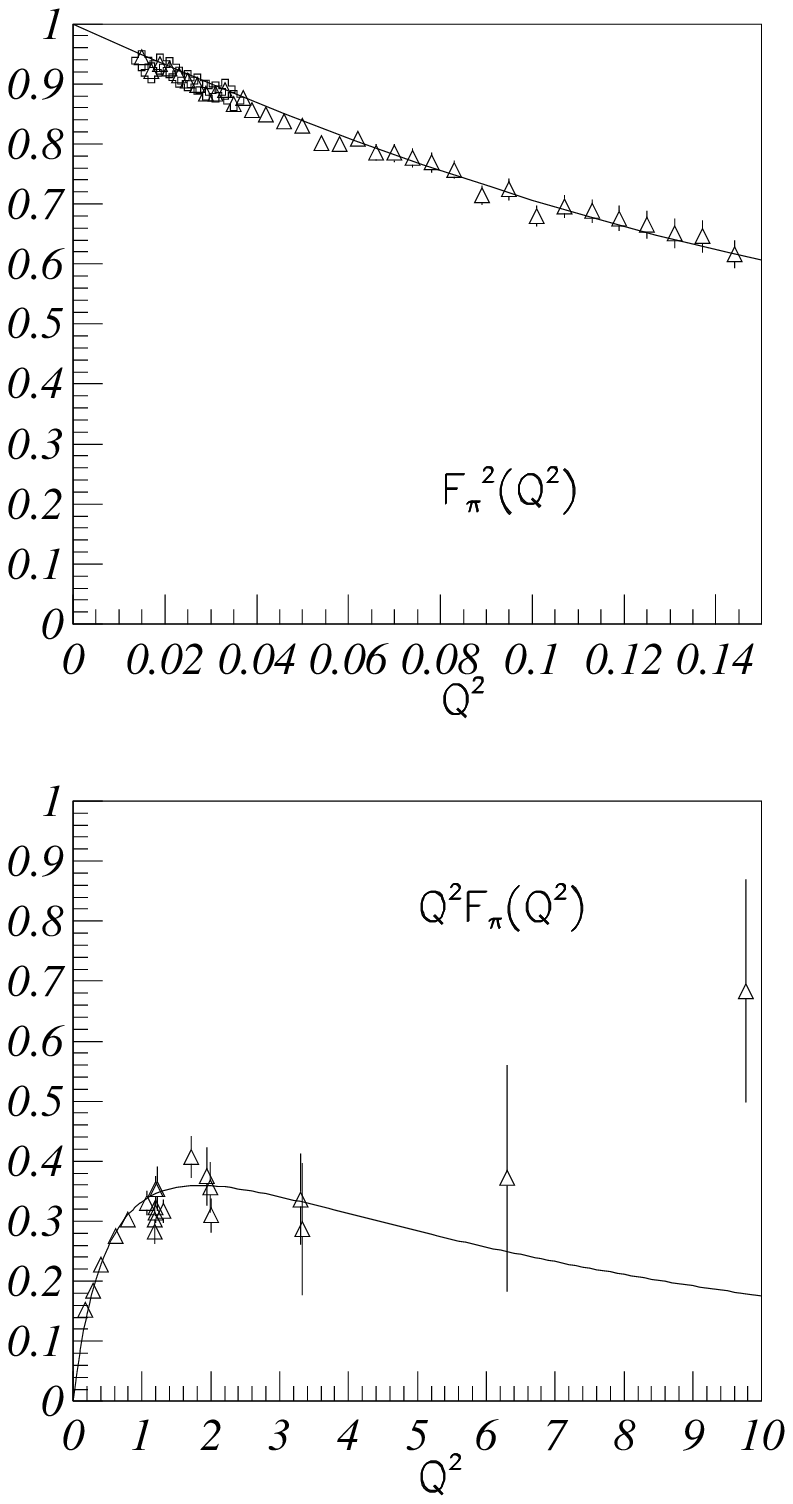,width=8.cm}}
& 
\mbox{\epsfig{file=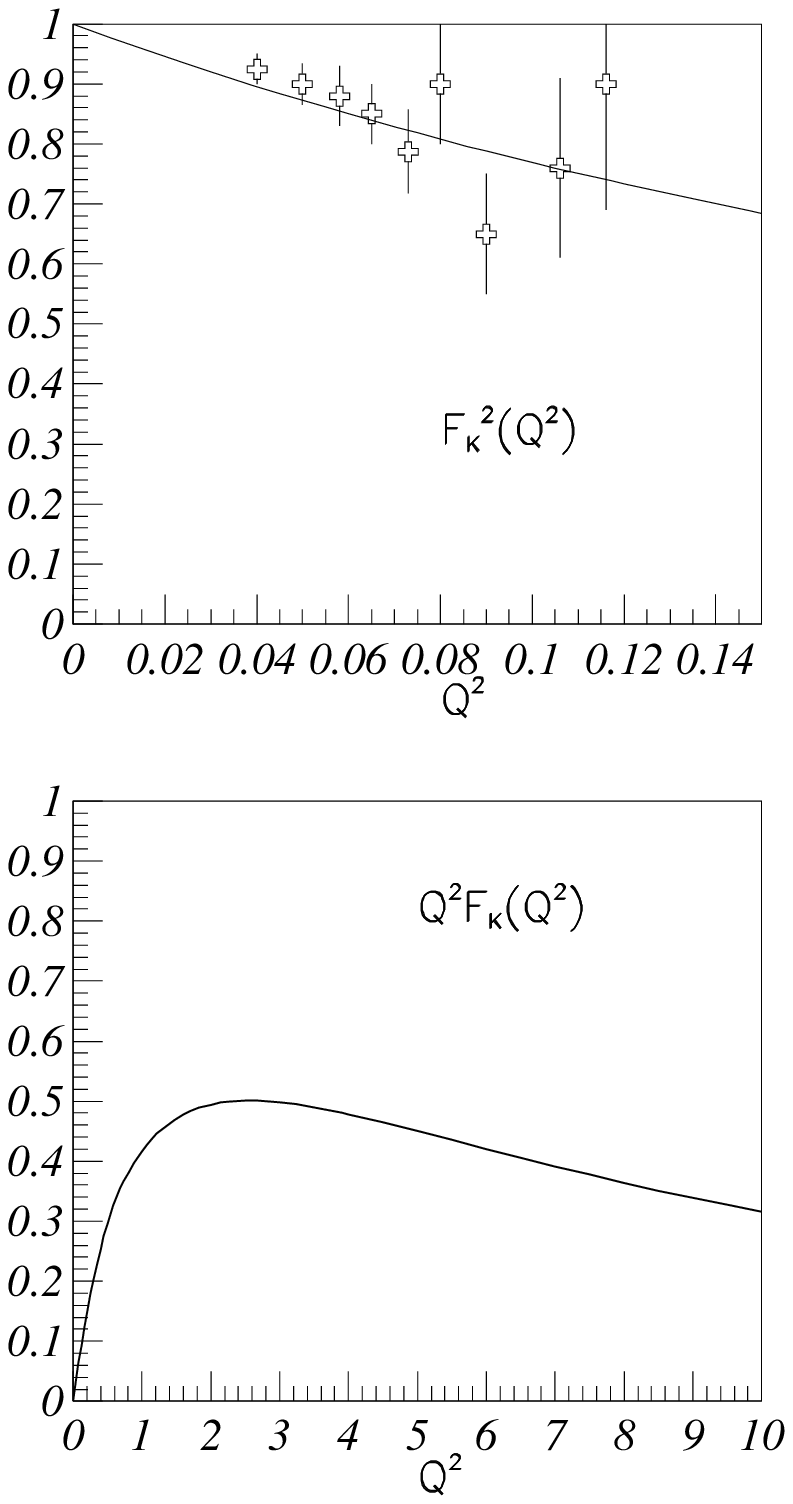,width=8.cm}}
\end{tabular}
\caption{\label{fig:ffpik}
The $\pi^+$ and $K^+$ elastic form factors evaluated for $\alpha_\pi=0.02$.}
\end{figure}
\end{center}

In the nonrelativistic quantum mechanics
a bound-state wave function is determined by the motion of the particle
with the mass $\mu_P$ in the potential independent of masses, and thus $\alpha$
does not depend on the masses as well. Relativistic effects
destroy this simple feature of the wave function. In QCD the situation is much
more
complicated because additional dimensional quantities such as $\Lambda_{QCD}$
and the
condensates appear.
So, $\alpha$ should be considered as some unknown function of the quark masses.
It is possible to obtain the information on the behavior of $\alpha$ as a
function of $m_Q$
at fixed $m_q=m_{u,d}=0.25\;GeV$ in the two regions: at small $m_Q$ and
$m_Q\to\infty$.
\begin{table}[h]
\caption{\label{table:parameters1}
The constituent quark masses and the calculated $f_P$ for $\alpha=0.02$.}
\centering
\begin{tabular}{|c|c||c|c|c|}
quark &quark mass,$GeV$& meson &meson mass,$GeV$ &$f_P,MeV$  \\
\hline\hline
u,d   & 0.25   &$\pi^+(u\bar d)  $& 0.14       &    130      \\
\hline
s     & 0.40   &$K^+(u\bar s)    $& 0.49       &    160      \\
\hline
c     & 1.80   &$D^+(c\bar d)    $& 1.87       &    234      \\
\hline
b     & 5.20   &$B^+(u\bar b)    $& 5.27       &    202      \\
\end{tabular}
\end{table}
At $m_Q\le 0.5\;GeV$  the value of $\alpha$ can be determined by
describing the data in the light-meson sector.
The light-quark masses given in Table \ref{table:parameters1} and
$\alpha_\pi=\alpha_K=0.02$ provide a good description of the data on
$f_\pi$, $f_K$, and the elastic form factors, Fig. \ref{fig:ffpik}. 
The meson decay constants and form factors are calculated with the values
$g_A^0(M^2)=1$ and $f_c(q^2)=f_c(0)$, respectively. 
\newpage
In the region $m_Q\to\infty$ the behavior of $\alpha$ can be found on the basis
of the heavy quark symmetry.
To this end, let us consider the amplitudes of the elastic and inelstic
transitions between pseudoscalar mesons consisting of heavy $Q$ and 
light $q$ quarks. 

For the case of the transition between two heavy mesons it is convenient 
to introduce instead of $q^2$ the dimensionless recoil variable $\omega=vv'$, where 
$v$ and $v'$ are the 4-velocities of the initial and final mesons, respectively, 
and to analyse the transition in terms of the velocity-dependent form factors. 

For the elastic-transition amplitude 
\begin{equation}
\label{5el}
\langle M,P'|\bar Q\gamma_\mu Q|M,P\rangle=(P'+P)_\mu\;F_{el}(q^2);\qquad q^2\le 0
\end{equation}
the recoil takes the form 
\begin{equation}
\omega=1-\frac{q^2}{2M^2}\ge 1, 
\end{equation}
and the velocity-dependent form factor just coincides with the elastic form factor 
\begin{equation}
h_{el}(\omega)=F_{el}(q^2). 
\end{equation}
The function $h_{el}$ can be expanded in powers of the variable $\omega-1$ 
near zero recoil point $\omega=1$ as follows 
\begin{equation}
h_{el}(\omega)=1-\rho^2_{el}(\omega-1)+O((\omega-1)^2). 
\end{equation}

For the amplitude of the inelastic transition between two pseudoscalar mesons 
\begin{eqnarray}
\label{5inel}
\langle M_{2},p_2|\bar Q_2\gamma_\mu Q_1|M_{1},p_1\rangle &=&
(p_1+p_2)_\mu F_{+}(q^2)+(p_1-p_2)_\mu F_{-}(q^2); 
\nonumber\\
0&<&q^2<(M_1-M_2)^2
\end{eqnarray}
the recoil reads 
\begin{equation}
\omega=\frac{M_1^2+M_2^2-q^2}{2M_1M_2}\ge 1, 
\end{equation}
and the velocity-dependent form-factors are related to the form factors $F_\pm$ 
according to the relation 
\begin{equation}
h_{\pm}(\omega)=\frac{M_1\pm M_2}{2\sqrt{M_1M_2}}F_{+}(q^2)+\frac{M_1\mp
M_2}{2\sqrt{M_1M_2}}F_{-}(q^2). 
\end{equation}
In the limit of infinitely heavy quarks $Q_{1,2}$,
both the elastic and inelastic amplitudes are expressed to a $1/m_Q$ accuracy in 
terms of the single universal Isgur-Wise (IW) function $\xi(\omega$) \cite{iwhh}
\begin{eqnarray}
\label{5hqlimit}
h_+(\omega)&=&h_{el}(\omega)=\xi(\omega)+O(1/m_Q), \nonumber\\
h_-(\omega)&=&O(1/m_Q). 
\end{eqnarray}
The Isgur-Wise function can be expanded near zero recoil as follows 
\begin{eqnarray}
\xi(\omega)=1-\rho^2(\omega-1)+O((\omega-1)^2).
\end{eqnarray}
It is important to stress, that the heavy-quark symmetry predicts the absolute 
normalization of the transition form factor in the heavy-quark limit. 

In addition, the heavy-quark symmetry gives the universal relation for
heavy-meson decay constants
\begin{equation}
\label{5fp}
\sqrt{M_P}f_Q={\rm const}.
\end{equation}
The asymptotic relations (\ref{5hqlimit}) and (\ref{5fp}) are the zero-order
terms of the $1/m_Q$-expansion which is calculable within the Heavy quark 
effective theory (HQET) \cite{hqet}. A particular form of the IW function 
depends on the heavy meson wave function. In next Sections  
we discuss in detail 
the structure of this expansion for the transition of a heavy pseudoscalr meson 
into pseudoscalar and vector mesons. 

\begin{figure}[thb]
\begin{center}
\mbox{\epsfig{file=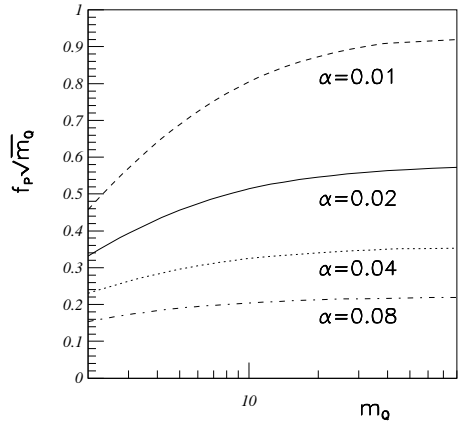,width=7.cm}}   
\end{center}
\caption{The quantity $m_Q^{0.5}f_P$ as the function of $m_Q$ at $m_q=0.25\;
GeV$.
\label{fig:fp}}
\end{figure}

The expressions (\ref{5hqlimit}) and (\ref{5fp}) mean that the HQ symmetry
restricts the possible behavior of the meson wave function at large $m_Q$.
%
%
\newpage
Table \ref{table:heavymeson} gives the results on
$f_P$ and $\rho^2_{el}$ vs $m_Q$ at $m_q=0.25\;GeV$, and
Fig.\ref{fig:fp} presents the quantity $\sqrt{m_Q}f_P$ as the function of
$m_Q$ for various values of $\alpha$.
\begin{table}
\caption{\label{table:heavymeson}
The decay constants $f_P$ of pseudoscalar mesons built up of quarks
with the masses $m_Q$ and $m_q$ and the slope of $h_{el}$ at $\omega=1$
calculated from $\langle M_Q|\bar Q\gamma_\mu Q|M_Q\rangle$ as functions of $m_Q$ at
$m_q=0.25\;GeV$.}
\centering
\begin{tabular}{|c||c|c||c|c||c|c||c|c|}
& \multicolumn{2}{c||} {$\alpha=0.01$} & \multicolumn{2}{c||} {$\alpha=0.02$} &
\multicolumn{2}{c||} {$\alpha=0.04$} & \multicolumn{2}{c|} {$\alpha=0.08$} \\
\hline
$m_Q, GeV$ & $f_P,MeV$ & $\rho^2_{el}$ &  $f_P,MeV$ & $\rho^2_{el}$ & $f_P,MeV$
& $\rho^2_{el}$ & $f_P,MeV$ &$\rho^2_{el}$ \\
\hline\hline
0.25 &   151 & 0.04 & 130  & 0.06 &  104 & 0.08 &  80 & 0.1  \\
\hline
0.4  &   190 & 0.25 & 160  & 0.35 &  128 & 0.5  &  97 & 0.65 \\
\hline
1.8  &   324 & 0.6  & 234  & 0.65 &  163 & 0.82  & 110 & 1.0  \\
\hline
5.2  &   308 & 0.75 & 202  & 1.0  &  132 & 1.05 &  85 & 1.1  \\
\hline
10   &   254 & 1.0  & 162  & 1.05 &  102 & 1.1  &  64 & 1.25 \\
\hline
20   &   195 & 1.0  & 122  & 1.1  &   76 & 1.23 &  48 & 1.45 \\
\hline
40   &   143 & 1.0  &  89  & 1.11 &   55 & 1.25 &  34 & 1.66 \\
\hline
80   &   103 & 1.0  &  63  & 1.11 &   39 & 1.25 &  24 & 1.66 \\
\end{tabular}
\end{table}
In the HQ limit, for a finite binding energy of the meson
the heavy meson and the heavy quark masses coincide, $M_Q/m_Q=1$.
So, the value of $\sqrt{m_Q}f_P$ should be independent of the heavy quark mass.
Clearly, the asymptotic relations (\ref{5hqlimit}) and (\ref{5fp}) are satisfied
if the parameter $\alpha$ of the wave function (\ref{5exppar})
tends to a constant $\alpha_\infty$ as $m_Q\to\infty$.

Thus, the function $\alpha(m_Q)$ has the following behavior: it is equal to
0.02 at $m_Q\le0.5\;GeV$ and
tends to a constant $\alpha_\infty$ as $m_Q\to \infty$. For investigating the
$B$ and $D$ mesons and
their decays we need the information on $\alpha$ in the region $m_Q=2\div
5\;GeV$.

The simplest way is to extract $\alpha$ at $m_Q=2\div5\;GeV$ from the
analysis of $f_D$ and $f_B$ as we have done for the light mesons.
In the absence of the experimental data we refer to the results of other
models.
As one can see, the decay constants $f_P$ calculated with $\alpha$ from
the range $0.02\le\alpha_D,\alpha_B\le 0.04$ cover the regions
$160\;MeV\le f_D \le 230\;MeV$ and $130\;MeV\le f_B \le 200\;MeV$ which include
the
predictions of most of the models.
Hence, the values of $\alpha_D$ and $\alpha_B$
related to the true wave functions of $D$ and $B$ mesons
are expected to be inside the interval $0.02\div 0.04$.

However, there is an attractive possibility to specify $\alpha_{D,B}$ more
precisely.
Namely, it seems reasonable to assume $\alpha$ to be
approximately constant in the region $m_Q\ge 1\div 2\;GeV$.
There are at least two arguments behind this assumption.
Firstly, a system consisting of a heavy and a
light particles behaves like a quasinonrelativistic system. And secondly, there
are no visible sources
within QCD to yield steep changes of $\alpha$ in this region.
Then for the $B$ and $D$ mesons one expects $\alpha_D=\alpha_B=\alpha_\infty$.
The next step is to estimate $\alpha_\infty$.
We consider the value $\alpha_\infty=0.02$ to be both
attractive and reasonable: on the one hand, the same parameter describes all
ground-state mesons, and on the other hand, one finds for $\alpha_\infty=0.02$
\newpage
\begin{eqnarray}
\nonumber
\sqrt{m_\infty}f_{P\infty}\simeq 5.8\;GeV^{3/2}
\end{eqnarray}
in agreement with the value $0.6\div 0.7$ estimated in \cite{sr4}.

Assuming $\alpha_D=\alpha_B=\alpha_\infty$, we can estimate the magnitude of
the
higher order $1/m_Q$ corrections which determine
the deviations of the calculated $f_P$ and $\rho^2_{el}$ at finite $m_Q$
from the asymptotic relations (\ref{5hqlimit}) and (\ref{5fp}).
Rather strong violation of the HQ symmetry for $b-$ and
$c-$quarks ($5\div15$\% at $m_Q=5\;GeV$ and $20\div30$\% at $m_Q=2\;GeV$) can
be observed both in $f_P$ and $\rho^2_{el}$ at $\alpha_\infty=0.02\div 0.04$.

We shall analyze the transition form factors obtained at $\alpha_{D,B}=0.02$
and $0.04$.
If our assumption $\alpha_D=\alpha_B=\alpha_\infty=0.02$ does not work
properly, the form factor calculations for $\alpha=0.02$
and $\alpha=0.04$ give an interval which is expected to include the true value.
Table \ref{table:parameters1} gives the numerical parameters of the model.

\vspace{.4cm}
The results on the axial-vector decay constant $f_P$ are shown
in Fig.\ref{fig:fp} and Table \ref{table:heavymeson}.
Assuming $\alpha(m_Q)=\alpha_\infty$ at $m_Q\ge2\;GeV$,
one can see the asymptotic relation $\sqrt{m_Q}f_P=const$ to work
perfectly at $m_Q>40-50\;GeV$,
and finds essential corrections to the asymptotic relations at lower $m_Q$.
For $\alpha_\infty=0.02$ one obtains $f_D=234\;MeV$ and $f_B=202\;MeV$ that
confirms the
expectation $f_D\simeq f_B$ \cite{sr4}. These values for the decay 
constants correspond to the constituent quark leptonic decay constant $g_A^0=1$. 

\begin{center}
\begin{figure}
\begin{tabular}{cc}
\mbox{\epsfig{file=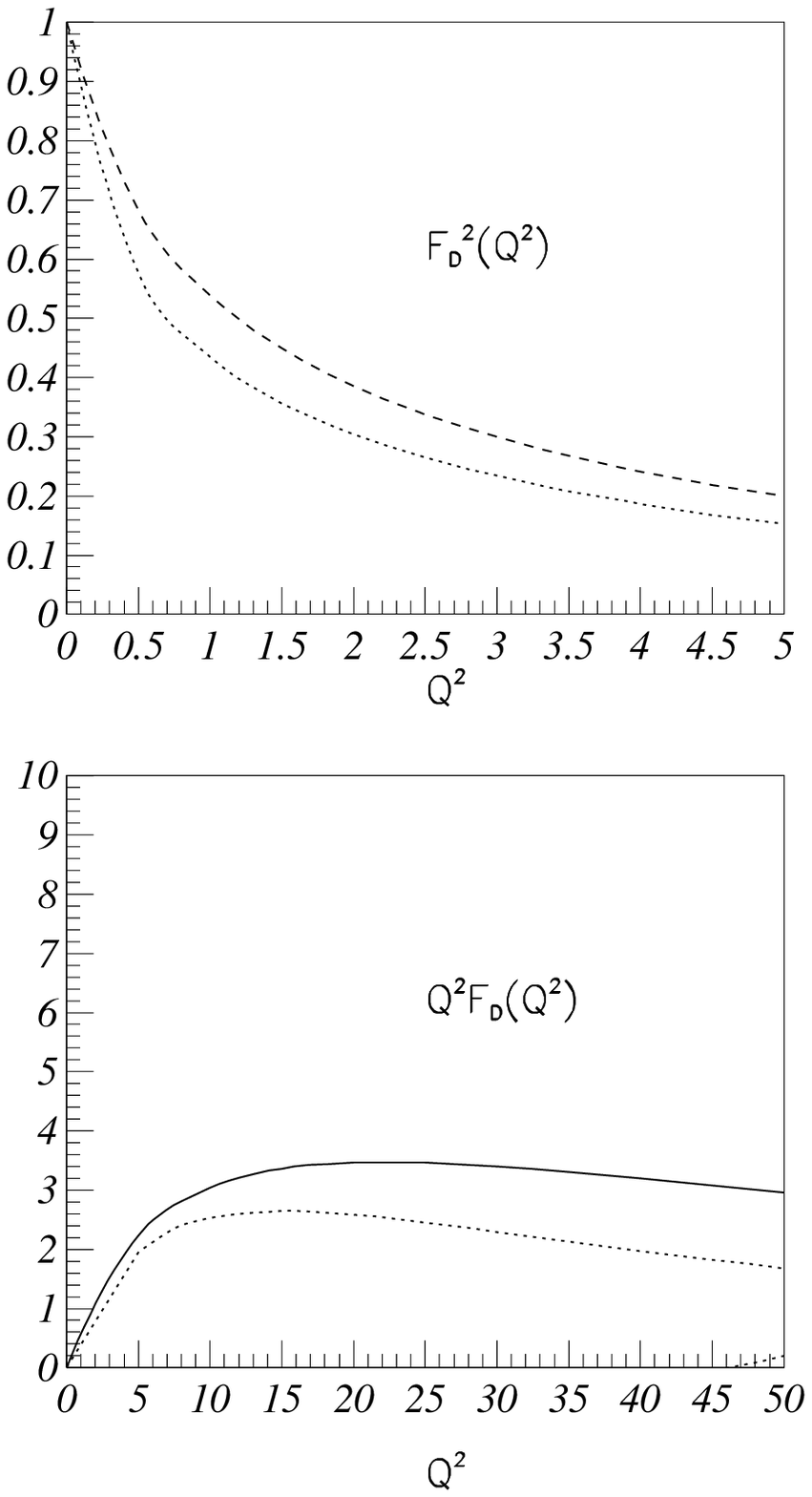,width=7cm}}  
& 
\mbox{\epsfig{file=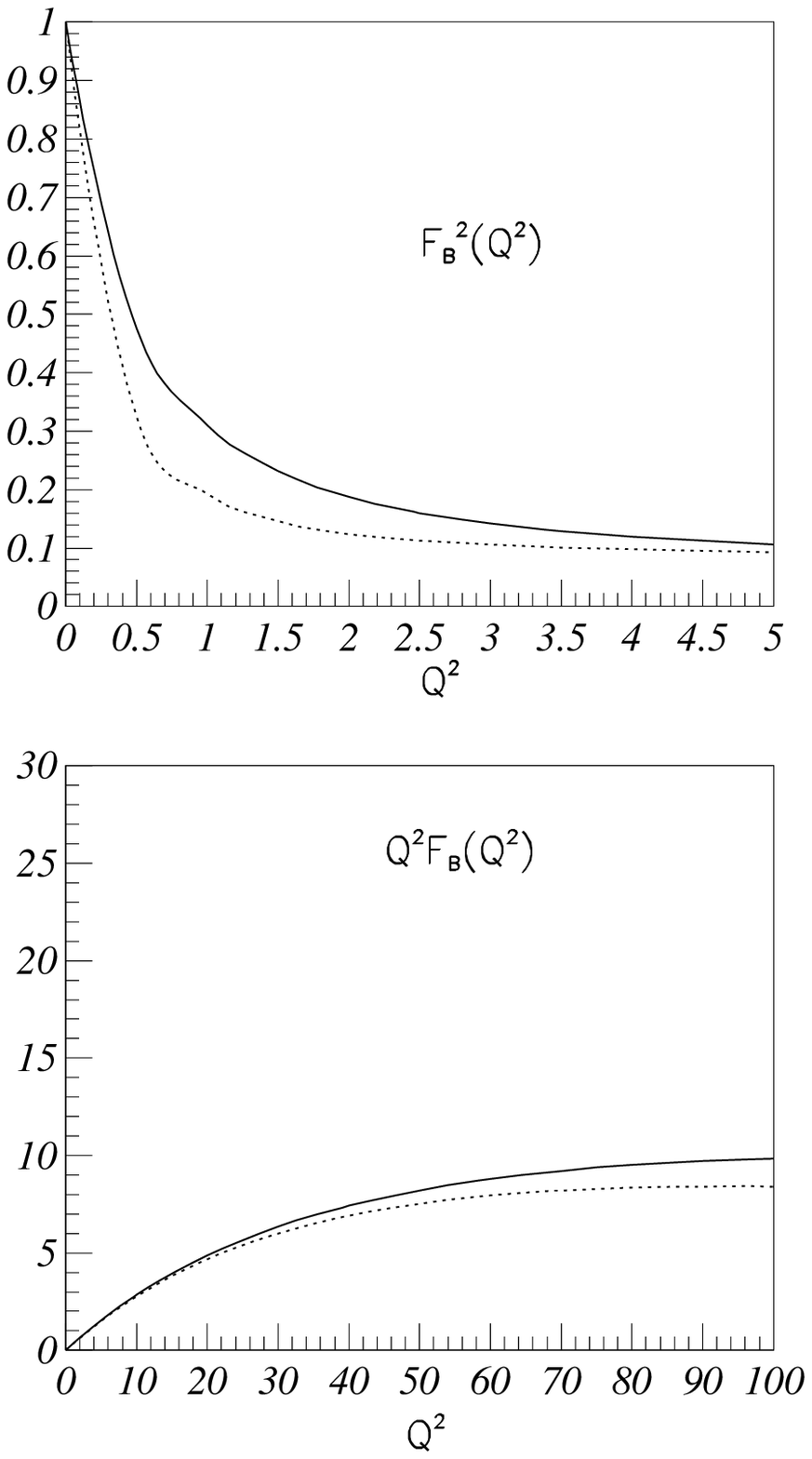,width=7cm}}  
\end{tabular}
\caption{\label{fig:ffdb}
The $D$ and $B$ elastic form factor $F_+$. 
Solid line $\alpha=0.02$, dashed line $\alpha=0.04$.}
\end{figure}
\end{center}
\begin{center}
\begin{figure}
\begin{tabular}{cc}
\mbox{\epsfig{file=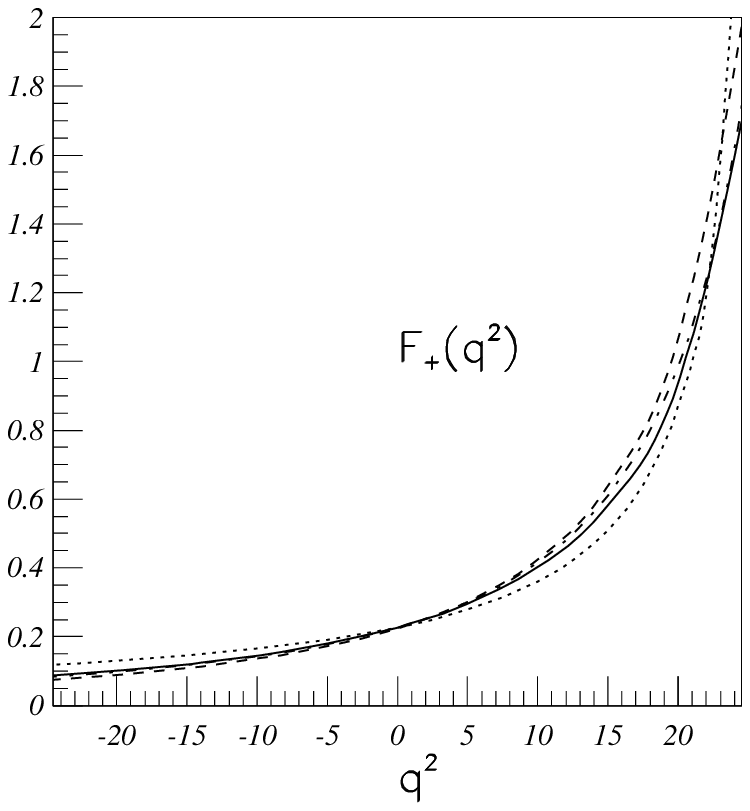,width=7.cm}}       
& 
\mbox{\epsfig{file=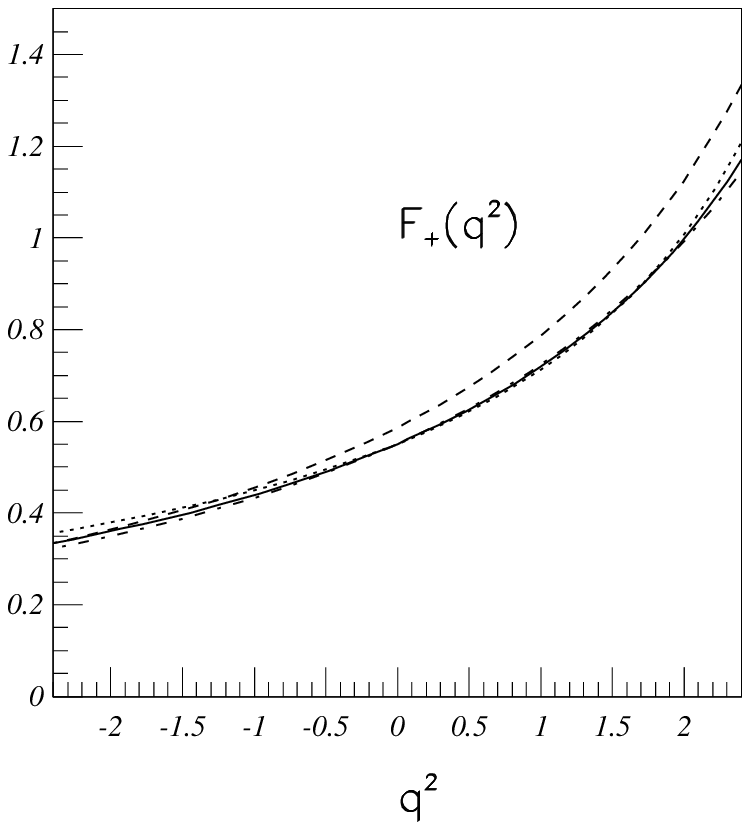,width=7.cm}}       
\end{tabular}
\caption{\label{fig:bd2pi}
The form factor $F+$ for the $B\to\pi$ (left) and $D\to\pi$ (right) transitions. 
Solid - $\alpha_{B(D)}=0.02$,
dotted - the monopole fit, dash-dotted - the dipole fit. Dashed -
$\alpha_{B(D)}=0.04$.}
\end{figure}
\end{center}
\begin{center}
\begin{figure}
\begin{tabular}{cc}
\mbox{\epsfig{file=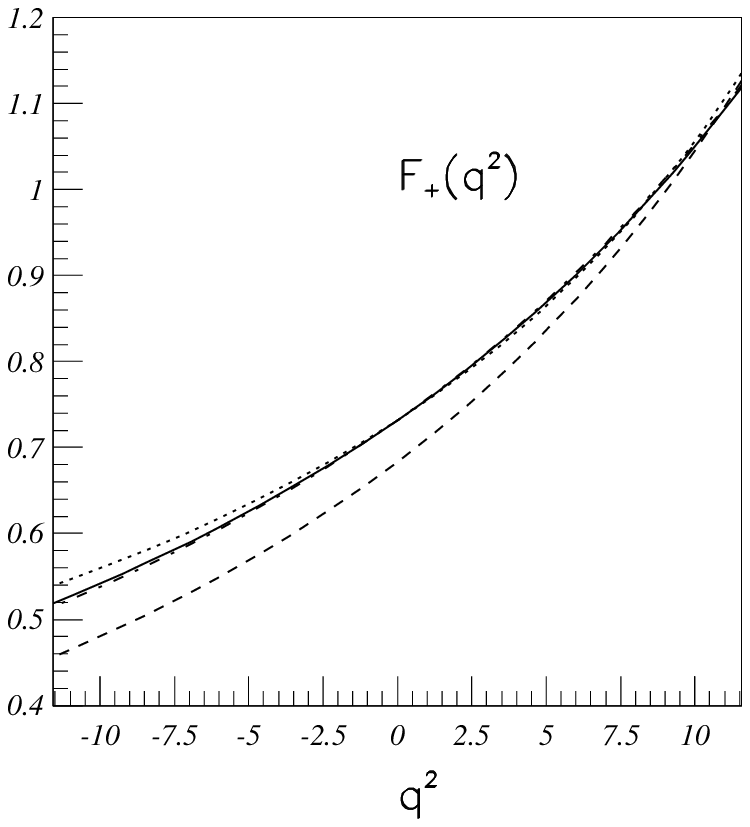,width=7.cm}}   
& 
\mbox{\epsfig{file=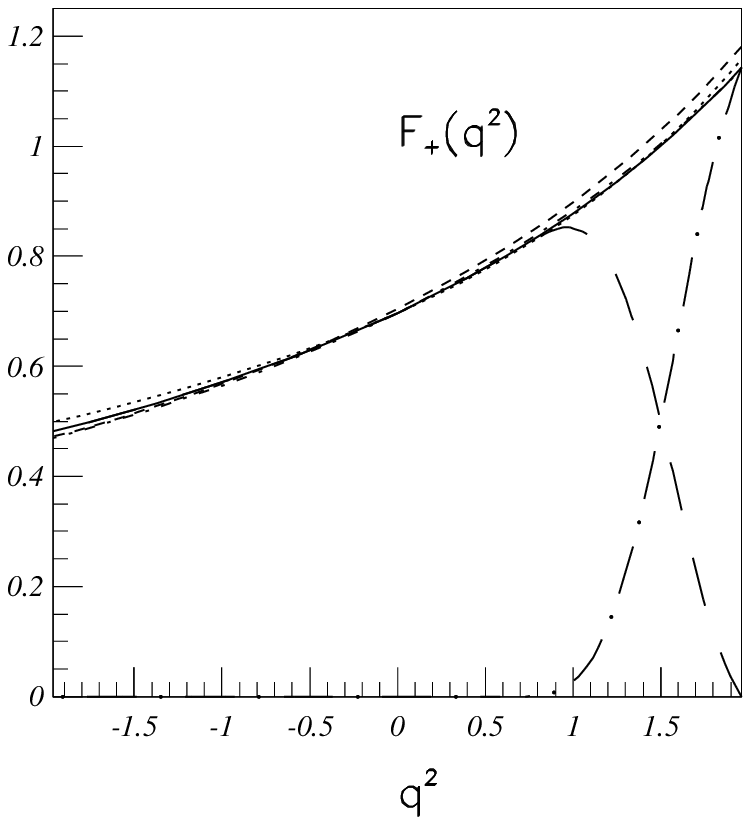,width=7.cm}}   
\end{tabular}
\caption{\label{fig:d2k}
(a) The form factor $F_+(q^2)$ for $B\to D$. Solid -
$\alpha_D=\alpha_B=0.02$,
dotted - the monopole fit, dash-dotted - the dipole fit. Dashed -
$\alpha_D=\alpha_B=0.04$;  
(b) The form factor $F_+$ for the $D\to K$ decay. Solid - $\alpha_D=0.02$,
dotted - the monopole fit, dash-dotted - the dipole fit.
Long-dashed - the Landau singularity contribution, long-dash-dotted - the
non-Landau term.
Dashed - $\alpha_D=0.04$.}
\end{figure}
\end{center}

Figures \ref{fig:ffdb}, \ref{fig:bd2pi} and \ref{fig:d2k} present the form factors 
calculated with $\alpha_{D,B}=0.02$ and $0.04$.

\vspace{.4cm}

Figure \ref{fig:d2k} (b) shows the relative magnitude of the various contributions to the 
form factor $F_+$, separately, as the function of $q^2$ for the $D\to K$ transition. 
Clearly, the full form factor is a monotoneously-rising function of $q^2$, 
whereas the behaviour of its normal and anomalous parts is rather specific: 

The normal part rises up to some value of $q^2$ where it takes the maximal value,  
then it goes down rather steeply and vanishes at the maximal $q^2=(m_2-m_1)^2$ 
corresponding to 'quark zero recoil point'. 

The anomalous part is identically zero at negative $q^2$ but comes into the game as $q^2$
goes into the positive region. It is small for small positive $q^2$, but 
increases steeply as $q^2$ approaches its maximal value. 

In other words, at $q^2\le 0$ the contribution of the non-Landau singularity 
(anomalous term) is absent, and the Landau-type singularity (normal term) 
determines the form factor; in the region $0<q^2<(m_2-m_1)^2$
both of them are essential; at the point $q^2=(m_2-m_1)^2$ the
contribution of the Landau singularity vanishes, and the non-Landau
singularity determines the decay form factor at this 'quark zero recoil' point.

For heavy-to-heavy meson transitions,
a specific relationship between the Landau and the
non-Landau contributions to the dispersion representation is observed: the
normal Landau contribution dominates the form factor at all $q^2<(m_2-m_1)^2$,
whereas the region where the anomalous singularity is essential 
shrinks to a very narrow vicinity of this point. 
So, effectively the transition form factor is determined by the
contribution of the Landau singularity only. 
Thus, the HQ symmetry can be formulated in the language of
the analytic properties of the transition form factors as the
dominance of the Landau singularity in the almost whole kinematical region.

In the case of the meson decay related to a heavy-to-light
quark transition, the anomalous non-Landau contribution
is important in a broad kinematical region. So the relations suggested by the
HQ symmetry would not work properly in this case. 

Table \ref{table:fits} shows parameters of the monopole 
$F_+(q^2)=F_+(0)/(1-q^2/M^2_{mon})$
and the dipole $F_+(q^2)=F_+(0)/(1-q^2/M^2_{dip})^2$ fits to the calculated 
transition form factors. 
The dipole formula approximates the
calculated form factors with better than 1\% accuracy.
Parameters of the monopole fit agree very well with the vector meson dominance.

The $K\to\pi$ transition form factor is well approximate by the linear function
$F_+(q^2)=F_+(0)+aq^2$, $F_+(0)=0.96$, $a=1.27\;GeV^{-2}$ in agreement with the
results of \cite{lr}.

\begin{table}[h]
\caption{\label{table:fits}
Parameters of the monopole and dipole fits to the $F_+$ form factor.
Masses of the lowest vector mesons which are expected to dominate the
form factors are given in brackets. }
\centering
\begin{tabular}{|c||c|c|c||c|c|}
  & \multicolumn{3}{c||} {$\alpha_B=\alpha_D=0.02$} & \multicolumn{2}{c|}
{$\alpha_B=\alpha_D=0.04$}  \\
\hline
Decay & $F_+(0)$ & $M_{mon},\;GeV$  & $M_{dip},\;GeV$ & $F_+(0)$ &
$M_{dip},\;GeV$  \\
\hline\hline
$B\to D$   & 0.73 & 5.7           & 7.7  & 0.68 & 7.20  \\
\hline
$B\to \pi$ & 0.23 & 5.2 [5.324]   & 6.2  & 0.22 & 6.08  \\
\hline
$D\to K$   & 0.70 & 2.22 [2.11]   & 3.0  & 0.70 & 2.95  \\
\hline
$D\to\pi$  & 0.55 & 2.1 [2.01]    & 2.8  & 0.59 & 2.68  \\
\end{tabular}
\end{table}

\subsection{\label{i.5}Discussion}
In this section we investigated form factors of hadron transitions within the relativistic
dispersion approach based on the constituent quark picture and proposed a formalism for 
a direct calculation of hadron decay form factors. The developed approach was applied to 
the analysis of the electroweak properties and transitions of pseudoscalar mesons.

The main results of this chapter are as follows:

\vspace{.4cm}

\noindent 
1. We analysed elastic and transition form factors 
for pseudoscalar mesons at spacelike momentum transfers, $q^2<0$, 
and obtained for them double spectral representations in the mass variables 
$s_1$ and $s_2$, squares of the invariant masses of the initial and final 
$q\bar q$ pairs, respectively. These representations involve double spectral 
densities of the corresponding triangle Feynman diagrams and wave functions 
of the participating mesons. 

\vspace{.4cm}
\noindent 
2. We performed the analytic continuation of the form factors to timelike momentum
transfers. We observed the appearance of the anomalous cut and, respectively, 
the appearance of the anomalous contribution to the form factor for $q^2>0$. 
The anomalous cut appears due to the motion of the singularities of the Feynman triangle 
graph from the second sheet onto the physical sheet through the normal cut. 
The normal cut is related to the Landau-Cutkosky singularities of the Feynamn graphs, 
whereas the anomalous cut is related to the non-Landau type singularities which come 
into the game in the region of timelike momentum transfers. 
It is important ot emphasise that both the normal and anomalous contributions to the form
factors are expressed in terms of the wave functions of the initial and final mesons 
only in the physical region above the threshold, $s_1>(m_2+m_3)^2$ and $s_2>(m_1+m_3)^2$.

\vspace{.4cm}
\noindent 
3. We demonstarted the equivalence of our approach based on spectral representations 
with the light-cone constituent quark model for the description of leptonic decays and 
for transition form factors between pseudoscalar mesons at 
{\it spacelike momentum transfers}.
It should be noticed that the dispersion approach has important advantages compared to 
the light-cone approach in the region $q^2>0$, where the direct application of the latter 
is hampered by the contribution of pair-creation subprocesses. 

\newpage

\vspace{.4cm}
\noindent 
4. For meson decays related to heavy-to-heavy quark transitions a dominance of 
the normal contribution over the anomalous contribution for almost all $q^2$ 
from the decay region except for the very vicinity of the zero-recoil point 
has been observed.
This allows a formulation of the heavy-quark symmetry in the language of the
analytic properties of the decay form factors as the dominance of the normal 
Landau contribution in the almost whole kinematic region of momentum transfers.

\vspace{.4cm}
\noindent  
5. Electroweak properties and form factors of pseudoscalar mesons
have been analysed using a simple parametrization of the meson wave function based on
the heavy quark symmetry. We have examined the dependence of the axial-vector
decay constant on the heavy-quark mass, and found $f_D\simeq 235\;MeV$ and 
$f_B\simeq200\;MeV$. 

Analysing the dependence of $f_P$ and the heavy meson form factor on the heavy
quark mass we have found that the violation of the HQ symmetry relations can be
expected at the 10-20\% level for the $b$ and $c$-quark masses. 

\vspace{.4cm}

In the next Chapter we consider the $1/m_Q$ expansion of the transition form factors 
given by the double spectral representations and match this expansion to the 
heavy quark expansion in QCD. In particular, we demonstrate that the heavy quark 
expansion of the $P\to P$ transition form factors is fully compatible with the $1/m_Q$ 
expansion in QCD in the leading and subleading $1/m_Q$ orders. 
\newpage
\section{\label{sec:ii}
Heavy quark expansion and universal form factors in the dispersion approach}

This Chapter presents a detailed discussion of form factors for weak meson decays 
within the dispersion approach following the analysis of Ref. \cite{m2}. 

We calculate the double spectral densities for the form factors describing 
the transition of a pseudoscalar ($P$) meson to pseudoscalar ($P$) and vector ($V$) mesons 
induced by the vector, axial-vector and tensor currents. These spectral densities are given 
in terms of the soft wave functions of the participating mesons and the double spectral 
densities of the corresponding triangle Feynman graphs. 

We then analyse the spectral representations for the form factors in the two 
specific cases of the {heavy-to-heavy} and {heavy-to-light} transitions. 

\vspace{.5cm}
\begin{itemize}
\item[(i)]
The {\it heavy-to-heavy} transition means that the masses of the initial quark $m_2$ 
and the final quark $m_1$ participating in the weak transition are much larger than the 
confinement scale $\Lambda$
\begin{eqnarray}
\label{h2h}  
m_2>m_1, \quad m_2\sim m_1\sim m_Q\gg\Lambda.  
\end{eqnarray} 
In this case the meson transition can be analysed using the formalism of Heavy quark 
effective theory (HQET) \cite{luke}, an effective theory obtained from QCD for heavy quarks. 
Expansions of the form factors in powers of $1/m_Q$ can be obtained in terms of 
the universal form factors which appear in each order of $1/m_Q$. These process-independent 
form factors contain the information about the long-distance dynamics in the heavy-quark limit 
and can be calculated only within some nonperturbative approach.  

We studied the form factors of the dispersion approach for the quark masses satisfying the
relation (\ref{h2h}). We perform the $1/m_Q$ expansion of our spectral representations 
and require its structure to match to the known structure of the $1/m_Q$ expansion 
in HQET.  
\item[(ii)]
The {\it heavy-to-light} transition means that 
\begin{eqnarray}
\label{h2l}  
m_2=m_Q\gg m_1\simeq\Lambda.  
\end{eqnarray} 
In this case the explicit structure of the $1/m_Q$ expansion cannot be obtained 
directly from QCD by the existing methods, but QCD provides relations between the 
form factors in the region of large $q^2$ near zero recoil \cite{iwhl}. 
We therefore require the form factors of our dispersion approach to obey these 
relations. 
\end{itemize}
The conditions (i) and (ii) allow us to determine the necessary subtraction terms in the spectral
representations for the form factors. As prompted by the structure of the $1/m_Q$ expansion of the dispersion form 
factors, no subtractions are necessary for $P\to P$ transition form factors, but subtractions 
for some of the $P\to V$ form factors are necessary. 

\vspace{.4cm}
Section \ref{ii.1} gives the definitions of all the necessary form factors for $P\to P$ and 
$P\to V$ weak transitions. 
The structure of the $1/m_Q$ expansion of the transition form factor induced by the vector and
axial-vector current has been calculated within HQET in \cite{luke}. 
We apply the formalism of HQET to transitions induced by the tensor current and report 
the $1/m_Q$ expansion of the form 
factors $h_{g_+}$, $h_{g_-}$, $h_{g_0}$, and $h_{s}$ (the definitions are given in
the next section) in the leading (LO) and next-to-leading (NLO) orders in $1/m_Q$. 
In particular, we find the $1/m_Q$ correction to the value of the form factor $h_{g_+}$ 
at zero recoil to vanish exactly as for the form factors $h_{f_+}$ and $h_f$ (the Luke theorem). 
 
We later use the $1/m_Q$ expansion of the weak form factors from HQET as a benchmark for 
testing the $1/m_Q$ expansion of the form factors of our dispersion approach. 

\vspace{.4cm}
 In section \ref{ii.2} we calculate the double spectral densities for form factors
and give results for the necessary subtraction terms which are explained later in the 
sections \ref{ii.3} and \ref{ii.4}. 

\vspace{.4cm}

In section \ref{ii.3} we consider the case of the {heavy-to-heavy} meson transition 
and perform the expansion of the spectral representations for the transition form factors in 
the LO and NLO in $1/m_Q$.  

The spectral representations without subtractions are found to agree with HQET 
in the LO for all $P\to P$ and $P\to V$ form factors. The next-to-leading order analysis 
shows the necessity of subtractions in the spectral representations for some of the $P\to V$ 
form factors in order to match to the structure of the HQET. 
The matching condition provides constraints on the subtraction terms.  

Assuming the strong peaking of the meson soft wave functions in terms of the relative 
quark momenta with a width of order of the confinement scale, we calculate the Isgur-Wise 
function and the NLO universal form factors in terms of the wave function of the infinitely 
heavy meson. Our dispersion approach leads to the following relations for the universal NLO 
form factors: 
\begin{eqnarray}
\label{nloffs}
\chi_2(\omega)&=&0;\qquad 
\nonumber\\
\chi_3(\omega)&=&0;\qquad 
\nonumber\\
\xi_3(\omega)&>&0,\; 
\qquad
\xi_3(1)={\langle z \rangle}/{3}, 
\end{eqnarray}
where $\langle z \rangle$ is an average kinetic energy of the light quark in the heavy meson rest frame. 

\vspace{.4cm}
\noindent 
In section \ref{ii.4} we discuss {heavy-to-light} quark transitions and 
consider the $1/m_Q$ expansion for the wave functions and the 
form factors in this case. 
We then consider the heavy-to-light meson transitions in which case a small parameter 
$\Lambda_{QCD}/m_Q$ emerges and analyse the form factors in the leading $\Lambda_{QCD}/m_Q$ order.
Requiring the fulfillment of the Isgur--Wise relatins for the heavy-to-light transitions 
\cite{iwhl} further constrains the subtraction terms providing explicit spectral 
representations with subtractions for the form factors of interest.

\vspace{.4cm} 
Section \ref{ii.5} illustrates the main results with numerical
estimates and evaluate the universal form factors for various quark model parameters. 

\subsection{\label{ii.1}
Meson transition amplitudes and heavy-quark expansion in QCD}
The amplitudes of meson decays induced by the quark transition $q_2\to q_1$ 
through the 
vector $V_{\mu}={\bar q_1}\gamma_{\mu}q_2$, 
axial-vector $A_{\mu}={\bar q_1}\gamma_{\mu}\gamma_{5}q_2$, 
tensor $T_{\mu\nu}={\bar q_1}\sigma_{\mu\nu}q_2$, 
and pseudo-tensor $T^5_{\mu\nu}={\bar q_1}\sigma_{\mu\nu}\gamma_5q_2$ 
currents have the following structure \cite{iwhl}
\begin{eqnarray}
\label{ffdef}
\langle P(M_2,p_2)|V_\mu(0)|P(M_1,p_1)\rangle &=&f_1(q^2)p_{1\mu}+f_2(q^2)p_{2\mu},  
\nonumber \\
\langle V(M_2,p_2,\epsilon)|V_\mu(0)|P(M_1,p_1)\rangle &=&2g(q^2)\epsilon_{\mu\nu\alpha\beta}
\epsilon^{*\nu}\,p_1^{\alpha}\,p_2^{\beta}, 
\nonumber \\
\langle V(M_2,p_2,\epsilon)|A_\mu(0)|P(M_1,p_1)\rangle &=&
i\epsilon^{*\alpha}\,[\,f(q^2)g_{\mu\alpha}+a_1(q^2)p_{1\alpha}p_{1\mu}+
               a_2(q^2)p_{1\alpha}p_{2\mu}\,],   
\nonumber \\
\langle P(M_2,p_2)|T_{\mu\nu}(0)|P(M_1,p_1)\rangle &=&-2i\,s(q^2)\,(p_{1\mu}p_{2\nu}-p_{1\nu}p_{2\mu}), 
\nonumber \\
\langle V(M_2,p_2,\epsilon)|T_{\mu\nu}(0)|P(M_1,p_1)\rangle &=&i\epsilon^{*\alpha}\,
               [\,g_{1}(q^2)\epsilon_{\mu\nu\alpha\beta}p^{1\beta}+
               g_{2}(q^2)\epsilon_{\mu\nu\alpha\beta}p^{2\beta}
\nonumber \\                
&&\qquad +g_0(q^2)p_{1\alpha}
\epsilon_{\mu\nu\beta\gamma}p_1^{\beta}p_2^{\gamma}\,], 
\nonumber \\
\langle P(M_2,p_2)|T^5_{\mu\nu}(0)|P(M_1,p_1)\rangle &=&
s(q^2)\epsilon_{\mu\nu\alpha\beta}P^\alpha q^\beta, 
\nonumber \\
\langle V(M_2,p_2,\epsilon)|T^5_{\mu\nu}(0)|P(M_1,p_1)\rangle &=&
g_{1}(q^2)(\epsilon^{*}_\nu p_{1\mu}-\epsilon^{*}_{\mu}p_{1\nu})
+g_{2}(q^2)(\epsilon^{*}_{\nu}p_{2\mu}-\epsilon^{*}_{\mu}p_{2\nu})
\nonumber \\                
&&\qquad 
+g_0(q^2)(\epsilon^{*}p_{1})(p_{1\nu}p_{2\mu}-p_{1\mu}p_{2\nu}), 
\end{eqnarray}
with $q=p_{1}-p_{2}$, $P=p_{1}+p_{2}$. 
Also the following linear combinations of the form factors will be used 
\begin{eqnarray}
f_{\pm}&=&\frac{1}{2}(f_1\pm f_2),\qquad 
\nonumber\\
a_{\pm}&=&\frac{1}{2}(a_1\pm a_2),\qquad 
\nonumber\\
g_{\pm}&=&\frac{1}{2}(g_1\pm g_2). 
\end{eqnarray}
The matrix element of the penguin operator relevant for rare decays 
has the following structure 
\begin{eqnarray}
&&\langle V(M_2,p_2,\epsilon)|\bar q_1\sigma_{\mu\nu}q^\nu(1+\gamma_5)q_2|P(M_1,p_1)\rangle
=-i\epsilon_{\mu\nu\alpha\beta}\epsilon^{*\nu}p_1^{\alpha}p_2^{\beta} 2g_{+}
\nonumber\\
&&\qquad-(g_{\mu\nu}\cdot qP-q_\nu P_\mu)\epsilon^{*\nu}\left(g_{+}-\frac{q^2}{qP}g_{-}\right)
+(\epsilon^{*}q)\left(q_\mu-\frac{q^2}{qP}P_\mu\right)\left(g_{-}-\frac{1}{2}qP\,g_0\right).
\end{eqnarray}
We use the following conventions: 
\begin{eqnarray}
\label{notations}
\gamma^{5}=i\gamma^{0}\gamma^{1}\gamma^{2}\gamma^{3},
\qquad \sigma_{\mu \nu}={\frac{i}{2}}[\gamma_{\mu},\gamma_{\nu}], \qquad
\epsilon^{0123}=-1.
\end{eqnarray}
Accordingly,  
\begin{eqnarray}
Sp(\gamma^{5}\gamma^{\mu}\gamma^{\nu}\gamma^{\alpha}\gamma^{\beta})=
4i\epsilon^{\mu\nu\alpha\beta}, \qquad 
\sigma_{\mu\nu}\gamma^5=-\frac{i}{2}\epsilon_{\mu\nu\alpha\beta}
\sigma^{\alpha\beta}.
\end{eqnarray}
The relativistic-invariant form factors contain the dynamical 
information on the process and should be calculated within a nonperturbative approach 
for any particular initial and final mesons. 

For analysing the transition in the case when both the parent and the daughter quarks
inducing the meson transition are heavy, i.e. $m_1\simeq m_2\gg \Lambda_{QCD}$  
it is convenient to introduce a new dimensionless variable 
\begin{eqnarray}
\omega=v_1v_2=\frac{M_1^2+M_2^2-q^2}{M_1M_2}
\end{eqnarray} 
and velocity-dependent form factors $h(\omega)$ connected with 4-velocities 
and not 4-momenta as in (\ref{ffdef}) in the following way 
\begin{eqnarray}
\label{hqffdef}
\langle P(M_2,p_2)|V_\mu(0)|P(M_1,p_1)\rangle &=&\sqrt{M_1M_2}\,[h_{f_+}(\omega)(v_1+v_2)_{\mu}
+h_{f_-}(\omega)(v_1-v_2)_{\mu}],  
\nonumber \\
\langle V(M_2,p_2,\epsilon)|V_\mu(0)|P(M_1,p_1)\rangle &=&
\sqrt{M_1M_2}\;h_g(\omega)\epsilon_{\mu\nu\alpha\beta}
\epsilon^{*\nu}\,v_1^{\alpha}\,v_2^{\beta}, 
\nonumber \\
\langle V(M_2,p_2,\epsilon)|A_\mu(0)|P(M_1,p_1)\rangle &=&
i\epsilon^{*\alpha}\,\sqrt{M_1M_2}\,[h_f(\omega)(1+\omega)g_{\mu\alpha}
\nonumber \\
&&\qquad\qquad\qquad
-h_{a_1}(\omega)v_{1\alpha}v_{1\mu}
\nonumber \\
&&\qquad\qquad\qquad
-h_{a_2}(\omega)v_{1\alpha}v_{2\mu}\,],   
\nonumber \\
\langle P(M_2,p_2)|T_{\mu\nu}(0)|P(M_1,p_1)\rangle &=&-2i\,\sqrt{M_1M_2}\;h_s(\omega)\,(v_{1\mu}v_{2\nu}-v_{1\nu}v_{2\mu}), 
\nonumber \\
\langle V(M_2,p_2,\epsilon)|T_{\mu\nu}(0)|P(M_1,p_1)\rangle &=&i\epsilon^{*\alpha}\sqrt{M_1M_2}\,
[\,h_{g_+}(\omega)\epsilon_{\mu\nu\alpha\beta}(v_1+v_2)^{\beta}
\nonumber\\
&&\qquad\qquad\qquad
+h_{g_-}(\omega)\epsilon_{\mu\nu\alpha\beta}(v_1-v_2)^{\beta}
\nonumber\\
&&\qquad\qquad\qquad
+h_{g_0}(\omega)v_{1\alpha}\epsilon_{\mu\nu\beta\gamma}v_1^{\beta}v_2^{\gamma}\,]. 
\nonumber\\
\langle P(M_2,p_2)|T^5_{\mu\nu}(0)|P(M_1,p_1)\rangle &=&
2\,\sqrt{M_1M_2}\;h_s(\omega)\,\epsilon_{\mu\nu\alpha\beta}
v_1^\alpha v_2^\beta, 
\nonumber\\
\langle V(M_2,p_2,\epsilon)|T^5_{\mu\nu}(0)|P(M_1,p_1)\rangle &=&
\sqrt{M_1M_2}\left[
h_{g_+}(\omega)(\epsilon^{*}_\nu(v_1+v_2)_{\mu}-\epsilon^{*}_{\mu}(v_1+v_2)_{\nu})
\right.
\nonumber \\                
&&\qquad\qquad
+h_{g_-}(\omega)(\epsilon^{*}_\nu(v_1-v_2)_{\mu}-\epsilon^{*}_\mu(v_1-v_2)_{\nu})
\nonumber \\                
&&\qquad\qquad 
\left.+h_{g_0}(\omega)(\epsilon^{*}v_{1})(v_{1\nu}v_{2\mu}-v_{1\mu}v_{2\nu})
\right].  
\end{eqnarray}
These form factors are related to the form factors introduced by the 
relations (\ref{ffdef}) as follows
\begin{eqnarray}
\label{ff2hqff}
f_1&=&\frac{M_2}{\sqrt{M_1M_2}}[h_{f_+}+h_{f_-}],\quad
\nonumber\\
f_2&=&\frac{M_1}{\sqrt{M_1M_2}}[h_{f_+}-h_{f_-}],\quad
\nonumber\\
s&=&\frac{1}{2\sqrt{M_1M_2}}h_s,\quad  
\nonumber\\
g&=&\frac{1}{2\sqrt{M_1M_2}}h_g,\quad  
\nonumber\\
a_1&=&-\frac{1}{\sqrt{M_1M_2}}\frac{M_2}{M_1}h_{a_1},\quad 
\nonumber\\
a_2&=&-\frac{1}{\sqrt{M_1M_2}}h_{a_2},\quad 
\nonumber\\
f&=&\sqrt{M_1M_2}(1+\omega)h_f,\quad
\nonumber\\
g&=&\frac{1}{2\sqrt{M_1M_2}}h_g,\quad  
\nonumber\\
g_1&=&  \frac{M_2}{\sqrt{M_1M_2}}[h_{g_+}+h_{g_-}],\quad
\nonumber\\
g_2&=&  \frac{M_1}{\sqrt{M_1M_2}}[h_{g_+}-h_{g_-}],\quad
\nonumber\\
g_0&=&\frac{1}{\sqrt{M_1M_2}}\frac{1}{M_1}h_{g_0}
\end{eqnarray}
The form factors $h$ are convenient quantities as in the leading $1/m_Q$ order 
all of them are expressed through a single universal function of the dimensionless 
variable $\omega$ -- the Isgur--Wise function. 
A consistent heavy--quark expansion of the form factors, i.e. expansion in inverse powers of the heavy--quark mass, 
can be constructed within the Heavy Quark Effective Theory based on QCD with heavy quarks. 

The general structure of the $1/m_Q$ expansion of the heavy quark form factors 
in QCD for the meson transition $M_1\to M_2$ induced by heavy quark transition $m_2\to m_1$ have the 
form (omitting corrections $O(\alpha_s,\,\alpha_s/m_Q,\,1/m_Q^2)$:
\begin{eqnarray}
\label{hqexp}
h_{f_+}=&\xi+     &\left(\frac{1}{m_1}+\frac{1}{m_2}\right)\rho_1,\nonumber\\
h_{f_-}=&         &\left(\frac{1}{m_1}-\frac{1}{m_2}\right)\left(-\frac{\bar\Lambda}2\xi+\xi_3\right),\nonumber\\
h_{g}=  &\xi+     &\left(\frac{1}{m_1}+\frac{1}{m_2}\right)\frac{\bar\Lambda}2\xi+\frac1{m_1}\rho_2
                            +\frac1{m_2}\left({\rho_1-\xi_3}\right),\nonumber\\
h_{f}=  &\xi+     &\left(\frac{1}{m_1}+\frac{1}{m_2}\right)\frac{\omega-1}{\omega+1}\frac{\bar\Lambda}2\xi+\frac1{m_1}\rho_2
                            +\frac1{m_2}\left({\rho_1-\frac{\omega-1}{\omega+1}\xi_3}\right),\nonumber\\
h_{a_1}=&         &\frac1{m_1}\frac1{\omega+1}\left(-\bar\Lambda\xi +2(\omega+1)\chi_2-\xi_3\right),\nonumber\\
h_{a_2}=&\xi+     &\left(\frac{\omega-1}{\omega+1}\frac{1}{m_1}+\frac{1}{m_2}\right)\frac{\bar\Lambda}2\xi+
\frac1{m_1}\left(\rho_2-2\chi_2-\frac1{\omega+1}\xi_3\right)+\frac1{m_2}(\rho_1-\xi_3).
\nonumber\\
h_{s}=&\xi+       &\left(\frac{1}{m_1}+\frac{1}{m_2}\right)\left({\frac{\bar\Lambda}2\xi-\xi_3+\rho_1}\right),\nonumber\\
h_{g_+}=&-\xi-    &\frac1{m_2}\rho_1-\frac1{m_1}\rho_2,\nonumber\\
h_{g_-}=&         &\left(\frac{1}{m_1}-\frac{1}{m_2}\right)\frac{\bar\Lambda}2\xi+\frac1{m_2}\xi_3,\nonumber\\
h_{g_0}=&         &\frac1{m_1}\left({    \frac{\bar\Lambda\xi+\xi_3}{\omega+1}+2\chi_2}\right).
\end{eqnarray}
In the leading $1/m_Q$ order (LO) all the form factors are represented through the single universal
Isgur--Wise function $\xi(\omega)$ , whereas in the next-to-leading order (NLO) the 4 new form factors 
$\rho_1, \rho_2, \chi_2$, and $\xi_3$ appear. 
The universal form factors are functions of a single variable $\omega$. 

The form factor $\xi_3$ originates from the expansion of the transition quark current, and the form factors 
$\rho_1,\rho_2,\chi_2$ are connected with the nontrivial relationship between the mesonic states in the full and the
effective theory. The universal form factors satisfy the conditions
\begin{eqnarray}
\label{hqnorm}
\xi(1)&=&1,\quad 
\nonumber\\
\rho_1(1)&=&0, 
\nonumber\\
\rho_2(1)&=&0,
\end{eqnarray}
whereas no constraints on $\xi_3$ and $\chi_2$ are imposed by the heavy quark symmetry. 
As found by Luke \cite{luke}, the $1/m_Q$ corrections to the form factors $h_f$ and $h_{f_+}$ vanish due to kinematical or
dynamical reasons:
\begin{eqnarray}
\label{luke1}
h_f(1)&=&1+O(1/m_Q^2),
\nonumber\\
h_{f_+}(1)&=&1+O(1/m_Q^2). 
\end{eqnarray}
The same is found to be true also for the form factor $h_{g_+}$: namely,
\begin{equation}
\label{luke2}
h_{g_+}(1)=-1+O(1/m_Q^2).
\end{equation}

The parameter $\bar\Lambda$ in (\ref{hqexp}) comes from the $1/m_Q$ expansion of the mass of a meson consisting of the 
heavy quark and light degrees of freedom
\begin{equation}
\label{lambdadef}
M_Q=m_Q+\bar\Lambda+O(1/m_Q). 
\end{equation}
In our notations for heavy quarks and mesons, this gives 
\begin{eqnarray}
M_1&=&m_2+\bar\Lambda+\ldots,\qquad 
\nonumber\\
M_2&=&m_1+\bar\Lambda+\ldots,
\end{eqnarray}
for the parent and daughter particles, respectively. 

It is straightforward to derive the following useful relations
\begin{eqnarray}
\label{M2m}
\frac{M_1+M_2}{\sqrt{M_1M_2}}&=&\frac{m_2+m_1}{\sqrt{m_1m_2}}\left[{1-\left(\frac1m_1+\frac1m_2\right)
\left(\frac{m_2-m_1}{m_2+m_1}\right)^2\frac{\bar\Lambda}2+\ldots}\right]\nonumber\\
\frac{M_1-M_2}{\sqrt{M_1M_2}}&=&\frac{m_2-m_1}{\sqrt{m_1m_2}}
\left[{1-\left( \frac1{m_1}+\frac1{m_2}\right)\frac{\bar\Lambda}2+\ldots }\right]\nonumber\\
\sqrt{M_1M_2}&=&\sqrt{m_1m_2}
\left[{1+\left( \frac1{m_1}+\frac1{m_2}\right)\frac{\bar\Lambda}2+\ldots }\right],
\end{eqnarray}
where the dots denote higher order terms. 

Using the relations (\ref{ff2hqff}) and (\ref{M2m}), we obtain for the form factors (\ref{ffdef}) the following
expansions  
\begin{eqnarray}
\label{f1exp}
f_1&=&\frac{m_1}{\sqrt{m_1m_2}}\left[{\xi+\frac1{m_1}\left(\rho_1+\xi_3\right) +\frac1{m_2}\left(\rho_1-\xi_3\right)}\right],\\ 
\label{f2exp}
f_2&=&\frac{m_2}{\sqrt{m_1m_2}}\left[{\xi+\frac1{m_1}\left(\rho_1-\xi_3\right) +\frac1{m_2}\left(\rho_1+\xi_3\right)}\right],\\
\label{sexp}
  s&=&  \frac{1}{2\sqrt{m_1m_2}}\left[{\xi+\frac1{m_1}\left(\rho_1-\xi_3\right) +\frac1{m_2}\left(\rho_1-\xi_3\right)}\right],\\
\label{g1exp}
g_1&=&-\frac{m_1}{\sqrt{m_1m_2}}\left[{\xi+\frac1{m_1}\rho_2+\frac1{m_2}\left(\rho_1-\xi_3\right)}\right],\\ 
\label{g2exp}
g_2&=&-\frac{m_2}{\sqrt{m_1m_2}}\left[{\xi+\frac1{m_1}\rho_2+\frac1{m_2}\left(\rho_1+\xi_3\right)}\right],\\ 
\label{gexp}
  g&=&  \frac{1}{2\sqrt{m_1m_2}}\left[{\xi+\frac1{m_1}\rho_2 +\frac1{m_2}\left(\rho_1-\xi_3\right)}\right],\\
\label{a1exp}
a_1&=&-\frac{1}{\sqrt{m_1m_2}}\frac1{m_2}\frac1{\omega+1}\left[{-\bar\Lambda\xi+2(\omega+1)\chi_2-\xi_3}\right],\\
\label{a2exp}
a_2&=&-\frac{1}{\sqrt{m_1m_2}}
\left[{\xi+\frac1{m_1}
\left(\rho_1-\xi_3\right) +\frac1{m_2}\left(\rho_1-\xi_3\right)
-\frac{\bar\Lambda}{m_1}\frac1{\omega+1}\xi}\right.
\nonumber\\
&&\qquad\qquad\qquad\left.{-\frac{2\chi_2}{m_1}+\frac1{m_1}\frac{\omega}{\omega+1}}\xi_3\right],\\
\label{fexp}
  f&=&{\sqrt{m_1m_2}}(\omega+1)\left[{\xi+\frac{\bar\Lambda\xi\omega}{\omega+1}\left(\frac1m_1+\frac1m_2\right)+\frac1m_1\rho_2
  +\frac1m_2\left(\rho_1-\frac{\omega-1}{\omega+1}\xi_3\right)    }\right]\\
\label{g0exp}
g_0&=&\frac{1}{(m_1m_2)^{3/2}}\left[\frac{\bar\Lambda\xi+\xi_3}{\omega+1}+2\chi_2\right]  
\end{eqnarray}
For the following analysis it is worth noting that the behavior of the combination $2p_1p_2\cdot g-m_1\cdot g_2$ 
and $f$ in LO and NLO coincide, namely  
\begin{equation}
\label{fmodified}
f\simeq 2p_1p_2\cdot g-m_1\cdot g_2.
\end{equation}
It is also convenient to introduce the form factor $a'_2$ such that 
\begin{equation}
a'_2=a_2+2s.
\end{equation}
In what follows we need the expansions of the following linear combinations of the form factors $a'_2$ 
and $a_1$
\begin{eqnarray}
\label{aminus}
a_1m_2-a'_2m_1&=&-\frac1{\sqrt{m_1m_2}}\left[ 4\chi_2-\xi_3 \right],\\ 
\label{aplus}
a_1m_2+a'_2m_1&=&-\frac1{\sqrt{m_1m_2}}\left[ -\frac{2\bar\Lambda\xi}{\omega+1}+\xi_3 \frac{\omega-1}{\omega+1} \right].
\end{eqnarray}

\subsection{\label{ii.2}Transition form factors in the dispersion approach}
The results presented in the previous section are strict consequences of QCD in the heavy-quark 
limit. The universal form factors $\xi$ and $\rho_1,\rho_2, \chi_2,\xi_3$ can be calculated 
within a nonperturbative dynamical approach. 

We study the form factors within the dispersion approach based of the constituent quark
picture which turns out to be an efficient framework for describing meson decays. 
We start with $q^2<0$ and represent the form
factors as double spectral representations in the invariant masses of the initial and final $q\bar q$ pairs. 
The form factors at $q^2>0$ are derived by performing the analytical continuation. 

\begin{figure}[hbt]
\begin{center}  
\mbox{\epsfig{file=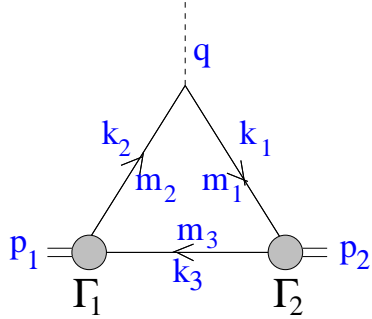,width=4.8cm}  }  
\end{center}
\caption{One-loop graph for the meson transition.\label{fig:trianglegraph}}
\end{figure}

The transition of the initial meson  $q(m_2)\bar q(m_3)$ with the mass $M_1$ 
to the final meson $q(m_1)\bar q(m_3)$ with the mass $M_2$ 
induced by the quark transition $m_2\to m_1$ through the current $\bar q(m_1) J_\mu q(m_2)$ is 
described by the triangle diagram of Fig.\ref{fig:trianglegraph}. 
As explained in the previous section, for constructing the 
double spectral representation we must  
consider a double--cut graph where all intermediate particles go on mass shell
but the initial and final mesons have the off--shell momenta 
$\tilde p_1$ and $\tilde p_2$ such that $\tilde p_1^2=s_1$ and $\tilde p_2^2=s_2$ with 
$(\tilde p_1-\tilde p_2)^2=q^2$ kept fixed. 

For the transition $B\to D,D^*$ we have $m_2=m_b, m_1=m_c$, and $m_3=m_u$. 
The constituent quark structure of the initial and final mesons is given in terms of the vertices $\Gamma_1$
and $\Gamma_2$, respectively. 
The vertex describing the initial pseudoscalar meson has the spinorial structure 
\begin{eqnarray}
\label{pseudoscalarvertex}
\Gamma_1=i\gamma_5G_1/\sqrt{N_c}.   
\end{eqnarray}
The vertex for the final pseudoscalar state reads 
\begin{eqnarray}
\Gamma_2=i\gamma_5G_2/\sqrt{N_c}.   
\end{eqnarray}
The final vector meson is described by the vertex 
\begin{eqnarray}
\label{vectorvertex}
\Gamma_{2\mu}=[A\gamma_\mu+B(k_1-k_3)_\mu]\,G_2/\sqrt{N_c}, 
\end{eqnarray}
where for an $S$--wave vector meson 
\begin{eqnarray}
A=-1, \qquad B=1/(\sqrt{s_2}+m_1+m_3).  
\end{eqnarray}
As explained above, the double spectral densities $\tilde f$ of the form factors are obtained by calculating the relevant traces and
isolating the Lorentz structures depending on $\tilde p_1$ and $\tilde p_2$. 
The invariant factors of such Lorentz structures provide the double spectral densities $\tilde f$ 
corresponding to contributions of the two--particle singularities in the Feynman graph. 
Let us point out that this procedure allows one to obtain 
double spectral densities, whereas subtraction terms should be determined independently. We 
determine the subtraction terms from matching the $1/m_Q$ expanded form factors of the quark model to the
heavy-quark expansion in QCD. 

Recall that at $q^2<0$ the spectral representations of the form factors have the form 
\begin{equation}
\label{drnormal}
f_i(q^2)=
\frac1{16\pi^2}\int\limits^\infty_{(m_1+m_3)^2}ds_2\varphi_2(s_2)
\int\limits^{s_1^{+}(s_2,q^2)}_{s_1^{-}(s_2,q^2)}ds_1\varphi_1(s_1)
\frac{\tilde f_i(s_1,s_2,q^2)}{\lambda^{1/2}(s_1,s_2,q^2)},
\end{equation}
where the wave function $\varphi_i(s_i)=G_i(s_i)/(s_i-M_i^2)$ and 
\begin{eqnarray}
s_1^\pm(s_2,q^2)&=&
\frac{s_2(m_1^2+m_2^2-q^2)+q^2(m_1^2+m_3^2)-(m_1^2-m_2^2)(m_1^2-m_3^2)}{2m_1^2}
\nonumber\\
&&\pm\frac{\lambda^{1/2}(s_2,m_3^2,m_1^2)\lambda^{1/2}(q^2,m_1^2,m_2^2)}{2m_1^2}. 
\nonumber
\end{eqnarray}
Here 
$\lambda(s_1,s_2,q^2)=(s_1+s_2-q^2)^2-4s_1s_2$
is the triangle function. 

Calculating the necessary traces and isolating the coefficients of the relevant
Lorentz structures, we come to the following expressions for the
double spectral densities $\tilde f_i(s_1,s_2,q^2)$ of the form
factors \cite{m2}:
\begin{eqnarray}
\label{s}
\tilde s&=&2\,[m_1\alpha_{2}+m_2\alpha_{1}+m_3(1-\alpha_{1}-\alpha_{2})],
\\
\label{f1}
\tilde f_1&=&2m_1\tilde s+4\alpha_2[s_2-(m_1-m_3)^2]-2m_3\tilde s,
\\
\label{f2}
\tilde f_2&=&2m_2\tilde s+4\alpha_1[s_1-(m_2-m_3)^2]-2m_3\tilde s,
\\
\label{g}
\tilde g&=&-A\tilde s-4B\beta,
\\
\label{g1}
\tilde g_{1}&=&A\tilde f_1-8\beta+8B(m_1+m_3)\beta,
\\
\label{g2}
\tilde g_{2}&=&A\tilde f_2+8B(m_2-m_3)\beta,
\\
\label{a2}
\tilde a_{2D}&=&-2\tilde s+ 4BC_2\alpha_1+\alpha_{12}C_0,
\\
\label{a1}
\tilde a_{1D}&=&-4A\,(2m_3+BC_1)\alpha_1+\alpha_{11}C_0,
\\
\label{f}
\tilde f_D&=&-4A[m_1m_2m_3+\frac{m_2}2(s_2-m_1^2-m_3^2)
+\frac{m_1}2(s_1-m_2^2-m_3^2)-\frac{m_3}2(q^2-m_1^2-m_2^2)]
\nonumber\\
&&+C_0\beta,
\\
\label{g0}
\tilde g_{0D}&=&-8A\alpha_{12}-8B\,[-m_3\alpha_{1}+(m_3-m_2)\alpha_{11}+(m_3+m_1)\alpha_{12}],
\end{eqnarray}
where 
\begin{eqnarray}
\label{cc}
C_0&=&-8A(m_2-m_3)+4BC_3,\quad 
\\
C_1&=&s_2-(m_1+m_3)^2, 
\\
C_2&=&s_1-(m_2-m_3)^2, \quad 
\\
C_3&=&q^2-(m_1+m_2)^2-C_1-C_2. 
\end{eqnarray}
The coefficients $\alpha_1$, $\alpha_2$, $\alpha_{11}$, $\alpha_{12}$, and $\beta$ 
are defined according to the relations
\begin{eqnarray}
&&\frac{1}{8\pi}\int dk_1 dk_2 dk_3 \delta(\tilde p_1-k_2-k_3)\delta(\tilde p_2-k_3-k_1)
\delta(k^2_1-m_1^2)\delta(k^2_2-m_2^2)\delta(k^3_2-m_3^2)k_{3\mu}
\nonumber\\
&&
=\left(\alpha_1 \tilde p_{1\mu}+\alpha_2 \tilde p_{2\mu}\right)
\Delta(s_1,s_2,q^2), 
\nonumber\\
&&\frac{1}{8\pi}\int dk_1 dk_2 dk_3 \delta(\tilde p_1-k_2-k_3)\delta(\tilde p_2-k_3-k_1)
\delta(k^2_1-m_1^2)\delta(k^2_2-m_2^2)\delta(k^3_2-m_3^2)k_{3\mu}k_{3\nu}
\nonumber\\
&&
=\left(
\alpha_{11} \tilde p_{1\mu}\tilde p_{1\nu}
+\alpha_{12} (\tilde p_{1\mu}\tilde p_{2\nu}+\tilde p_{1\nu}\tilde p_{2\mu})
+\alpha_{22} \tilde p_{2\mu}\tilde p_{2\nu}
+\beta g_{\mu\nu}
\right)
\Delta(s_1,s_2,q^2), 
\nonumber
\end{eqnarray}
with $\Delta(s_1,s_2,q^2)$ given by Eq. (\ref{thetafunction}). Explicitly, they read
\begin{eqnarray}
\label{alpha1}
\alpha_1&=&\left[(s_1+s_2-q^2)(s_2-m_1^2+m_3^2)-2s_2(s_1-m_2^2+m_3^2)\right]
/{\lambda(s_1,s_2,q^2)},
\\
\label{alpha2}
\alpha_2&=&
\left[(s_1+s_2-q^2)(s_1-m_2^2+m_3^2)-2s_1(s_2-m_1^2+m_3^2)\right]/{\lambda(s_1,s_2,q^2)},
\\
\label{beta}
\beta&=&\frac14\left[2m_3^2-\alpha_1(s_1-m_2^2+m_3^2)-\alpha_2(s_2-m_1^2+m_3^2)\right],
\\
\label{alpha11}
\alpha_{11}&=&\alpha_1^2+4\beta {s_2}/{\lambda(s_1,s_2,q^2)}, \quad 
\\
\label{alpha12}
\alpha_{12}&=&\alpha_1\alpha_2-2\beta(s_1+s_2-q^2)/\lambda(s_1,s_2,q^2),
\end{eqnarray}
In Eqs (\ref{a2}-\ref{g0}) we label with a subscript 'D' the true double spectral densities 
obtained from the Feynman graph for those form factors which require subtractions. 
We fix this subtraction procedure by requiring the $1/m_Q$ expansion of the form factors to have a 
proper form in accordance with QCD in leading and next--to--leading orders for the case 
of a meson transition caused by heavy--to--heavy quark
transition. 
It is convenient to introduce for such form factors the modified double spectral densities
which include properly defined subtraction terms as follows 
\begin{eqnarray}
\label{f+sub}
\tilde f&=&\tilde f_D+[(M_1^2-s_1)+(M_2^2-s_2)]\tilde g, \\
\label{a1+sub}
\tilde a_1&=&\tilde a_{1D}
+\frac{1}{(\bar\omega+1)m_2}\left(\frac{M_1^2-s_1}{\sqrt{s_1}}+\frac{M_2^2-s_2}{\sqrt{s_2}}\right)
\frac{\tilde g}2,\\
\label{a2+sub}
\tilde a_2&=&\tilde a_{2D}
+\frac{1}{(\bar\omega+1)m_1}\left(\frac{M_1^2-s_1}{\sqrt{s_1}}+\frac{M_2^2-s_2}{\sqrt{s_2}}\right)
\frac{\tilde g}2,\\
\label{g0+sub}
\tilde g_0&=&\tilde g_{0D}
+\frac{1}{(\bar\omega+1)m_1m_2}\left(\frac{M_1^2-s_1}{\sqrt{s_1}}+\frac{M_2^2-s_2}{\sqrt{s_2}}\right)
\frac{\tilde g}2.
\end{eqnarray}

As the analytical continuation to the timelike region is performed, 
in addition to the normal contribution which is just the expression (\ref{drnormal}) taken
at $q^2>0$, anomalous contribution emerges.  
As we have shown in the previous section, the origin of the anomalous contribution is connected 
with the motion of the zero of the 
triangle function $\lambda(s_1,s_2,q^2)$ located at $s_1^R=(\sqrt{s_2}-\sqrt{q^2})^2$ from the 
unphysical sheet at $q^2<0$ onto the physical 
sheet at $q^2>0$ through the normal cut between the points $s_1^-(s_2,q^2)$ and $s_1^+(s_2,q^2)$.
Pinching of the points $s_1^R$ and $s_1^-$ occurs at $s_2^0$ given by Eq. (\ref{s20}). 
The spectral representation of the form factor at $q^2>0$ takes the form 
\begin{eqnarray}
\label{final}
f(q^2)&=&\frac{1}{16\pi^2}
\int\limits_{(m_1+m_3)^2}^\infty\varphi_2(s_2)
\int\limits_{s_1^-}^{s_1^+}
\varphi_1(s_1)
\frac{\tilde f(s_1,s_2,q^2)}{\lambda^{1/2}(s_1,s_2,q^2)}\nonumber\\
&+&\theta(q^2)\frac{1}{8\pi^2}\int\limits_{s_2^0}^\infty\varphi_2(s_2)
\int\limits_{s_1^R}^{s_1^-}
ds_1\varphi_1(s_1)\frac{\tilde f(s_1,s_2,q^2)}{\lambda^{1/2}(s_1,s_2,q^2)},  
\end{eqnarray}
One should also take into account that if the double spectral density $\tilde f$ is singular at the point 
$s_1^R$, which is the case for the form factors describing transitions to vector meson, 
then a properly defined anomalous part contains subtraction terms which   
appear as the contribution of a small circle around $s_1^R$. 
Obviously, these subtractions have quite different nature from subtractions  
in the spectral representations at $q^2<0$ performed to match the structure of the heavy quark expansion in
QCD. The corresponding expression was derived and can be found in \cite{mplb2}.

The normal contribution dominates the form factor at small timelike and vanishes as 
$q^2=(m_2-m_1)^2$ while the anomalous contribution is negligible at small $q^2$ and steeply rises
as $q^2\to(m_2-m_1)^2$. 
 
Both for pseudoscalar and vector mesons with the mass $M$ built up of the 
constituent quarks $m_Q$ and $m_{q}$, the function $\varphi$ is normalized 
according to the relation 
\begin{equation}
\label{norma}
\frac1{8\pi^2}\int ds \varphi^2(s)\frac{\lambda^{1/2}(s,m_Q^2,m_{q}^2)}{s}
[s-(m_Q-m_{q})^2]=1. 
\end{equation}
This equation corresponds to the normalization of the elastic charge form factor at $q^2=0$. 

Recall the spectral representation for the leptonic decay constant of the 
pseudoscalar meson $f_P$  
\begin{equation}
\label{fp}
f_P=\sqrt{N_c}(m_Q+m_{q})\int
ds\;\varphi(s)\frac{\lambda^{1/2}(s,m_Q^2,m_{q}^2)}
{8\pi^2 s}\frac{s-(m_Q-m_{q})^2}{s}.
\end{equation}

In the subsequent sections we analyse the form factors given by the dispersion representation 
(\ref{final}) with the spectral densities (\ref{s}--\ref{g2}) and (\ref{f+sub}--\ref{g0+sub}) 
and demonstrate them to have the following properties in the case of a heavy parent meson: 
for the transition induced 
by the heavy--to--heavy quark transition they satisfy the LO and NLO relations \cite{luke} 
of the $1/m_Q$ expansion in accordance with QCD 
provided the functions $\varphi_i$ are localized near the $q\bar q$ threshold with the width of order
$\Lambda_{QCD}$. For the meson decay induced by the heavy--to--light quark transition they satisfy 
the LO relations between the form factors of the vector, axial--vector and tensor currents \cite{iwhl}.  

\subsection{\label{ii.3}Heavy-quark expansion in the dispersion approach 
for heavy-to-heavy transitions}
In this section we consider the form factors of the dispersion quark model in 
the case when both $m_2$ and $m_1$ are large. We calculate the universal form factors and 
demonstrate that requiring the structure of the $1/m_Q$ expansion in 
the quark model to be consistent with the structure of such expansion in QCD allows us to determine the subtraction 
terms. 
 
\vspace{.5cm} 
\underline{\it Soft wave function and normalization condition} 
\vspace{.5cm} 

First, we need to specify the properties of the soft wave function of a heavy meson. 
A basic property of such soft wave function $\varphi(s,m_Q,m_{q},\bar\Lambda)$ is a strong peaking near the $q\bar q$ 
threshold. 
For elaborating the $1/m_Q$ expansion, it is convenient to formulate such peaking in terms of the 
variable $z$ such that $s=(m_Q+m_3+z)^2$ (hereafter we denote the mass of the light quark as $m_3$).
The region above the $q\bar q$ threshold which contributes to the 
spectral representation corresponds to $z>0$. A localization of the soft wave function in terms of $z$ means that the 
wave function is nonzero as $z\le \Lambda_{QCD}$. In the heavy meson case $m_Q \gg m_{3}\simeq z\simeq \bar\Lambda$.
Let us notice that for a heavy meson the localization in terms of $z$ is equivalent to the localization in terms of the 
relative momentum in the meson rest frame 
\begin{equation}
\label{z2k}
{\vec k}^2=z(z+2m_3)+O(1/m_Q).
\end{equation}
The normalization condition (\ref{norma}) which is 
a consequence of the vector current conservation in the full theory
provides an (infinite) chain of relations in the effective theory.  
Namely, expanding the soft wave function in $1/m_Q$ as follows 
\begin{equation}
\label{phi}
\varphi(s,m_Q,m_{3},\bar\Lambda)=\frac\pi{\sqrt{m_Q}}\phi_0(z,m_{3},\bar\Lambda)\left[{1+\frac {m_3}{4m_Q}
\chi_1(z,m_{3},\bar\Lambda)+O(1/m_Q^2)}\right],
\end{equation}
we come to the normalization condition in the form 
\begin{equation}
\int dz \phi^2_0(z)\sqrt{z}(z+2m_3)^{3/2}\left[1+\frac{m_3}{2m_Q}\chi_1(z)-\frac{m_3}{2m_Q}+\ldots\right]=1. 
\end{equation}
This exact relation is equivalent to an infinite chain of equations in different $1/m_Q$ orders. 
Lowest order relations take the form  
\begin{eqnarray}
\label{normaphi}
&\int &dz\;\phi^2_0(z)\sqrt{z}(z+2m_3)^{3/2}=1, \\
&\int & dz\;\phi^2_0(z)\sqrt{z}(z+2m_3)^{3/2}\chi_1(z)=1. 
\end{eqnarray}

\vspace{.5cm}
\underline{\it The variables $\omega$ and $\bar\omega$}
\vspace{.5cm}

In the description of the transition processes the dispersion formulation 
of the quark model has the following feature: 
since the underlying process is the quark transition, the relevant kinematical variable for the description of the
dynamics of the process is the quark recoil $\bar\omega$ which is defined as follows
\begin{equation}
q^2=(m_2-m_1)^2-2m_1m_2(\bar\omega-1).
\end{equation}
The relationship between $\omega$ and $\bar\omega$ is given by the condition that the spectral representation for the
form factor is written at fixed value of $q^2$ and hence 
\begin{eqnarray}
q^2=(M_1-M_2)^2-2M_1M_2(\omega-1)
=(m_2-m_1)^2-2m_1m_2(\bar\omega-1).
\end{eqnarray}
In the case of heavy particle transitions these quantities are related to each other as
\begin{eqnarray}
\bar\omega=\omega+\bar\Lambda\left(\frac{1}{m_1}+\frac{1}{m_2}\right)(\omega-1)+O(1/m_Q^2).
\end{eqnarray}
We shall obtain the representations of the form factors as functions of the variable $\bar\omega$. 
The variables $\omega$ and $\bar\omega$ are different by the terms of order $1/m_Q$ at nonzero recoil. 
On the other hand, the quark and meson zero recoil points coincide with $1/m_Q^2$ accuracy. 
This means that in the analyses of the $1/m_Q$ expansion at nonzero recoil 
the difference between the $\omega$ and $\bar\omega$ might be ignored in the Isgur--Wise function, but gives nontrivial
contribution to the NLO form factors. At the same time, at zero recoil the difference between $\omega$ and $\bar\omega$ 
might be neglected both in the leading and next--to--leading orders. 
Namely, the quark model provides the expansion of the form factor in the following form 
\begin{eqnarray}
h&=&h_0(\bar\omega)+\frac{1}{m_1}h^{(1)}_1(\bar\omega)+\frac{1}{m_2} h^{(2)}_1(\bar\omega)+\ldots\\
 &=&h_0(\omega)+h'_0(\omega)\bar\Lambda\left(\frac{1}{m_1}+\frac{1}{m_2}\right)(\omega-1)
   +\frac{1}{m_1}h^{(1)}_1(\bar\omega)+\frac{1}{m_2}\bar h^{(2)}_1(\bar\omega)+\ldots.
\end{eqnarray}
As we shall see later, among the NLO form factors only $\rho_{1,2}$ are affected by the 
the difference between $\omega$ and $\bar\omega$ whereas $\xi, \xi_3, \chi_2$ are not. 

\vspace{.5cm}
\underline{\it Relative magnitudes of the normal and the anomalous contributions}
\vspace{.5cm}

We are going now to demonstrate that the anomalous contribution comes into the game only in close vicinity 
of the zero recoil point whereas beyond this region is negligible. 

Let us study the behavior of the anomalous contribution in the region  
\begin{eqnarray}
\bar\omega-1\simeq m_Q^{-(2+\varepsilon)}.
\end{eqnarray}
Introducing the variables $z_1$ and $z_2$ such that $s_1=(m_2+m_3+z_1)^2$ and $s_2=(m_1+m_3+z_2)^2$ we find that 
the magnitude of the anomalous contribution is controlled by the value of $z_2^0(\bar\omega)$ such that 
$s_2^0=(m_1+m_3+z_2^0(\bar\omega))^2$ 
which is the lower boundary of the $z_2$ integration. If $z_2^0(\bar\omega)$ becomes large, i.e. of the order $m_Q$, 
the anomalouis contribution is suppressed by the fall--down of the soft wave function.
This suppression is at least stronger than $1/m_Q^2$. This means that the anomalous contribution is nonvanishing only if  
\begin{eqnarray}
z_2^0(\bar\omega)=\frac{m_1m_2\sqrt{\bar\omega^2-1}+m_1m_2\bar\omega-m_1^2}{\sqrt{m_1^2+m_2^2-2m_1m_2\bar\omega}}
-m_1+O(m_3)
\simeq\bar\Lambda.
\end{eqnarray}
In the region $\bar\omega-1\simeq m_Q^{-(2+\varepsilon)}$, 
one finds $z_2^0(\bar\omega)\simeq m_Q^{-\varepsilon/2}$. 
Hence the anomalous contribution comes actually into the game only in the $O(1/m_Q^2)$ vicinity of the 
zero recoil point but otherwise might be neglected. On the other hand, at the quark zero recoil point $\bar\omega=1$, 
the normal contribution vanishes and the form factor is given by the anomalous contribution. 

We shall calculate the form factors in the region $\bar\omega-1=O(1)$ where only the normal contribution should be 
taken into account in leading and subleading orders. 

\vspace{.5cm}
\underline{\it The LO analysis}
\vspace{.5cm}

To perform the LO analysis of the form factors let us start with the integration measure. 
With $1/m_Q$ accuracy it can be represented in the form 
\begin{eqnarray}
\frac1{16\pi^2}\frac{ds_1ds_2\theta
\left(s_2\ge(m_1+m_3)^2\right)\theta(s_1^-\le s_1\le s_1^+)}
{\lambda^{1/2}(s_1,s_2,q^2)}
\simeq
\frac1{4\pi^2}dz_2\sqrt{z_2(z_2+2m_3)}\frac{d\eta}{2}\theta(z_2)\theta(\eta^2<1), 
\end{eqnarray}
and the expression for $z_1$ reads 
\begin{eqnarray}
\label{z1}
z_1=z_2\bar\omega +m_3(\bar\omega -1)+\eta\sqrt{z_2(z_2+2m_3)}\sqrt{\bar\omega ^2-1}+O(1/m_Q).
\end{eqnarray}
Let us point out that the LO integration measure is symmetric in $z_1$ and $z_2$. 

Next, we need expanding the spectral densities (\ref{s})-(\ref{g0}). To this end we must take
into account that under the integral sign
$z_1$ and $z_2$ are localized in the region $z\le\bar\Lambda$ due to the
soft wave functions $\phi(z)$.

In LO the kinematical coefficients (\ref{cc})-(\ref{alpha1}) in the region $\bar\omega-1=O(1)$ simplify to 
\begin{eqnarray}
\lambda(s_1,s_2,q^2)&=&4m_1^2m_2^2(\bar\omega^2-1),\\
\alpha_1&=&\frac1{m_2(\bar\omega+1)}\left[ m_3+z_2+\frac{z_2-z_1}{\bar\omega-1}\right],\\
\alpha_2&=&\frac1{m_1(\bar\omega+1)}\left[ m_3+z_1+\frac{z_1-z_2}{\bar\omega-1}\right],\\
\beta&=&\frac12\left[m_3^2-\frac2{\omega+1}(m_3+z_1)(m_3+z_2)+ \frac{(z_1-z_2)^2}{\bar\omega^2-1}\right],\\
\alpha_{11}&=&\alpha_1^2+\frac\beta{m_2^2(\bar\omega^2-1)},\quad 
\\
\alpha_{12}&=&\alpha_1\alpha_2-\frac{\beta\;\bar\omega}{m_1m_2(\bar\omega^2-1)},\\
B&=&\frac1{2m_1}, 
\\
C_0&=&-4m_2(\bar\omega-1),
\\
C_1&=&2m_1z_2,
\\
C_2&=&2m_2(z_1+2m_3).  
\end{eqnarray}
One finds the LO behavior of the form factor densities (\ref{s}--\ref{a2}) is determined by the term proportional to
$\tilde s$. The latter reads in the LO 
\begin{eqnarray}
\tilde s\simeq 2\left( m_3+\frac{z_1+z_2+2m_3}{\bar\omega+1} \right). 
\end{eqnarray}
The LO expression for $\tilde f$ takes the form  
\begin{eqnarray}
\tilde f_D=(\bar\omega+1)\left(m_3+\frac{z_1+z_2+2m_3}{\bar\omega+1}\right).
\end{eqnarray}
The spectral densities $\tilde a_1$ and $\tilde g_{0D}$ vanish in the leading order. 

Hence the LO relations (\ref{f1exp}--\ref{g0exp}) are fulfilled with the Isgur--Wise (IW) function 
\begin{eqnarray}
\label{xi}
\xi(\omega )=\int
dz_2 \phi_0(z_2)\sqrt{z_2(z_2+2m_3)}\int\limits_{-1}^{1}\frac{d\eta}2
\phi_0(z_1)\left({m_3+\frac{2m_3+z_1+z_2}{1+\omega }}\right). 
\end{eqnarray}
In (\ref{xi}) we used the equality of $\bar\omega$ and $\omega$ with $1/m_Q$ accuracy. 
The normalization condition of the LO wave functions (\ref{normaphi}) yields $\xi(1)=1$. 
For the slope of the IW function at zero recoil, $\rho^2=-\xi'(1)$, one finds 
\begin{eqnarray}
\label{rho2}
\rho^2=\frac13\int
dz \sqrt{z}(z+2m_3)^{3/2}{\left(\phi_0'(z)\right)}^2z(z+2m_3)
\end{eqnarray}
 
Let us point out that the subtraction terms in the spectral densities do not contribute in the LO 
relations. As we shall see later, they are important in the NLO analysis. 

\vspace{.5cm}
\underline{\it The NLO analysis of the form factors $s$, $f_1$, $f_2$, $g_1$, $g_2$, and $g$}
\vspace{.5cm}

First, let us concentrate on the NLO relations (\ref{f1exp}--\ref{gexp}). 
It is convenient to analyze the linear combinations of the form factors 
which do not contain the LO contribution. These combinations are 
\begin{eqnarray}
\label{1exp}
g-s&=&\frac{1}{2\sqrt{m_1m_2}}\frac{1}{m_1}\left[\rho_2-\rho_1+\xi_3\right],\\ 
\label{2exp}
g_1+f_1&=&\frac{1}{\sqrt{m_1m_2}}\left[-\rho_2+\rho_1+\xi_3\right],\\
\label{3exp}
g_2+f_2&=&-\frac{m_2}{m_1}\frac{1}{\sqrt{m_1m_2}}\left[\rho_2-\rho_1+\xi_3\right],\\
\label{4exp}
f_1-2m_1s&=&\frac{2\xi_3}{\sqrt{m_1m_2}}. 
\end{eqnarray}

The spectral densities of the form factor combinations in the l.h.s. of eqs. (\ref{1exp}--\ref{3exp}) read  
\begin{eqnarray}
\label{1dexp}
\tilde g-\tilde s&=&-\frac{2}{m_1}\beta,\\ 
\label{2dexp}
\tilde g_1+\tilde f_1&=&-4\beta,\\
\label{3dexp}
\tilde g_2+\tilde f_2&=&4\frac{m_2}{m_1}\beta,
\end{eqnarray}
Comparison with the eqs. (\ref{1exp}--\ref{3exp}) leads to the relation 
\begin{eqnarray}
\rho_1(\omega)=\rho_2(\omega), 
\end{eqnarray}
For the form factor $\xi_3$ we come to the representation 
\begin{eqnarray}
\label{xi3a}
\xi_3(\omega )&=&-\int 
dz_2 \phi_0(z_2)\sqrt{z_2(z_2+2m_3)}
\nonumber\\
&&\times\int\limits_{-1}^{1}\frac{d\eta}2\phi_0(z_1) 
\frac12\left[m_3^2-\frac2{\omega+1}(m_3+z_1)(m_3+z_2)+ \frac{(z_1-z_2)^2}{\omega^2-1}\right], 
\end{eqnarray}
with $z_1$ given by the expression (\ref{z1}). In (\ref{xi3a}) we have neglected the $O(1/m_Q)$ difference between 
$\omega$ and $\bar\omega$. 

On the other hand, the equation (\ref{4exp}) yields the representation for the form factor $\xi_3$ in a different
form 
\begin{eqnarray}
\label{xi3b}
\xi_3(\omega)&=&\int 
dz_2 \phi_0(z_2)\sqrt{z_2(z_2+2m_3)}\int\limits_{-1}^{1}\frac{d\eta}2\phi_0(z_1)\nonumber \\
&\times&\left[\frac{z_2+2m_3}{\omega+1}\left(z_2+m_3+\frac{(z_1-z_2)\;\omega}{\omega-1}\right)
-\frac{m_3}{2}\left({m_3+\frac{2m_3+z_1+z_2}{1+\omega }}\right)\right]. 
\end{eqnarray}
One can check that for the soft wave functions providing convergency of the integrals and nonsingular at $z=0$ 
the representations (\ref{xi3a}) and (\ref{xi3b}) are equivalent. 
At zero recoil one finds 
\begin{eqnarray}
\xi_3(1)&=&\int dz_2\;\sqrt{z_2}(z_2+2m_3)^{3/2}\;\phi^2_0(z_2)\frac{z_2}3
\nonumber\\
&\equiv&\frac{\langle{z}\rangle}{3}, 
\end{eqnarray}
where $\langle z \rangle$ is an average kinetic energy of the light constituent quark inside an (infinitely) 
heavy meson in its rest frame.  
It is worth noting that the function $\xi_3$ is positive for all $\omega$. 

The universal form factor $\rho_1=\rho_2$ can be found from the expansion of $s(\omega)$ with a NLO accuracy. 
In this case the
NLO terms in the $1/m_Q$ expansions of the integration measure, the wave function, and the spectral density $\tilde s$
contribute. 
Namely, we can write for the form factor $h_s$ the expression 
\begin{eqnarray}
h_s(\omega)&=&\int\left[d\mu_0+\frac{d\mu_1^{(1)}}{m_1}+\frac{d\mu_1^{(2)}}{m_2}\right]
\left[\tilde s(\bar\omega)+\frac{S_1}{m_1}
+\frac{S_2}{m_2}\right]
\nonumber\\
&&\times\phi_0(z_1)
\left(1+\frac{m_3}{4m_1}\chi_1(z_1)\right) \phi_0(z_2)
\left(1+\frac{m_3}{4m_2}\chi_1(z_2)\right)\nonumber\\
&=&\xi(\bar\omega)+\left( \frac1{m_1} + \frac1{m_2}\right)\xi^{(1)}(\bar\omega)+\ldots
\nonumber\\
&=&\xi(\omega)+\left( \frac1{m_1} + \frac1{m_2}\right)[\xi^{(1)}(\omega)+\xi'(\omega)\bar\Lambda(\omega-1)]+\ldots,
\end{eqnarray}
and hence $\rho_1(\omega)=\xi^{(1)}(\omega)+\xi'(\omega)\bar\Lambda(\omega-1)-\xi_3(\omega)-\frac{\bar\Lambda}2\xi(\omega)$. 
We do not present explicit expression for $\rho_1$. However, let us consider $\rho_1$ at zero recoil. 
The analysis of the anomalous contribution at $\bar\omega=1$ gives
\begin{eqnarray}
h_s(\bar\omega=1)=1+\left( \frac1{m_1} + \frac1{m_2}\right) \left( \frac{\bar\Lambda}2-\frac{\langle z \rangle}{3}\right)+\ldots
\end{eqnarray}
Using the relations $h_s(\omega=1)=h_s(\bar\omega=1)+O(1/m_Q^2)$ and $\xi_3(1)=\langle z \rangle/3$ 
we obtain $\rho_1(1)=0$ just as required by HQET.  

\newpage
\vspace{.5cm}
\underline{\it The NLO analysis of the form factor $f$ }
\vspace{.5cm}

First, let us demonstrate that the dispersion representation of the form factor $f$ requires subtraction.
To this end consider the anomalous contribution at $\bar\omega=1$. For the form factor $f_D$ 
constructed from the spectral density $\tilde f_D$ through the double dispersion representation 
without subtractions we find
\footnote{Hereafter we denote
as $a_D$ a form factor reconstructed from its double discontinuity 
through dispersion representation without subtractions.}
\begin{eqnarray}
f_D(1)=2\sqrt{M_1M_2}\left[1+\frac12\left(\frac1{m_1}+\frac1{m_2} \right)
\left(\langle z \rangle+m_3-\bar\Lambda\right)+\ldots \right]
\end{eqnarray}
This value contradicts the Luke theorem which requires the $1/m_Q$ corrections to vanish at zero recoil.
Let us demonstrate that the form factor $f$ constructed from the 
spectral density with the included subtraction term
\begin{eqnarray}
\label{tildeffull}
\tilde f&=&\tilde f_D+(M_1^2-s_1+M_2^2-s_2)\tilde g
\nonumber\\
&=&\tilde f_D+(2p_1p_2-2\tilde p_1 \tilde p_2)\tilde g, 
\end{eqnarray}
satisfies the NLO relation (\ref{fexp}).
Here $2\tilde p_1 \tilde p_2= s_1+s_2-q^2$. 

We have noticed above that to LO and NLO the expansion of the form factor $f$ and the combination 
$2p_1p_2\cdot g-m_1\cdot g_2$ coincide (\ref{fmodified}). 
Hence checking the NLO relations for the form factor $f$ is equivalent to checking 
with the NLO accuracy the relation 
\begin{eqnarray}
\label{fnew}
f_D\simeq f_{2\tilde p_1\tilde p_2\cdot \tilde g} -m_1 \cdot g_2. 
\end{eqnarray}
The spectral density of the r.h.s. of eq. (\ref{fnew}) can be written as
\begin{eqnarray}
\label{fsym}
2\tilde p_1\tilde p_2\cdot \tilde g -m_1 \cdot \tilde g_2&\simeq& 
2m_1m_2(\bar\omega+1)\tilde s+ [2m_1(z_2+m_3)+2m_2(z_1+m_3)]\tilde s
\nonumber\\
&&-8(m_1+m_2)\beta-4m_2(\bar\omega-1)\beta. 
\end{eqnarray}
For checking the expression (\ref{fnew}) in NLO we need the expansion of the spectral density $\tilde s$ in LO and NLO
which has the structure
\begin{eqnarray}
\frac12\tilde s&=&m_1\alpha_2+ m_2\alpha_1+m_3(1-\alpha_1-\alpha_2)
\nonumber\\
&\equiv& m_3+\frac{z_1+z_2+2m_3}{\bar\omega+1}
+\frac{S_1}{m_1}+\frac{S_2}{m_2}
+\ldots,
\end{eqnarray}
with 
\begin{eqnarray}
S_1=\frac12 z_2(z_2+2m_3)-z_1m_3-(z_2+2m_3)\frac{z_1-z_2}{\bar\omega-1}-\frac{(z_1+z_2+2m_3)^2}{\bar\omega+1}, 
\end{eqnarray}
and $S_2$ is obtained from $S_1$ by replacing  $z_1$ and $z_2$. 
The spectral density $\tilde f_D$ reads
\begin{eqnarray}
\tilde f_D&=&4m_1m_2(\bar\omega+1)\left(m_3+\frac{z_1+z_2+2m_3}{\bar\omega+1}\right)
+2m_2z_2(z_2+2m_3)+2m_1z_1(z_1+2m_3)-4m_2(\bar\omega-1)\beta.
\end{eqnarray}
Notice that for checking the NLO relation (\ref{fnew}) between the form factors 
we do not need explicit expression for the integration measure in the NLO: 
in the LO the spectral densities are equal 
and hence the NLO contributions from the integration measure into both sides of the eq (\ref{fnew}) are equal too. 
Finally, Eq. (\ref{fnew}) is satisfied if the following relation is valid
\begin{eqnarray}
\label{fconsisitency}
&
\int& dz_2\;\sqrt{z_2(z_2+2m_3)}\phi_0(z_2)\int\limits_{-1}^{1}\frac{d\eta}2\phi_0(z_1)
\nonumber \\ &&\times
\left[
-z_2(z_2+2m_3)+2(z_1+m_3)\left( m_3+\frac{z_1+z_2+2m_3}{\bar\omega+1}  \right)-4\beta +2(\bar\omega+1)S_1\right]=0,
\end{eqnarray}
with $z_1$ given by (\ref{z1}). One can check this relation to be true for any 
function $\phi_0(z)$ regular at $z=0$. Hence, the form factor $f$ calculated with the 
subtracted double spectral density 
(\ref{tildeffull}) satisfies the HQET relations in LO and NLO at all $\omega$.  

Strictly speaking, the NLO analysis does not allow us to uniquely specify the subtraction term: namely,
any spectral density of the form 
\begin{eqnarray}
\tilde f=\tilde f_D +(M_1^2-s_1+M_2^2-s_2) \tilde \rho_f 
\end{eqnarray}
has a proper NLO behavior in accordance with (\ref{fexp}) provided the spectral density $\tilde \rho_f$ 
behaves in the LO as 
$$
\tilde \rho_f\simeq 2\left(m_3+\frac{z_1+z_2+2m_3}{\bar\omega+1}\right).
$$
As we demonstrate in the next section, the analysis of the heavy--to--light LO relations requires the 
identification 
$\tilde\rho_f=\tilde g.$  

Let us notice that the form factor obtained within the light-cone approach \cite{jaus} can be represented 
as dispersion representation with the spectral density \cite{m2}
\begin{eqnarray}
\tilde f_{LC}=\frac{M_2}{\sqrt{s_2}}\tilde f_{D}+
M_2\left({\frac{s_1-s_2-q^2}{2\sqrt{s_2}}-\frac{M_1^2-M_2^2-q^2}{2M_2}}\right)
\frac{\tilde a_1+\tilde a_2}2. 
\end{eqnarray}
At zero recoil one finds 
\begin{eqnarray}
f_{LC}(1)=2\sqrt{M_1M_2}\left[1-\frac12\left(\frac1{m_1}-\frac1{m_2} \right)
\left(\langle z \rangle+m_3-\bar\Lambda
\right)+\ldots \right]
\end{eqnarray}
that contradicts the Luke theorem. 

\vspace{.5cm}
\underline{\it The NLO analysis of the form factors $a_{1,2}$ and $\chi_2$}
\vspace{.5cm}

The double spectral densities of the form factors $a_1$ and $a_2'\equiv a_2-2s$ in the LO read 
\begin{eqnarray}
\tilde a_{1D}&=&4\left[(z_2+2m_3)\alpha_1-m_2\alpha_{11}(\bar\omega-1)\right],\\
\tilde a'_{2D}&=&4\left[\frac{m_2}{m_1}(z_2+2m_3)\alpha_1-m_2\alpha_{12}(\bar\omega-1)\right]. 
\end{eqnarray}
The quantity $a_2'$ is more convenient than $a_2$ for calculations because  
the LO term of the heavy quark expansion of $a_2'$ is zero. 

The unsubtracted spectral representation for 
$a_1(\omega)m_2-a_2(\omega)m_1$ in combination with eq (\ref{aminus}) gives 
\begin{eqnarray}
\chi_2(\omega)=\xi_3(\omega)-\frac14
\int
dz_2 \phi_0(z_2)\sqrt{z_2(z_2+2m_3)}\int\limits_{-1}^{1}\frac{d\eta}2
\phi_0(z_1)\left[\tilde a_1m_2-\tilde a_2m_1\right]. 
\end{eqnarray}
Substituting the representation (\ref{xi3a}) for $\xi_3$ we find 
\begin{eqnarray}
\chi_2(\omega)=0.
\end{eqnarray}

Let us now consider the linear combination $a_1(\omega)m_2+a_2'(\omega)m_1$. As a first step, show that the unsubtracted 
dispersion representation is not compatible with HQET. To this end calculate the unsubtracted form factors 
$a_{1D}$ and $a_{2D}'$ at zero recoil: 
\begin{eqnarray}
a_{1D}(1)&=&\frac1{\sqrt{m_1m_2}}\frac1m_2\left[ \frac{2}{3}\langle z \rangle+\frac {m_3}{2}  \right],\\
a'_{2D}(1)&=&\frac1{\sqrt{m_1m_2}}\frac1m_1\left[ \frac{1}{3}\langle z \rangle+\frac {m_3}{2}  \right],
\end{eqnarray}
and hence 
\begin{eqnarray}
a_{1D}(1)m_2+a'_{2D}(1)m_1=\frac1{\sqrt{m_1m_2}}[\langle z \rangle+{m_3}]
\end{eqnarray}
in contradiction with the HQET result eq. (\ref{aplus}) 
\begin{eqnarray}
\frac{\bar\Lambda}{\sqrt{m_1m_2}}.
\end{eqnarray}
This fact suggests a necessity of subtraction in the quantity $a_1m_2+a_2'm_1$. 
Let us write the spectral density with subtraction in the form 
\begin{eqnarray}
\tilde a_{1}m_2+\tilde a'_{2}m_1=\tilde a_{1D}m_2+\tilde a'_{2D}m_1+\frac\kappa{\bar\omega+1}
\left(\frac{M_1^2-s_1}{2\sqrt{s_1}}+\frac{M_2^2-s_2}{2\sqrt{s_2}}\right)\frac{\tilde \rho_a}2
\end{eqnarray}
with 
\begin{eqnarray}
\label{rhoa}
\tilde \rho_a\simeq 2\left(m_3+\frac{z_1+z_2+2m_3}{\bar\omega+1}\right).
\end{eqnarray}
Then corresponding representation for the form factor reads
\begin{eqnarray}
&a_1(\omega)m_2+a_2'(\omega)m_1=
\frac1{4\sqrt{m_1m_2}}
\int
dz_2 \phi_0(z_2)\sqrt{z_2(z_2+2m_3)}\int\limits_{-1}^{1}\frac{d\eta}2\phi_0(z_1)
\\
\nonumber
&\times\left[ \tilde a_{1D}m_2+\tilde a'_{2D}m_1-\frac{2\kappa}{\omega+1}(z_1+z_2+2m_3)
\left(m_3+\frac{z_1+z_2+2m_3}{\omega+1}    \right)     
\right]+\frac{\kappa\bar\Lambda}{2\sqrt{m_1m_2}}\frac{\xi(\omega)}{\omega+1}.
\end{eqnarray}
According to the HQET relation (\ref{aplus}) this quantity should be equal to 
\begin{eqnarray}
\frac{1}{\sqrt{m_1m_2}}\left[\frac{2\bar\Lambda}{\omega+1}\xi(\omega) -\xi_3(\omega)\frac{\omega-1}{\omega+1}  \right]
\end{eqnarray}
The term proportional $\xi(\omega)$ yields $\kappa=4$. One can check that this value also makes 
other parts of both expressions equal. 
Hence we arrive at the subtracted spectral density 
\begin{eqnarray}
\tilde a_{1}m_2+\tilde a'_{2}m_1=\tilde a_{1D}m_2+\tilde a'_{2D}m_1+\frac{1}{\bar\omega+1}
\left(\frac{M_1^2-s_1}{\sqrt{s_1}}+\frac{M_2^2-s_2}{\sqrt{s_2}}\right)\tilde \rho_a.
\end{eqnarray}
The resulting spectral densities of the form factors $a_1$ and $a_2$ with the built--in subtraction terms take 
the form 
\begin{eqnarray}
\tilde a_1&=&\tilde a_{1D}
+\frac{1}{(\bar\omega+1)m_2}\left(\frac{M_1^2-s_1}{\sqrt{s_1}}+\frac{M_2^2-s_2}{\sqrt{s_2}}\right)
\frac{\tilde \rho_a}2,\\
\tilde a_2&=&\tilde a_{2D}
+\frac{1}{(\bar\omega+1)m_1}\left(\frac{M_1^2-s_1}{\sqrt{s_1}}+\frac{M_2^2-s_2}{\sqrt{s_2}}\right)
\frac{\tilde \rho_a}2. 
\end{eqnarray}

\vspace{.5cm}
\underline{\it The NLO analysis of $g_0$. }
\vspace{.5cm}

The unsubtracted spectral density $\tilde g_{0D}$ has the form 
\begin{eqnarray}
\tilde g_{0D}\simeq\frac4{\sqrt{m_1m_2}}\left[m_2\alpha_1(m_3+m_2\alpha_1+m_1\alpha_2)-\frac\beta{\bar\omega+1}\right].
\end{eqnarray}
At zero recoil one finds
\begin{eqnarray}
g_{0D}(1)=\frac{1}{(m_1m_2)^{3/2}}\left[ \frac{\langle z \rangle+m_3}{2}+\frac{\langle z \rangle}6\right].
\end{eqnarray}
On the other hand, taking into account our earlier finding $\chi_2=0$, the HQET result (\ref{g0exp}) reads 
\begin{eqnarray}
g_{0D}(1)=\frac{1}{(m_1m_2)^{3/2}}\left[ \frac{\bar\Lambda}{2}+\frac{\langle z \rangle}6\right].
\end{eqnarray}
Our experience obtained in considering $a_{1}m_2+a'_2m_1$ hints that the subtraction procedure adds a term
proportional at zero recoil to $\bar\Lambda-\langle z \rangle-m_3$, and hence we expect subtraction to work properly also
in the case of $g_0$. 

As a matter of fact, the spectral density 
\begin{eqnarray}
\label{tildegsub}
\tilde g_0=\tilde g_{0D}+\frac{1}{(\bar\omega+1)m_1m_2}
\left(\frac{M_1^2-s_1}{\sqrt{s_1}}+\frac{M_2^2-s_2}{\sqrt{s_2}}\right)\frac{\tilde \rho_{g_0}}2
\end{eqnarray}
satisfies the HQ expansion (\ref{g0exp})
provided the function $\tilde \rho_{g_0}$ behaves in the LO as 
\begin{eqnarray}
\label{rhog0}
\tilde \rho_{g_0}\simeq 2\left(m_3+\frac{z_1+z_2+2m_3}{\bar\omega+1}\right).
\end{eqnarray}

\vspace{.5cm}
\underline{\em The scalar form factor}
\vspace{.5cm}

Let us consider the form factor of the transition between two pseudoscalar
mesons induced by the scalar current
\begin{eqnarray}
\label{scalar1}
\langle P(M_2,p_2)|\bar q_1 q_2|P(M_1,p_1)\rangle &=&f_s(q^2)
\end{eqnarray}
For the analysis of the heavy-quark transition we introduce the velocity-dependent
form factor as follows
\begin{eqnarray}
\langle P(M_2,p_2)|\bar q_1 q_2|P(M_1,p_1)\rangle &=&\sqrt{M_1
M_2}(1+\omega)h_{f_s}(\omega)
\end{eqnarray}
such that
\begin{eqnarray}
f_s(q^2)=\sqrt{M_1 M_2}(1+\omega)h_{f_s}(\omega).
\end{eqnarray}
From HQET we find the following expansion
\begin{eqnarray}
\label{scalar2}
h_{f_s}=\xi+\left(\frac{1}{m_1}+\frac{1}{m_2}
\right)
\left[
\frac{\omega-1}{\omega+1}
\left(\frac{\bar\Lambda}2\xi-\xi_3\right)+\rho_1\right].
\end{eqnarray}
An important consequence of this expansion is the relation
\begin{eqnarray}
h_{f_s}(1)=1+O(1/m_Q^2),
\end{eqnarray}
which shows that the deviation of $h_{f_s}(1)$ from unity emerges only
at the $1/m_Q^2$ order, as well as for the form factors
$h_{f}$, $h_{f_+}$, and $h_{g_+}$ (\ref{luke1}).

Notice that in the leading and subleading orders in $1/m_Q$
the following quantities have the same expansion
\begin{eqnarray}
f_s&\sim& 2(p_1p_2-m_1m_2)s+m_2 f_1+m_1 f_2
\nonumber\\
&\sim& 2(p_1p_2-M_1M_2)s+M_1 f_1+M_2 f_2.
\end{eqnarray}

The form factor can be obtained according to the relation
\begin{eqnarray}
\label{class1}
f_s(q^2)=\frac{1}{m_2-m_1}
q^\mu\langle P(M_2,p_2)|\bar q_1 \gamma_\mu q_2|P(M_1,p_1)\rangle,
\end{eqnarray}
which follows from the equations of motion for quark fields. We then find
\begin{eqnarray}
\label{scalar7}
f_s(q^2)=\frac{1}{m_2-m_1}\left[(M_1^2-M_2^2)f_+(q^2)+q^2\,f_-(q^2)\right]
\end{eqnarray}
and
\begin{eqnarray}
h_{f_s}=\frac{1}{m_2-m_1}\left[
 h_{f_+}(M_1-M_2)(\omega+1)
-h_{f_-}(M_1+M_2)(\omega-1)\right].
\end{eqnarray}
In this case the structure of the expansion (\ref{scalar2}) is satisfied
automatically.

In the dispersion approach the double discontinuity of the form factor $f_s$
is found from the Feynman graph
\begin{eqnarray}
\tilde f_s^D=
 2m_3\left((m_1-m_2)^2-q^2\right)
+2m_2\left(s_2-(m_1-m_3)^2\right)
+2m_1\left(s_1-(m_2-m_3)^2\right)
-4m_1m_2m_3.
\end{eqnarray}
The double discontinuities of the form factors $f_+$, $f_-$ and $f_s$ obey the classical equation of motion
\begin{eqnarray}
\label{class2}
\tilde f^D_s(s_1,s_2,q^2)=\frac{1}{m_2-m_1}\left[(s_1-s_2)\tilde f_+(s_1,s_2,q^2)+q^2 \tilde f_-(s_1,s_2,q^2)\right].
\end{eqnarray}
As becomes clear from comparing Eqs.(\ref{class1}) and (\ref{class2}),
in order to satisfy the stucture of the expansion (\ref{scalar2}) the form factor
requires subtraction. The spectral density which includes the proper subtraction
term reads
\begin{eqnarray}
\tilde f_s=\tilde f_s^D+\left[(M_1^2-s_1)-(M_2^2-s_2)\right]\frac{\tilde f_+}{m_2-m_1}.
\end{eqnarray}
Equivalently, the behavior of the form factor $f_s$ can be brought in accordance
with HQET in the NLO by defining the subtraction as follows
\begin{eqnarray}
\tilde f_s=\tilde f_s^D+\left[(M_1^2-s_1)+(M_2^2-s_2)\right]\tilde \rho_{f_s},
\end{eqnarray}
provided $\tilde \rho_{f_s}\sim \tilde \rho_{g_0} \sim \tilde \rho_{a}$, Eqs.
(\ref{rhoa}) and (\ref{rhog0}).

\vspace{1.cm}

Concluding this section let us summarize our main results: we have calculated the universal form factors
and demonstrated the dispersion representations with relevant subtractions in the case of 
heavy--to--heavy transitions to reproduce the structure of the heavy--quark expansion in QCD in the leading 
and next--to--leading orders. 
However, we have not been able to fix uniquely these subtractions. As we shall see in the next section, 
the heavy--to--light transitions provide further
restrictions on the form of the subtraction terms. 

\subsection{\label{ii.4}Heavy--to--light meson transitions}
In this section we discuss meson decays induced by the heavy--to--light quark transitions in which case 
$M_1=m_2+O(1)$ is large, while $M_2\simeq m_1$ is kept finite. 
As found by Isgur and Wise \cite{iwhl}, in the region $\omega-1=O(1)$ 
the form factors of the tensor current can be expressed through the form factors of the vector and 
axial--vector currents in the leading $1/m_2$ order as follows 
\begin{eqnarray}
\label{hls}
s(q^2)&=&\frac{1}{2M_1}f_1(q^2),  \\
\label{hlg2}
g_{2}(q^2)&=&-2M_1\,g(q^2), \\
\label{hlg0}
g_0(q^2)&=&\frac{1}{M_1}[2g(q^2)+a_{2}(q^2)], \\
\label{hlg1}
g_{1}(q^2)&=&\frac{1}{M_1}[-f(q^2)+2p_1p_2\cdot g(q^2)]. 
\end{eqnarray}
 
We address the two issues: 
(i) perform the leading order $1/m_2$ expansion of the form factors and show the fulfillment of the relations
(\ref{hls}--\ref{hlg1}) and 
(ii) discuss the scaling behavior of the 
form factors of the transition of different heavy mesons into a fixed final light state. 

\vspace{.5cm}
\underline{\it The LO $1/m_Q$ expansion of the form factors in the quark model}
\vspace{.5cm}
 
In the case of heavy--light transitions one observes the appearance of the two scales: 
the light scale $M_2\simeq m_1\simeq m_3 \simeq \bar\Lambda$, and the heavy scale $M_2\simeq m_1$, and we may 
expand the form factors in inverse powers of the small parameter $\bar\Lambda/m_2$. 
The kinematical coefficients in the leading $\bar\Lambda/m_2$ order in the region $\bar\omega-1=O(1)$ simplify to
\begin{eqnarray}
\label{hlcoefficients}
\lambda(s_1,s_2,q^2)&\simeq &4m_2^2\left[(z_1+m_3+\bar\omega m_1)^2-s_2 \right],
\\
\quad \beta&=&O(1),
\\
\alpha_1&\simeq&\frac{1}{m_2}
\frac{(z_1+m_3+\bar\omega m_1)(s_2-m_1^2+m_3^2)/2-s_2(z_1+m_3)}
{(z_1+m_3+\bar\omega m_1)^2-s_2}
\nonumber\\
&=&O\left(\frac{1}{m_2}\right),
\\
\alpha_2 &\simeq &
\frac{(z_1+m_3+\bar\omega m_1)(z_1+m_3)-(s_2-m_1^2+m_3^2)/2}
{(z_1+m_3+\bar\omega m_1)^2-s_2}
\nonumber\\
&=&O\left(1\right),
\\
\alpha_{12} &\simeq &\alpha_1\alpha_2-\frac{\beta(z_1+m_3+\bar\omega m_1)}{2m_2\left[(z_1+m_3+\bar\omega m_1)^2-s_2
\right]}
\nonumber\\
&=&O\left(\frac{1}{m_2}\right),
\\
\alpha_{11} &\simeq  &\alpha_1^2+\frac{1}{m_2^2}\frac{\beta s_2}{\left[(z_1+m_3+\bar\omega
m_1)^2-s_2\right]}
\nonumber\\
&=&O\left(\frac{1}{m_2^2}\right).
\end{eqnarray}
For the double spectral densities $\tilde f_i$ these expressions yield the following LO relations
\begin{eqnarray}
\label{hlds}
\tilde s&=&\frac{1}{2m_2}\tilde f_2, \\
\label{hldg2}
\tilde g_{2}&=&2m_2\tilde g,\\
\label{hldg0}
\tilde g_{0D}&=&
\frac{1}{m_2}(2\tilde g+\tilde a_{2D}), \\
\label{hldg1}
\tilde g_{1}&=&\frac{1}{m_2}[-\tilde f_D+2\tilde p_1\tilde p_2\cdot\tilde g],
\end{eqnarray}
with $2\tilde p_1\tilde p_2=s_1+s_2-q^2$. 
Notice that these relations hold also for $\bar\omega=O(m_2)$. 

First two equations directly give the LO equality of the corresponding form factors. 
The relation (\ref{hldg0}) between the unsubtracted spectral densities is more interesting and requires 
the LO identity of the subtraction terms 
\begin{eqnarray}
\tilde \rho_a\simeq \tilde \rho_{g_0}.
\end{eqnarray}
The choice $\tilde \rho_a=\tilde \rho_{g_0}=\tilde g$ is acceptable although there are no firm backgrounds 
to justify this very choice. 

The most informative is the relation (\ref{hldg1}): this relation not only suggests a necessity of subtraction
in the form factor $f$ but also determines the subtraction term. When considering the heavy--to--heavy
transitions we have found that the subtracted spectral density of the form (\ref{tildeffull})
\begin{eqnarray}
\tilde f=\tilde f_D+(M_1^2-s_1+M_2^2-s_2)\tilde \rho_f
\end{eqnarray}
matches the HQET expansion if $\tilde\rho_f\simeq \tilde g$ in LO. The eq. (\ref{hldg1}) prescribes 
$\tilde\rho_f=\tilde g$. 

\vspace{.5cm}
\underline{\it Scaling of the form factors with $m_Q$}
\vspace{.5cm}

Let us point out that in the case of the heavy--to--light transition there is no parametrical suppression of the
anomalous contribution to the form factors compared with the normal one as it was in heavy--to--heavy transitions
and both parts contribute on equal footings. We shall present all the results for the normal part but the same
holds also for the anomalous part. 

In the region $\bar\omega=O(1)$ one finds for the normal contribution
\begin{eqnarray}
\label{scaling}
f_i(\bar\omega)
=
\frac{1}{\sqrt{m_2}}
\int
\frac{ds_2\varphi(s_2)\lambda^{1/2}(s_2,m_1^2,m_3^2)}
{16\pi\,m_1}
\int\limits_{-1}^{1}
\frac
{d\eta \sqrt{\bar\omega^2-1}\;\phi_0(z_1)\tilde f_i(z_1,z_2,m_1,m_2,m_3,\bar\omega)}
{\sqrt{(m_1(\bar\omega+1)+z_1+z_2+2m_3)(m_1(\bar\omega-1)+z_1-z_2)}}
\end{eqnarray}
where
\begin{eqnarray}
&&s_2=(m_1+m_3+z_2)^2,\nonumber\\
&&z_1=z_2\bar\omega\left({1+\frac{2m_3+z_2}{m_1}}\right)+m_3(\bar\omega-1)+\eta\sqrt{\bar\omega^2-1}
\frac{\lambda^{1/2}(s_2,m_1^2,m_3^2)}{2m_1}+O(1/m_2). \nonumber
\end{eqnarray}
Using the relations (\ref{hlcoefficients}) one finds that 
the spectral densities scale at large $m_2$ as  
\begin{equation}
\tilde f_i=m_2^{n_i}\rho_i(\bar\omega,m_1,m_3,z_1,z_2).
\end{equation}
Namely, 
\begin{eqnarray}
\tilde f_1&=&O(1),\quad \nonumber\\
\tilde f_2&=&O(m_2),\quad \nonumber\\
\tilde g&=&O(1),\quad \nonumber\\
\tilde a_1&=&O(1/m_2),\quad \nonumber\\
\tilde a_2&=&O(1),\nonumber\\
\tilde f&=&O(m_2),\quad \nonumber\\
\tilde g_1&=&O(1),\quad \nonumber\\
\tilde g_2&=&O(m_2),\quad \nonumber\\
\tilde g_0&=&O(1/m_2),\quad \nonumber\\
\tilde s&=&O(1).
\nonumber
\end{eqnarray}
Hence, the form factors have the scaling behaviour of the form
\begin{equation}
f_i=m_2^{n_i-1/2}r_i(\bar\omega;m_1,m_3;\varphi_2,\phi_0).
\end{equation}
The variables $\omega$ and $\bar\omega$ are connected with each other as follows
\begin{equation}
\bar\omega=\frac{M_2}{m_1}\omega-\frac{\bar\Lambda}{m_1}+O(1/m_2),
\end{equation}
and hence the ratio
\begin{equation}
f_i(\omega)/m_Q^{n_i+1/2}
\end{equation}
is universal for the transition of any heavy meson into a fixed  light state. 
This behaviour reproduces the results of \cite{iwhl} up to the
logarithmic corrections which arise from the anomalous scaling of the quark currents in QCD. 

To summarize, matching the LO $1/m_2$ relations between the form factors in the quark model to 
the corresponding relations of \cite{iwhl} allowed us to determine the subtraction term in the form factor 
$f$ and equated to each other subtraction terms in the form factors $a_{1,2}$ and $g_0$. 

\subsection{\label{ii.5}Numerical estimates for the universal form factors}
In this section we apply the derived results to the model--dependent estimates of the universal form factors. 

\begin{figure}[hbt]
\begin{center}  
\mbox{\epsfig{file=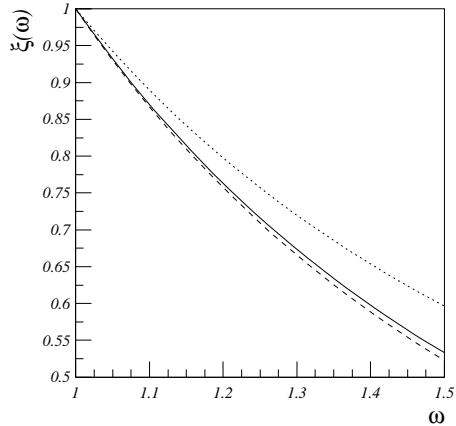,width=7.cm}  }
\end{center}
\caption{The Isgur-Wise function calculated with various quark--model parameters:
dashed line -- set 1, solid line -- set 2, dotted line -- set 3. 
\label{fig:iw}}
\end{figure}

1. First, let us study the dependence of the IW function on the parameters of the quark model. 
We adopt the exponential parametrization of 
the radial wave functions in the form $w(\vec k^2)\simeq \exp(-\vec k^2/2\beta^2)$. 
For obtaining the LO wave function $\phi_0$ we must take into
account the relationship between the parameters $z$ and $\vec k^2$ which is found from the following equation 
\begin{eqnarray}
\sqrt{s}=\sqrt{\vec k^2+m_Q^2}+\sqrt{\vec k^2+m_3^2}=m_Q+m_3+z.
\end{eqnarray}
This yields the relation 
\begin{eqnarray}
\vec k^2=z(z+2m_3)+O(1/m_Q).
\end{eqnarray}
Using the eq. (\ref{4vertex}) we come to the following form of the LO wave function 
\begin{eqnarray}
\phi_0(z)\simeq\sqrt{\frac{z+m_3}{z+2m_3}}\exp\left(-\frac{z(z+2m_3)}{2\beta_0^2}\right),
\end{eqnarray}
where $\beta_0$ is the LO harmonic oscillator parameter 
\begin{eqnarray}
\beta(m_Q)=\beta_0+O(1/m_Q).
\end{eqnarray}
For calculations we need to specify $\beta_0$. 
The exact value of $\beta_0$ is not known but extrapolating the known values of $\beta_D$, $\beta_B$, and
$\beta_{D^*}$ we find $\beta_0\simeq 0.4$ both in the WSB \cite{wsb} and the ISGW2 \cite{isgw2} models. 
In the model for pseudoscalar mesons discussed in the previous Chapter we have found 
the following approximate relation 
\begin{eqnarray}
\beta(m_Q)\simeq 2.5\frac {m_Qm_3}{m_Q+m_3},
\end{eqnarray}
which gives $\beta_0\simeq 2.5\; m_3$. 
Table \ref{table:parametersiw} presents the relevant numerical parameters. 
The results of calculating the IW function through the eq. (\ref{xi}) are shown in Fig. \ref{fig:iw}. 
Table \ref{table:parametersiw} demonstrates the values $\xi'(1)$ calculated with the eq. (\ref{rho2}) 
and the parameters of the quadratic fit of the form 
\begin{eqnarray}
\xi(\omega)=1-\rho_i^2(\omega-1)+\delta_i(\omega-1)^2,
\end{eqnarray}
obtained by interpolating the results of calculations in the range $1\le\omega\le1.5$. One can observe the value of $-\xi'(1)$ 
to be considerably larger than the parameter $\rho^2$ obtained by the interpolation
procedure. 
\begin{table}[hbt]
\caption{\label{table:parametersiw}
Parameters of the quadratic fit to the IW function and the NLO form factor $\xi_3$ in various quark--model versions.} 
\centering
\begin{tabular}{c||c|c||c|c|c||c|c|c|}
Ref.                & $m_3$ & $\beta_0$ & $-\xi'(1)$ & $\rho^2$ & $\delta$ & $\xi_3(1)$ & $\rho^2_{\xi_3}$ & $\delta_{\xi_3}$ \\
\hline\hline
Set 1 \cite{wsb}    & 0.35  &    0.4    &   1.47     &   1.34     &   0.78  &   0.077    &   1.49     &   0.88  \\
Set 2 \cite{isgw2}  & 0.33  &    0.4    &   1.43     &   1.31     &   0.76  &   0.08     &   1.4      &   0.84  \\
Set 3 \cite{m1}     & 0.25  &   0.63    &   1.37     &   1.12     &   0.62  &   0.17     &   1.2      &   0.65  
\end{tabular}
\end{table}

The IW function obtained from the dispersion quark model agrees with the results 
of the light-cone quark model applied to the elastic $P\to P$ transition for (infinitely) heavy mesons \cite{simula} 
$\xi(1.5)=0.62$ and turns out to be a bit smaller than the SR result $\xi(1.5)=0.66$ \cite{sr7}.

The NLO form factor $\xi_3(\omega)$ is found to be very sensitive to the light--quark mass. 
The parameters of the quadratic fit to $\xi_3(\omega)$ calculated through eq. (\ref{xi3a}) in the form 
\begin{eqnarray}
\label{fit}
h_i(\omega)=h_i(1)\;[1-\rho_i^2(\omega-1)+\delta_i(\omega-1)^2]
\end{eqnarray}
for various sets of the quark--model parameters are shown in Table {\ref{table:parametersiw}. 
One can observe an approximate relation $\xi_3(\omega)/\xi(\omega)\simeq 0.08\;GeV$ for the light quark mass
$m_3\simeq0.35\;GeV$ and $\xi_3(\omega)/\xi(\omega)\simeq 0.17\;GeV$ for the light quark mass $m_3\simeq0.25\;GeV$ 
in the whole region $1\le\omega\le1.5$. 
This ratio is to be compared with the SR result with the $O(\alpha_s)$ corrections omitted \cite{neubert} 
\begin{eqnarray}
\xi_3(\omega)/\xi(\omega)=\bar\Lambda/3\simeq 0.16\; GeV,\quad \bar\Lambda\simeq 0.5\;GeV. 
\end{eqnarray}

\newpage
2. We are in a position to estimate the higher order corrections as we can calculate the form factors at finite masses  
and the leading--order contribution separately. For the ISGW2 parameter set ($m_c=1.82$, $m_b=5.2$, 
$\beta_{D}=0.45$,  $\beta_{B}=0.43$, and $\beta_{D^*}=0.38$), the results on $h_{f_+}$ and $h_f$ 
are shown in Fig. \ref{fig:xi}, and Table \ref{table:xi} presents the results of interpolating in the range 
$1\le\omega\le1.5$ with a quadratic fit (\ref{fit}).
\begin{table}[hbt]
\caption{\label{table:xi}
The form factors calculated with the parameters of ISGW2 model.}
\centering
\begin{tabular}{|c||c|c|c|}
   & $h_{f_+}$ &$h_{f}$  & $\xi$ \\
\hline
\hline
$h(1)$   & 0.93  & 0.96 & 1.0  \\
$\rho^2$ & 0.87  & 1.06 & 1.25 \\
$\delta$ & 0.37  & 0.53 & 0.7  
\end{tabular}
\end{table}

\begin{figure}[hbt]
\begin{center}  
\mbox{\epsfig{file=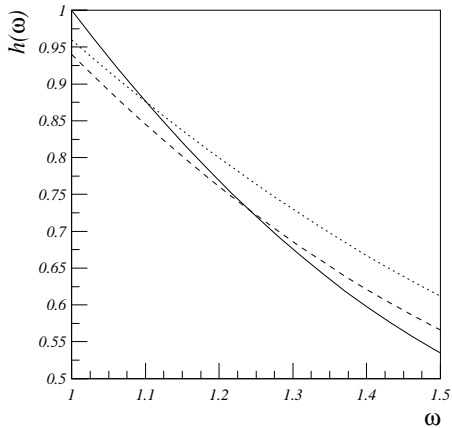,width=7.cm}  }    
\end{center}
\caption{The velocity--dependent form factors calculated for the ISGW2 parameters:
solid line -- the IW function, dashed line -- $h_{f}(\omega)$, dotted
line -- $h_{f_+}(\omega)$. 
\label{fig:xi}}
\end{figure}
The form factor $h_f$ can be seen to agree with the results of a combined fit to 
the ALEPH, ARGUS, CLEO, DELPHI, and OPAL data \cite{m2}
\begin{eqnarray}
\rho^2_{f}=1.07\pm 0.20. 
\end{eqnarray}
One observes sizeable difference between the form factors $h_f$ and $h_{f_+}$ and the Isgur--Wise 
function both in the value at zero recoil and the slope parameter. 
This indicates sizeable higher order corrections. At zero recoil the $1/m_Q$ corrections 
to $h_f$ vanish and writing \cite{neubert} 
\begin{eqnarray}
\label{deltam2}
h_f(1)=1+\delta_{1/m^2}
\end{eqnarray}
we find $\delta_{1/m^2}=-0.04$ that agrees well with the value $-0.055\pm0.025$ \cite{neubert}. 
At the nonzero recoil the higher--order corrections yield a more flat $\omega$--dependence of the form factor 
$h_f$ compared with the IW function such that 
the slope of the IW function in our model turns out to be considerably larger compared with the slope 
of $h_f$: An approximate relation $\rho^2_{IW}\simeq \rho^2_{h_f}+0.2$ is found. 
This is different from the sum-rules result $\rho^2_{IW}\simeq \rho^2_{h_f}-0.2$ \cite{neubert}. 
Let us also point out that $h'(1)$ obtained by the direct calculation 
(the theoretical value) turns out to be considerably 
larger than the parameter $\rho^2$ of the quadratic fit obtained by the interpolation 
over the range $1\le\omega\le1.5$ which could be measured experimentally. 
The same happens for the Isgur-Wise function. 

\newpage
\subsection{Discussion}
A detailed analysis of our dispersion approach to meson transition form factors was performed in
this chapter. We calculated the double spectral densities of the form factors at spacelike 
momentum transfers from the Feynman graphs, and determined the necessary subtractions 
by considering the two following cases: 
\begin{itemize}
\item[(i)]
by performing the $1/m_Q$ expansion of the form factors in the case 
of heavy-to-heavy transitions and matching them to HQET, and 
\item[(ii)]
by considering the relations between the form factors of the vector, axial-vector
and tensor currents in the case of the heavy-to-light transition and matching them to 
the general relations in QCD obtained in ref. \cite{iwhl}. 
\end{itemize}
This procedure led to the spectral representations with appropriate subtractions at 
spacelike momentum transfers. 

The form factors at timelike momentum transfers were then obtained by performing the 
analytical continuation in $q^2$. The analytical continuation yielded the appearance 
of the anomalous contribution to the form factors at $q^2>0$. As a result, 
we have come to an explicit model of the form factors describing the weak 
meson transitions which develops the correct structure of the heavy-quark expansion in agreement
with HQET. 
 
Let us emphasize that the obtained representations allow to calculate form factors both at spacelike and
timelike momentum transfers in terms of the wave functions of the participating mesons. 

\vspace{.4cm}
\noindent 
The following results were obtained in this chapter:

\vspace{.4cm}
\noindent 
1. Applying the method of ref. \cite{luke} to meson amplitudes induced by the 
heavy-quark tensor current we have found the expansions of the relevant form 
factors in LO and NLO in HQET. The $1/m_Q$ corrections in the form factor $h_{g_+}$ 
vanish at zero recoil just as in $h_{f_+}$ and $h_f$. 

\vspace{.4cm}
\noindent 
2. The $1/m_Q$ expansion of the soft wave function was constructed. The normalization condition 
of the wave function yields an infinite chain of normalization 
conditions for wave function components emerging in various $1/m_Q$ orders. 
The LO normalization condition provides the proper normalization of the Isgur-Wise function 
at zero recoil. 

\vspace{.4cm}
\noindent 
3. Form factors of {\it heavy-to-heavy} meson transitions were considered.  

First, the leading $1/m_Q$-order analysis of the spectral representations for the form factors 
was performed. It was shown that to this accuracy all the transition form factors are 
represented through the Isgur-Wise function in accordance with HQET. 
The Isgur-Wise function was calculated in terms of the leading-order component of the heavy 
meson wave function. Subtraction terms in the spectral representations for the form factors 
do not contribute to this accuracy. 

The anomalous contribution which emerges when the analytical continuation to the timelike 
region is performed comes into the game only in the $1/m_Q^2$ vicinity of the zero 
recoil point. It is otherwise suppressed by at least the second inverse power of the 
heavy-quark mass. 

\vspace{.4cm}
\noindent 
We then analysed spectral representations for the form factors with next-to-leading order
accuracy. 

We found that the $P\to P$ form factors do not require any subtractions to this accuracy 
and calculated the universal form factor $\xi_3$ in terms of the LO wave function. 

In the case of the $P\to V$ transition the representations without subtractions 
perfectly match HQET 
for the form factors $g$, $g_1$, $g_2$, and 
$m_1a_2-m_2a_1$ and leads to the relation $\rho_1=\rho_2$ (or $\chi_3=0$) 
and $\chi_2=0$. 
At the same time, spectral representations without subtractions for the form factors 
$f$, $g_0$, and $m_1a_2+m_2a_1$ do not agree with HQET to this order. 

Matching of the dispersion representations for these form factors to HQET shows the necessity of
subtractions and restricts the form of the subtraction terms. 

For performing the heavy-quark expansion only a strong localization of 
the meson momentum distribution with a width of order of the confinement scale 
was necessary. 
No other constraints on the soft wave functions or numerical parameters of the 
model have emerged. 

\vspace{.4cm}
\noindent 
4. We analysed {\it heavy-to-light} meson transitions and 
found that the Isgur-Wise relations
between the form factors of the tensor and vector and axial--vector currents \cite{iwhl} 
further constrain subtraction terms in the spectral representations for the form factors. 

\vspace{.4cm}
\noindent 
5. We observed a discrepancy between the predictions of the various versions of the quark model 
on the NLO universal form factors : Namely, the analysis of the Wirbel-Stech-Bauer model 
\cite{wsb} performed in \cite{luke} resulted in $\rho_1\ne
\rho_2\ne 0$ and $\rho_3\ne 0$, but $\rho_4=0$; the quasipotential quark model \cite{faustov} predicts all the NLO form factors to be nonzero; 
a consistent relativistic 
treatment of the $q\bar q$ intermediate states in Feynman graphs performed in this paper gives 
$\rho_1=\rho_2$, $\rho_3$=0, and  $\rho_4\ne 0$. 

The diagreement between our relativistic approach and the WSB model \cite{wsb} 
can be traced back to the not fully consistent treatment of the quark spins in the latter. 
On the other hand, the origin of the discrepancy between the relativistic quasipotential 
approach of \cite{faustov} and our dispersion approach is not fully understood and should be considered in
more detail. 

\vspace{.4cm}
\noindent 
6. Numerical results from our dispersion approach using several parameter sets 
(i.e. constituent quark masses and wave functions) demonstrate only a moderate 
dependence of the Isgur-Wise functions on the value of the light-quark mass and 
the shape of the soft wave function. 

Namely, the light-quark mass $m_3=0.25\div 0.35$ and the wave-function width $\beta_0=0.4\div 0.65$ 
give $\xi(1.5)=0.55\div 0.6$ that is a bit smaller than the SR result $\xi(1.5)=0.66$ \cite{sr7}. 

The form factor $h_f$ is found to behave in agreement with predictions of other models and 
experimental results. 

We observed a sizeable difference between the form factor $h_f$ and the Isgur-Wise 
function both in the value at zero recoil and the slope parameter. This difference might be 
important for the analysis of the experimental data for the $B\to (D,D^*)l\nu$ decays.  

The size of the higher-order $1/m_Q$-corrections to $h_f(1)$, $\delta_{1/m^2}=-0.04$, 
obtained using the ISGW2 parameters agrees favorably with the sum rule estimates. 

The slope of the Isgur-Wise function in our model turns out to be considerably larger compared with the slope 
of $h_f$: An approximate relation $\rho^2_{IW}\simeq \rho^2_{h_f}+0.2$ (see Fig. \ref{fig:xi})
is found. This is opposite to the sum-rules 
result $\rho^2_{IW}\simeq \rho^2_{h_f}-0.2$ \cite{neubert}. 
We would like to notice that the slope parameter
is very sensitive to the interpolation procedure: namely, the value $h'(1)$ turns out to be considerably 
larger than the result of the quadratic interpolation over the range $1\le\omega\le1.5$.

Our dispersion approach to the transition form factors can be further improved by
performing the heavy--quark expansion in higher orders and matching to HQET order by order. 
It should be taken into account that the quark model effectively describes the whole $1/m_Q$ series, 
but the short-distance corrections are not contained in the model and should be considered separately. 
\newpage
\section{\label{sec:iii}
Calculation of the weak form factors}

In this Chapter the dispersion approach developed in previous
sections is applied to form factors of weak transitions of heavy mesons. 
This issue was studied in \cite{mplb1,mplb2,mb1,mb2,ms}. 
We present form factors for weak decays of $B_{(s)}$ and $D_{(s)}$
mesons to light pseudoscalar and vector mesons obtained by the recent 
detailed analysis \cite{ms}. 

The use of this dispersion approach based on the constituent quark picture allows 
us to reveal the intimate connection between different decay modes and to perform the
calculations in the full physical $q^2$-region. In fact quark models provide for the only 
approach which leads to relations between the decays of various
mesons through the meson wave functions and gives the form factors in the full
$q^2$ range. However, quark models are not closely related to the QCD
Lagrangian (or at least this relationship is not well understood yet) and
therefore have input parameters which are not directly measurable and may
not be of fundamental significance.  

We reduce the usual disadvantages of quark models related to ill-defined
effective quark masses and not precisely known meson wave functions
by fitting the quark model parameters to lattice QCD results for the 
$B\to \rho$ transition form factors at large momentum transfers and to 
the measured total $D\to (K,K^*)l\nu$ decay rates. 
This allows us to predict numerous form factors for various decay channels 
and for all kinematically accessible $q^2$ values.

As demonstrated in the previous Section, 
the calculated form factors have the correct structure of the long-distance
corrections in accordance with QCD in the leading and next-to-leading $1/m_Q$
orders, if the radial wave functions $w(k^2)$ are localized in a region of
the order of the confinement scale, $k^2\le\Lambda^2$.
We assume a simple gaussian parametrization of the radial wave function
\begin{equation}
\label{gauss}
w(k^2)\propto\exp(-k^2/2\beta^2),
\end{equation}
which satisfies the localization requirement for $\beta\simeq
\Lambda_{QCD}$.

\subsection{Parameters of the model}
We consider the slope parameter $\beta$ of the meson wave function
(\ref{gauss})
and the constituent quark masses
as fit parameters. The relevant values for the $B$, $\rho$, and $\pi$ mesons
are determined \cite{mb1} from the requirement to reproduce the observed value 
of the pion decay constant $f_\pi=130\; MeV$ and from the fit to the lattice results
on the form factors $T_2(q^2)$ and $A_1(q^2)$ (see Eq. (\ref{wsbff}) below)
at $q^2=19.6$ and $17.6\;$GeV$^2$ \cite{lat}. 
The spectral representations 
for the decay constants (\ref{fp}) and for the form factors (\ref{drnormal}) 
are used for calculations. 
 
The values obtained for $m_b$, $m_u$, and $\beta_B$,
$\beta_\rho$, $\beta_\pi$ are displayed in Table \ref{table:parameters}.
\begin{table}[hbt]
\caption{\label{table:parameters}
Constituent quark masses and slope parameters of the exponential wave
function
(in $GeV$).}
\centering
\begin{tabular}{|cc|c|c|c||c|c|c|c|c|c|c|c|c|c|c|}
& $m_u$ & $m_s$ & $m_c$ & $m_b$
&  $\beta_\pi$  &$\beta_K$&  $\beta_D$ &  $\beta_B$ &
$\beta_{\eta_s}$ & $\beta_{D_s}$ & $\beta_{B_s}$ &
$\beta_\rho$ & $\beta_{K^*}$ & $\beta_{D^*}$ & $\beta_\phi$ \\
& 0.23 & 0.35  & 1.45  & 4.85  &
0.36 & 0.42  & 0.46  & 0.54  &
0.45 & 0.48  & 0.56  &
0.31 & 0.42  & 0.46  &  0.45
\end{tabular}
\end{table}

A few comments are in order:
\begin{itemize}
\item
As it was already noticed, the spectral representation (\ref{drnormal}) 
takes into account the long-distance contributions connected with the structure of the
initial and final mesons. To describe additional long-distance effects in the $q^2$ 
channel, Eq. (\ref{drnormal}) should be multiplied by the constituent quark form factor 
$f_{q_2\to q_1}(q^2)$ which contributes to the resonance structure in the $q^2$ channel. 
However, in the region of calculation $q^2<(m_2-m_1)^2$, the wave functions provide already
for a rise of the form facotrs with $q^2$ which is well compatible with a properly located
meson pole. Thus an additional quark form factor is not needed there, but we shall use a 
proper extrapolation formula when considering the vicinity of the poles. 
\item
The quark model double spectral representations take into account long-range
QCD effects but not the short-range perturbative corrections. However,
by fitting the wave functions and masses to reproduce the lattice points,
these
corrections are effectively taken care of:
Corrections to the quark propagators correspond to the appearance of
the effective quark masses. Corrections to the vertices at the relevant
values of the recoil variable
$\omega=(M_B^2+m_\pi^2-q^2)/2M_BM_\pi$ should be small as found in form
factors of other meson transitions \cite{amn1}.
\item
The value obtained for the $b$-quark mass $m_b=4.85\;GeV$ is close to the
one-loop pole mass which in fact is the relevant mass for quark model
calculations.
\end{itemize}

\newpage
We now need to fix the parameters describing the strange and charmed mesons.
The charm quark mass can be determined from the well-known $1/m_Q$ expansion
of the heavy meson mass in terms of the heavy quark
mass and the hadronic parameters
$\bar \Lambda$, $\lambda_1$ and $\lambda_2$. Using the recent estimates
of these parameters \cite{bigi} one finds
\begin{eqnarray}
m_b-m_c\simeq 3.4\; GeV.
\end{eqnarray}
This provides $m_c\simeq 1.45\;GeV$. For $m_s$ one expects $m_s\simeq
350-370\;MeV$
taking into account that $m_u=230\;MeV$.

A stringent way to constrain the parameters $m_s$, $\beta_K$, $\beta_{K^*}$,
and $\beta_D$ is provided by the measured integrated rates of the
semileptonic decays $D\to (K,K^*)l\nu$. In addition we apply the relation
(\ref{fp})
which connects $\beta_K$ with $m_s$ by using the known value of the
$K$-meson
leptonic decay constant $f_K=160\;MeV$.
The parameter values found this way are displayed in
Table \ref{table:parameters}\footnote{In \cite{m2}
a different set of the parameters was used which also provided a good
description
of the available experimental data on semileptonic $B$ and $D$ decays.
However, the corresponding form factors have a rather flat $q^2$-dependence
and do not match the lattice results at large $q^2$.}.
The corresponding form
factors and decay rates are given in Tables \ref{table:fitsd2k} and
\ref{table:ratesd2k}.

The polarization of the $K^*$ in the $D\to K^*l\nu$
decays turns out to be in good agreement with the experimental result
(Table \ref{table:ratesd2k}), and
the  calculated $D$ meson decay constant $f_D=200\; MeV$ corresponds to the
expectation for the magnitude of this quantity.

The parameter $\beta_{D^*}$ cannot be found this way, but it should be
close to $\beta_D$ because of the heavy quark symmetry requirements.
We therefore set $\beta_{D^*}=\beta_D$.

Also listed in Table \ref{table:parameters} are the parameters which
describe strange heavy mesons. They are discussed in subsection D.

\begin{table}[hbt]
\caption{\label{table:decconst}
Leptonic decay constants of pseudoscalar mesons in $MeV$ calculated via
(\protect\ref{fp}) for the quark-model parameters of Table 
\protect\ref{table:parameters}.}
\centering
\begin{tabular}{|cc|c|c|c|c|c|c|}
&$f_\pi$  &$f_K$&  $f_D$ &  $f_B$ & $f_{\eta_s}$ & $f_{D_s}$ & $f_{B_s}$ \\
&132 & 160 & 200 & 180 &  183  &  220  &  200 \\
\end{tabular}
\end{table}
\begin{table}[hbt]
\caption{\label{table:masses}
Meson masses in $GeV$ from PDG \protect\cite{pdg}}
\centering
\begin{tabular}{|cc|c|c|c|c|c|c|c|c|c|c|c|c|}
&$M_\pi$&$M_K$ & $M_\eta$ &$M_{\eta'}$ & $M_D$& $M_{D_s}$ & $M_B$ &
$M_{B_s}$ &
$M_\rho$&$M_{K^*}$& $M_\phi$ & $M_{D^*}$ & $M_{D^*_s}$\\
&0.14   & 0.49 & 0.547 & 0.958  & 1.87  & 1.97 & 5.27  &  5.37
&  0.77 & 0.89    &  1.02 &    2.01   &    2.11
\end{tabular}
\end{table}
The knowledge of the wave functions and the quark masses allows the
calculation of the
form factors in Eq. (\ref{ffdef}). It is however more convenient to present our
results in terms
of the dimensionless form factors
$F_+$, $F_0$, $F_T$, $V$, $A_0$, $A_1$, $A_2$, $T_1$, $T_2$, $T_3$
\cite{wsb} which are the
following linear combinations of the form factors given in Eq. (\ref{ffdef}):
\begin{eqnarray}
\label{wsbff}
F_+&=&f_+,\quad 
\nonumber \\
F_0&=&f_{+}+\frac{q^2}{Pq}f_{-},\quad  
\nonumber \\
F_T &=& (M_1+M_2)s,\quad 
\nonumber \\
V&=&(M_1+M_2)g,  
\nonumber \\
A_1&=& \frac{1}{M_1+M_2}f,\quad  
\nonumber \\
A_2&=&-(M_1+M_2)a_+, \quad
\nonumber \\
A_0 &=& \frac{1}{2M_2}\left( f + q^2\cdot a_{-}+ Pq\cdot a_+\right), \nonumber
\nonumber\\
T_1&=&-g_+,\quad
\nonumber \\
T_2&=&-g_{+}-\frac{q^2}{Pq}g_{-},\quad
\nonumber \\
T_3&=& g_{-}-\frac{Pq}{2} g_0.
\end{eqnarray}
\newpage

Each of these form factors contains 
contributions of $q^2$-channel resonances only of the same spin. 
This is the advantage of the form factors of the set (\ref{wsbff}) 
compared with the form factors defined by the Eq. (\ref{ffdef}),  
some of which contain contributions of resonances with different spins.

\subsubsection{The form factors $F_+$, $F_T$, $V$, $T_1$, and $A_0$}
The form factors $F_+$, $F_T$, $V$, $T_1$ contain a pole at
$q^2=M_{V}^2\equiv M_{1^-}^2$
and $A_0$ contains a pole at $q^2=M_B\equiv M_{0^-}^2$. 
The residues of the form factors at these poles are given in terms of the 
product of the weak and strong coupling constants. 
These coupling constants therefore determine the
behavior of the form factors at $q^2$ near the resonance poles (beyond the decay
region). We consider as an example the case of the $B\to \pi,\rho$ transition.

\vspace{.3cm}
\underline{\it Weak decay constants}
\vspace{.3cm}

Weak decay constants of mesons are defined by the following relations
\begin{eqnarray}
\langle 0|\bar q\gamma_\mu\gamma_5 b |B(q) \rangle &=&i
f_P^{(B)}q_\mu\nonumber\\
\langle 0|\bar q\gamma_\mu b |B^*(q) \rangle
&=&\epsilon_\mu^{*(B^*)}M_{B^*}f_V^{(B^*)}
\nonumber\\
\langle 0|\bar q\sigma_{\mu\nu} b |B^*(q) \rangle &=&
i(\epsilon_\mu^{*(B^*)}q_\nu-\epsilon_\nu^{*(B^*)}q_\mu)f_T^{(B^*)},
\end{eqnarray}
where $\epsilon_\mu^{(B^*)}$ is the $B^*$ polarization vector.
In the heavy quark limit one has 
$$
f_P^{(B)}=f_V^{(B^*)}=f_T^{(B^*)}.
$$

\vspace{.3cm}
\underline{\it Strong coupling constants}
\vspace{.3cm}

Strong coupling constants are connected with the three-meson amplitudes as
follows
\begin{eqnarray}
\langle \pi(p_2)B^*(q)|B(p_1) \rangle
&=&-\frac{1}{2}g_{B^*B\pi}P_\mu\epsilon_\mu^{(B^*)}
\nonumber\\
\langle \rho(p_2)B^*(q)|B(p_1)\rangle
&=&\frac{1}{2}\epsilon_{\alpha\beta\mu\nu}
\epsilon_\alpha^{(\rho)}\epsilon_\beta^{(B^*)}P_\mu q_\nu
\frac{g_{BB\rho}}{M_B^*}
\nonumber\\
\langle \rho(p_2)B(q)|B(p_1)\rangle
&=&\frac{1}{2}g_{BB\rho}P_\mu\epsilon_\mu^{(\rho)},
\end{eqnarray}
where $q=p_1-p_2$, $P=p_1+p_2$ and $\epsilon_\mu$ is the polarization vector
of the
vector meson.

In the heavy quark limit 
$$
g_{B^*B\rho}=g_{BB\rho}.
$$

The form factors $F_+$, $F_T$, $V$, $T_1$
contain pole at $q^2=M^2_{B^*}$ due to the contribution of the intermediate
$B^*$ ($1^-$ state) in the $q^2$ channel. The residue of this pole is given
in terms of the
product of the weak and strong coupling constants such that in the region
$q^2\simeq M_{B^*}^2$ the form factors read
\begin{eqnarray}
F_+&=&\frac{g_{B^*B\pi}f_V^{(B^*)}}{2M_B^*}\frac{1}{1-q^2/M^2_{B^*}}+...,
\nonumber\\
F_T&=&\frac{g_{B^*B\pi}f_T^{(B^*)}}{2M_B^*}\frac{M_B+M_\pi}{M_B^*}\frac{1}{1
-q^2/M^2_{B^*}}+...,\nonumber\\
V&=&\frac{g_{B^*B\rho}f_V^{(B^*)}}{2M_B^*}\frac{M_B+M_\rho}{M_{B^*}}\frac{1}
{1-q^2/M^2_{B^*}}+...,\nonumber\\
T_1&=&\frac{g_{B^*B\rho}f_T^{(B^*)}}{2M_B^*}\frac{1}{1-q^2/M^2_{B^*}}+...
\nonumber\\
\end{eqnarray}
Here $\dots$ stand for the terms non-singular at $q^2=M_{B^*}^2$.

Similarly, $A_0$ contains the contribution of the $B$ ($0^-$ state).
In the region of
$q^2\simeq M_B^2$  it can be represented as follows
\begin{eqnarray}
A_0=\frac{g_{BB\rho}f_{P}^{(B)}}{2M_B}\frac{M_B}{2M_\rho}\frac{1}{1-q^2/M^2_
{B}}+...
\end{eqnarray}
Let us notice that the residues of the form factors are not all
independent and are
connected with each other as follows:
\begin{eqnarray}
\label{constraint}
\frac{Res(F_T)Res(V)}{Res(F_+)Res(T_1)}=\frac{M_B+M_\rho}{M_{B^*}}\frac{M_B+
M_\pi}{M_{B^*}}.
\end{eqnarray}
This relation can be used as a cross-check of the consistency of the
extrapolation for the form factors.

The coupling constants are related to the residues of
the form factors according to the relations
\begin{eqnarray}
\frac{g_{B^*B\pi}f_V^{(B^*)}}{2M_{B^*}}&=&Res(F_+),
\nonumber\\
\frac{g_{BB\rho}f_P^{(B)}}{2M_{\rho}}&=&2Res(A_0),
\nonumber\\
\frac{f_T^{(B^*)}}{f_V^{(B^*)}}&=&
\frac{Res(F_T)}{Res(F_+)}\frac{M_{B^*}}{M_B+M_\pi},
\nonumber\\
\frac{g_{B^*B\rho}}{g_{B^*B\pi}}&=&
\frac{Res(V)}{Res(F_+)}\frac{M_{B^*}}{M_B+M_\rho}.
\end{eqnarray}

\subsubsection{The form factors $F_0$, $A_1$, $A_2$, $T_2$ and $T_3$}

The remaining form factors, $F_0$, $A_1$, $A_2$, $T_2$ and $T_3$, do not
contain
contributions of the lowest lying negative parity states (for instance,
$F_0$ contains a contribution
of the $0^+$ state, and $A_1$ contains that of the $1^+$ which have
considerably
higher masses). As a result they have a rather flat $q^2$ behaviour in the
decay region, whereas
the form factors $F_+$, $F_T$, $V$, $T_1$, $A_0$ are rising  more steeply.

From the spectral representations (\ref{drnormal}) together with the parameter values
of Table I the form factors are obtained numerically.  For the applications
it is convenient, however, to represent our results by simple fit formulas 
which interpolate these numerical values within a 1\% accuracy for all $q^2$
values in the region $0<q^2 < (m_2-m_1)^2$. Also, they should be appropriate
for a simple extrapolation to the resonance region.

Let us start with the form factors $F_+$, $F_T$, $V$, $T_1$, $A_0$.
If we interpolate the results of the calculation with the simple
three-parameter fit formula
\begin{eqnarray}
\label{fit0}
f(q^2)=\frac{f(0)}{(1-q^2/M^2)(1-q^2/(\alpha M)^2)},
\end{eqnarray}
 the least-$\chi^2$
interpolation procedure leads in all cases to a value of the parameter $M$
which is within 3\% equal to the lowest resonance mass.
We consider this fact to be an important indication for the proper choice of 
the quark-model parameters and for the
reliability of our calculations. 
We therefore prefer to fix the pole mass $M$ to its physical value. 
The fit functions (\ref{fit0}) represent the results now with an accuracy of less
than 2\%. To achieve the accuracy of less than 1\% in all cases we take the form
\cite{mb1}:
\begin{eqnarray}
\label{fit1}
f(q^2)=\frac{f(0)}{(1-q^2/M^2)[1-\sigma_1 q^2/M^2+\sigma_2q^4/M^4]},
\end{eqnarray}
where $M=M_P$ for the form factor $A_0$ and $M=M_V$ for the form factors
$F_+$, $F_T$, $V$, $T_1$.
In the Tables below we quote numerical values of $\sigma_2$ only if an accuracy 
of better than 1\% cannot be achieved with $\sigma_2=0$, and 
take $\sigma_2=0$ if this accuracy can already be achieved with the two 
parameters $f(0)$ and $\sigma_1$. 

For the heavy-to-light meson transitions the masses of the lowest resonances
are not very much different from the highest $q^2$ values in the decay.
Eq. (\ref{fit1}) then allows an estimate of the residues of these poles. These
residues can be expressed in terms of products of weak and strong coupling
constants. The errors for these constants induced by 
changing $\sigma_1 $ and $\sigma_2$ in our fitting procedure 
(keeping to the 1\% requirement) do not exceed 10\%. Moreover,
the residues of the form factors at the meson pole are not
independent and satisfy certain constraint (\ref{constraint}), which provides a consistency check of the extrapolations. 
The mismatch in (\ref{constraint}) is always below 10\%, and in most of the
cases much lower. 

For the form factors $F_0$, $A_1$, $A_2$, $T_2$ and $T_3$ the contributing
resonances ($0^+$, $1^+$, etc)
lie farther away from the physical decay region and  the effect of any
particular resonance is smeared out.
For these form factors the interpolation formula taken is\footnote{ 
One should note that the parameters $\sigma_1$ and $\sigma_2$ in 
the fit formula (\ref{fit3}) for the form factors $F_0$, $A_1$, $A_2$, $T_2$, and $T_3$
are introduced in a different way than in the fit formula (\ref{fit1}) for the form factors 
$F_+$, $F_T$, $V$, $T_1$, and $A_0$.} 
\begin{eqnarray}
\label{fit3}
f(q^2)= f(0) / [1-\sigma_1 q^2/M_V^2+\sigma_2q^4/M_V^4].
\end{eqnarray}
If setting $\sigma_2=0$ allows us to describe the calculation results
with better than
1\% accuracy for all $q^2$, a simple monopole two-parameter formula is used.

The values of $f(0)$, $\sigma_1$, and $\sigma_2$ are given for each decay
mode in the relevant sections.

\newpage
\subsection{Charmed meson decays}

\subsubsection{$D\to K,K^*$}

The $D\to K,K^*$ decays are induced by the charged current $c\to s$
quark transition. As described in the previous section, the measured total
rates of these decays are
used for a precise fit of the parameters of our model. With the parameters
of Table \ref{table:parameters}
we obtain the form factors listed in Table \ref{table:fitsd2k}.
Table \ref{table:compd2k} compares the form factors at $q^2=0$ with the
results
of other approaches and Table \ref{table:ratesd2k} presents the decay rates.

\begin{table}[hbt]
\caption{\label{table:fitsd2k}The $D \to K, K^*$ transition form
factors. $M_V=M_{D_s^*}=2.11\; GeV$, $M_P=M_{D_s}=1.97\; GeV$. 
For the form factors $F_+, F_T, V, A_0, T_1$ the fit formula Eq. (\protect\ref{fit1}) is used, 
for the other form factors - Eq. (\protect\ref{fit3})}
\centering
\begin{tabular}{|c||c|c|c||c|c|c|c|c|c|c|}
 & \multicolumn{3}{c||}{$D \to K$} & \multicolumn{7}{c|}{$D \to K^*$}\\
\hline
          &$F_{+}$&$F_{0}$& $F_T$ & $V$   & $A_0$ & $A_1$  & $A_2$ & $T_1$
& $T_2$  & $T_3$   \\
\hline
$f(0)$     & 0.78 & 0.78  & 0.75  & 1.03  & 0.76  &  0.66  & 0.49  &  0.78
& 0.78   & 0.45    \\
$\sigma_1$ & 0.24 & 0.38  & 0.27  & 0.27  & 0.17  &  0.30  & 0.67  &  0.25
& 0.02   & 1.23    \\
$\sigma_2$ &      & 0.46  &       &       &       &  0.20  & 0.16  &
& 1.80   & 0.34     \\
\end{tabular}
\end{table}
\begin{table}[hbt]
\caption{\label{table:compd2k}
Comparison of the results of different approaches for the semileptonic
$D\to K,K^*$ form factors at $q^2=0$.}
\centering
\begin{tabular}{|c||c|c|c|c|c|}
 Ref.                &$F_+(0)$   & $F_T(0)$  &  $V(0) $  & $A_1(0)$  &
$A_2(0)$    \\
\hline
This work
                     &  0.78     &   0.75    & 1.03      &  0.66     &  0.49
\\
WSB \cite{wsb}
                     &  0.76     &    $-$    & 1.3       &  0.88     &  1.2
\\
Jaus'96 \cite{jaus}
                     &  0.78     &    $-$    & 1.04      &  0.66     &  0.43        
\\
SR \cite{bbd}
                     &  0.60(15) &    $-$    & 1.10(25)  &  0.50(15) &0.60(15)    
\\
Lat(average) \cite{lattice}
                     &  0.73(7)  &    $-$    & 1.2(2)    &  0.70(7)  &0.6(1)    
\\
Lat \cite{ape}
                     &  0.71(3)  &  0.66(5)  &    $-$    &    $-$    &$-$     
\\
\hline 
Exp \cite{ryd}
                     &  0.76(3)  &    $-$       &  1.07(9)  &  0.58(3)  &
0.41(5)
\end{tabular}
\end{table}
\begin{table}[hbt]
\caption{\label{table:ratesd2k}
The $D\to (K,K^*)l\nu$ decay rates in $10^{10}\;s^{-1}$ obtained
within different approaches, $|V_{cs}|=0.975$.}
\centering
\begin{tabular}{|c||c|c|c|c|}
 Ref.               &$\Gamma(D\to K)$ & $\Gamma(D\to K^*)$ &
$\Gamma(K^*)/\Gamma(K)$ & $\Gamma_L/\Gamma_T$ \\
\hline
This work           &  9.7      & 6.0       &  0.63     &  1.28       \\
Jaus'96 \cite{jaus} &  9.6      & 5.5       &  0.57     &  1.33       \\
SR \cite{bbd}       &  6.5(1.5) & 3.8(1.5)  &  0.50(15) &  0.86(6)    \\
\hline
Exp \cite{cleod2k}  &  9.3(4)   & 5.7(7)    &  0.61(7)  &  1.23(13)
\cite{pdg}
\end{tabular}
\end{table}
Extrapolating the form factors to $q^2=M^2_{D^*}$ (or $q^2=M_D^2$ for $A_0$)
gives the following estimates of the coupling constants 
\begin{eqnarray}
\frac{g_{D_s^*DK}f_V^{(D_s^*)}}{2M_{D_s^*}}&=&1.05\pm0.05, \qquad
\nonumber\\
\frac{g_{D_sDK^*}f_P^{(D_s)}}{2M_{K^*}}&=&1.7\pm 0.1, \qquad
\nonumber\\
\frac{f_T^{(D_s^*)}}{f_V^{(D_s^*)}}&=&0.95\pm0.05, \qquad
\nonumber\\
\frac{g_{D_s^*DK^*}}{g_{D_s^*DK}}&=&1.1\pm0.1. \nonumber
\end{eqnarray}

\newpage
\subsubsection{$D\to \pi,\rho$}

These decays are induced by the $c\to d$ charged current. Since all the
necessary parameters
have already been fixed, this mode allows for parameter-free predictions.
Table  \ref{table:fitsd2pi} presents the results of our calculations. In
Tables \ref{table:compd2pi} and
\ref{table:ratesd2pi} we compare our results with different approaches and
with experimantal data.

\begin{table}[hbt]
\caption{\label{table:fitsd2pi}The calculated $D \to \pi,\rho$ transition
form
factors. $M_V=M_{D^*}=2.01\; GeV$, $M_P=M_{D}=1.87\; GeV$.
For the form factors $F_+, F_T, V, A_0, T_1$ the fit formula Eq. (\protect\ref{fit1}) is used, 
for the other form factors - Eq. (\protect\ref{fit3}).}
\centering
\begin{tabular}{|c||c|c|c||c|c|c|c|c|c|c|}
 & \multicolumn{3}{c||}{$D \to \pi$} & \multicolumn{7}{c|}{$D \to \rho$}\\
\hline
          &$F_{+}$&$F_{0}$& $F_T$ & $V$   & $A_0$ & $A_1$  & $A_2$ & $T_1$
& $T_2$  & $T_3$   \\
\hline
$f(0)$     & 0.69 & 0.69  & 0.60  & 0.90  & 0.66  &  0.59  & 0.49  &  0.66
& 0.66   & 0.31    \\
$\sigma_1$ & 0.30 & 0.54  & 0.34  & 0.46  & 0.36  &  0.50  & 0.89  &  0.44
& 0.38   & 1.10    \\
$\sigma_2$ &      & 0.32  &       &       &       &        &       &
& 0.50   & 0.17
\end{tabular}
\end{table}

\begin{table}[hbt]
\caption{\label{table:compd2pi}
Comparison of the results of different approaches for the  semileptonic
$D\to \pi,\rho$ form factors at $q^2=0$.}
\centering
\begin{tabular}{|c||c|c|c|c|c|}
 Ref.                    &$F_+(0)$   & $F_T(0)$  &$V(0) $   & $A_1(0)$  &
$A_2(0)$  \\
\hline
This work                &  0.69     & 0.60      &  0.90     &  0.59      &
0.49     \\
WSB  \cite{wsb}          &  0.69     & $-$       & 1.23      &  0.78     &
0.92     \\
Jaus'96\cite{jaus}       &  0.67     & $-$       & 0.93      &  0.58     &
0.42     \\
SR  \cite{bbd}           &  0.50(15) & $-$       & 1.0(2)    &  0.5(2)   &
0.4(2)   \\
Lat(ave) \cite{lattice}  &  0.65(10) & $-$       & 1.1(2)    &  0.65(7)  &
0.55(10)    \\
Lat \cite{ape}         &  0.64(5)  &  0.60(7)  &    $-$    &    $-$    &
$-$     \\
\end{tabular}
\end{table}

\begin{table}[hbt]
\caption{\label{table:ratesd2pi}
The $D\to (\pi,\rho)l\nu$ decay rates in $10^{10}\;s^{-1}$,
$|V_{cd}|=0.22$.}
\centering
\begin{tabular}{|c||c|c|c|c|}
 Ref.             &$\Gamma(D\to \pi)$ & $\Gamma(D\to \rho)$ &
$\Gamma(\rho)/\Gamma(\pi)$ & $\Gamma_L/\Gamma_T$ \\
\hline
This work                &  0.95     & 0.42      &  0.45     &  1.16
\\
WSB \cite{wsb}           &  0.68     & 0.67      &  1.0      &  0.91
\\
Jaus'96\cite{jaus}       &  0.8      & 0.33      &  0.41     &  1.22
\\
SR \cite{bbd}            &  0.39(8)  & 0.12(4)   &  $-$      &  1.31(11)
\\
Melikhov'97 \cite{m2}    &  0.62     & 0.26      &  0.41     &  1.27
\\
\hline
Exp \cite{expd2pi,cleopik,e653} 
                         &  0.92(45) & 0.45(22)  &  0.50(35) &  $-$
\end{tabular}
\end{table}
The form factors at $q^2=0$ are close to the
predictions of the relativistic quark model of Ref. \cite{jaus}, but the
$q^2$ dependence is different such that our model and \cite{jaus}
predict different decay rates.
Although the experimental errors are very
large and nearly all theoretical results agree with experiment, we notice
perfect agreement of our decay rates with the central values.

For the coupling constants we get the following relations
\begin{eqnarray}
\frac{g_{D^*D\pi}f_V^{(D^*)}}{2M_{D^*}}&=& 1.05\pm0.05, \qquad 
\frac{g_{DD\rho}f_P^{(D)}}{2M_\rho}=2.1\pm0.2,  \qquad
\frac{f_T^{(D^*)}}{f_V^{(D^*)}}=0.9\pm0.1,  \qquad
\frac{g_{D^*D\rho}}{g_{D^*D\pi}}=1.3\pm0.2.  \nonumber
\end{eqnarray}
Taking $f_V^{(D^*)}\simeq 220\;MeV$, we find
$$
g_{D^*D\pi}=18\pm 3
$$
in good agreement with a calculation of $g_{D^*D\pi}$ based on combining
PCAC with the dispersion approach \cite{mb2}. This value also comapres very well 
with the first recent measurement of $g_{D^*D\pi}$ reported by CLEO 
\cite{cleo-ddpi}
$$
g_{D^*D\pi}=17.9\pm 0.3\pm 1.9.
$$ 
Notice a serious disagreement between our result and the sum rule estimate 
\cite{bbkr}
$$
g^{SR}_{D^*D\pi}=12.5\pm 1.0. 
$$ 
The experimental result clearly speaks in favour of our prediction.   
\newpage

\newpage
\subsection{Beauty meson decays}

\subsubsection{$B\to D,D^*$}

These decays arise from the heavy-quark $b\to c$ transition. Here one has
rigorous
predictions for the expansion of the form factors in terms of the
heavy-quark
mass \cite{luke}. Namely, the main part of the form factors can be expressed
through the
universal form factor - the Isgur-Wise function. However, different models
provide
different $q^2$-dependences of the Isgur-Wise function as well as
different subleading $1/m_Q$ corrections.

We recall that our spectral representations of the form factors explicitly
respect
the structure of the long-distance QCD corrections
in the leading and the subleading orders of the heavy-quark expansion. Thus,
we expect reliable predictions for the form factors. Our numerical
results are
summarized in Tables \ref{table:fitsb2d}, \ref{table:compb2d}, and
\ref{table:ratesb2d}.
\begin{table}[hbt]
\caption{\label{table:fitsb2d} The $B \to D,D^*$ transition form
factors. $M_V=M_{B_c^*}\simeq M_P=M_{B_c}=6.4\; GeV$. 
For the form factors $F_+, F_T, V, A_0, T_1$ the fit formula Eq. (\protect\ref{fit1}) is used, 
for the other form factors - Eq. (\protect\ref{fit3}).}
\centering
\begin{tabular}{|c||c|c|c||c|c|c|c|c|c|c|}
 & \multicolumn{3}{c||}{$B \to D$} & \multicolumn{7}{c|}{$B \to D^*$}\\
\hline
          &$F_{+}$&$F_{0}$& $F_T$ & $V$   & $A_0$ &  $A_1$  & $A_2$ & $T_1$
& $T_2$  & $T_3$   \\
\hline
$f(0)$     & 0.67 & 0.67  & 0.69  & 0.76  & 0.69  &  0.66  & 0.62  &  0.68
& 0.68   & 0.33    \\
$\sigma_1$ & 0.57 & 0.78  & 0.56  & 0.57  & 0.58  &  0.78  & 1.40  &  0.57
& 0.64   & 1.46    \\
$\sigma_2$ &      &       &       &       &       &        & 0.41  &
&        & 0.50     \\
\end{tabular}
\end{table}
\begin{table}[hbt]
\caption{\label{table:compb2d}
Comparison of the results of different approaches for the semileptonic
$B\to D,D^*$ form factors at $q^2=0$.}
\centering
\begin{tabular}{|c||c|c|c|c|}
 Ref.                 &$F_+(0)$    & $A_1(0)$  & $R_1(0)=V(0)/A_1(0)$ &
$R_2(0)=A_2(0)/A_1(0)$  \\
\hline
This work             &  0.67      &  0.66     &  1.15               &0.94  
\\
Jaus'96 \cite{jaus}   &  0.69      &  0.69     &  1.17               & 0.93
\\
Neubert \cite{neubert}&            &           &  1.3                &  0.8
\\
\hline
Exp \cite{cleoR1}     &            &           &  1.18$\pm$0.15$\pm$0.16 &
0.71$\pm$0.22$\pm$0.07
\end{tabular}
\end{table}
\begin{table}[hbt]
\caption{\label{table:ratesb2d}
The $B\to (D,D^*)l\nu$ decay rates in $|V_{cb}|^2\;ps^{-1}$. }
\centering
\begin{tabular}{|c||c|c|c|c|}
 Ref.      &$\Gamma(B\to D)$ & $\Gamma(B\to D^*)$ & $\Gamma(D^*)/\Gamma(D)$
& $\Gamma_L/\Gamma_T$ \\
\hline
This work               &  8.57    & 22.82 & 2.66     &  1.11   \\
Jaus'96 \cite{jaus}     &  9.6     & 25.33 & 2.64     &         \\
Mel'97 \cite{mplb2}    
&  8.7     & 21.0  & 2.65     &  1.28   \\
\hline
Exp        &  $(1.34\pm0.15)10^{-2}\; ps^{-1}$ \cite{cleob2d}  &
$(2.98\pm0.17)10^{-2}\; ps^{-1}$ \cite{cleob2dstar}  & 2.35(1.3) &
$1.24(0.16)$  \cite{cleogr}
\end{tabular}
\end{table}
For the coupling constants we find
\begin{eqnarray}
\frac{g_{B_c^*BD}f_V^{(B_c^*)}}{2M_{B_c^*}}&=& 1.56\pm0.15, \qquad
\nonumber\\
\frac{g_{B_cBD^*}f_P^{(B_c)}}{2M_{D^*}}&=&3.3\pm0.3, \qquad
\nonumber\\
\frac{f_T^{(B_c^*)}}{f_V^{(B_c^*)}}&=&0.9\pm0.1, \qquad
\nonumber\\
\frac{g_{B_c^*BD^*}}{g_{B_c^*BD}}&=&1.05\pm0.05. 
\nonumber
\end{eqnarray}

\newpage

\subsubsection{$B\to K,K^*$}

These decays are induced by the $b\to s$ Flavor Changing Neutral Current
(FCNC).
We recall that the $B\to \pi,\rho$ form factors at large $q^2$ have been
used to fix
the parameters of the model. Thus we expect that the predictions for the
$B\to K,K^*$ form factors which in fact differ from the former mode only by
SU(3) violating effects should be particularly reliable.
Table \ref{table:fitsb2k} presents the calculated form factors and Fig
\ref{fig:b2kstar}
exhibits our predictions together with the available lattice results at large
$q^2$. The
good agreement shows that the size and the sign of the SU(3) violating
effects
are correctly accounted for.
\begin{table}[hbt]
\caption{\label{table:fitsb2k}The calculated $B \to K, K^*$ transition form
factors.
$M_V=M_{B_s^*}=5.42\; GeV$, $M_P=M_{B_s}=5.37\; GeV$.
For the form factors $F_+, F_T, V, A_0, T_1$ the fit formula Eq. (\protect\ref{fit1}) is used, 
for the other form factors - Eq. (\protect\ref{fit3}).}
\centering
\begin{tabular}{|c||c|c|c||c|c|c|c|c|c|c|}
 & \multicolumn{3}{c||}{$B \to K$} & \multicolumn{7}{c|}{$B \to K^*$}\\
\hline
          &$F_{+}$&$F_{0}$& $F_T$ & $V$   & $A_0$ &  $A_1$  & $A_2$ & $T_1$
& $T_2$  & $T_3$   \\
\hline
$f(0)$     & 0.36 & 0.36  & 0.35  & 0.44  & 0.45  &  0.36  & 0.32  &  0.39
& 0.39   & 0.27    \\
$\sigma_1$ & 0.43 & 0.70  & 0.43  & 0.45  & 0.46  &  0.64  & 1.23  &  0.45
& 0.72   & 1.31    \\
$\sigma_2$ &      & 0.27  &       &       &       &  0.36  & 0.38  &
& 0.62   & 0.41
\end{tabular}
\end{table}
\begin{table}[hbt]
\caption{\label{table:compb2k}
Comparison of the results of different approaches for the
$B\to K,K^*$ form factors at $q^2=0$.}
\centering
\begin{tabular}{|c||c|c|c|c|c|c|c|c|}
 Ref.      &$F_+(0)$  &$F_T(0)$ & $V(0) $   & $A_1(0)$  &  $A_2(0)$  &
$A_0(0)$ &$T_1(0)$ & $T_3(0)$  \\
\hline
This work  &  0.36    &  0.35   & 0.44      &  0.36     &  0.32      & 0.45
& 0.39   &  0.27   \\
SR \cite{colangelo}
           &  0.25    &  -      & 0.47      &  0.37     &  0.40      & 0.30
& 0.38    &  -       \\
Lat+Stech \cite{lat}
           &  -       &  -      & 0.38      &  0.28     &  -         & 0.32
& 0.32    &  -       \\
LCSR'98 \cite{lcsr1}
           &  0.34    &  0.374  & 0.46      &  0.34     &  0.28      & 0.47
& 0.38    &  0.26  \\
Lat \cite{ape}
           &  0.30(4) &  0.29(6)&-          &  -        &  -         & -
& -       &  -
\end{tabular}
\end{table}
\begin{figure}[b]
\begin{center}
\epsfig{file=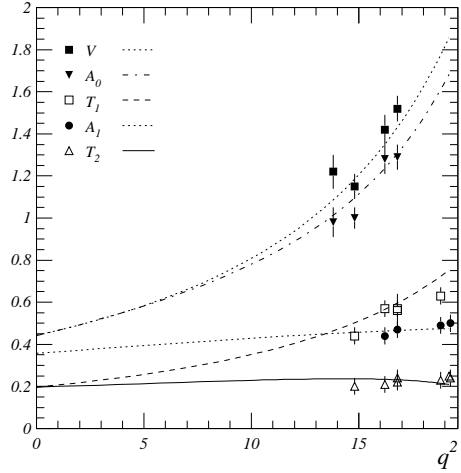,width=7cm}     
\caption{Form factors of the $B\to K^*$ transition vs the lattice results.
\label{fig:b2kstar}}
\end{center}
\end{figure}
For the coupling constants we obtain
\begin{eqnarray}
\frac{g_{B_s^*BK}f_V^{(B_s^*)}}{2M_{B_s^*}}&=& 0.65\pm0.05, \qquad
\frac{g_{B_sBK^*}f_P^{(B_s)}}{2M_{K^*}}=1.65\pm0.1, \qquad
\nonumber\\
\frac{f_T^{(B_s^*)}}{f_V^{(B_s^*)}}&=&0.95\pm0.05\qquad
\frac{g_{B_s^*BK^*}}{g_{B_s^*BK}}=1.15\pm0.05. \nonumber
\end{eqnarray}
\newpage
\subsubsection{$B\to \pi,\rho$}

The $B\to\rho$ transition has been used for determining
the parameters of our quark model in the $u,d$ and $b$ sectors by fitting
the quark-model form factors to
available lattice results on $T_2$ and $A_1$ at large $q^2$ \cite{mb1}, see 
Figure \ref{fig:ffbpirho}. 
The parametrizations for the $B\to \pi,\rho$ form factors  
are given in Table \ref{table:fitsb2pi}. 

The form factor $F_0$ at large $q^2$ lies   
below the central lattice values but nevertheless agrees with lattice results 
within the given error bars.
Notice however that in our model the form factor $F_0$ is calculated as a
difference
of $f_+$ and $f_-$ and at large $q^2$ turns out to be much more sensitive to
the subtle details
of the pion wave function, than $f_+$ and $f_-$ separately. A simple
Gaussian wave function
which works quite well for $f_+$ and $f_-$, might not be sufficiently
accurate for $F_0$.
\begin{table}[hbt]
\caption{\label{table:fitsb2pi}The calculated $B \to \pi, \rho$ transition
form factors.
$M_V=M_{B^*}=5.32\; GeV$, $M_P=M_{B}=5.27\; GeV$. 
For the form factors $F_+, F_T, V, A_0, T_1$ the fit formula Eq. (\protect\ref{fit1}) is used, 
for the other form factors - Eq. (\protect\ref{fit3}). }
\centering
\begin{tabular}{|c||c|c|c||c|c|c|c|c|c|c|}
 & \multicolumn{3}{c||}{$B \to \pi$} & \multicolumn{7}{c|}{$B \to \rho$}\\
\hline
          &$F_{+}$&$F_{0}$& $F_T$ & $V$   & $A_0$ &  $A_1$  & $A_2$ & $T_1$
& $T_2$  & $T_3$   \\
\hline
$f(0)$     & 0.29 & 0.29  & 0.28  & 0.31  & 0.30  &  0.26  & 0.24  &  0.27
& 0.27   & 0.19    \\
$\sigma_1$ & 0.48 & 0.76  & 0.48  & 0.59  & 0.54  &  0.73  & 1.40  &  0.60
& 0.74   & 1.42    \\
$\sigma_2$ &      & 0.28  &       &       &       &  0.10  & 0.50  &
& 0.19   & 0.51
\end{tabular}
\end{table}

\begin{center}
\begin{figure}[h]
\begin{tabular}{cc}
\mbox{\epsfig{file=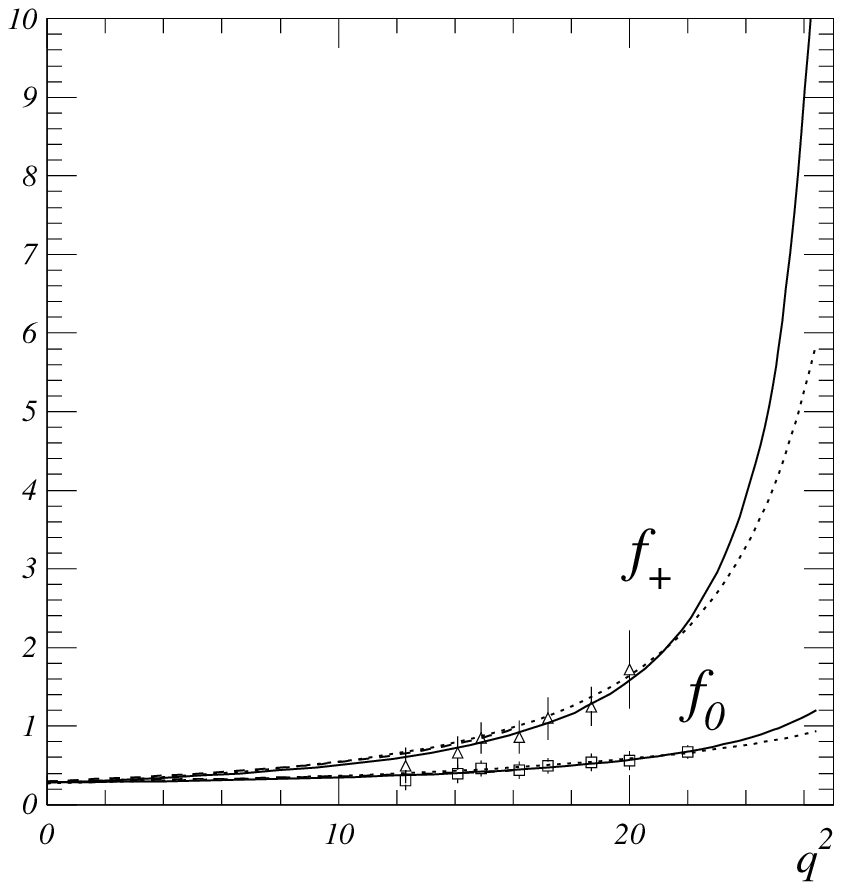,width=6.cm}}     
& 
\mbox{\epsfig{file=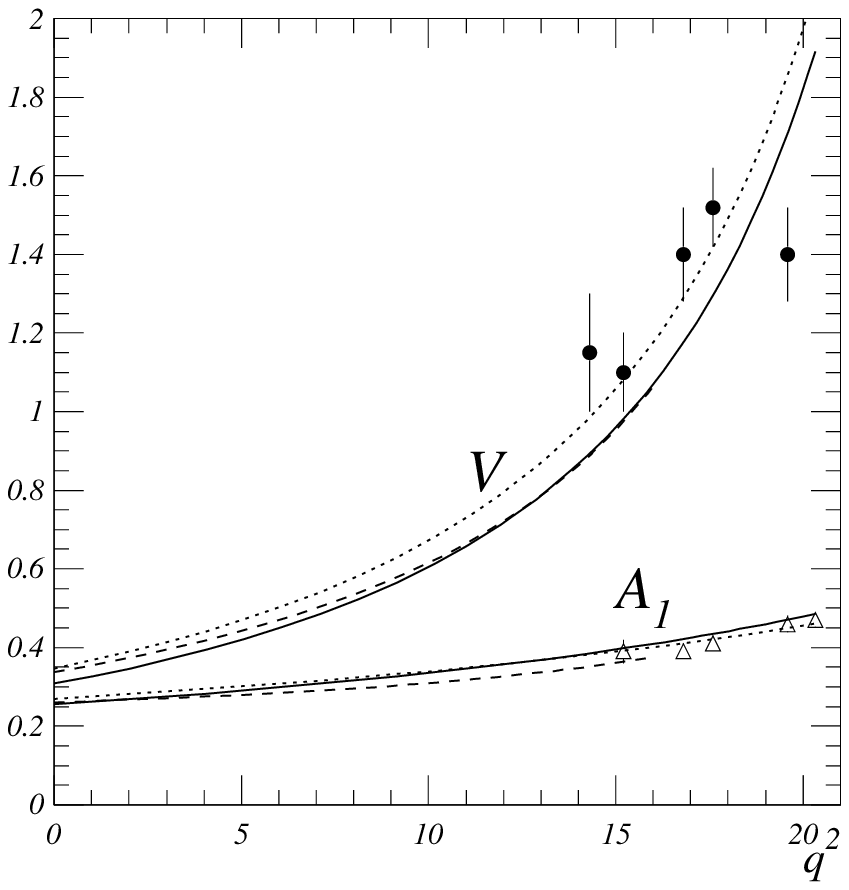,width=6.cm}}   
\\
\mbox{\epsfig{file=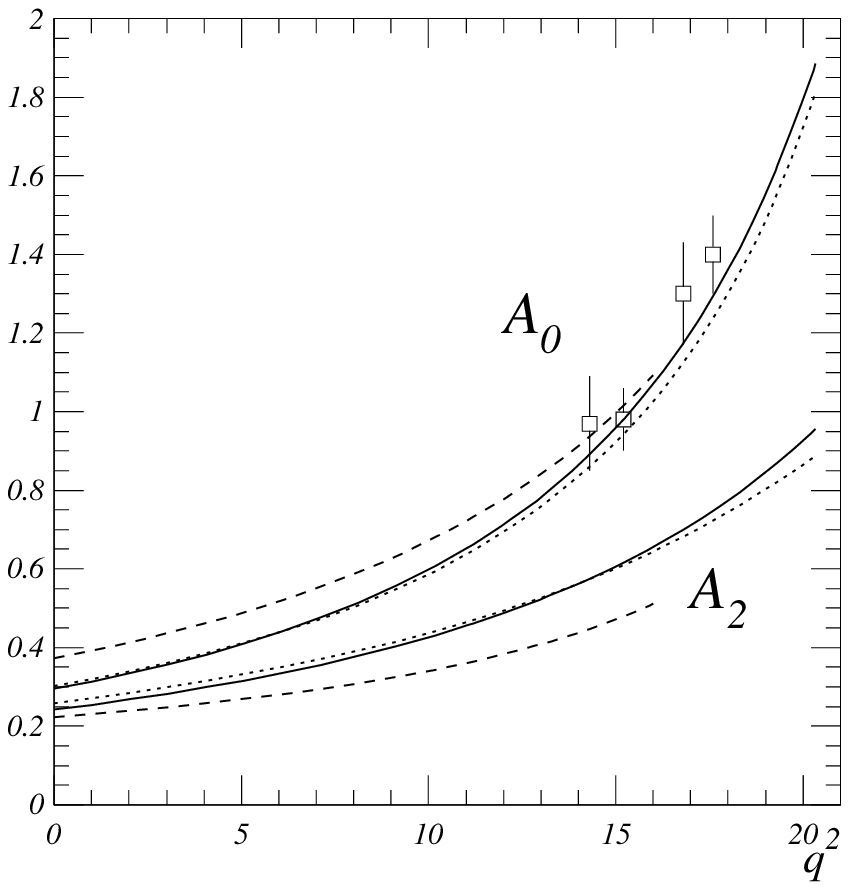,width=6.cm}}   
& 
\mbox{\epsfig{file=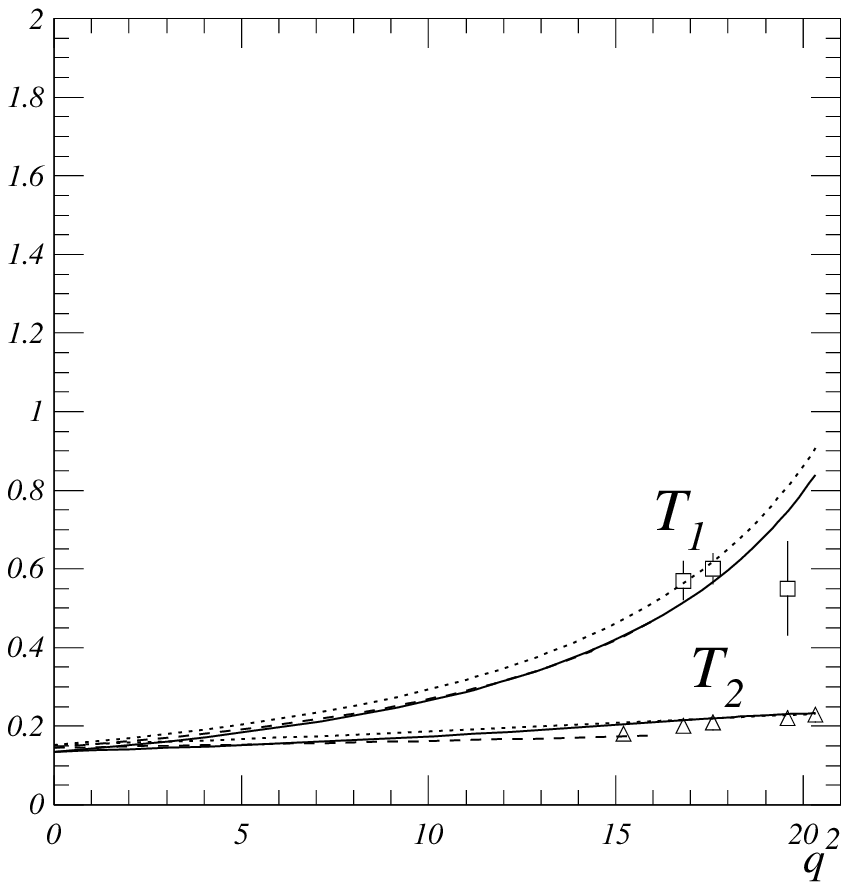,width=6.cm}}   
\end{tabular}
\caption{\label{fig:ffbpirho}
The $B\to \pi$ and $B\to \rho$ form factors vs lattice results.
Solid lines - results from the dispersion approach obtained in 
Ref. \protect\cite{mb1}, 
dotted lines - lattice-constrained parametrizations of Ref. 
\protect\cite{lat}, 
dashed lines - results from light-cone sum rules 
\protect\cite{lcsr1}}. 
\end{figure}
\end{center}
\newpage

\begin{table}[hbt]
\caption{\label{table:compb2pi}
Comparison of the results of different approaches on weak
$B\to \pi,\rho$ form factors at $q^2=0$.}
\centering
\begin{tabular}{|c||c|c|c|c|c|c|c|c|}
 Ref.      &$F_+(0)$  &$F_T(0)$ & $V(0) $   & $A_1(0)$  &  $A_2(0)$  &
$A_0(0)$ &$T_1(0)$ & $T_3(0)$ \\
\hline
This work  &  0.29    &  0.28   & 0.31      &  0.26     &  0.24      & 0.29
& 0.27   &  0.19   \\
Jaus'96 \cite{jaus}
           &  0.27    &  -      & 0.35      &  0.26     &  0.24      & -
&  -     &         \\
LCSR'98 \cite{lcsr1}
           &  0.305   &  0.296  & 0.34      &  0.26     &  0.22      & 0.37
& 0.29   &  0.20   \\
Lat \cite{ape}
           &  0.28(4) &  0.28(7)& -         &  -        &  -         & -
& -      &  -
\end{tabular}
\end{table}
Table \ref{table:compb2pi} compares the results obtained from the quark
model of Ref. \cite{mb1} with results from the quark model of Jaus
\cite{jaus} and latest light-cone sum rule (LCSR) results \cite{lcsr1}.
One observes very good agreement between the
quark model of Jaus, LCSRs, and our approach. The only visible
difference with the LCSR method occurs in the form factor $A_0(0)$, which is
caused by small differences of the two methods in $A_1(0)$ and $A_2(0)$
(recall that $A_0(0)=\left((M_1+M_2)A_1(0)-(M_1-M_2)A_2(0)\right)/2M_2$)
This discrepancy  exceeds the 15\% error bar quoted for the LCSR results
only marginally.
If the LCSR results at small $q^2$ and lattice results at large
$q^2$ are correct, our approach surely provides a realistic description of
the
form factors at all kinematically accessible $q^2$ values.

Extrapolating the form factors to the poles, we obtain
\begin{eqnarray}
\frac{g_{B^*B\pi}f_V^{(B^*)}}{2M_{B^*}}&=& 0.6\pm0.05, \qquad
\nonumber\\
\frac{g_{BB\rho}f_P^{(B)}}{2M_\rho}&=&1.4\pm0.2, \qquad
\nonumber\\
\frac{f_T^{(B^*)}}{f_V^{(B^*)}}&=&0.97\pm0.03, \qquad
\nonumber\\
\frac{g_{B^*B\rho}}{g_{B^*B\pi}}&=&1.2\pm0.1. \nonumber
\end{eqnarray}

It is convenient to define the quantity $\hat g$ according to the relation 
\begin{eqnarray}
\hat g=\frac{f_\pi}{2\sqrt{M_BM_{B^*}}}g_{BB^*\pi}. 
\end{eqnarray}
Using $f_V^{B^*}\simeq 200\;MeV$ gives the estimate
\begin{eqnarray}
g_{BB^*\pi}&=&32\pm5,\qquad 
\nonumber\\
\hat g&=&0.4\pm0.06  
\end{eqnarray}
The latter value is in good agreement with the lattice result $\hat
g=0.42\pm 0.04\pm0.08$
\cite{latg} and is only slightly smaller than $\hat g=0.5\pm 0.02$ \cite{mb2}
based on combining PCAC with our dispersion approach.

Table \ref{table:ratesb2pi} compares the calculated semileptonic decay rates 
from different theoretical approaches. 
\begin{table}
\caption{\label{table:ratesb2pi}
The $B\to(\pi\rho)l\nu$ Decay rates in units $|V_{ub}|^2{\rm ps^{-1}}$.}
\centering
\begin{tabular}{|c|c|c|c|}
Ref. & $\Gamma(B\to\pi\ell\nu)$  &  $\Gamma(B\to\rho\ell\nu)$  &
$\Gamma_L/\Gamma_T$   \\ 
\hline
This work &	  $8.0^{+0.8}_{-0.2}$ & $15.8 \pm 2.3$ & $0.88\pm0.08$  \\
ISGW2 QM \cite{isgw}  & 9.6 & 14.2 & 0.3   \\
Lat \cite{lat} & $8.5^{+3.4}_{-0.9}$ & $16.5^{+3.5}_{-2.3}$ &
$0.80^{+0.04}_{-0.03}$ \\ 
LCSR \cite{braun} & $-$    &   13.5$\pm$ 4.0 & 0.52$\pm$ 0.08  
\\
LCSR \cite{ball}  & 8.85    &	17.7	    &  1.12    
\end{tabular}
\end{table}
Combining the experimental data reported by CLEO  
${\cal B}(B^0\to\rho^- l^+\nu)=
(2.57\pm 0.29^{0.33}_{-0.46}\pm 0.41)\cdot 10^{-4}$ \cite{cleob2pi}
with our prediction for $\Gamma(B\to\rho l\nu)$ 
leads to the following estimate for the central value of $V_{ub}$: 
\begin{eqnarray}
|V_{ub}|=3.25\cdot 10^{-3}. 
\end{eqnarray}
Using the $B^0\to\pi^- l^+\nu$ measurement from Belle 
${\cal B}(B^0\to\pi^- l^+\nu)=
(1.28\pm 0.20\pm 0.26)\cdot 10^{-4}$
\cite{belle} 
and our prediction for $\Gamma(B\to\pi l\nu)$
leads to the central value 
\begin{eqnarray}
|V_{ub}|=3.22\cdot 10^{-3}.  
\end{eqnarray}
The perfect agreement between these values speaks in favour 
of the reliability of our predictions for the form factors. 

\vspace{3cm}
Summing up our results on the decays of the nonstrange heavy mesons,
we found no disagreement neither with the exisiting
experimental data nor with the available results of the lattice QCD or
sum rules in their specific regions of validity. The only exception is the
form factor $F_0$
at large $q^2$ in $B\to \pi$ and $D\to \pi$ decays, where our results are lying 
slightly below the lattice points. However, this disagreement can be
related to a strong sensitivity of $F_0$ at large $q^2$ to the details of
the pion wave function.
Small changes in the pion wave function, which only marginally affect $f_+$
and $f_-$, can change the form factor $F_0$. But such subtle effects are beyond the scope
of our present analysis.

In the next section we apply our model to the decays of strange heavy
mesons for which a few new parameters have to be introduced which are
specific to the description of  weak decays of strange heavy mesons
to light mesons.

\newpage
\subsection{Decays of the strange mesons $D_s$ and $B_s$ \label{d}}
Before dealing with these decays, we must first specify the slope parameters
of the
$B_s$ and the $D_s$ wave function. We obtain these parameters by applying
(\ref{fp})  and using
$f_{B_s}/f_B=1.1$ and $f_{D_s}/f_D=1.1$ in agreement with the lattice
estimates for these quantities \cite{lattice}.
The resulting values of the slope parameters are listed in Table
\ref{table:parameters}.
Since all other parameters have already been fixed the
calculation of the form factors is straight forward. The only exceptions are
the decays into the
$\eta, \eta', \phi$ final states. For these decays we need to know the
$\phi$
wave function, the mixing angle and the slope of the radial wave function
of the $s\bar s$ component in $\eta$ and $\eta'$. Our procedure of fixing
these parameters
are discussed in the relevant subsection.

\subsubsection{$D_s\to K, K^*$ }

These meson transition are driven by the charged-current $c\to d$ quark
transition. The results of the calculation are given in Table \ref{table:fitsds2k}.
The predictions
for the semileptonic decay rates are displayed in
Table \ref{table:ratesds2k}.
\begin{table}[hbt]
\caption{\label{table:fitsds2k}The calculated $D_s \to K,K^*$ transition
form
factors. $M_V=M_{D^*}=2.01\; GeV$, $M_P=M_{D}=1.87\; GeV$. 
For the form factors $F_+, F_T, V, A_0, T_1$ the fit formula Eq. (\protect\ref{fit1}) is used, 
for the other form factors - Eq. (\protect\ref{fit3}).}
\centering
\begin{tabular}{|c||c|c|c||c|c|c|c|c|c|c|}
 & \multicolumn{3}{c||}{$D_s \to K$} & \multicolumn{7}{c|}{$D_s \to K^*$}\\
\hline
          &$F_{+}$&$F_{0}$& $F_T$ & $V$   & $A_0$ &  $A_1$  & $A_2$ & $T_1$
&   $T_2$  & $T_3$   \\
\hline
$f(0)$     & 0.72 & 0.72  & 0.77  & 1.04  & 0.67  &  0.57  & 0.42  &  0.71
&   0.71   & 0.45    \\
$\sigma_1$ & 0.20 & 0.41  & 0.24  & 0.24  & 0.20  &  0.29  & 0.58  &  0.22
&$-$0.06   & 1.08    \\
$\sigma_2$ &      & 0.70  &       &       &       &  0.42  &       &
&   0.44   & 0.68
\end{tabular}
\end{table}

\begin{table}[hbt]
\caption{\label{table:ratesds2k}
The $D_s\to (K,K^*)l\nu$ decay rates in $10^{10}\;s^{-1}$, $|V_{cd}|=0.22$.}
\centering
\begin{tabular}{|c||c|c|c|c|}
Ref & $\Gamma(D_s\to K)$ & $\Gamma(D_s\to K^*)$ & $\Gamma(K^*)/\Gamma(K)$ &
$\Gamma_L/\Gamma_T$ \\
\hline
This work     &  0.63   &  0.38     &   0.6  &  1.21
\end{tabular}
\end{table}
For the coupling constants we obtain
\begin{eqnarray}
\frac{g_{D^*D_sK}f_V^{(D^*)}}{2M_{D^*}}&=&0.95\pm0.05, \qquad
\nonumber\\
\frac{g_{DD_sK^*}f_P^{(D)}}{2M_{K^*}}&=&1.85\pm 0.15,  \qquad
\nonumber\\
\frac{f_T^{(D^*)}}{f_V^{(D^*)}}&=&0.9\pm0.1,\qquad
\nonumber\\
\frac{g_{D^*D_sK^*}}{g_{D^*D_sK}}&=&1.15\pm0.15.\nonumber
\end{eqnarray}

\newpage
\subsubsection{$D_s\to \eta,\eta',\phi$}

These decay modes are induced by the charged current $c\to s$ quark
transition.
The pseudoscalar mesons $\eta$ and $\eta'$ are mixtures of the nonstrange
and the strange components with
the flavour wave functions
$\eta_n\equiv \frac{\bar u u +\bar d d}{\sqrt{2}}$ and $\eta_s=\bar s s$,
respectively,
\begin{eqnarray}
\eta &=&\cos(\varphi)\,\eta_n-\sin(\varphi)\,\eta_s \nonumber\\
\eta'&=&\sin(\varphi)\,\eta_n+\cos(\varphi)\,\eta_s,
\end{eqnarray}
with the angle $\varphi\simeq 40^o$ \cite{amn1}.

The decay rates of interest are
\begin{eqnarray}
\label{etas}
\Gamma(D_s\to \eta l\nu) &=& \sin^2(\varphi)
\Gamma(D_s\to \eta_s(M_\eta) l\nu)\nonumber\\
\Gamma(D_s\to \eta' l\nu) &=& \cos^2(\varphi)
\Gamma(D_s\to \eta_s(M_{\eta'}) l\nu).
\end{eqnarray}
Let us give a brief explanation of these formulas:
The semileptonic decay rates are determined by the form factor $f_+$.
The spectral representation of this form factor does not involve the
final meson mass explicitly. This means that for the $\bar s s$
component of both
$\eta$ and $\eta'$ we have to deal with the same form factor, which can be
expressed
through the radial wave function of this component.
On the other hand, the phase-space volume of the decay process is determined
by the physical meson masses, as indicated in (\ref{etas}).
It should be clear, however, that the $\eta_s$ is not an
eigenstate of the Hamiltonian and does not have a definite mass.

Assuming universality of the wave functions of the ground state pseudoscalar
$0^-$ nonet,
the radial wave function of the nonstrange component $\Psi_{\eta_n}$
coincides with the
pion radial wave function \cite{amn1}. The radial wave function
$\Psi_{\eta_s}$ should be determined independently.
From the
analysis of a broad set of processes the leptonic decay constant $f_s$ of
the strange component
$\eta_s$,
has been found to lie in the interval $f_s=(1.36\pm 0.04)f_\pi$
\cite{kroll}. This allows us to determine
the slope
parameter $\beta_{\eta_s}$ in such a way that the calculated value of $f_s$
lies in this interval, and the calculated ratio
$\Gamma(D_s \to \eta)/\Gamma(D_s \to \eta')$ agrees with the experimental
data for $\varphi=40^o$.
This procedure yields for the slope parameter $\beta_{\eta_s}=0.45$
\footnote{Another procedure of taking into account the SU(3) breaking
effects to
obtain $\Psi_{\eta_s}$ from $\Psi_{\eta_n}$ has been proposed in
\cite{amn1}.}.
For the slope parameter $\beta_{\phi}$ of the wave function of the
$\phi$-meson,
which is the vector $\bar ss$ state, we expect a value close to
$\beta_{\eta_s}$.

In fact, $\beta_\phi=0.45\;GeV$ leads to the $B_s\to\phi$ transition form
factors which agree well with the LCSR results at $q^2=0$ (see subsection
4).
With all other quark model parameters fixed from the description of the
nonstrange
heavy meson decays and by taking a simple Gaussian form of the radial wave
function,
the decay rate $\Gamma(D_s\to \phi l\nu)$ is a function of $\beta_\phi$.
This function has a minimum at the value $\beta_\phi=0.45\;GeV$;
nevertheless, the corresponding value of the decay rate is $1\sigma$ above
the central experimental value.

The results of our calculations are given in Tables \ref{table:fitsds2phi}
and \ref{table:ratesds2ss}.
\begin{table}[hbt]
\caption{\label{table:fitsds2phi}The calculated $D_s \to \eta_s,\phi$
transition form
factors. $M_V=M_{D_s^*}=2.11\; GeV$, $M_P=M_{D_s}=1.97\; GeV$. 
For the form factors $F_+, F_T, V, A_0, T_1$ the fit formula Eq. (\protect\ref{fit1}) is used, 
for the other form factors - Eq. (\protect\ref{fit3}).}
\centering
\begin{tabular}{|c||c|c|c|c|c||c|c|c|c|c|c|c|}
 &   \multicolumn{3}{r}{$D_s \to \eta_s(M_\eta)$}
 &   \multicolumn{2}{r||}{$D_s \to \eta_s(M_{\eta'})$}
 &   \multicolumn{7}{c|}{$D_s \to\phi$}\\
\hline
            &$F_{+}$&$F_{0}$&$F_{T}$&$F_{0}$& $F_T$
     & $V$   & $A_0$ &  $A_1$  & $A_2$ & $T_1$ &   $T_2$  & $T_3$   \\
\hline
$f(0)$      & 0.78  & 0.78  & 0.80  & 0.78  & 0.94
            &  1.10 &  0.73 &  0.64 & 0.47  &  0.77   &   0.77   & 0.46
\\
$\sigma_1$ & 0.23  & 0.33  & 0.24  & 0.21  & 0.24
            &  0.26 &  0.10 &  0.29 &  0.63   & 0.25  &   0.02   & 1.34
\\
$\sigma_2$ &       & 0.38  &       & 0.76  &
            &       &       &       &       &         &   2.01   & 0.45
\end{tabular}
\end{table}
\begin{table}[hbt]
\caption{\label{table:ratesds2ss}
The $D_s\to (\eta,\eta',\phi)l\nu$ decay rates in $10^{10}\;s^{-1}$,
$|V_{cs}|=0.975$.
The experimental rates are obtained from the corresponding branching ratios
using the
$D_s$ lifetime $\tau_{D_s}=0.495\pm 0.013\;ps$ from the 1999 update
\protect\cite{pdg}}
\centering
\begin{tabular}{|c||c|c|c|}
Ref & $\Gamma(D_s\to \eta)$ & $\Gamma(D_s\to \eta')$ & $\Gamma(D_s\to \phi)$
\\
\hline
This work     &  5.0            &  1.85           &   5.1    \\
Exp \cite{cleods2eta}&  5.2$\pm$1.3   &  2.0$\pm$0.8   &   4.04$\pm$1.01
\end{tabular}
\end{table}
For the coupling constants we obtain
\begin{eqnarray}
\frac{g_{D_s^*D_s\eta_s}f_V^{(D_s^*)}}{2M_{D_s^*}}&=&1.0\pm0.1, \qquad
\nonumber\\
\frac{g_{D_sD_s\phi}f_P^{(D_s)}}{2M_{\phi}}&=&1.6\pm 0.3,  \qquad
\nonumber\\
\frac{f_T^{(D_s^*)}}{f_V^{(D_s^*)}}&=&0.93\pm0.03,\qquad
\nonumber\\
\frac{g_{D_s^*D_s\phi}}{g_{D_s^*D_s\eta_s}}&=&1.08\pm0.04.
\nonumber
\end{eqnarray}

\newpage

\subsubsection{$B_s\to K,K^*$}

This mode is driven by the $b\to u$ charged current transition.
The only additional new parameter needed here is the slope of the $B_s$ wave
function. We
obtain it by using (\ref{fp}) and taking $f_{B_s}/f_B=1.1$. The results of
our
calculation are given in Table \ref{table:fitsbs2k}.
\begin{table}[hbt]
\caption{\label{table:fitsbs2k}
The calculated $B_s \to K, K^*$ transition form factors. $M_V=M_{B^*}=5.32\;
GeV$, $M_P=M_{B}=5.27\; GeV$. 
For the form factors $F_+, F_T, V, A_0, T_1$ the fit formula Eq. (\protect\ref{fit1}) is used, 
for the other form factors - Eq. (\protect\ref{fit3}).}
\centering
\begin{tabular}{|c||c|c|c||c|c|c|c|c|c|c|}
 & \multicolumn{3}{c||}{$B_s \to K$} & \multicolumn{7}{c|}{$B_s \to K^*$}\\
\hline
          &$F_{+}$&$F_{0}$& $F_T$ & $V$   & $A_0$ &  $A_1$  & $A_2$ & $T_1$
&   $T_2$  & $T_3$   \\
\hline
$f(0)$     & 0.31 & 0.31  & 0.31  & 0.38  & 0.37  &  0.29  & 0.26  &  0.32
&   0.32   & 0.23    \\
$\sigma_1$ & 0.63 & 0.93  & 0.61  & 0.66  & 0.60  &  0.86  & 1.32  &  0.66
&   0.98   & 1.42    \\
$\sigma_2$ & 0.33 & 0.70  & 0.30  & 0.30  & 0.16  &  0.60  & 0.54  &  0.31
&   0.90   & 0.62
\end{tabular}
\end{table}

These form factors lead to the following relations
\begin{eqnarray}
\frac{g_{B^*B_sK}f_V^{(B^*)}}{2M_{B^*}}&=&0.44\pm0.04, \qquad
\nonumber\\
\frac{g_{BB_sK^*}f_P^{(B)}}{2M_{K^*}}&=&1.3\pm 0.1, \qquad
\nonumber\\
\frac{f_T^{(B^*)}}{f_V^{(B^*)}}&=&0.95\pm0.05,\qquad
\nonumber\\
\frac{g_{B^*B_sK^*}}{g_{B^*B_sK}}&=&1.2\pm0.1.
\nonumber
\end{eqnarray}
The form factors at $q^2=0$ are compared with the LCSR predictions in Table
\ref{table:compbs2k}.
We observe some disagreement between our predictions and the LCSR
calculation which
gives smaller values for all the form factors. A closer look at the origin
of this discrepancy shows
that its source is the strength and sign of the SU(3)-breaking effects.
They lead to opposite
corrections in the two approaches.
\begin{table}[h]
\caption{\label{table:compbs2k}
Comparison of the QM and LCSR results on the $B_s\to K,K^*$ form factors at
$q^2=0$.}
\centering
\begin{tabular}{|c||c|c|c|c|c|c|}
 Ref.        & $V(0) $   & $A_1(0)$  &  $A_2(0)$  & $A_0(0)$ &$T_1(0)$ &
$T_3(0)$ \\
\hline
This work    & 0.38      &  0.29     &  0.26      & 0.37     & 0.32    &
0.23   \\
LCSR'98 \cite{lcsr1}
             & 0.262     &  0.19     &  0.164     & 0.254    & 0.22    &
0.16
\end{tabular}
\end{table}
To discuss these  SU(3) breaking effects, let us start with
$B\to \rho$, which in fact differs from the $B_s\to K^*$ only
by the flavour of the spectator quark, and move to $B_s\to K^*$ by
accounting
for the SU(3) violating effects:

Within the LCSR method there are two changes which affect the form factors:
first, the change $f_{B_s}\to f_B$
leads to an increase of the $B_s\to K^*$ form factors;
second, the change of the symmetric twist-two distribution amplitude of the
$\rho$-meson
to the asymmetric one of the $K^*$ meson
leads to a decrease of the form factors. The second
effect turns out to be much stronger than the first one with the result of
an overall decrease of the
form factors.

In the quark model, the same SU(3) breaking effects take place:
The change of the spectator mass (it determines the increase of
$f_{B_s}/f_{B}$)
and the change of the $K^*$ meson wave function (due to the change of both
the quark mass
and the slope parameter of the light meson wave function). Here the
influence of the slope of the heavy
meson wave function upon the form factor is only marginal. Therefore, the
resulting effect of these
changes leads to an increase of the form factors.

We want to point out that the difference between the results of the two
approaches
does not arise from specific effects (higher twists, higher radiative
corrections etc) which are present in the LCSRs but absent in the quark
model.
The observed difference is only due to the different strength of the SU(3)
violating
effects at the level of the twist-2 distribution amplitude. As was discussed
in \cite{amn1},
this distribution amplitude can be expressed through the radial soft wave
function of the meson.
The change of the quark-model wave function caused by SU(3) violating
effects does not induce a strong
asymmetry in the leading twist-2 distribution amplitude.

In view of the discrepancy between our results and the LCSR it would be
interesting
to have independent calculations of the $B_s\to K^*$ form factors at small
$q^2$ from the
3-point sum rules, as well as a lattice calculation for large $q^2$.

\newpage
\subsubsection{$B_s\to \eta,\eta',\phi$}

These weak meson transitions are induced by the FCNC $b\to s$ quark
transition.
The results of the form factor calculation are given in Table
\ref{table:fitsbs2phi} and compared with the
LCSR predictions at $q^2=0$ in Table \ref{table:compbs2phi}. The agreement
between the two values is satisfactory at least within the declared 15\%
accuracy of the LCSR predictions.
This allows us to expect that also the $D_s\to \phi, \eta,\eta'$ form
factors and the corresponding
decay rates (given earlier in subsection 2) are calculated reliably.
\begin{table}[hbt]
\caption{\label{table:fitsbs2phi}The calculated $B_s \to \eta_s,\phi$
transition form
factors. $M_V=M_{B_s^*}=5.42\; GeV$, $M_P=M_{B_s}=5.37\; GeV$. 
For the form factors $F_+, F_T, V, A_0, T_1$ the fit formula Eq. (\protect\ref{fit1}) is used, 
for the other form factors - Eq. (\protect\ref{fit3}).}
\centering
\begin{tabular}{|c||c|c|c|c|c||c|c|c|c|c|c|c|}
 &   \multicolumn{3}{r}  {$B_s \to \eta_s(M_\eta)$}
 &   \multicolumn{2}{r||}{$B_s \to \eta_s(M_{\eta'})$}
 &   \multicolumn{7}{c|} {$B_s \to\phi$}\\
\hline
            &$F_{+}$&$F_{0}$&$F_{T}$&$F_{0}$& $F_T$
     & $V$   & $A_0$ &  $A_1$  & $A_2$ & $T_1$ &   $T_2$  & $T_3$   \\
\hline
$f(0)$      & 0.36  & 0.36  & 0.36  & 0.36  & 0.39
            &  0.44 &  0.42 &  0.34 & 0.31  &  0.38   &   0.38   & 0.26
\\
$\sigma_1$  & 0.60  & 0.80  & 0.58  & 0.80  & 0.58
            &  0.62 &  0.55 &  0.73 &  1.30   & 0.62  &   0.83   & 1.41
\\
$\sigma_2$  & 0.20  & 0.40  & 0.18  & 0.45  & 0.18
            &  0.20 &  0.12 &  0.42 &  0.52 &  0.20   &   0.71   & 0.57
\end{tabular}
\end{table}
\begin{table}[h]
\caption{\label{table:compbs2phi}
Comparison of the QM and LCSR results on the $B_s\to \phi$ form factors at
$q^2=0$.}
\centering
\begin{tabular}{|c||c|c|c|c|c|c|}
 Ref.        & $V(0) $   & $A_1(0)$  &  $A_2(0)$  & $A_0(0)$ &$T_1(0)$ & $T_3(0)$ \\
\hline
This work    & 0.44      &  0.34     &  0.31      & 0.42     & 0.38    &  0.26   \\
LCSR'98 \cite{lcsr1}
             & 0.433     &  0.296    &  0.255     & 0.382    & 0.35    &  0.25
\end{tabular}
\end{table}
For the coupling constants we obtain
\begin{eqnarray}
\frac{g_{B_s^*B_s\eta_s}f_V^{(B_s^*)}}{2M_{B_s^*}}&=&0.6\pm0.05, \qquad
\nonumber\\
\frac{g_{B_sB_s\phi}f_P^{(B_s)}}{2M_{\phi}}&=&1.5\pm 0.1, \qquad
\nonumber\\
\frac{f_T^{(B_s^*)}}{f_V^{(B_s^*)}}&=&0.95\pm0.05, \qquad
\nonumber\\
\frac{g_{B_s^*B_s\phi}}{g_{B_s^*B_s\eta_s}}&=&1.13\pm0.06.\nonumber
\end{eqnarray}

\newpage
\subsection{Discussion}
In this Chapter we have calculated numerous form factors of heavy meson transitions to light
mesons which are relevant for the semileptonic (charged current) and penguin
(flavor-changing neutral current) decay processes. 
Our approach is based on evaluating the triangular decay graph within a relativistic
quark model which has the correct analytical properties and satisfies all known general 
requirements of long-distance QCD.

The model connects different decay channels in a unique way and gives the
form factors for all relevant $q^2$ values. The disadvantage of the constituent quark model
connected with its dependence on ill-defined parameters such as the effective constituent 
quark masses, have been reduced by
using several constraints: the quark masses and the slope parameters of the
wave functions are 
chosen such that the calculated form factors reproduce the lattice results
for the $B\to \rho$
form factors at large $q^2$ and the observed integrated rates of the
semileptonic $D\to K,K^*$ decays.

Our main results from this Chapter are as follows:
\begin{itemize}
\item
In spite of the rather different masses and properties of mesons
involved in weak transitions, all existing data on the form factors, both
from
theory and experiment, can be understood in our quark picture.
Namely, all the form factors are essentially described by the few degrees of
freedom of constituent quarks, i.e. their wave functions and their effective
masses. Details of the soft wave functions are not crucial; only the
spatial extention  of these wave functions of order of the confinement scale
is important.
In other words, only the meson radii are essential.

\item
The calculated transition form factors are in all cases in good agreement
with the results available from lattice QCD and from sum rules
in their specific regions of validity. The only exception is a disagreement
with the LCSR results for the $B_s\to K^*$ transition where
we predict larger form factors. This disagreement is caused by a different
way of taking into account the
SU(3) violating effects when going from $B\to\rho$ to $B_s\to K^*$ and is
not related to specific
details of the dynamics of  the decay process.
We suspect that the LCSR method overestimates the SU(3) breaking
in the long-distance region
but this problem deserves further clarification. 

\item
We have estimated the products  
of the meson weak and strong coupling constants by using the fit formulas for 
the form factors for the extrapolation to the meson pole. 
The error of such estimates connected with the errors in the extrapolation 
procedure is found to be around 5-10\%. 
\end{itemize}
We cannot provide for definite error estimates of our predictions for the form factors 
because of the approximate
character of the constituent quark model. However from the fine agreement
obtained in cases where checks are possible,
we believe that the actual accuracy of our predictions for the form factors
is around 10\%.
Since some parameters have been fixed by using lattice results and have also
been tested
using the sum rule predictions, further improvements of the accuracy of our
predictions will follow
if these approaches attain smaller errors.
Of course, each precisely measured decay will also allow a more accurate
determination of the
parameters of the model and thus can be used to diminish the errors at least
for closely related decays.

\newpage
\section{\label{sec:iv}Weak annihilation in the rare radiative $B\to \rho\gamma$ decay 
and the $B\to\gamma \ell\nu$ form factors}
Results obtained in the previous Chapter allow us to 
fully describe semileptonic $B_{(s)}$ and $D_{(s)}$ decays.  
We now discuss rare radiative decays of heavy mesons 
induced by the flavour-changing neutral currents (FCNC) such as 
$B\to K^*\gamma$ and $B\to\rho\gamma$ decays. This Chapter is based on Refs. 
\cite{bmns,ms1}.

The investigation of rare semileptonic $B$ decays induced by the 
flavour-changing neutral current transitions $b\to s$ and $b\to d$ represents an important 
test of the standard model or its extentions. 
Rare decays are forbidden at the tree level in the standard model and occur
at the lowest order only through 
loop diagrams. This fact opens the possibility to probe the structure of the 
electroweak sector at large mass scales due to contributions of virtual particles 
in the loop diagrams. Interesting
information about the structure of the theory is contained in the Wilson coefficents in the 
effective Hamiltonian 
which describes the $b\to s,d$ transition at low energies. 
These Wilson coefficients take different values in different theories with testable 
consequences in rare $B$ decays 
\cite{hewett,rizzo,mmoriond,burdman_asym,mnsplb2,mnsplb3,mnsplb4,aliball}.  

Among rare $B$ decays the radiative decays $b\to s\gamma$ and $b\to d\gamma$ have the 
largest probabilites. The $b\to s\gamma$ transitions are CKM favoured and have larger 
branching ratios than the $b\to d\gamma$ transitions.  
The $b\to s\gamma$ transition has been observed by CLEO in the exclusive channel    
$B\to K^*\gamma$ in 1993 \cite{cleo_rare_excl} and measured inclusively in 1995 
\cite{cleo_rare_incl}. The $B\to \rho\gamma$ decay will be extensively studied by BaBar and BELLE. 

Any reliable extraction of the perturbative
(short-distance) effects encoded in the Wilson coefficients of the effective
Hamiltonian requires an accurate separation of the nonperturbative
(long-distance) contributions, which therefore should be known with
high accuracy. 

In rare exclusive semileptonic decays one faces several types of 
long-distance contributions which arise in the meson transition amplitude of the 
effective Hamiltonian, which contains penguin and 4-quark operators 
(see \cite{gp} for more details).  

The dominant long-distance effects for the radiative decays arise from 
the electromagnetic penguin operator and is given in terms of the form factor 
$T_1(q^2=0)$. 

The 4-quark operators generate many different contributions: 

One of them is the so-called long-range charming penguins, i.e. intermediate 
$c\bar c$ states in the $q^2$-channel which include the $c\bar c$ resonances 
($\psi, \psi'$, ...) and the charmed hadronic continuum. 
The charming penguins are crucial for rare semileptonic decays $b\to (s,d)l^+l^-$, 
but provide only a small contribution for the radiative decays $b\to (s,d)\gamma$ 
\cite{kruger,nonfact,mplb3}: In the factorization approximation 
the $c\bar c$ contribution precisely vanishes due to the gauge invariance \cite{mplb3}, 
whereas the nonfactorizable effects turn out to be only at the level of few percent
\cite{nonfact}. 

Another effect of the 4-quark operators is the so-called weak annihilation which 
corresponds to the annihilation of quarks in the channel of the initial or 
final mesons.  

In the $B\to K^*\gamma$ decay the weak annihilation is negligible compared to the 
penguin effect: it is suppressed by two powers of the small parameter $\lambda\simeq 0.2$ 
of the Cabibbo-Kobayashi-Maskawa (CKM) matrix. In $B\to \rho\gamma$ 
both effects have the same order in $\lambda$ and must be taken into account. 

The penguin contribution is known rather well. 
On the other hand, the WA in $B\to \rho\gamma$ has been studied in less detail: the relevant 
form factors were analysed within sum rules \cite{wa-sr1,wa-sr2} 
and perturbative QCD \cite{korch}. However, some contributions to these form factors were 
neglected. These contributions may be relevant if precise measurements become available. 

We now apply the dispersion approach to the weak annihilation in the $B^-\to \rho^-\gamma$ decay.  

In the factorization approximation \cite{neubertstech} the weak-annihilation amplitude can be represented as the product 
of meson leptonic decay constants and matrix elements of the weak current between 
meson and photon. The latter contain the meson-photon transition form factors and 
contact terms which are determined by the equations of motion. The photon can be emitted from the 
loop containing the $b$ quark which is described by $B\gamma$ transition form factors. It can also 
be emitted from the loop containing only light quarks described by the $\rho\gamma$ 
transition form factors (Fig \ref{fig:diag}).  

\begin{center}
\begin{figure}[hb]
\mbox{\epsfig{file=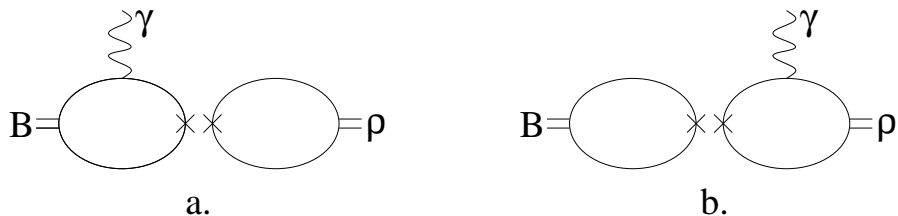,width=12cm}}
\caption{\label{fig:diag}
Diagrams describing the weak annihilation process for $B\to\rho\gamma$ in the 
factorization approximation: 
(a) The photon is emitted from the loop containing the $b$ quark, 
(b) The photon is emitted from the loop containing only light quarks.}
\end{figure}
\end{center}

We calculate the $B\gamma$ form factors within our relativistic dispersion approach which 
expresses these form factors in terms of the $B$ meson wave function. 
We demonstrate that the form factors calculated by the dispersion approach 
behave in the limit $m_b\to\infty$ in agreement with perturbative QCD. 
The $B$ meson wave function was previously tested in the $B\to$ light meson weak decays and 
is known quite well, allowing us to provide reliable numerical estimates for the $B\gamma$ form factors.    

The $\rho\gamma$ transition form factor is related to the divergence of the vector and 
axial-vector currents. In the case of the axial-vector current it is   
proportional to the light-quark masses if the classical equation of motion is applied. 
Because the quark momenta in the loop are high, these masses have to be identified with 
current quark masses. For this reason the corresponding $\rho\gamma$ form factor was neglected in previous 
analyses \cite{wa-sr1,wa-sr2}.    
We find however that this argument is not correct: this form factor 
remains finite in the limit of vanishing light quark mass $m\to 0$ and behaves like 
$\sim M_\rho f_\rho/M_B^2$ which means the violation of the classical equation of motion.  
Here we present the result for the $\rho\gamma$ transition form factor. 
A detailed discussion of the anomaly appearing for the matrix element 
$\langle \rho\gamma|\partial_\nu A_\nu|0\rangle$ can be found in Ref. \cite{ms1}

Finally, we provide numerical estimates of the weak-annihilation amplitude 
taking into account the $B\gamma$, $\rho\gamma$ transition form factors, and the 
contact term contributions.   

In section \ref{sec:iv.1} the effective Hamiltonian for the $b\to d$ transition and the general 
structure of the amplitude are presented. In Section \ref{sec:iv.2} we discuss the photon emission from the 
$B$ meson loop and obtain the $B\gamma$ transition form factors within the dispersion 
approach.  
Section \ref{sec:iv.3} contains results for the $\rho\gamma$ transition form factors.  
In Section \ref{sec:iv.4} the numerical estimates are given. The concluding Section \ref{sec:iv.5}
summarises our results for the weak annihilation. 

\subsection{\label{sec:iv.1}The effective Hamiltonian, the amplitude and the decay rate}
The amplitude of the weak radiative 
$B\to \rho$ transition is given by the matrix element of the effective Hamiltonian 
for the $b\to d$ transition 
\begin{eqnarray}
A(B\to\rho\gamma)=\langle \gamma(q_1)\rho(q_2)|H_{\rm eff}(b\to d)|B(p) \rangle,  
\end{eqnarray}
where $p$ is the $B$ momentum, $q_2$ is the $\rho$ momentum, and $q_1$ is the photon
momentum, $p=q_1+q_2$, $q_1^2=0$, $q_2^2=M_\rho^2$, $p^2=M_B^2$. 
The effective weak Hamiltonian has the structure\cite{nunu,gws,burasmuenz,ali}:
\begin{eqnarray}
\label{Heff}
H_{\rm eff}(b\to d) &=& \frac{G_F}{\sqrt{2}}\,\xi_t
C_{7\gamma}(\mu){\cal O}_{7\gamma}
-\frac{G_F}{\sqrt{2}}{\xi_u}
\left (C_1(\mu){\cal O}_1+C_2(\mu){\cal O}_2\right), 
\end{eqnarray}
where only operators relevant for the penguin and weak ahnnihilation effects are listed. 
$G_F$ is the Fermi constant, $\xi_q=V^*_{qd}V_{qb}$, $C_i$'s are the Wilson 
coefficients and ${\cal O}_i$'s are the basis operators 
\begin{eqnarray}
{\cal O}_{7\gamma} &=& \frac{e}{8\pi^2}\,
\bar d_{\alpha}\sigma_{\mu\nu}
m_b(\mu)\left (1+\gamma_5\right )\, b_{\alpha}\, F_{\mu\nu}, \\
{\cal O}_1 &=& \bar d_{\alpha}\gamma_{\nu}(1-\gamma_5)u_{\alpha}\;
\bar u_{\beta}\gamma_{\nu}(1-\gamma_5) b_{\beta},\nonumber
\\
{\cal O}_2 &=& \bar d_{\alpha}\gamma_{\nu}(1-\gamma_5) u_{\beta}\; 
\bar u_{\beta}\gamma_{\nu}(1-\gamma_5) b_{\alpha}.  
\end{eqnarray}
with $e=\sqrt{4\pi\alpha_{\rm em}}$. Other notations are given in Eq. (\ref{notations}).  
 
The amplitude can be parametrized as follows  
\begin{eqnarray}
A(B^-\to\rho^-\gamma)=\frac{eG_F}{\sqrt{2}}
\left[
\epsilon_{q_1\epsilon^\ast_1 q_2 \epsilon_2^\ast}F_{\rm PC}
+i \epsilon_2^{\ast\nu}\epsilon_1^{\ast\mu} \left(g_{\nu\mu}\,pq_1-p_\mu q_{1\nu}\right)F_{\rm PV}
\right], 
\end{eqnarray}
where $F_{\rm PC}$ and $F_{\rm PV}$ are the parity-conserving and 
parity-violating invariant amplitudes, respectively. 
$\epsilon_2$($\epsilon_1$) is the 
$\rho$-meson (photon) polarization vector. We use the short-hand notation 
$\epsilon_{abcd}=\epsilon_{\alpha\beta\mu\nu}a^{\alpha}b^{\beta}c^{\mu}d^{\nu}$ 
for any 4-vectors $a,b,c,d$. 

For the decay rate one finds 
\begin{eqnarray}
\label{rate}
\Gamma(B^-\to\rho^-\gamma)=\frac{G^2_F\,\alpha_{em}}{16}M_B^3
\left(1-{M^2_\rho}/{M_B^2}\right)^3
     \left( |F_{\rm PC}|^2+|F_{\rm PV}|^2 \right). 
\end{eqnarray}

\subsubsection{The penguin amplitude}
The main contribution to the amplitude is given by the  
electromagnetic penguin operator $O_{7\gamma}$: 
\begin{eqnarray}
A_{\rm peng}(B\to\rho\gamma)&=&
-\frac{eG_F}{\sqrt{2}}\xi_t C_7\frac{m_b}{2\pi^2} T_1(0)
\left(
\epsilon_{q_1\epsilon_1^* q_2 \epsilon_2^*}
+i\epsilon_2^{\ast\nu}\epsilon_1^{\ast\mu}(g_{\nu\mu}\,pq_1-p_\mu q_{1\nu})
\right),
\end{eqnarray}
where $T_1$ is the form factor of the $B\to\rho$ transition through 
the tensor current \cite{wsb,m1,m2,ms}.
The corresponding contribution to the invariant amplitudes is therefore 
\begin{eqnarray}
\label{fpeng}
F^{\rm peng}_{\rm PC}=F^{\rm peng}_{\rm PV}=-\xi_t C_7\frac{m_b}{2\pi^2}T_1(0). 
\end{eqnarray}

\subsubsection{The weak annihilation amplitude}
The radiative $B\to\rho \gamma$ transition also receives contribution from the 
4-fermion operators ${\cal O}_1$ and ${\cal O}_2$.  
For the charged $B^-\to\rho^-(q_2)\gamma(q_1)$ transition the corresponding amplitude reads 
\begin{eqnarray}
\label{WA}
A_{\rm WA}(B^-\to\rho^-\gamma)&=&-\frac{G_F}{\sqrt{2}}\xi_u 
\langle \rho(q_2)\gamma(q_1)|\bar d\gamma_\nu(1-\gamma_5)u\cdot 
\bar u\gamma_\nu(1-\gamma_5)b|B(p)\rangle,  
\end{eqnarray}  
In what follows we suppress the lable WA.   
Neglecting the nonfactorizable soft-gluon exchanges, i.e. assuming vacuum saturation, 
the complicated matrix element in Eq. (\ref{WA}) is reduced to simpler quantities - 
the meson-photon matrix elements of the bilinear quark currents and the 
meson decay constants. The latter are defined as usual  
\begin{eqnarray}
\langle \rho(q_2)|\bar d\gamma_\nu u|0\rangle &=& \epsilon_{2\nu}^\ast M_\rho f_\rho, \qquad  f_\rho>0,
\nonumber \\
\langle 0|\bar u\gamma_\nu \gamma_5 b|B(p)\rangle &=& ip_\nu f_B,\qquad  f_B>0.
\end{eqnarray}
It is convenient to isolate the parity-conserving contribution which emerges from the
product of the two equal-parity currents, and the parity-violating contribution 
which emerges from the product of the two opposite-parity currents. 

\vspace{.4cm}
\underline{\it The parity-violating contribution}
\vspace{.4cm}

The parity-violating contribution to the weak annihilation amplitude 
has the form  
\begin{eqnarray}
\label{apva}
A_{\rm PV}(B\to\rho\gamma)&=&\frac{G_F}{\sqrt{2}}\xi_u a_1  
\left\{
\langle \rho\gamma|\bar d \gamma_\nu u|0 \rangle 
\langle 0|\bar u \gamma_\nu\gamma_5 b|B \rangle  
+
\langle \rho|\bar d \gamma_\nu u|0 \rangle 
\langle \gamma|\bar u \gamma_\nu\gamma_5 b|B \rangle \right\}. 
\end{eqnarray}
Here $a_1$ is an effective Wilson coefficient, which we take as $a_1=C_1+C_2/N_c$ 
at the scale $\sim$ 5 GeV.  

It is convenient to denote 
\begin{eqnarray}
A^{\rm PV}_1=\langle \rho^-(q_2)|\bar d \gamma_\nu u|0 \rangle 
\langle \gamma(q_1)|\bar u \gamma_\nu\gamma_5 b|B^-(p) \rangle 
\end{eqnarray}
and 
\begin{eqnarray}
A^{\rm PV}_2=
\langle \rho^-(q_2)\gamma(q_1)|\bar d \gamma_\nu u|0 \rangle 
\langle 0|\bar u \gamma_\nu\gamma_5 b|B^-(p) \rangle.  
\end{eqnarray}

\noindent 1. Let us start with $A^{\rm PV}_1$. 
One can write  
\begin{eqnarray}
\langle \gamma(q_1)|\bar u \gamma_\nu\gamma_5 b|B^-(p) \rangle=
e\,\epsilon_1^{\ast\mu} T^B_{\mu\nu} 
\end{eqnarray}
where 
\begin{eqnarray}
T^B_{\mu\nu}(p,q_1)&=&i\int dx e^{iq_1x}\langle 0|T(J_\mu(x),\bar u\gamma_\nu \gamma_5 b)|B^-(p)\rangle,
\end{eqnarray}
and 
\begin{eqnarray}
J_\mu(x)=\frac23\bar u\gamma_\mu u
-\frac13 \bar d\gamma_\mu d
-\frac13 \bar b\gamma_\mu b  
\end{eqnarray}
is the electromagnetic quark current. 

The amplitude $T^B_{\mu\nu}$ in general contains 5 independent Lorentz structures 
and can be parametrised in various ways.  

A possible way is to write $T^B_{\mu\nu}$ as follows
\begin{eqnarray}
\label{qq}
T^B_{\mu\nu}=T^{\perp}_{\mu\nu}+\frac{ip_{\mu}p_\nu}{pq_1} C_1
+\frac{ip_{\mu}q_\nu}{pq_1} C_2, 
\end{eqnarray}
where 
\begin{eqnarray}
\label{tperpL}
T^{\perp}_{\mu\nu}&=&i\left(g_{\mu\nu}pq_1-p_{\mu}q_{1\nu}\right)F_{1A}(q_1^2)
\nonumber\\ 
&+&i\left(q_1^2\,p_{\mu}-{pq_1}\,q_{1\mu}\right)q_{1\nu}F_{2A}(q_1^2)
\nonumber\\ 
&+&i\left(q_1^2\,p_{\mu}-{pq_1}\,q_{1\mu}\right)p_{\nu}F_{3A}(q_1^2).  
\end{eqnarray}
is transverse with respect to $q_{1\mu}$, 
\begin{eqnarray}
\label{tperp}
q_1^\mu \,T^{\perp}_{\mu\nu}=0.  
\end{eqnarray} 
The Lorentz structures in this expansion are chosen such that they contain no singularity for
$q_1^2\to 0$. 

The invariant amplitudes $C_1$ and $C_2$ in the longitudinal structure can be determined using 
the conservation of the 
electromagnetic current $\partial_\mu J_\mu=0$, which leads to the relation 
\begin{eqnarray}
\label{eq3}
q_{1\mu} T^B_{\mu\nu}(p,q_1)&=&-\langle 0|[\hat Q, \bar d\gamma_\nu \gamma_5 u]|B^-(p)\rangle
\nonumber\\
&=&iQ_Bf_B p_\nu.  
\end{eqnarray}
This gives 
\begin{eqnarray}
C_1&=&Q_Bf_B, 
\nonumber\\
C_2&=&0. 
\end{eqnarray}
In particular, $C=-f_B$ for the $B^-$ meson. Notice however, that the longitudinal part of the amplitude 
in this form is {\it not a contact term}.\footnote{We recall that according to the usual definition, 
a contact term is a quantity which is represented by the delta-function and its dirivatives in the coordinate 
space. One can then easily check, that the Fourier transform of the expression $\frac{ip_{\mu}p_\nu}{pq_1}$ 
is not a contact term.} This means that some of the invariant amplitudes in the transverse part 
are mixtures of the true form factors (i.e. quantities with definite and known analytic properties) and 
remnants of the contact terms. 

After setting $q_1^2=0$ and multiplying by $\epsilon_1^{\ast\mu}$ 
(this very combination is needed for $A_1^{\rm PV}$) only some of the Lorentz structures 
contribute and we find 
\begin{eqnarray}
\label{ttt}
\epsilon_1^{\ast\mu}T^B_{\mu\nu}|_{q_1^2=0}=i\epsilon_1^{\ast\mu}
\left\{(g_{\mu\nu}pq_1-p_{\mu}q_{1\nu})F_{1A}(0)-p_\mu p_{\nu} 
\frac{2f_B}{M_B^2-M_\rho^2}\right\},
\end{eqnarray}
where we have used the relation $q_1p=\frac{1}{2}(M_B^2-M_\rho^2)$ for $q_1^2=0$. 

The parametrization of Eq. (\ref{qq}) is of course not unique and there are other 
possibilities. There is however only one choice when the longitudinal part is 
a pure contact term. Such a choice, which is also the most natural one, is prompted by the structure of 
the Feynman diagram:  

Let us rewrite the usual electromagnetic coulping of the quark as follows ($q_1=k'-k$):
\begin{eqnarray}
(m+\hat k')\gamma_\mu(m+\hat k)
=(m+\hat k')\left\{\gamma_\mu-\hat q_1\,\frac{q_{1\mu}}{q_1^2} \right\}(m+\hat k)+
\frac{q_{1\mu}}{q_1^2}\,\left[(k'^2-m^2) (m+\hat k)-(k^2-m^2) (m+\hat k')\right].
\end{eqnarray} 
The first term is explicitly transverse with respect to $q_{1\mu}$ and leads to $T^\perp_{\mu\nu}$. 

\noindent
The second term, containing the factors $(k^2-m^2)$ and $(k'^2-m^2)$, leads to the contact term 
$$ip_\nu\frac{q_{1\mu}}{q_1^2}f_B.$$ 

Finally, we come to the following parametrization  
\begin{eqnarray}
\label{qqprime}
T^B_{\mu\nu}=T^{'\perp}_{\mu\nu}+\frac{iq_{1\mu}p_\nu}{q^2_1} C'_1
+\frac{iq_{1\mu}q_{1\nu}}{q^2_1} C'_2, 
\end{eqnarray}
with 
\begin{eqnarray}
\label{tperpLprime}
T^{'\perp}_{\mu\nu}&=&i\left(g_{\mu\nu}-\frac{q_{1\mu}q_{1\nu}}{q_1^2}\right)pq_1\,F'_{1A}(q_1^2)
\nonumber\\ 
&+&i\left(p_{\mu}-\frac{pq_1}{q_1^2}\,q_{1\mu}\right)q_{1\nu}F'_{2A}(q_1^2)
\nonumber\\ 
&+&i\left(p_{\mu}-\frac{pq_1}{q_1^2}\,q_{1\mu}\right)p_{\nu}F'_{3A}(q_1^2).  
\end{eqnarray}
Notice, that this is the only parametrization of the amplitude, which provides a distinct separation 
of the amplitude: 
form factors in the gauge-invariant transverse part of the amplitude, and contact terms in the 
longitudinal part of the amplitude. 

Clearly, $F'_{1A}=F_{1A}$, whereas other form factors of the sets $F$ and $F'$ are different.  
Now the Lorentz structures contain explicit singularities for $q_1^2=0$ which must cancel each
other in the amplitude $T^B_{\mu\nu}$. This leads to the constraints on the form factors at
$q_1^2=0$ 
\begin{eqnarray}
-q_1p\,F_{1A}'(0)-q_1p\,F_{2A}'(0)+C'_2&=&0, 
\nonumber\\
\label{relationsffs}
-q_1p\,F_{3A}'(0)+C'_1&=&0. 
\end{eqnarray}
Equation (\ref{eq3}) gives 
\begin{eqnarray}
C'_1&=&-f_B, 
\nonumber\\
C'_2&=&0, 
\end{eqnarray}
and hence 
\begin{eqnarray}
F'_{1A}(0)&=&-F'_{2A}(0),
\nonumber\\
F'_{3A}(0)&=&-\frac{2f_B}{M_B^2-M_\rho^2}. 
\end{eqnarray}
By virtue of the relations (\ref{relationsffs}), for the amplitude $A^{\rm PV}_1$ at $q_1^2=0$ we find  
\begin{eqnarray}
A^{\rm PV}_1&=&
ie\,f_\rho M_\rho \epsilon_1^{\ast\mu}\epsilon_2^{\ast\nu}\left\{
g_{\mu\nu}pq_1 F_{1A}(0)
+p_{\mu}q_{1\nu}F_{2A}(0)
-p_\mu p_{\nu}F_{3A}(0)\right\}
\nonumber\\
&=&
ie\,f_\rho M_\rho \epsilon_1^{\ast\mu}\epsilon_2^{\ast\nu}\left\{
(g_{\mu\nu}pq_1-p_{\mu}q_{1\nu})F_{1A}(0)-p_\mu q_{1\nu}\frac{2f_B}{M_B^2-M_\rho^2}\right\}. 
\end{eqnarray}
Notice that the contact term does not contribute to the amplitude {\it directly}, but 
{\it indirectly} determines the value of the form factor $F_{3A}(0)$.

\noindent 2. Let us now turn to $A_2^{\rm PV}$. 

Using the equation of motion for the quark fields ($Q_u=2/3\;e$, $Q_d=-1/3\;e$)
\begin{eqnarray}
i\gamma_\nu\partial^\nu q(x)&=& m q(x) - Q_q A_\nu \gamma^\nu q(x), 
\nonumber\\
i\partial^\nu \bar q(x)\gamma_\nu&=& -m \bar q(x) + Q_q A_\nu \bar q(x)\gamma^\nu , 
\end{eqnarray}
one obtains for $A_2^{\rm PV}$ 
\begin{eqnarray}
A^{\rm PV}_2&=&ip_\nu f_B \langle \rho^-\gamma|\bar d \gamma_\nu u|0 \rangle
\nonumber\\
&=&
f_B \langle \rho^-\gamma|\partial_\nu(\bar d \gamma_\nu u)|0 \rangle
\nonumber\\
&=&-(Q_d-Q_u)e\,\langle \rho^-\gamma|A^\nu \bar d \gamma_\nu u|0 \rangle
\nonumber\\
&=&e\,\epsilon_1^{\ast\mu}\epsilon_2^{\ast\nu} g_{\mu\nu}f_\rho M_\rho f_B +O(m_u,m_d). 
\end{eqnarray}  
Now the full contribution is due to the contact terms only, since the 
transverse form-factor part of the amplitude  $\langle \rho^-\gamma|\bar d \gamma_\nu u|0 \rangle$ 
vanishes after the multiplication by $p_\nu$.  

\noindent 3. For the sum we find 
\begin{eqnarray}
A^{\rm PV}_1+A^{\rm PV}_2&=&ie\,f_\rho M_\rho \epsilon_1^{\ast\mu}\epsilon_2^{\ast\nu}
\left(g_{\mu\nu}pq_1-p_{\mu}q_{1\nu}\right)\left[F_{1A}(0)+\frac{2f_B}{M_B^2-M_\rho^2}\right]
\end{eqnarray}  
Summing the contributions of the photon emission from the $B$-meson loop and the $\rho$-meson 
loop gives the amplitude $A_{\rm PV}$ which can be represented in the form 
$A_{\rm PV}=\epsilon_1^{\ast\mu} T_\mu$ with  
$q_1^\mu T_\mu=0$ as required by gauge invariance. 
The weak-annihilation contribution to the form factor $F_{\rm PV}$ 
for the $B^-\to\rho^-\gamma$ decay takes the form (after redefining $F_{1A}=2F_A/M_B$)  
\begin{eqnarray}
\label{fpv}
F^{\rm WA}_{\rm PV}=\xi_u a_1 f_\rho M_\rho\;\left[\frac{2F_{A}}{M_B}+\frac{2f_B}{M_B^2-M_\rho^2}\right].
\end{eqnarray}
Notice that in addition to the form factor contribution $\sim F_{A}$, there is a contribution 
proportional to $f_B$. The latter is determined by the contact terms: namely,  
the contact terms which are present in the amplitudes of the photon
emission from the $B$ meson loop and from the $\rho$-meson loop do not cancel each other, 
but lead to a nonvanishing contribution.\footnote{ 
Here we disagree with the recent statement of \cite{khodjct} that contact terms cancel each other 
and do not contribute to the final amplitude. The conclusion of \cite{khodjct} is based on choosing 
the longitudinal part of the amplitude in the form 
$\left\{g_{\mu\nu}+\frac{q_{2\mu}q_{2\nu}}{q_2q_1}\right\}f_B$ 
which is {\it not a contact term} according to the usual definition of the latter. 
As a result, the invariant amplitude of \cite{wa-sr1,wa-sr2,khodjct} at $q_1^2=0$ corresponds to the combination 
$F_A+\frac{M_Bf_B}{M_B^2-M_\rho^2}$.  
We take this relationship into account when comparing numerical results from different approaches 
later in this section.} 
In the amplitude of the $B^+\to\rho^+\gamma$ decay  
the contact term proportional to $f_B$ changes sign, and also 
$F_A^{B^+}=-F_A^{B^-}$ as can be obtained from charge conjugation of 
the amplitude $A_{\rm PV}$. For the $B^0\to\rho^0\gamma$ decay the contact term 
proportional to $f_B$ is absent. 

\vspace{.4cm}
\underline{\it The parity-conserving amplitude}
\vspace{.4cm}

This amplitude reads  
\begin{eqnarray}
\label{apc}
A_{\rm PC}(B\to\rho\gamma)&=&-\frac{G_F}{\sqrt{2}}\xi_u a_1 
\left\{\langle \rho|\bar d\gamma_\nu u|0 \rangle \langle \gamma|\bar u \gamma_\nu b|B \rangle 
+
\langle \gamma\rho|\bar d\gamma_\nu \gamma_5 u |0 \rangle 
\langle 0| \bar u\gamma_\nu \gamma_5 b|B \rangle \right\}. 
\end{eqnarray}
The $B\to\gamma$ amplitude from the first term in the brackets contains the form factor
$F_V(q_2^2=M_\rho^2)$:   
\begin{eqnarray}
\langle \gamma(q_1)|\bar u\gamma_\nu b|B(p)\rangle = 
e\,\epsilon_{\nu\mu p q_1} \epsilon_1^{\ast\mu} \;\frac{2F_V}{M_B}.
\end{eqnarray}
The second term in (\ref{apc}) can be reduced to the divergence of the 
axial-vector current and contains another form factor, $G_V$: namely,   
\begin{eqnarray}
\langle 0|\bar u\gamma_\nu \gamma_5 b|B\rangle 
\langle \gamma\rho|\bar d\gamma_\nu\gamma_5 u|0 \rangle&=&
f_B \langle \gamma\rho|\partial_\nu \bar d\gamma_\nu\gamma_5 u |0 \rangle
\nonumber\\
&=&
e\,f_B G_V \epsilon_{q_1\epsilon_1^\ast q_2\epsilon_2^\ast}. 
\end{eqnarray}
Thus, the weak annihilation contribution to $F_{\rm PC}$ reads  
\begin{eqnarray}
\label{fpc}
F^{\rm WA}_{\rm PC}=\xi_u a_1 M_\rho f_\rho\left[\frac{2F_V}{M_B}-\frac{f_B\;G_V}{M_\rho f_\rho}\right]. 
\end{eqnarray}
Summing up, within the factorization approximation the weak annihilation amplitude 
can be expressed in terms of the three form factors $F_A$, $F_V$, and $G_V$. 

\newpage   
\subsection{\label{sec:iv.2}Form factors for the $B\to\gamma\ell\nu$ transition}
In this section we derive formulas for the form factors $F_{A,V}$ 
describing the weak transition $B\to\gamma l\nu$
within the dispersion approach 
to the transition form factors. 
Recall that the 
form factors $F_{A,V}$ describe the transition of the $B$-meson to the photon with the momentum 
$q_1$, $q_1^2=0$, induced by the axial-vector (vector) current with the momentum transfer 
$q_2$, $q_2^2=M_\rho^2$. For technical reasons, it is convenient to treat the form factor $F_{A(V)}$ as describing the 
amplitude of the photon-induced transition of the $B$-meson into a $b\bar u$ axial-vector 
(vector) virtual particle with the corresponding factor $1/(s-q_2^2)$ in the dispersion integral. 
Then we can directly apply the equations obtained in Chapter II for 
meson-to-meson transition form factors. 

\subsubsection{The form factor $F_A$}
The form factor $F_A$ is given by the diagrams of Fig \ref{fig:Fa}. 
Fig \ref{fig:Fa}a shows $F_A^{(b)}$, 
the contribution to the form factor of the process when the $b$ quark interacts with the 
photon; Fig \ref{fig:Fa}b describes the contribution of the process when the quark $u$ interacts while 
$b$ remains a spectator. One has    
\begin{eqnarray}
F_A=F_A^{(b)}+F_A^{(u)}. 
\end{eqnarray} 
\begin{center}
\begin{figure}[hb]
\mbox{\epsfig{file=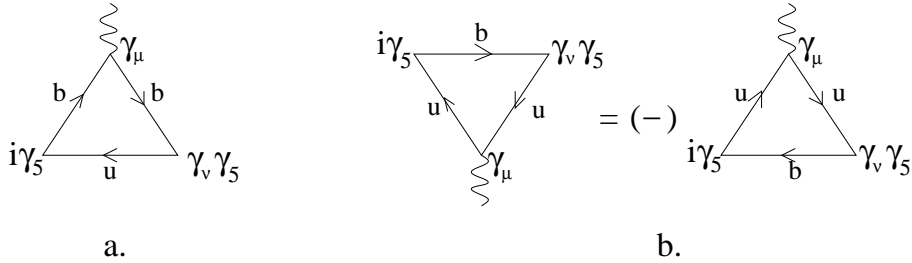,width=12cm}}
\caption{\label{fig:Fa}
Diagrams for the form factor $F_A$: a) $F_A^{(b)}$, b) $F_A^{(u)}$.}
\end{figure}
\end{center}
The $B^-$ meson is described 
by the vertex $\bar b(k_b)\; i\gamma_5 u(k_u)\;G(s)/{\sqrt{N_c}}$, with 
$G(s)=\varphi_B(s)(s-M_B^2)$. The $B$-meson wave function $\varphi_B$ is normalized according to 
the relation (\ref{norma}). 

It is convenient to change the direction of the quark line in the loop diagram of 
Fig \ref{fig:Fa}b. This is done by performing the charge conjugation of the matrix element 
and leads to a sign change for the $\gamma_\nu\gamma_5$ vertex. 

Now both diagrams in Fig \ref{fig:Fa} a,b are reduced to the diagram of Fig \ref{fig:Fat} 
which defines the form factor $F_A^{(1)}(m_1,m_2)$: Setting $m_1=m_b$, $m_2=m_u$ gives 
$F_A^{(b)}$: 
$$
F_A^{(b)}=Q_b F_A^{(1)}(m_b,m_u).
$$ 
Similarly, setting $m_1=m_u$, $m_2=m_b$ 
gives $F_A^{(u)}$, 
$$F_A^{(u)}=-Q_u F_A^{(1)}(m_u,m_b).
$$ 
\begin{center}
\begin{figure}[hb]
\mbox{\epsfig{file=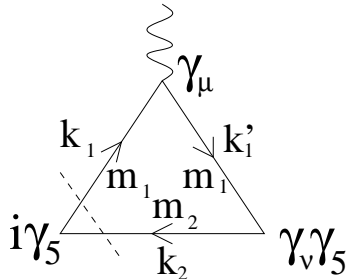,width=4.5cm}}
\caption{\label{fig:Fat}
The triangle diagram for $F_A^{(1)}(m_1,m_2)$. The cut corresponds to calculating the 
imaginary part in the variable $p^2$. }
\end{figure}
\end{center}
For the diagram of Fig \ref{fig:Fat} (quark 1 emits the photon, quark 2 is the spectator)    
the trace reads  
\begin{eqnarray}
\nonumber 
&&-{\rm Sp}\;i\gamma_5(m_2-\hat k_2)\gamma_\nu\gamma_5(m_1+\hat k'_1)\gamma_\mu(m_1+\hat k_1)
\nonumber\\
&&\qquad=4i(k_1+k_1')_\mu(m_1 k_2+m_2 k_1)_\nu+
4i(g_{\mu\nu}q_\alpha-g_{\mu\alpha}q_\nu)(m_1 k_2+m_2 k_1)_\alpha. 
\end{eqnarray}
The spectral density of the form factor $F^{(1)}_A(m_1,m_2)$ in the variable $p^2$, $p=k_1+k_2$,  
is the coefficient of the structure $g_{\mu\nu}$ obtained after the integration of the trace over 
the quark phase space. Performing necessary calculations, we arrive at the following single 
dispersion integral  

\begin{eqnarray}
\label{fadisp}
\frac{2}{M_B}F_{A}^{(1)}&=&\frac{\sqrt{N_c}}{4\pi^2}\int\limits_{(m_b+m_u)^2}^\infty
\frac{ds\;\varphi_B(s)}{(s-M_\rho^2)}
\nonumber\\
&&\times
\left\{
\left(m_1\log\left(\frac{s+m_1^2-m_2^2+\lambda^{1/2}(s,m_1^2,m_2^2)}
{s+m_1^2-m_2^2-\lambda^{1/2}(s,m_1^2,m_2^2)}\right)
+(m_2-m_1)\frac{\lambda^{1/2}(s,m_b^2,m_u^2)}{s}
\right)\right.
\nonumber\\
&&+\left. \frac{1}{pq_1}\left(
\frac{\lambda^{1/2}(s,m_b^2,m_u^2)}{2s}-
m_1^2\log\left(\frac{s+m_1^2-m_2^2+\lambda^{1/2}(s,m_1^2,m_2^2)}
{s+m_1^2-m_2^2-\lambda^{1/2}(s,m_1^2,m_2^2)}\right)\right)
\right\}.  
\end{eqnarray}
For $q_1^2=0$ one has $pq_1=(M_B^2-M_\rho^2)/2$. 

Now, let us analyse the behaviour of the form factor in the limit $m_b\to\infty$. 
To this end it is convenient to rewrite the spectral representation (\ref{fadisp}) 
in terms of the light-cone variables as follows (see \cite{amn1} for details) 
\begin{eqnarray}
\label{fa-lc}
\frac{2}{M_B}F_{A}^{(1)}(m_1,m_2)&=&\frac{\sqrt{N_c}}{4\pi^2}\int 
\frac{dx_1 dx_2 dk_\perp^2}{x_1^2 x_2}\delta(1-x_1-x_2)
\frac{\varphi_B(s)}{s-M_\rho^2}
\nonumber\\
&&\qquad\qquad\times\left\{m_1x_2+m_2 x_1-(m_1-m_2)k_\perp^2/pq_1\right\}.  
\end{eqnarray}
Here $x_i$ is the fraction of the $B$-meson light-cone momentum carried by the 
quark $i$, and 
$$
s=m_1^2/x_1+m_2^2/x_2+k_\perp^2/x_1x_2.
$$ 
For the form factors $F_{A}^{(u)}$ and $F_{A}^{(b)}$ one obtains 
\begin{eqnarray}
\label{fau}
\frac{2}{M_B}F_{A}^{(u)}&=&-Q_u\frac{\sqrt{N_c}}{4\pi^2}\int \frac{dx dk_\perp^2}{x_u^2 x_b}
\frac{\varphi_B(s)}{s-M_\rho^2}\left\{m_u\,x_b+m_b\,x_u+\frac{2(m_b-m_u)k_\perp^2}{M_B^2-M_\rho^2}
\right\}, \nonumber
\\
\frac{2}{M_B}F_{A}^{(b)}&=&\;\;\;Q_b\frac{\sqrt{N_c}}{4\pi^2}\int \frac{dx dk_\perp^2}{x_b^2 x_u}
\frac{\varphi_B(s)}{s-M_\rho^2}\left\{m_b\,x_u+m_u\,x_b+\frac{2(m_u-m_b)k_\perp^2}{M_B^2-M_\rho^2}
\right\}, \nonumber
\end{eqnarray}
with 
\begin{eqnarray}
s=\frac{m_b^2}{x_b}+\frac{m_u^2}{x_u}+\frac{k_\perp^2}{x_u x_b}.     
\end{eqnarray}
Recall that the $B$-meson decay constant is given in terms of the 
wave function by Eq. (\ref{fp}). 
Due to the wave function $\varphi_B(s)$, the integral in the heavy quark limit is dominated 
by the region $x_u=\bar\Lambda/m_b$, $x_b=1-\bar\Lambda/m_b$, 
where $\bar\Lambda$ is a constant of order $M_B-m_b$. This leads to the following expansion 
of the form factors in the $1/m_b$ series 
\begin{eqnarray}
\label{hqlimit}
\frac{2}{M_B}F_{A}^{(u)}&=&-Q_u\frac{f_B}{\bar\Lambda m_b}+...\nonumber\\
\frac{2}{M_B}F_{A}^{(b)}&=&\;\;\,\,Q_b\frac{f_B}{m_b^2}+...
\end{eqnarray}
Clearly, the dominant contribution in the heavy quark limit comes from the process 
when the light quark emits the photon, whereas the emission of the photon from the heavy 
quark gives only a $1/m_b$ correction. The expressions (\ref{hqlimit}) for the form factor 
$F_{A}^{(u)}$ agrees with the result of Ref. \cite{korch}, 
while we have found a different sign for $F_{A}^{(b)}$. 

\subsubsection{The form factor $F_V$} 
The consideration of the form factor $F_{V}$ is very similar to the form factor $F_{A}$. 
$F_V$ is determined by the two diagrams shown in Fig \ref{fig:Fv}:  
Fig \ref{fig:Fv}a gives $F_V^{(b)}$, the contribution of the process when the $b$ quark interacts with the 
photon; Fig \ref{fig:Fv}b describes the contribution of the process when the quark $u$ interacts. 
One has    
\begin{eqnarray}
F_V=F_V^{(b)}+F_V^{(u)}. 
\end{eqnarray}
\begin{center}
\begin{figure}[hb]
\mbox{\epsfig{file=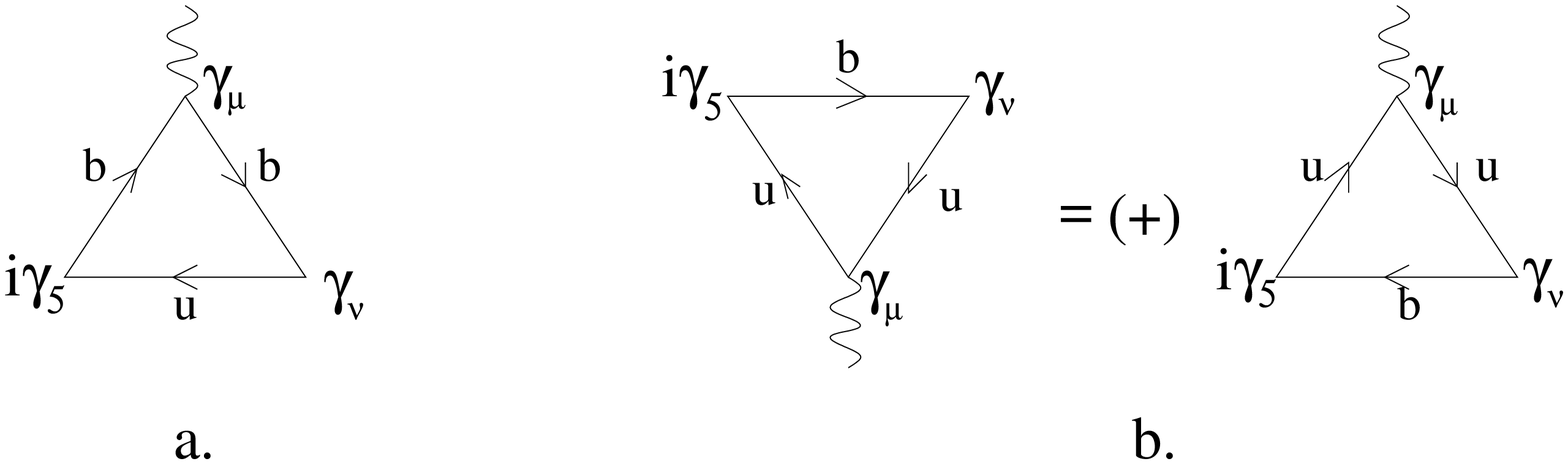,width=12cm}}
\caption{\label{fig:Fv}Diagrams for the form factor $F_V$: a) $F_V^{(b)}$, b) $F_V^{(u)}$.} 
\end{figure}
\end{center}
It is again convenient to change the direction of the quark line in the loop diagram of 
Fig \ref{fig:Fv}b 
describing $F_V^{(u)}$ 
by performing the charge conjugation of the matrix element. For the vector current $\gamma_\nu$ 
in the vertex the sign does not change (in contrast to the $\gamma_\nu\gamma_5$ case 
considered above). 

Then both diagrams in Fig \ref{fig:Fv} a, b are 
reduced to the diagram of Fig \ref{fig:Fvt} which gives the 
form factor $F_V^{(1)}(m_1,m_2)$: Setting $m_1=m_b$, $m_2=m_u$ gives $F_V^{(b)}$: 
$$
F_V^{(b)}=Q_b F_V^{(1)}(m_b,m_u).
$$
Setting $m_1=m_u$, $m_2=m_b$ gives $F_V^{(u)}$ 
$$F_V^{(u)}=Q_u F_V^{(1)}(m_u,m_b).
$$ 
\begin{center}
\begin{figure}[hb]
\mbox{\epsfig{file=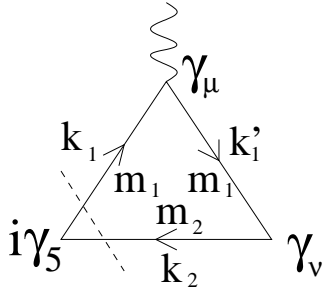,width=4.2cm}}
\caption{\label{fig:Fvt} The triangle diagram for $F_V^{(1)}(m_1,m_2)$. 
The cut corresponds to calculating the imaginary part in the variable $p^2$.}
\end{figure}
\end{center}
The trace corresponding to the diagram of Fig 4 (1 - active quark, 2 - spectator) reads 
\begin{eqnarray}
-{\rm Sp}\;
\left[i\gamma_5(m_2-\hat k_2)\gamma_\nu (m_1+\hat k'_1)\gamma_\mu(m_1+\hat k_1)
\right]=
-4\epsilon_{\nu\mu \alpha q_1}(m_1 k_2+m_2 k_1)_\alpha. 
\nonumber
\end{eqnarray}
The spectral representation for the form factor takes the form
\begin{eqnarray}
\frac{2}{M_B}F_V^{(1)}&=&\frac{\sqrt{N_c}}{4\pi^2}
\int\limits_{(m_b+m_u)^2}^\infty
\frac{ds\;\varphi_B(s)}{(s-M_\rho^2)}
\\
\nonumber
&&\times\left\{
(m_2-m_1)\frac{\lambda^{1/2}(s,m_b^2,m_u^2)}{s}+
m_1\log\left(\frac{s+m_1^2-m_2^2+\lambda^{1/2}(s,m_1^2,m_2^2)}
{s+m_1^2-m_2^2-\lambda^{1/2}(s,m_1^2,m_2^2)}\right)
\right\}. 
\end{eqnarray}
To analyse the heavy quark limit $m_b\to\infty$ we again represent the form factor 
in terms of the light-cone variables 
\begin{eqnarray}
\label{fv-lc}
\frac{2}{M_B}F_{V}^{(1)}&=&-\frac{\sqrt{N_c}}{4\pi^2}
\int \frac{dx_1 dx_2 dk_\perp^2}{x_1^2 x_2}\delta(1-x_1-x_2)
\frac{\varphi_B(s)}{s-M_\rho^2}\left(m_1x_2+m_2 x_1\right).  
\end{eqnarray}
For $F_{V}^{(u)}$ and $F_{V}^{(b)}$ the corresponding expressions read 
\begin{eqnarray}
\label{fvu}
\frac{2}{M_B}F_{V}^{(u)}&=&-Q_u\frac{\sqrt{N_c}}{4\pi^2}\int \frac{dx dk_\perp^2}{x_u^2 x_b}
\frac{\varphi_B(s)}{s-M_\rho^2}\left\{m_u\,x_b+m_b\,x_u\right\}, \nonumber
\\
\frac{2}{M_B}F_{V}^{(b)}&=&-Q_b\frac{\sqrt{N_c}}{4\pi^2}\int \frac{dx dk_\perp^2}{x_b^2 x_u}
\frac{\varphi_B(s)}{s-M_\rho^2}\left\{m_b\,x_u+m_u\,x_b\right\}, \nonumber
\end{eqnarray}
By the same procedure as used for $F_A$, we obtain in the limit $m_b\to\infty$ 
\begin{eqnarray}
\label{hqlimit1}
\frac{2}{M_B}F_{V}^{(u)}&=&-Q_u\frac{f_B}{\bar\Lambda m_b}+...\nonumber\\
\frac{2}{M_B}F_{V}^{(b)}&=&-Q_b\frac{f_B}{m_b^2}+...
\end{eqnarray}
The dominant contribution in the heavy quark limit again comes from the process 
when the light quark emits the photon. Now both form factors $F_{V}^{(u)}$ and $F_{V}^{(b)}$
in (\ref{hqlimit1}) perfectly agree with the expansions obtained in \cite{korch}.  

As seen from Eqs. (\ref{hqlimit}) and (\ref{hqlimit1}), one finds $F_A=F_V$ in the heavy quark limit, 
in agreement with the large-energy effective theory \cite{leet}. 

\newpage
\subsection{\label{sec:iv.3}The form factor $G_V$}
The form factor $G_V$ is determined by the divergence of the matrix element of the charged current 
between the vacuum and the $\rho^-\gamma$ state,
\begin{eqnarray}
ip_\nu\langle \gamma(q_1)\rho^-(q_2)|\bar d\gamma_\nu\gamma_5 u|0\rangle=
eG_V \epsilon_{q_1\epsilon_1^\ast q_2\epsilon_2^\ast}. 
\end{eqnarray}
The corresponding diagrams are shown in Fig \ref{fig:rhogamma}.  
\begin{center}
\begin{figure}[hb]
\mbox{\epsfig{file=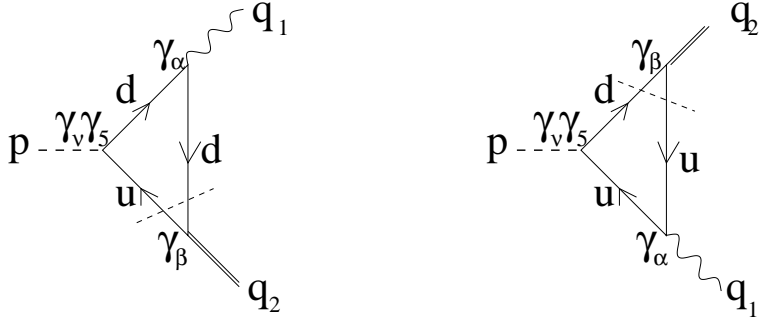,width=10cm}}
\vspace{.25cm}
\caption{\label{fig:rhogamma} Diagrams describing the amplitude 
$\langle \gamma(q_1)\rho^-(q_2)|\bar d \gamma_\nu\gamma_5 q|0\rangle$, $p=q_1+q_2$, $p^2=M_B^2$. 
The diagram (a) is multiplied by the $d$-quark charge, and 
the diagram (b) is multiplied by the $u$-quark charge. 
The cut corresponds to the calculation of the imaginary part in the variable $q_2^2$. }
\end{figure}
\end{center}
If the classical equations of motion are applied, the form factor $G_V$ is proportional to the 
light-quark masses.   
This is the reason why this form factor was neglected in previous 
analyses \cite{sr1,gp}. However, a proper calculation shows that this argument is not correct: 
in fact the classical equations 
of motions do not hold and the divergence contains the anomaly. 

The anomalous behavior of the divergence of the axial-vector current in the chiral limit 
is a well-known phenomenon discovered in the two-photon amplitude 
$\langle \gamma\gamma|\partial_\nu \bar q\gamma_\nu\gamma_5 q|0\rangle$ \cite{abj}. 
A very clear way to demonstrate the anomaly is to start with the matrix element of the axial vector 
current and to calculate the spectral representations for the relevant form factors. 
The anomaly is then obtained by performing the divergence at the final stage of the calculation \cite{dz}.  
A similar treatment applied to the matrix element $\langle \gamma\rho^-|\bar d\gamma_\nu\gamma_5 u|0\rangle$
leads to the form factor $G_V$ which does not vanish in the limit $m\to 0$. 
We will present a detailed discussion of the anomaly in radiative $B$ decays in a separate 
publication \cite{ms1}. Here we only quote the final result: considering the spectral 
representation in the variable $q_2^2$ and introducing the appropriate $\rho$-meson radial wave function 
$\psi_\rho(s)$, one finds the following expression for the form factor $G_V$ 
\begin{eqnarray}
\label{gvd}
G_V=\sqrt{N_c}(Q_u+Q_d)\left[-\frac{M_B^2}{4\pi^2}\int\frac{ds\;\psi_\rho(s)\,(s-M_\rho^2)}{(s-M_B^2-i0)^2}\right].  
\end{eqnarray} 
$\psi_\rho(s)$ is normalized according to the relation (for massless quarks) 
\begin{eqnarray}
\label{norma1}
\frac{1}{8\pi^2}\int\limits_{0}^\infty ds\; s\;|\psi_\rho(s)|^2=1.  
\end{eqnarray}
Clearly, $G_V$ is finite for $m=0$ which means a violation of the classical equations
of motion for the axial-vector current. Eq. (\ref{gvd}) describes the anomaly which 
takes place for the $\gamma\rho$ final state as well as for the $\gamma\gamma$ one. 
There is however an important difference between the two cases: 
For the $\gamma\gamma$ final state the divergence remains finite in the limit $m\to 0$ and 
$p^2=M_B^2\to\infty$. 
For the $\rho\gamma$ final state the divergence is finite for $m\to 0$ but 
decreases as $1/p^2$ for $p^2\to\infty$. 
It is convenient to introduce the parameter $\kappa$ such that 
\begin{eqnarray}
\label{kappa}
G_V= (Q_u+Q_d)\;\kappa\frac{M_\rho f_\rho}{M_B^2},  
\end{eqnarray} 
with $\kappa$ staying finite for $m=0$ and $M_B\to\infty$.

\subsection{\label{sec:iv.4}Numerical estimates}

Table \ref{table:wa-comparison} presents formulas for the penguin and the weak-annihilation 
invariant amplitudes for the $B^-\to\rho^-\gamma$ and $B^0\to\rho^0\gamma$ decays. 

Before giving numerical estimates, let us look at the scaling behavior of the form factors 
in the limit $M_B\to\infty$: 
\begin{eqnarray}
T_1(0)\sim M_B^{-3/2} \cite{lcsr}, 
\quad F_{A,B}\sim M_B^{-1/2},  \quad G_V\sim M_B^{-2} 
\end{eqnarray}
such that 
\begin{eqnarray}
F^{\rm peng}\sim M_B^{-1/2}, \quad F^{\rm WA}_{\rm PV,PC}\sim M_B^{-3/2}. 
\end{eqnarray}
The terms proportional to $f_B$ ($f_B\sim 1/\sqrt{M_B}$) are $1/M_B$-suppressed 
compared with the terms containing the form factors $F_{A,V}$. 
As we see below numerically this leads to a suppression by a factor of $4-5$. 
\begin{table}[htb]
\caption{\label{table:wa-comparison} The penguin and the weak-annihilation invariant amplitudes 
for the $B^-\to\rho^-\gamma$ and $B^0(b\bar d)\to\rho^0\gamma$ decays. }
\centering
\begin{tabular}{|l|l|l|}
    Quantity & $B^-\to\rho^-\gamma$   & $B^0\to\rho^0\gamma$   \\
\hline
$F^{\rm peng}$   &
$-\xi_t C_7\frac{m_b}{2\pi^2}T^{B^-\to\rho^-}_1(0)$   & 
$-\xi_t C_7\frac{m_b}{2\pi^2}T^{B^0\to\rho^0}_1(0)$  \\
$F^{\rm WA}_{\rm PV}$   &
$\xi_u a_1 M_\rho f_\rho\;\left(\frac{2F^{B^-}_{A}}{M_B}+\frac{2f_B}{M_B^2-M_\rho^2} \right)$   & 
$\xi_u a_2 M_\rho f_{\rho^0}\;\left(\frac{2F^{B^0}_{A}}{M_B}\right)$   \\
$F^{\rm WA}_{\rm PC}$    &
$\xi_u a_1 M_\rho f_\rho\;\left(\frac{2F^{B^-}_V}{M_B}-(Q_u+Q_d)\;\kappa\frac{f_B}{M^2_B}\right)$    &
$\xi_u a_2 M_\rho f_{\rho^0}\;\left(\frac{2F^{B^0}_V}{M_B}-2\;Q_u\;\kappa\frac{f_B}{M^2_B}\right)$  
\end{tabular}
\end{table} 
We now proceed to numerical estimates for the $B$-meson decay. 
The scale-dependent Wilson coefficients $C_i(\mu)$ and $a_1(\mu)$ take the following values
at the renormalization scale $\mu\simeq 5$ GeV \cite{ali}: 
\begin{eqnarray}
C_1=1.1,\quad  C_2=-0.241, \quad  C_{7\gamma}=-0.312, \quad a_1=C_1+C_2/N_c\simeq 1.02.   
\quad a_2=C_2+C_1/N_c\simeq 0.27\pm 0.1.
\end{eqnarray}
The penguin form factor was previously calculated within the dispersion approach 
with the result $T_1^{B^-\to\rho^-}(0)=0.27\pm 0.3$, see Table \ref{table:fitsb2pi}
in Chapter IV.   
Using the same parameters and the $B$ meson wave function as in Chapter IV, we obtain the 
form factors $F_{A,B}$ shown in Table \ref{table:wa-results}. 
Our result for the form factor $F_V$ is in good agreement with the estimates from other approaches. 
The form factor $F_A$ agrees well with the constraints from perturbative QCD and turns out to be 
considerably larger than the corresponding sum rule estimate. 

The value of the $G_V$ is sensitive to the details of the $\rho$ meson wave function  
$\psi_\rho$. The reason for that is the presence of the term $(s-M_\rho^2)$ in the integrand in Eq. 
(\ref{gvd}) which changes sign in the integration region. Assuming $\psi_\rho(s)\simeq {\beta^2}/{(\beta^2+s)^2}$ and setting 
$\beta=0.8$ GeV, which gives a good description of the $\rho$ meson radius, leads to 
$\kappa=-1.8$.   
Conservatively, we take $-\kappa<2.0$ and use this result for further estimates. 
The form factor $G_V$ does not contribute more than a few \% to the full amplitude, but 
can sizeably correct the weak-annihilation part, especially for the 
$B^0\to\rho^0\gamma$ decay.  
\begin{table}[htb]
\caption{\label{table:wa-results} The weak-annihilation form factors 
$F_A$, $F_V$ and $G_V$. The accuracy of our estimates 
is about 10\%. The sum rule results are recalculated from \protect\cite{wa-sr2} 
according to the relation 
$-F^{SR}_{A}=F^{\rm SR}_{1}{M_B}/{f_{\rho^-}}+f_B/M_B$ 
and 
$-F^{SR}_{V}=F^{\rm SR}_{2}{M_B}/{f_{\rho^-}}$. 
The results from \protect\cite{korch} 
are recalculated according to 
$-F_{A,V}=\frac{1}{2}f^{\protect\cite{korch}}_{A,V}$ 
for 
$\bar \Lambda=0.5$ GeV. According to our estimates, $\kappa$ related to $G_V$ 
is found in the range $-\kappa<2.0$. }
\centering
\begin{tabular}{|r|r|r|r|}
    & Disp Approach & SR \cite{wa-sr2}  &  pQCD\cite{korch}  \\
\hline
$-F^{B^-}_{A}$                   & 0.120   & 0.110   & $\ge$0.09   \\
$-F^{B^-}_{V}$                   & 0.092   & 0.091   & $\ge$0.09   \\
$G^{\rho^-}$                     & $0.002\,\kappa$   &            \\
\hline
$F^{B^0}_{A}$   & 0.055   &  0.037  &  \\
$F^{B^0}_{V}$   & 0.043   &  0.046  &  \\
$G^{\rho^0}$       & $0.0055\,\kappa$   &  &       
\end{tabular}
\end{table} 
Using the obtained form factors and the decay constants $f_B=180$ MeV, 
$f_\rho=210$ MeV, $f_{\rho}^0=210$ MeV, 
we arrive at the results for the penguin and weak annihilation
invariant amplitudes shown in Table \ref{table:wa-resamp}. 

Our results for $F^{\rm peng}$ and $F^{\rm WA}_{\rm PV}$ amplitudes agree with the sum rule results 
\cite{wa-sr2}. We would like to notice, however, that the anatomy of $F^{\rm WA}$ in our analysis is 
different. Namely, we have found $F_A$ considerably larger than the sum rule result. 
But after including the contact term $\sim f_B$ which 
we have come to $F^{\rm WA}_{\rm PV}$ close to the sum rule estimate \cite{wa-sr2}. 

If the form factor $G_V$ is not taken into account, our estimate for 
$F^{\rm WA}_{\rm PC}$ agrees with the sum rules. Notice, however, that the contribution of 
$G_V$ is enhanced for the $B^0$ decay compared to the $B^-$ decay. 
Therefore, if the value $\kappa\simeq -1\div -2$ is confirmed by the future analysis, 
the $F^{\rm WA}_{\rm PC}$ for the $B^0\to\rho^0\gamma$ decay might appear sizeably increased.  
For a definite conclusion a more accurate study of the form factor $G_V$ is necessary. 

\begin{table}[htb]
\caption{\label{table:wa-resamp} Results for the invariant amplitudes and their ratios for the 
$B^-\to\rho\gamma$ and $B^0\to\rho^0\gamma$ decays. In the Standard Model $|\xi_u/\xi_t|\simeq 0.4$. 
According to our estimates $-\kappa\le 2$.} 
\centering
\begin{tabular}{|l|r|r|}
    Quantity & $B^-\to\rho^-\gamma$  & $B^0\to\rho^0\gamma$  \\
\hline
$F^{\rm peng}$, in MeV         & 
$20\,\xi_t$        & 
$14\,\xi_t$        \\
$F^{\rm WA}_{\rm PC}$, in MeV [this work]& 
$-5.6\,a_1\,\xi_u$  & 
$ 1.9\,a_2\,\xi_u$  \\
$F^{\rm WA}_{\rm PV}$, in MeV [this work]& 
$-6.0\,(1+0.06\,\kappa)\,a_1\,\xi_u$   & 
$ 2.5\,(1-0.3\,\kappa)\,a_2\,\xi_u$  \\
\hline
$F^{\rm WA}_{\rm PV}/F^{\rm peng}$ [this work]  & 
$-0.28\,a_1\,\xi_u/\xi_t$            &  
$0.14\,a_2\,\xi_u/\xi_t$         \\
$F^{\rm WA}_{\rm PC}/F^{\rm peng}$ [this work]  & 
$-0.33\,(1+0.06\,\kappa)\,a_1\,\xi_u/\xi_t$     & 
$0.18\,(1-0.33\,\kappa  )\,a_2\,\xi_u/\xi_t$          \\
\hline 
$F^{\rm WA}_{\rm PV}/F^{\rm peng}$ \cite{ap,wa-sr2} &   
$-0.3\,a_1\,\xi_u/\xi_t$   & 
$0.15\,a_2\,\xi_u/\xi_t$   \\
$F^{\rm WA}_{\rm PC}/F^{\rm peng}$ \cite{ap,wa-sr2} &   
$-0.3\,a_1\,\xi_u/\xi_t$   & 
$0.15\,a_2\,\xi_u/\xi_t$
\end{tabular}
\end{table}

The calculated form factors and the ratios of the weak-annihilation to the penguin
amplitudes is one of the necessary ingredients for the calculation of the 
branching ratios of the $B\to\rho\gamma$ decays, Isospin and CP Asymmetries. 
In addition to the above ratios, these quantities contain the phase induced by the strong 
interactions and the CP-violating phase of the CKM matrix (see \cite{ap} for details). 
The corresponding analysis was done recently in \cite{ap} using the value  
$F^{\rm WA}/F^{\rm peng}\simeq -0.3\;{\xi_u}/{\xi_t}$ and 
$F^{\rm WA}/F^{\rm peng}\simeq 0.03\;{\xi_u}/{\xi_t}$ for $B^-\to\rho^-\gamma$ and 
$B^0\to\rho^0\gamma$ decays, respectively. 

\subsection{\label{sec:iv.5}Discussion}
We have analysed the weak annihilation for the radiative decay $B\to\rho\gamma$ 
in the factorization approximation. 

\noindent 1. Making use of the dispersion approach,  
we have calculated the form factors 
$F_A$ and $F_V$ describing the radiative semileptonic weak 
transition $B\to\gamma l\nu$.   
They are given by the diagrams with photon emitted from the $B$ meson 
loop. We have performed the $1/m_b$ expansion of the spectral representations for 
these form factors and demonstrated them to exhibit a behaviour in agreement with the 
large-energy limit of QCD. 

\noindent 2. We have analysed the contribution to the weak annihilation amplitude from the diagram 
when the photon is emitted from the loop containing only light quarks. 
For the parity-conserving process this quantity is related to the divergence of 
the axial-vector current 
\begin{eqnarray}
\langle \gamma(q_1)\rho^-(q_2)|\partial_\nu \bar d \gamma_\nu \gamma_5 u|0\rangle=
e\epsilon_{q_1\epsilon_1^\ast q_2\epsilon^\ast_2}(Q_u+Q_d)\;\kappa\;{f_\rho M_\rho}/{M_B^2}. 
\end{eqnarray}
with $\kappa$ staying finite in the chiral limit $m\to 0$ and estimated to be $-\kappa\leq 2$. 
The anomalous behavior of this amplitude has the similar origin as the anomalous behaviour of 
the matrix element $\langle \gamma(q_1)\gamma(q_2)|\partial_\mu A_\mu|0\rangle$.
Let us stress this important result: the axial-vector current 
{\it is not conserved in the chiral limit} opposite to the common belief. 

\noindent 3. We have also included contact terms which were missed in some of the 
previous analyses. Numerical estimates for the 
weak annihilation contribution to the $B\to\rho\gamma$ amplitude are given 
in Table \ref{table:wa-results}.  

\newpage
\newpage
\section{\label{sec:mixing} Nonfactorizable effects in the $B^0-\bar B^0$ mixing}
In this Chapter we analyse corrections to factorization for 
the $B-\bar B$ mixing amplitude assuming that non-factorizable soft gluon exchanges 
can be described in terms of the local gluon condensate. 
The Chapter is based on Ref. \cite{mn}. 

Within this approximation the $\langle\alpha_s GG\rangle$-correction to the mixing amplitude can be expressed 
through specific $B$-meson transition form factors at zero momentum transfer. 
First, the correction is demonstarted to be strictly {\it negative} 
independent of the particular values of the form factors. 

As a next step, we study these form factors making use of  
the relativistic dispersion approach. 

The study of the oscillations in the system of neutral $B$ mesons provides important 
information on the pattern of the CP violation in the Standard model and its extentions 
(a detailed discussion can be found in \cite{mixingburas}). 
The main uncertainty in the theoretical description of the process arises from  
the nonperturbative long-distance contributions to the mixing amplitude.  
In the $B-\bar B$ mixing there are two kinds of such contributions:  
first, effects related to the presence of the $B$-mesons in the initial and final states, 
and, second, corrections to the weak 4-quark amplitude due to the soft gluon exchanges 
between quarks of the initial $B$ and the final $\bar B$ meson. 
By neglecting the second contribution the amplitude factorizes.   
In this case all nonperturbative effects connected with the meson formation are reduced 
to only one quantity - the leptonic decay constant $f_B$. 

Soft-gluon corrections to the weak 4-quark amplitude 
give rise to non-factorizable contributions, such that the influence of the 
$B$-meson structure is no longer described by the decay constant only. 
The understanding of the actual size of the non-factorizable effects and their 
adequate description in the heavy quark systems is an important and challenging task. 

The theoretical analysis of $B-\bar B$ mixing was first performed within 
the QCD sum rules \cite{mixingsr1,mixingsr2,mixingsr3} and later using lattice QCD, see 
\cite{mixinglat1,mixinglat2} and 
refs therein. The results from the sum rules are in all cases well compatible with 
factorization (vacuum saturation), within rather large errors. 
Recent lattice analyses definitely reported negative non-factorizable effects, 
around 5-10\% for the central values at the scale $\mu\simeq m_b$.  
Still, the errors of the calculations have a similar size. 
Results from these approaches are shown in Table \ref{table:mixingresults}. 
\begin{table}[htb]
\caption{\label{table:mixingresults}
Comparison of results from various approaches}
\centering
\begin{tabular}{|l|l|l|l|l|}
 Ref.     &$B_{B_d}(m_b)$&$B_{B_s}(m_b)$&$\hat B_{B_d}$&$\hat B_{B_s}/\hat B_{B_d}$\\
\hline
SR\cite{mixingsr1,mixingsr2}&  $1.0\pm 0.15$	&		&	       &	      \\
SR  \cite{mixingsr3}  &  $0.95\pm 0.1$	&		&	       &	      \\
\hline
Lat \cite{mixinglat1} &0.92(4)$^{+3}_{-0}$&0.91(2)$^{+3}_{-0}$&  1.41(6)$^{+5}_{-0}$&0.98(3)\\
Lat \cite{mixinglat2} &   0.93(8)    &   0.92(6)    &   1.38(11)   &  0.98(5)     \\
\hline
This work  &   $0.94\pm 0.035$  & $0.95\pm 0.03$ & $1.4\pm 0.05$ & $1.01\pm 0.01$    
\end{tabular}
\end{table}
We analyse corrections to factorization in the $B-\bar B$ mixing 
due to soft gluon exchanges in the leading $\alpha_s$-order, assuming   
the main effect of such exchanges to be described by the local gluon 
condensate introduced in Ref. \cite{svz}. 
Then the parameter $\Delta B_B$ which describes correction to factorization, is determined by the meson 
transition form factors at zero momentum transfer. 

We demonstrate that in the local gluon condensate approximation 
the value of $\Delta B_B$ turns out to be {\it negative}, independently of the particular values 
of these form factors. 
 
As a next step we calculate the relevant form factors making use of the 
dispersion approach. 
We obtain spectral representations for the form factors in terms of the $B$-meson 
wave function and the propagator of the color-octet $b\bar q$-system. 

Parameters of the model, such as the quark masses and the wave function have ben fixed, 
but for numerical estimates we also need the propagator of the 
color-octet $q\bar b$-system in the nonperturbative region, so we discuss the possible 
form of this propagator. The lack of reliable information about its details in the 
nonperturbative region leads to the main uncertainties in our numerical estimates. 

The Chapter is organised as follows: 
The next Section presents basic formulas for the $B-\bar B$ mixing. In
Section \ref{sec:mixb} we calculate the non-factorizable contribution to the mixing amplitude 
induced by the local gluon condensate and show that the correction is negative. 
In Section \ref{sec:mixc} we calculate the $B$-meson transition form 
factors within the dispersion approach. Section \ref{sec:mixd} gives numerical estimates. 

\newpage
\subsection{\label{sec:mixa} Effective Hamiltonian and the structure of the amplitude}
The effective Hamiltonian which summarizes the perturbative QCD corrections for the 
$B^0-\bar B^0$ mixing has the following form \cite{mixingheff}
\begin{equation}
\label{heff1}
H^{\Delta B=2}_{eff}=\frac{G^2_F\, M^2_W}{\sqrt{2}}\,
\left (V^*_{tb}V_{td}\right )^2\,C(\mu)\;
\bar d O_{\sigma} b \cdot \bar d O_{\sigma} b + h.c.
\end{equation}
where $G_F$ is the Fermi constant and $O_{\sigma}=\gamma_{\sigma}(1-\gamma_5)$.

The parameter $\mu$ stands for the renormalization scale which separates the 
hard region 
taken into account perturbatively and the soft region. 
The Wilson coefficient $C(\mu)$ includes perturbative corrections above the scale $\mu$.
The explicit expression for $C(\mu)$ can be 
taken from \cite{mixingheff}. The soft contributions are contained in the 
$B$-meson matrix elements of the operators in the effective Hamiltonian 
\begin{eqnarray}
\label{me1}
A=\langle\bar B^0 |\bar d O_{\sigma} b \cdot \bar d O_{\sigma} b| B^0\rangle.  
\end{eqnarray}
The diagrammatic representation for the amplitude $A$ of Eq. (\ref{me1}) is shown in 
Fig. \ref{fig:fig1}. 
\begin{figure}[htb]
\begin{center}
\vspace{.25cm}
\mbox{\epsfig{file=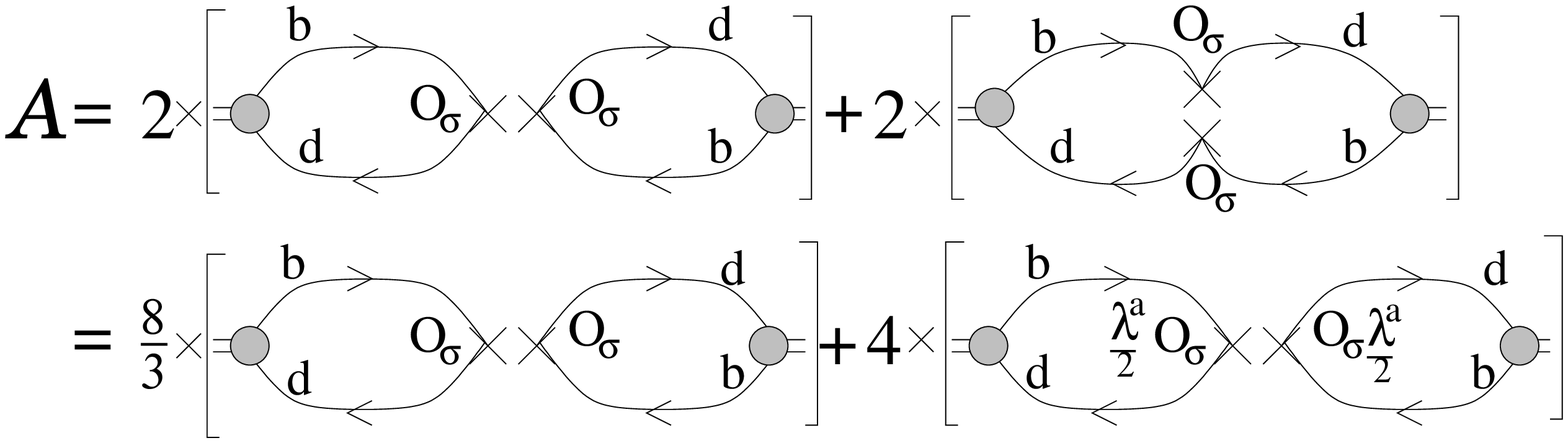,width=12cm}}
\vspace{.5cm}
\caption{\label{fig:fig1}
Different representations for the $B^0-\bar B^0$ mixing amplitude. 
The enclosed quark lines show the order of contraction of the spinorial indices. 
The circles stand for the initial (final) $B$ ($\bar B$) meson. The second line 
is obtained from the first line by performing the combined color and spinorial 
Fierz rearrangements.}
\end{center}
\end{figure}
Using the language of the hadronic intermediate states one finds that the 
contribution of the hadronic vacuum to this amplitude gives 
\begin{equation}
\label{vac}
\left < \bar B^0 | H^{\Delta B=2}_{eff} | B^0 \right > =
\frac{8}{3}\,\frac{G^2_F\, M^2_W\, M^2_B}{\sqrt{2}}\,
(V^*_{tb}V_{td})^2\, C(\mu)\, f^2_B, 
\end{equation}
with $f_B$ the leptonic decay constant of the $B$ meson defined according to the relation  
\begin{eqnarray}
\langle 0|\bar d\gamma_{\mu}\gamma_5 b | B(p)\rangle = i\,f_B\,p_{\mu}, \qquad  
\langle \bar B(p)  |\bar d\gamma_{\mu}\gamma_5 b | 0\rangle &=& -i\,f_B\,p_{\mu}. 
\end{eqnarray}

It is convenient to parametrize the full amplitude as follows 
\begin{equation}
\label{hadr}
\left < \bar B^0 | H^{\Delta B=2}_{eff} | B^0 \right > =
\frac{8}{3}\,\frac{G^2_F\, M^2_W\, M^2_B}{\sqrt{2}}\,
(V^*_{tb}V_{td})^2\, C(\mu)\, f^2_B\, B_B,
\end{equation}
such that the quantity $B_B-1\equiv \Delta B_B$  
measures contributions of the non-vacuum intermediate hadronic states. 

Using the language of quarks and gluons, the corrections to the amplitude of 
Fig. \ref{fig:fig1} due to 
soft gluon exchanges 
are obtained by inserting the (soft) gluons between the quark lines 
in Fig. \ref{fig:fig1}. 
Soft gluon exchanges between the quarks of the same loop 
lead to the $\alpha_s$-corrections either to the meson vertices or to the quark propagators. 
They only contribute to the leptonic decay constant $f_B$ but do not lead to 
non-factorizable effects. 

The non-factorizable effects originate from the soft gluon exchanges between 
quarks of different loops. To lowest $\alpha_s$-order these effects are described by the 
4 diagrams shown in Fig. \ref{fig:fig2}. 
\begin{figure}[tb]
\begin{center}
\begin{tabular}{c}
\mbox{\epsfig{file=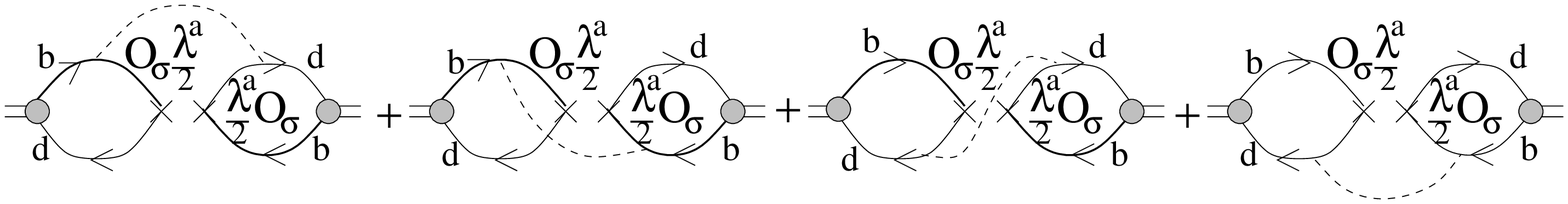,width=16cm}}
\end{tabular}
\caption{\label{fig:fig2} Soft-gluon exchanges leading to the non-factorizable effects.}
\end{center}
\end{figure}
We assume that the main effect of the soft gluon exchange can be described by
the local gluon condensate. In the next section we demonstrate that 
in this approximation the $\left<\alpha_sGG\right>$-correction to the factorization 
is {\it negative}. 

\subsection{\label{sec:mixb}$\Delta B$ in terms of the local gluon condensate} 
Let us consider the exchange of a soft gluon between different quark loops 
assuming the dominance of the local gluon condensate \cite{svz}. 
In this case a typical graph of Fig. \ref{fig:fig2} describing the soft-gluon contribution is reduced 
to the product of the three-point diagrams as shown in Fig. \ref{fig:fig3}. 
\begin{figure}[tb]
\begin{center}
\begin{tabular}{cc}
\mbox{\epsfig{file=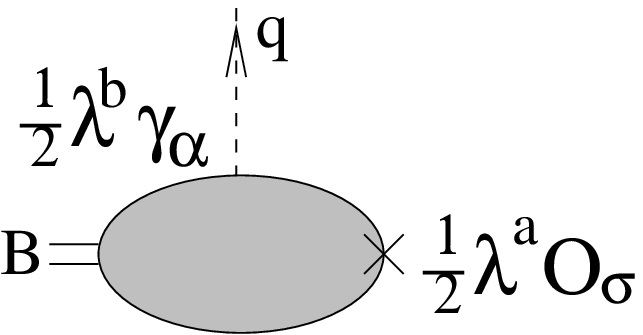,width=4cm}}\qquad\quad
&
\qquad\quad\mbox{\epsfig{file=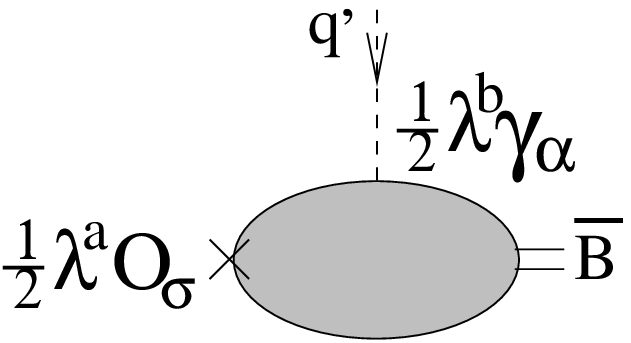,width=4cm}}
\end{tabular}
\caption{\label{fig:fig3} Diagrams for 
$T^{ab(l)}_{\sigma\alpha}(p,q)$ (a) and 
$T^{ab(r)}_{\sigma\alpha}(p,q)$ (b).}
\end{center}
\end{figure}
The corresponding amplitude has the form
\begin{eqnarray}
\label{aa1}
A^{(1)}=g^2
\int dq dq' T^{ab(l)}_{\sigma\alpha}(p,q)
            T^{ab'(r)}_{\sigma\alpha'}(p,q)
\frac{dx}{(2\pi)^4} \frac{dx'}{(2\pi)^4} {\rm e}^{-iqx+iq'x'}\langle A^b_\alpha(x)A^{b'}_{\alpha'}(x')\rangle,   
\end{eqnarray}
where $T^{ab(l)}_{\sigma\alpha}(p,q)$ with the superscript $l$ stands for the amplitude  
of Fig. \ref{fig:fig3}a with the $B$ in the initial state,  
and $T^{ab'(r)}_{\sigma\alpha'}(p,q')$ with the superscript $r$ for the amplitude of 
Fig. \ref{fig:fig3}b with the $\bar B$ in the final state.  

It is convenient to use the fixed-point gauge $x_\mu A^b_\mu(x)=0$ \cite{mixingsrff}. In this case 
the gluon potential reads 
\begin{eqnarray}
\label{gauge}
A^b_\alpha(x)=\frac{1}{2}x_\beta G^b_{\alpha\beta}+...,
\end{eqnarray}
where dots stand for terms with higher order derivatives.  
The average over the hadronic vacuum is performed according to the relation 
\begin{eqnarray}
\langle G^b_{\alpha\beta}G^{b'}_{\alpha'\beta'}\rangle=\frac{\delta^{bb'}}{96}
(g_{\alpha\alpha'}g_{\beta\beta'}-g_{\alpha\beta'}g_{\alpha'\beta})\langle GG \rangle,  
\end{eqnarray}
with $\delta^{aa}=N_c^2-1$. Here $\langle GG \rangle$ is the positive-valued gluon condensate 
$\langle\frac{\alpha_s}{\pi} GG \rangle=0.012\;{\rm GeV}^4$ \cite{svz}. 

The amplitude of Fig. \ref{fig:fig3} then takes the form 
\begin{eqnarray}
\label{aa2}
A^{(1)}&=&\frac{g^2}{4}
\int dq dq'  
T^{ab(l)}_{\sigma\alpha}(p,q)   \frac{\partial}{\partial q_\beta}\delta(q)
T^{ab'(r)}_{\sigma\alpha'}(p,q')\frac{\partial}{\partial q'_{\beta'}}\delta(q') 
\langle G^b_{\alpha\beta} G^{b'}_{\alpha'\beta'}\rangle
\nonumber\\
&=&
\frac{g^2}{4}\langle G^b_{\alpha\beta} G^{b'}_{\alpha'\beta'}\rangle
\frac{\delta^{ab }}{2\sqrt{N_c}}T^{(l)}_{\sigma\alpha, \beta }(p)
\frac{\delta^{ab'}}{2\sqrt{N_c}}T^{(r)}_{\sigma\alpha',\beta'}(p), 
\end{eqnarray}
where we have denoted 
\begin{eqnarray}
\label{relation}
\frac{\partial}{\partial q_\beta}T^{ab(l,r)}_{\sigma\alpha}(p,q)|_{q=0}\equiv
\frac{\delta^{ab}}{2\sqrt{N_c}}T^{(l,r)}_{\sigma\alpha,\beta}(p). 
\end{eqnarray}
Notice that only the parts of the amplitudes 
$T^{(l,r)}$ antisymmetric in indices $\alpha$ and $\beta$ give a nonvanishing contribution. 
These antisymmetric parts have the following general Lorentz structure: 
\begin{eqnarray}
\label{aaa} 
T^{A(l,q)}_{\sigma\alpha,\beta}&=&
4p^\nu\left[f_1^q \epsilon_{\alpha\beta\sigma\nu}
+i f_2^q (g_{\alpha\nu}g_{\beta\sigma}-g_{\beta\nu}g_{\alpha\sigma})\right], 
\nonumber\\
T^{A(l,\bar q)}_{\sigma\alpha,\beta}&=&
4p^\nu\left[f_1^{\bar q} \epsilon_{\alpha\beta\sigma\nu}
+i f_2^{\bar q} (g_{\alpha\nu}g_{\beta\sigma}-g_{\beta\nu}g_{\alpha\sigma})\right],
\nonumber\\ 
T^{A(r,q)}_{\sigma\alpha,\beta}&=&T^{A(l,\bar q)}_{\sigma\alpha,\beta}, \nonumber\\
T^{A(r,\bar q)}_{\sigma\alpha,\beta}&=&T^{A(l,q)}_{\sigma\alpha,\beta}, 
\nonumber\\ 
&&f_1^q=f_1^{\bar q}, \qquad f_2^q=-f_2^{\bar q}.  
\end{eqnarray}
The superscript $q(\bar q)$ in these expressions denotes the flavour of the 
quark (antiquark) to which the soft external gluon is attached in diagrams of 
Fig. \ref{fig:fig3}.  
The quantities $f_{1,2}^{q,\bar q}$ are real constants. The last three lines 
in Eq. (\ref{aaa}) are the consequence of the $C$-invariance of 
the strong interaction $S$-matrix. As we discuss in the next section, 
the constants $f_{1,2}^{q,\bar q}$ can be represented as specific $B$-meson 
transition form factors at zero momentum transfer.   

We have to collect now contributions of the four diagrams in Fig. \ref{fig:fig2}. 
For instance, the contribution to the amplitude of the subprocess in which the soft gluons 
are attached to the $b$ quark in the left loop and the $\bar b$ quark in the right loop 
reads 
\begin{eqnarray}
\label{correction}
T^{A(l,b)}_{\sigma\alpha;\beta}(p)T^{A(r,\bar b)}_{\sigma\alpha';\beta'}(p)
(g_{\alpha\alpha'}g_{\beta\beta'}-g_{\alpha\beta'}g_{\alpha'\beta})
=-192 M_B^2 \left((f^b_1)^2+(f^b_2)^2\right). 
\end{eqnarray}
Similar expressions are easily obtained for other diagrams using the relations (\ref{aaa}). 
Taking into account the relevant symmetry factors as shown in Fig. \ref{fig:fig1},  
we finally arrive at the following expressions 
for the factorizable and nonfactorizable contributions to the mixing amplitude:   
\begin{eqnarray}
\label{mixinga1}
A^{(0)}&\simeq&\frac{8}{3}M_B^2f_B^2\nonumber\\
A^{(1)}&\simeq& -8C_F M_B^2 \left< \frac{\alpha_s}{\pi}GG\right> \pi^2
\left[(f_1^{b}+f_1^{d})^2+(f_2^{b}-f_2^{d})^2
\right], 
\end{eqnarray}
with the color factor $C_F=\frac{N_c^2-1}{4N_c}$. 
Finally, for $\Delta B_B$ we find the expression 
\begin{eqnarray}
\label{deltab}
\Delta B_B(\mu)=-\frac{\left< \frac{\alpha_s}{\pi}GG\right>}{f_B^2}2\pi^2
\left[
\left(f_1^{b}+f_1^{d}\right)^2+\left(f_2^{b}-f_2^{d}\right)^2
\right]. 
\end{eqnarray}
The quantities $f_1$ and $f_2$ depend on the renormalization scale $\mu$. 

Obviously, the correction to the factorizable amplitude due to the 
local gluon condensate is {\it negative}. Let us point out that the 
relation (\ref{deltab}) 
is the direct consequence of the only assumption of the locality of the gluon 
condensate, dominating the nonfactorizable correction. It is 
completely independent of any details of the $B$-meson structure. 

\subsection{\label{sec:mixc} Form factors in the dispersion approach} 
We now proceed to the calcuation of the form  factors $f_1$ and $f_2$.  
We start with the amplitude $T^{(l)ab}_{\sigma\alpha}(p,q)$. It is diaginal in color
indices, so we write 
$T^{(l)ab}_{\sigma\alpha}(p,q)=\frac{\delta^{ab}}{\sqrt{N_c}}T^{(l)}_{\sigma\alpha}(p,q)$. 
The triangle diagram for $T^{(l)}_{\sigma\alpha}(p,q)$ is shown in 
Fig. \ref{fig:fig4}a. Throughout this section we shall omit the superscript $l$. 
The quark structure of the 
$B$-meson is described by the vertex $i\bar q_1\gamma_5 q_2 G(s)/\sqrt{N_c}$. 
One should set 
$m_1=m_b$, $m_2=m_d$ for the calculation of $f^{(b)}$, and $m_1=m_d$, $m_2=m_b$ for $f^{(d)}$. 
\begin{figure}[htb]
\begin{center}
\begin{tabular}{lr}
\mbox{\epsfig{file=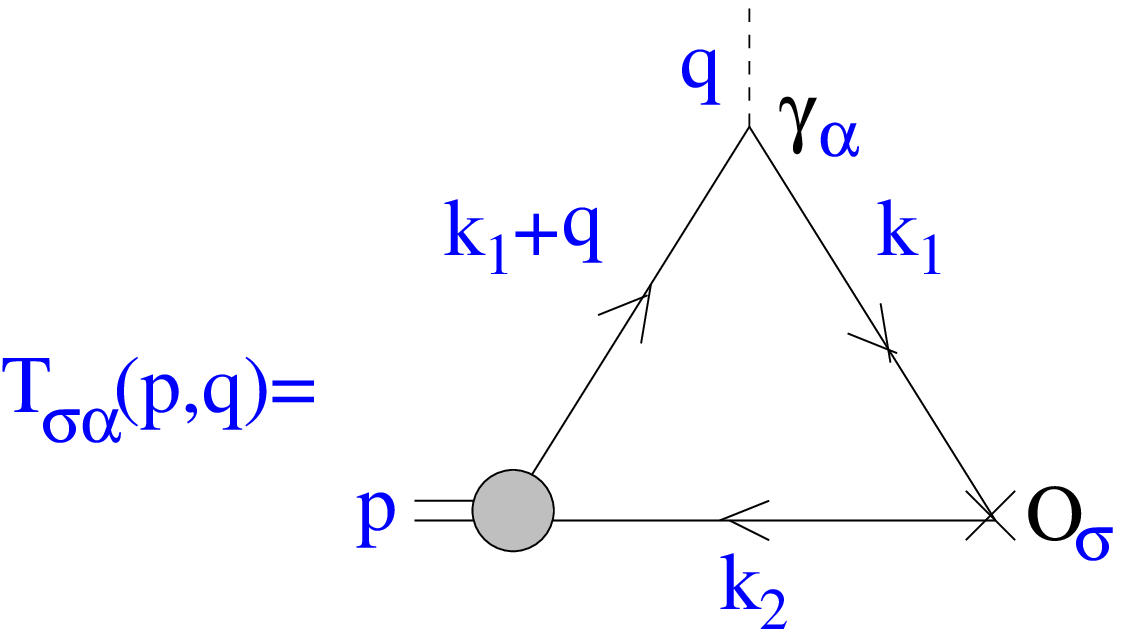,width=6.cm}} $\qquad$ & $\qquad$
\mbox{\epsfig{file=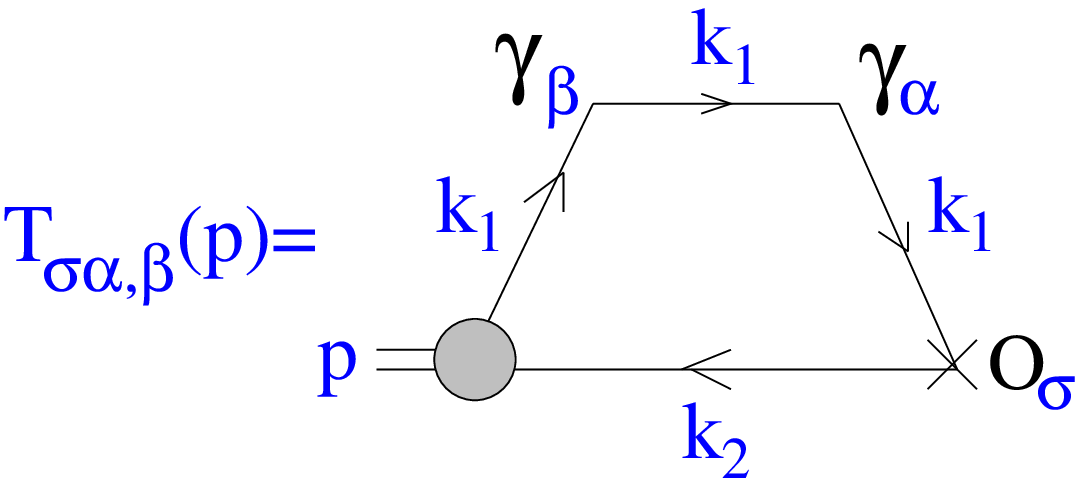,width=6.cm}}
\end{tabular}
\caption{\label{fig:fig4}
Diagrams for 
$T_{\sigma\alpha}(p,q)$ (a) and 
$T_{\sigma\alpha,\beta}(p)=\frac{\partial }{\partial q^\beta} T_{\sigma\alpha}(p,q)|_{q=0}$
(b). }
\end{center}
\end{figure}

We need to calculate $T_{\sigma\alpha,\beta}(p)=
\frac{\partial }{\partial q^\beta} T_{\sigma\alpha}(p,q)|_{q=0}$ (\ref{relation}).  
For the $T_{\sigma\alpha}(p,q)$ shown in Fig. \ref{fig:fig4}a, its derivative 
$T_{\sigma\alpha,\beta}(p)$ is given by the diagram in Fig. \ref{fig:fig4}b. 
The trace corresponding to the Feynman diagram of Fig. \ref{fig:fig4}b reads:   
\begin{eqnarray}
S_{\sigma\alpha;\beta}=
{\rm Sp}\left[ i\gamma_5 (m_2-\hat k_2)O_\sigma (m_1+\hat k_1)
\gamma_\alpha(m_1+\hat k_1)\gamma_\beta(m_1+\hat k_1)\right], 
\end{eqnarray}
where $O_\sigma=\gamma_\sigma(1-\gamma_5)$. Since this expression is further multiplied by 
$G_{\alpha\beta}$, only its antisymmetric part  in $\alpha$ and $\beta$ 
is necessary. The latter reads 
\begin{eqnarray}
S^A_{\sigma\alpha;\beta}&=&(m_1^2-k_1^2)
{\rm Sp} \left(i\gamma_5 (m_2-\hat k_2)\gamma_\sigma(1-\gamma_5) 
\gamma_\beta(m_1-\hat k_1) \right)\nonumber\\
&=&(m_1^2-k_1^2)
\left[4\epsilon_{\beta\alpha\sigma\nu}(m_2 k_1^\nu+m_1 k_2^\nu)
+4im_1 k_2^\nu
(g_{\alpha\sigma}g_{\beta\nu}-g_{\beta\sigma}g_{\alpha\nu})\right].  
\end{eqnarray}
The corresponding expression for the Feynman amplitude takes the form 
\begin{eqnarray}
\label{ta}
T^A_{\sigma\alpha;\beta}(p)&=&\frac{1}{(2\pi)^4i}\int dk_1 dk_2 \delta (p-k_1-k_2)
\frac{-1}{(m_1^2-k_1^2)^2(m_2^2-k_2^2)}
\nonumber\\&&\times
\left[4\epsilon_{\beta\alpha\sigma\nu}(m_2 k_1^\nu+m_1 k_2^\nu)
+4im_1 k_2^\nu(g_{\alpha\sigma}g_{\beta\nu}-g_{\beta\sigma}g_{\alpha\nu})\right]. 
\end{eqnarray}
We define $f_v$ and $f_s$ by the relations 
\begin{eqnarray}
\label{fv}
\frac{1}{(2\pi)^4i}\int dk_1 dk_2\delta (p-k_1-k_2)
\frac{k_1^\nu}{(m_1^2-k_1^2)^2(m_2^2-k_2^2)}&=&p^\nu f_v(p^2),\nonumber\\
\label{fs} 
\frac{1}{(2\pi)^4i}\int dk_1 dk_2\delta (p-k_1-k_2)\frac{1}{(m_1^2-k_1^2)^2(m_2^2-k_2^2)}&=& f_s(p^2). 
\end{eqnarray}
After the $k$-integration in Eq. (\ref{ta}) one recovers the structure of the amplitude from Eq. (\ref{aaa})
\begin{eqnarray}
\label{formula}
T^A_{\sigma\alpha;\beta}(p)=-4p^\nu
\left[f_1 \; \epsilon_{\beta\alpha\sigma\nu} 
+if_2\;(g_{\alpha\sigma}g_{\beta\nu}-g_{\beta\sigma}g_{\alpha\nu})\right],   
\end{eqnarray}
with 
\begin{eqnarray}
f_1&=&(m_2-m_1)f_v+m_1f_s,\nonumber\\
f_2&=& m_1 (f_s-f_v). 
\end{eqnarray}
Let us obtain now spectral representations for the form factors $f^F_v$ and $f^F_s$. 
We first notice that the expressions (\ref{fv}) correspond to the 
triangle Feynman diagram at zero momentum transfer $q=0$. This condition means that both 
$q^2=0$ and $p^2=p'^2$, where $p'=p-q$, so we can write 
$f^F_{i}(p^2)=f^F_{i}(q^2=0,p^2,p'^2=p^2)$, with $i=v,s$.   

It is convenient to start with the case $q^2\ne 0$ and $p^2\ne p'^2$. 
Then the form factors $f_i$ can be written in the form of 
the double spectral representation:   
\begin{eqnarray}
\label{mixingdouble}
f^F_i(q^2,p^2,p'^2)=\int \frac{ds}{s-p^2}\frac{ds'}{s'-p'^2}\Delta_i (s,s',q^2), 
\end{eqnarray}
where the double spectral density $\Delta_i$ can be calculated for any of the form factors 
from the Feynman integral. 

As the next step, we set $q^2=0$, but still treat $p^2$ and $p'^2$ as independent variables.  
In this case the double spectral representations of the form (\ref{mixingdouble}) simplify 
to the following single spectral representations 
\begin{eqnarray}
\label{dfs1}
f^F_s(p^2,p'^2)&=&\frac{1}{16\pi^2}
\int ds \frac{1}{s-p^2}\frac{1}{s-p'^2}
\log\left(\frac{s+m_1^2-m_2^2+\lambda^{1/2}(s,m_1^2,m_2^2)}
{s+m_1^2-m_2^2-\lambda^{1/2}(s,m_1^2,m_2^2)}\right),\nonumber\\ 
\label{dfv1}
f^F_v(p^2,p'^2)&=&\frac{1}{16\pi^2}
\int ds \frac{1}{s-p^2}\frac{1}{s-p'^2}
\frac{\lambda^{1/2}(s,m_1^2,m_2^2)}{s}.
\end{eqnarray}
The representations for the form factors in the form (\ref{dfv1}) 
are valid however only in the region of $p^2$ and $p'^2$ (far) below the threshold. 
The reason for that is the following: 
the representations (\ref{mixingdouble}) and (\ref{dfv1}) are based on the Feynman form of the 
quark propagators. This form is however valid only for highly virtual particles, while in 
the soft region it is strongly distorted by nonperturbative effects. In particular, 
the pole at $k^2=m^2$ in the propagator of a color object, like quark and gluon, is absent. 
Recently, this was confirmed in a lattice study of the gluon propagator \cite{mixinglatprop}. 

The modification of the quark propagator in the soft region leads to the 
change of the spectral representation for the form factors in the region 
of $p^2$ and $p'^2$ near and above the $b\bar q$ threshold. 
 
Notice that the quantities $1/(s-p^2)$ and $1/(s'-p'^2)$ in Eq. (\ref{mixingdouble}) are  
the propagators of the initial and final $b\bar q$ states with virtualities $s$ and $s'$, 
and the squared masses $p^2$ and $p'^2$, respectively. 
The nonperturbative effects which modify the quark and gluon 
propagators in the soft region, modify the propagators of the $b\bar q$ states as well. 

We know the character of these changes in the $p^2$-channel corresponding to the $B$-meson: 
soft interaction of quarks effectively 
replace the factor $1/(s-M_B^2)$ with a regular $B$-meson soft wave function $\varphi_B(s)$ in 
the integrand of (\ref{dfv1}).  

In the color-octet $p'^2$-channel the Feynman propagator $1/(s-p'^2)$ in Eq (\ref{dfv1}) 
is replaced by the propagator $D(s,p'^2)$ which involves proper modifications in the 
soft region. We therefore find the following representation 
for the form factors $f_i(p'^2)\equiv f_i(q^2=0, p^2=M_B^2, p'^2)$  
\begin{eqnarray}
\label{fvd}
f_v(p'^2)&=&\frac{1}{16\pi^2}
\int ds\; \phi_B(s)D(s,p'^2)
\frac{\lambda^{1/2}(s,m_1^2,m_2^2)}{s}, 
\nonumber
\\
\label{fsd}
f_s(p'^2)&=&\frac{1}{16\pi^2}
\int ds \:\phi_B(s)D(s,p'^2)
\log\left(\frac{s+m_1^2-m_2^2+\lambda^{1/2}(s,m_1^2,m_2^2)}
{s+m_1^2-m_2^2-\lambda^{1/2}(s,m_1^2,m_2^2)}\right). 
\end{eqnarray}
We know that $D(s,p'^2)\sim 1/(s-p'^2)$ at large values of $s-p'^2$, and also that 
$D(s,p'^2)$ 
is finite at $s=p'^2$. However, we do not know details of $D(s,p'^2)$ in the 
soft region.   
Motivated by the discussion in \cite{mixingzakharov}, we assume here that the nonperturbative 
effects can be described by the following modification of the propagator 
\begin{eqnarray}
\label{prop} 
D(s,p'^2)=\frac{1}{s-p'^2+M_0^2}
\end{eqnarray}
where $M_0$ is a mass parameter. In order to guarantee the absence of the pole 
in $D(s,p'^2)$ in the heavy quark limit, $M_0$ should scale with the heavy quark mass as 
follows: $M_0^2=O(\Lambda_{QCD}\,m_Q)$. So it is convenient to write $M_0$ in the form 
\begin{eqnarray}
\label{m0} 
M_0^2=w\, m_d\,m_b, 
\end{eqnarray}
where 
$m_d\simeq \Lambda_{QCD}$ is the constituent mass of the light quark, and 
$w$ is a parameter of order unity.\footnote{
Assuming that for the color-octet light $q\bar q$ system Eq (\ref{prop}) remains valid with 
$M_0^2=w m_d^2$, we find $D(k^2=0)\simeq 15\; {\rm GeV}^{-2}$ for $w=1$. 
This is close to the value of the gluon propagator $D(k^2=0)\simeq 18\;{\rm  GeV}^{-2}$ 
as found in \cite{mixinglatprop}. This agreement seems to be reasonable since one can expect 
the propagator of the light $q\bar q$ color-octet system to have a structure in the 
nonperturbative region similar to the structure of the gluon propagator.} 

In the next section we make use of Eqs. (\ref{fvd}) and (\ref{prop}) to analyse $f_{v,s}(p'^2)$ 
for $B_d$ and $B_s$ mesons.

\subsection{\label{sec:mixd}Numerical results.}
The parametrization for the $B$-meson wave function 
and values of the quark masses have been already fixed.  
Recall that we have determined the wave function by fitting the lattice 
data on the weak transition form factors for large momentum transfers at the normalization point 
$\mu\simeq 5\;$ GeV. Therefore, the $B$-meson soft wave function and the 
form factors $f_{s,v}(p'^2)$ at this scale are determined. 

Whereas the $B$-meson wave function is known quite well, good information about the details 
of $D(s,p'^2)$ is lacking. We therefore use the simple Ansatz (\ref{prop}) for $D(s,p'^2)$ 
in the full range of $s$ and $p'^2$ and treat $w$ as a free parameter of order unity.  
We expect that the variation of $w$ in the interval $w=0.5\div 2$ provides reasonable 
error estimates for the form factors and $\Delta B$. 
Notice that the value of $\Delta B$ corresponding to 
$w=1$ agrees favourably with the recent lattice estimates, see Table \ref{table:mixingresults}.  

The form factor $f_s^{(d)}$ which gives the main 
contribution to $\Delta B_B$ is shown in Fig. \ref{fig:oscbd} 
vs $p'^2$.   

Table \ref{table:ourresults} gives numerical results for  the form factors 
at $q^2=0$.  
The relative magnitudes of these form factors can be easily understood taking into 
account their scaling propertiers in the heavy quark limit: 
$f_s^{(d)}\sim m_b^{-1/2}$, $f_s^{(b)}\sim m_b^{-3/2}$, $f_v^{(b)}/f_s^{(b)}\sim 1$. 
\begin{table}[htb]
\centering
\caption{\label{table:ourresults}
Form factors and $\Delta B_{B_q}$ with $q=d$ for $B_d$, and  $q=s$ for $B_s$. The factors 
$f_{1,2}(p'^2)$ are calculated at $p'^2=M^2_{B_d}$ for $B_d$ and at $p'^2=M^2_{B_s}$ for $B_s$.
The errors correspond to the interval $w$=0.5$\div$2.}    
\begin{tabular}{|l|r|r|}
                      &                      $B_d$  &          $B_s$  \\
\hline
$f_v^{(b)}=f_v^{(q)}$ &  $9\pm 3\times10^{-3}$      & $8\pm 3\times10^{-3}$  \\
$f_s^{(b)}$           &  $1.1\pm 0.2 \times10^{-2}$ & $9\pm 3 \times10^{-3}$  \\
$f_s^{(d)}$           &  $1.4\pm 0.5\times10^{-1}$  & $1.0\pm 0.4\times10^{-1}$\\
\hline
$f_1^{(b)}$           &  $7\pm1.5\times10^{-3}$     & $8\pm3\times10^{-3}$  \\
$f_2^{(b)}$           &  $5.5\pm 1.5\times10^{-3}$  & $5.5\pm 2.0\times10^{-3}$ \\
$f_1^{(d)}$           &  $7.5\pm 2.5\times10^{-2}$  & $7.5\pm 2.5\times10^{-2}$ \\
$f_2^{(d)}$           &  $3.5\pm 1.0\times10^{-2}$  & $3.0\pm 1.5\times10^{-2}$ \\
\hline
$\Delta B_{B_q}$      &  $ -0.06\pm0.035$           &   $-0.05\pm0.03$        
\end{tabular}
\end{table}

\begin{figure}[htb]
\begin{center}
\mbox{\epsfig{file=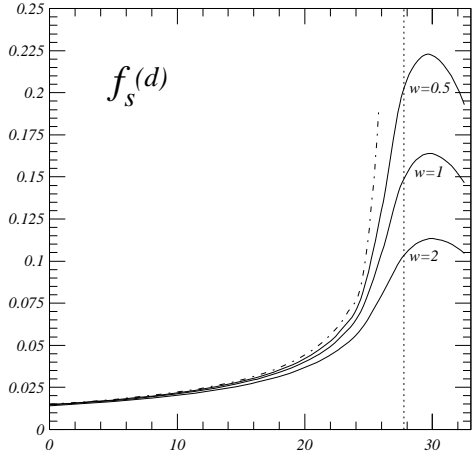,width=7.cm}}
\caption{\label{fig:oscbd}
The form factor $f_s{(p'^2)}$ for the $B_d$ meson vs $p'^2$.  
Solid curves are results of the calculation via the Eqs. \protect(\ref{fvd}) and 
\protect(\ref{prop}) for various values of the parameter $w$: upper curve - $w=0.5$, 
middle curve - $w=1.0$, lower curve - $w=2$. The dash-dotted curve 
in the region $p'^2<(m_b+m_d)^2$ corresponds to $w=0$.  
The vertical dotted line corresponds to $p'^2=M_B^2$. }
\end{center}
\end{figure}
Fig. \ref{fig:oscbd} shows the form factor $f_s^{(d)}$ which gives the main 
contribution to $\Delta B_B$. 

The value of $\Delta B$ is obtained from Eq. (\ref{deltab}) taking 
$\left <\frac{\alpha_s}{\pi}GG\right> = 0.012\; {\rm GeV}^4$ from \cite{svz}.   
Using the Wilson coefficent at the scale $\mu=m_b$ we obtain the  
renorm-invariant factors $\hat B_{B_d}$ and $\hat B_{B_s}$ (the definition are given in 
\cite{mixinglat2}) listed in Table \ref{table:mixingresults}.

Our results are in good agreement with the lattice results, 
except for a slightly different estimate of the SU(3) violating effects in the 
$B_d-\bar B_d$ and $B_s-\bar B_s$ cases. This nice agreement gives support to our 
Ansatz for the description of the nonperturbative effects in the propagator of the 
color-octet $q\bar q$ system. 

We would like to notice that taking the Ansatz 
(\ref{prop}) and (\ref{m0}) for the propagator of the $s\bar q$ system 
and setting $M_0^2=w\, m_s\,m_{\bar q}$, with $w\simeq 1$ allows a 
parameter-free estimate of the $K^0-\bar K^0$ mixing. 
In this case the result is much more stable with respect to the particular value of $w$ 
and therefore to the details of the propagator in the nonperturbative region.  
We obtained this way $\Delta B_K(1\; {\rm GeV})=-0.21 \pm 0.04$. 

\vspace{1cm}

\subsection{\label{sec:mixe}Discussion.}
In this Chapter we have considered corrections to factorization in the $B\bar B$ mixing 
amplitude due to soft-gluon exchanges, assuming that the main effect of such exchanges 
can be described in terms of the local gluon condensate. 

\vspace{.1cm}

\noindent 1. We have proved that within this
approximation the correction of order $\alpha_s\langle GG\rangle$ to factorization 
is {\it negative}. It can be expressed through the specific $B$-meson transition 
form factors at zero momentum transfer. 

\vspace{.1cm}

\noindent 2. We calculated these form factors within our dispersion approach. 
Spectral representations for the form factors 
in terms of the $B$-meson soft wave function and the propagator of the color-octet 
$q\bar b$ system were obtained. The behaviour of this propagator in the nonperturbative 
region was discussed. The obtained numerical estimates for $\Delta B_B$ 
favourably compare with the recent lattice calculations.

\vspace{.1cm}

Let us point out that the proposed approach can be extended to the description of more 
complicated problems. In particular, it can be applied to the analysis of the non-factorizable 
effects in non-leptonic $B$ decays. 
\newpage
\section{\label{sec:v}Quark-binding effects in inclusive decays of heavy mesons}
As we have discussed in Chapter II, the dispersion approach allows us to represent amplitudes 
of the hadron interactions as relativistic spectral representations in terms of the 
hadron wave function. In previous Chapters we analysed in detail form factors of meson decays and 
meson wave functions. 

In this Chapter we apply the dispersion approach to the 
analysis of quark-binding effects in 
inclusive semileptonic decays of heavy mesons. We derive spectral representations 
for various differential distributions, such as  
electron energy spectrum, $q^2$- and $M_X$-distributions, 
in terms of the same meson wave function \cite{msimula}.

Using the quark-model parameters determined in the previous Chapters, 
we provide numerical estimates for various distributions describing inclusive 
$B \to X_c \ell \bar{\nu}_{\ell}$ decays. 

Inclusive decays provide a promising possibility to determine 
those CKM matrix elements which describe the mixing of $b$ quark. This is due to the fact that 
a rigorous theoretical treatment of these decays, including nonperturbative effects, is possible. 
The consideration based on the Operator Product Expansion (OPE) 
and the Heavy Quark (HQ) expansion \cite{cgg}
allows to connect the rate of the inclusive $B$ meson decay with
the rate of the $b$ quark decay. An important
consequence of the OPE is the appearance of quark-binding effects 
in the integrated rates (both total and semileptonic)  
of heavy meson decays only in the second order of the $1/m_Q$ expansion \cite{bsuv93}. 
These second order corrections are expressed in terms of the two hadronic 
parameters, $\lambda_1$ and $\lambda_2$. The latter are the mesonic matrix 
elements of operators of dimension 5 which appear in the OPE of the 
product of two weak currents. The differential distributions can be 
obtained in the form of expansions in inverse powers of the heavy-quark 
mass $m_Q$ \cite{mw,bsuvd94,fls}. 

Whereas providing quite reliable results for the integrated semileptonic decay rate, 
the OPE method encounters difficulties in calculating various 
differential distributions. For instance, before comparing the OPE-based results for the 
differential distributions with the true distributions a proper smearing 
over duality interval is necessary. 

There are several reasons for complications arising in the calculation of differential 
distributions in the resonance region near zero recoil, namely:

\vspace{.25cm}
\noindent 1. duality-violating $1/m_Q$ effects (i.e. the difference between the true distributions 
and the smeared OPE results) in the differential distributions 
near zero recoil which originate from the delay in opening different 
hadronic channels, as noticed in \cite{isgur}. Although these effects 
are cancelled in the integrated semileptonic rate, they can considerably influence 
the kinematical distributions near the zero recoil point;

\vspace{.25cm}
\noindent 2. the convergence of the OPE series for the differential  
distributions persists only in the region where the quark 
produced in the semileptonic decay is sufficiently fast. This means that the OPE 
cannot directly predict distributions in some kinematical regions, such as:  

\begin{itemize}
\item
the photon energy spectrum $d\Gamma/dE_\gamma$ in the radiative 
$B\to X_s\gamma$: the window in the photon 
energy between $m_Q/2$ and $M_Q/2$ turns out to be completely 
inaccessible within the OPE formalism \cite{fls}; 
\item
the lepton energy spectrum $d\Gamma/dE_{\ell}$ at large values of 
$E_{\ell}$ in semileptonic or rare semileptonic decays; 
\item
the lepton $q^2$-distributions in semileptonic $B \to X_c ~(X_u)$ and rare 
$B \to X_s$ decays at 
large $q^2$ near zero recoil; in this region one encounters both the  
quark-binding and duality-violating effects. 
\end{itemize}

Problems related to the quark-binding effects can be solved in principle
by performing proper resummation of the nonperturbative corrections 
which in practice however leads to the appearance 
of {\it a priori} unknown distribution functions \cite{neubertmannel,bsuv94}. 

The inclusion of  
the quark-binding effects in heavy meson decays was 
first done in \cite{altarelli}, where an unknown distribution function of a 
heavy quark inside the heavy meson was introduced. Evidently, this 
distribution function is connected with the wave function of the heavy meson 
which also determines the exclusive transition form factors. To put this 
connection on a more solid basis, it is reasonable to consider the inclusive 
and exclusive processes within the same approach. 

We show in this Chapter that the dispersion approach based on the constituent quark picture 
can be used as an efficient tool for calculating differential distributions in inclusive 
decays of heavy mesons, covering also kinematical regions where 
OPE cannot provide a rigorous treatment. 
Our dispersion approach allows one to take into account quark-binding effects 
in inclusive heavy meson decays in terms of the meson soft wave function. 
The latter describes the heavy meson properties both in exclusive and inclusive 
processes and thus allows one to consider on the same ground 
long-distance effects in various kinds of hadronic processes. 

Quark-model calculations of inclusive distributions are essentially based 
on the evaluation of the box diagram (see Fig. \ref{fig:inclfig1} later on) by introducing the heavy meson
wave function in one way or another. To illustrate the basic features 
of such an approach as well as its advantages and limitations it suffices to 
consider the case of a nonrelativistic potential model with scalar 
currents. Inclusion of relativistic effects can be then performed. 

Let us consider a weak transition induced by the scalar current $J=\bar c b$, where 
both $b$ and $c$ are heavy. To make the nonrelativistic treatment consistent we 
assume that 
\begin{eqnarray}
m_b, m_c\gg \delta m\equiv m_b-m_c \gg \Lambda,
\end{eqnarray}
where $\Lambda$ is the typical scale of the quark binding effects in the heavy meson.   
In the nonrelativistic theory the general expression for the hadronic tensor 
\begin{eqnarray}
\label{w}
W(q_0,\vec q)&=&\frac{1}{\pi}{\rm Im}\int \langle B|T(J(x)J^+(0))|B\rangle e^{-iqx}dx  
\end{eqnarray}
is reduced to the form 
\begin{eqnarray}
\label{wnr}
W(q_0,\vec q)=\frac{1}{\pi}{\rm Im}\langle B|G_{c \bar{d}}(M_B-q^0-i0,\vec q)|B \rangle.  
\end{eqnarray}
Here 
\begin{eqnarray}
G_{c \bar{d}}(E,\vec q)=(\hat{H}_{c \bar{d}}(\vec q)-E)^{-1}
\end{eqnarray} 
is the full Green function corresponding to the full Hamiltonian operator 
of the $c\bar d$ system with the total momentum $\vec q$ 
\begin{eqnarray}
\hat H_{c \bar{d}}(\vec q)=m_c+m_d+\frac{(\hat{\vec k}+\vec q)^2}{2m_c}+\frac{\hat{\vec k^2}}{2m_d}
+V_{c \bar{d}}(\hat r). 
\end{eqnarray}
Thus, the hadronic tensor is the average of the full $c\bar d$ Green function 
over the ground state of the full $b\bar d$ Hamiltonian 
\begin{eqnarray}
\label{eigen}
\hat{H}_{b \bar{d}}|B\rangle=M_B|B\rangle=(m_b+m_d+\epsilon_B)|B\rangle.
\end{eqnarray}
In the rest frame of the $B$-meson one has  
\begin{eqnarray}
\hat H_{b \bar{d}}&=&m_b+m_d+\frac{\hat{\vec k^2}}{2m_b}+\frac{\hat{\vec k^2}}{2m_d}
+V_{b \bar{d}}(\hat r)
\nonumber\\
&\equiv& m_b+m_d+\hat h_{b \bar{d}}   
\end{eqnarray}
The following relation provides a basis for performing the 
OPE in the nonrelativisitc potential model \cite{ourduality} 
\begin{eqnarray}
\label{basic-ope}
\hat H_{c \bar{d}}-(M_B-q^0)&=&
-(\delta m-\frac{\vec q^2}{2m_c}-q^0)+\left(\hat h_{b \bar{d}}-\epsilon_B\right)
\nonumber\\
&&+
\frac{\hat{\vec k^2}+\hat V_1}{2}\left(\frac{1}{m_c}- \frac{1}{m_b}\right)
-\frac{\hat{\vec k}\vec q}{m_c}+O\left(\frac{\gamma^3\delta m}{m_c^3}\right)
\end{eqnarray}
where $\gamma \sim \Lambda_{QCD}$ and we have assumed the following expansion of the potential 
\begin{eqnarray}
\label{Vpot}
\hat V_{Q \bar{q}}=\hat V_0+\frac{\hat V_1}{2m_Q}+\frac{\hat V_2}{2m_Q^2}+...  
\end{eqnarray}
Starting with (\ref{basic-ope}) one constructs an OPE series using the 
amplitude of the 
free $b\to c$ quark 
transition as a zero-order approximation (hereafter referred to as the standard OPE). 
By virtue of the equations of motion, 
\begin{eqnarray}
\left(\hat h_{b \bar{d}}-\epsilon_B\right)|B\rangle=0, 
\end{eqnarray}
one observes the absence of $1/m_Q$ corrections 
to the leading order (LO) $b\to c$  amplitude,  
so that the corrections emerge only at the $1/m_Q^2$ order. 
A detailed analysis of the duality in the nonrelativistic potential model can be 
found in \cite{ourduality}. 
Being completely reliable for the calculation of the integrated decay rate, 
the choice of the free-quark decay as the zero-order approximation turns out to 
be inconvenient however for calculating differential distributions. 
In particular, the distribution 
in the invariant mass of the produced hadronic system, $M_X$, becomes very 
singular and is represented via $\delta(M_X - m_c)$ and its derivatives, 
such that the $1/m_Q$ corrections are even more singular than the leading-order result. 
This is the price one pays for the choice of the zero-order term. 

It is clear that the free-quark transition amplitude is not the 
unique choice of the zero-order approximation, at least in quantum mechanics. 
For instance, another structure of the expansion 
can be obtained if the free $c\bar d$ Green function is used as the zero-order approximation.  

In the nonrelativisitc quantum mechanics the relation between the full $G(E)$ and the free 
$G_0(E)$ Green functions is well known and reads
\begin{eqnarray}
G^{-1}(E)=H-E, \quad G^{-1}_0(E)=H_0-E, \quad G^{-1}(E)-G^{-1}_0(E)=V, 
\end{eqnarray}
or, equivalently,  
\begin{eqnarray}
\label{exp2}
G(E)&=&G_0(E)-G_0(E)VG(E)
\nonumber\\
&=&G_0(E)-G_0(E)VG_0(E)+G_0(E)VG_0(E)VG_0(E)+...
\end{eqnarray} 
For the heavy quark decay, in most of the kinematical $q^2$-region except for a 
vicinity of the zero recoil point, the Green function $G_0$ behaves as $1/m_Q$, and, since the matrix elements of the operator $V$ remain 
finite as $m_Q \to \infty$, the series 
(\ref{exp2}) is an expansion in powers of $1/m_Q$. 
Notice that in the nonrelativistic potential model the expansion (\ref{exp2}) is fully equivalent 
to the OPE series obtained from (\ref{basic-ope}). 
Inserting the expansion (\ref{exp2}) into the expression for the hadronic tensor $W$ given by 
eq. (\ref{wnr}), we 
come to the series shown in Fig \ref{fig:inclfig1}. 
\begin{figure}[htb]
\vspace{.5cm}
\begin{center}
\epsfig{file=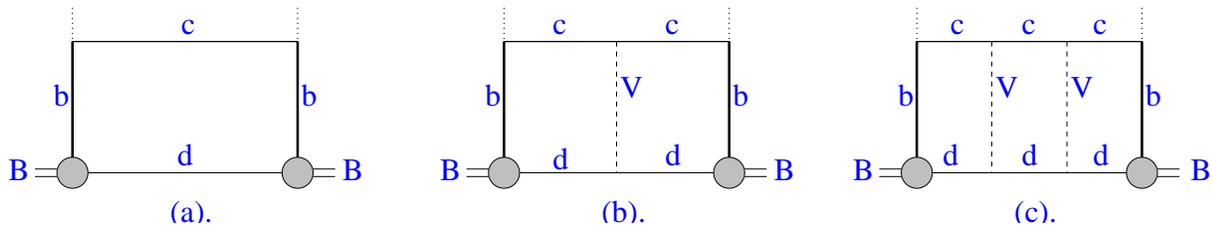,width=16cm}
\vspace{0.75cm}
\caption{ Expansion of the hadronic tensor in the quark model: 
(a) - the box diagram which provides the LO free-quark contribution, 
(b,c) - diagrams contributing in subleading $1/m_Q$ orders and containing the final state interaction $V \equiv V_{c \bar{d}}$. 
The short-dashed lines represent W bosons. 
\label{fig:inclfig1}}
\end{center}
\end{figure}
The leading contribution in the heavy quark limit is given by the box diagram 
of Fig. \ref{fig:inclfig1}a  
with the free $c$ and $\bar{d}$ quarks in the 
intermediate state. The corresponding analytical expression reads 
\begin{eqnarray}
W^{\rm QM}=\frac{1}{\pi}{\rm Im}\langle B|G^0_{c \bar{d}}(M_B-q^0,\vec q)|B\rangle .   
\end{eqnarray}
This is the quantity usually taken into account in quark-model calculations. 
It is easy to see, that the semileptonic decay rate calculation based on the box diagram of Fig. \ref{fig:inclfig1} only,  
reproduces the free quark semileptonic decay rate in the heavy-quark limit, but should contain also $1/m_Q$ 
corrections (see also the general structure of the quark-model results of Ref. \cite{termarti}). Namely, the next-to-leading order (NLO) term in the expansion (\ref{exp2}) is the diagram 
with a single insertion of the potential between the free $c$ and $\bar{d}$ quarks 
(the diagram of Fig. \ref{fig:inclfig1}b). 
It has the order $1/m_Q$ and precisely cancels the $1/m_Q$ contribution of the quark-model box diagram, 
yielding in this way the absence of the $1/m_Q$ correction in the difference between the decay 
rates of bound and free heavy quarks.   

In other words, the quark-model box-diagram calculation is just the first  
term in an alternative expansion of the full Green function: unlike the 
standard OPE series which starts with a single $c$-quark in the intermediate 
state, 
the quark-model starts with the free $c\bar d$ pair which is the eigenstate 
of the Hamiltonian
\begin{eqnarray}
H^0_{c \bar{d}}(\vec q)=m_c+m_d+\frac{(\vec k+\vec q)^2}{2m_c}+\frac{\vec k^2}{2m_d}. 
\end{eqnarray} 
Hereafter we refer to the expansion of the decay rate based on the expansion 
(\ref{exp2}) of the Green function as the quark-model (QM) expansion. 

\newpage
Summing up, the quark model provides an alternative $1/m_Q$ expansion with the following properties: 
\begin{itemize}

\item[1.] the box diagram of Fig. \ref{fig:inclfig1}a provides the LO $1/m_Q$ term and reproduces the 
free-quark decay in the limit $m_Q\to\infty$. All other terms contribute only 
in subleading $1/m_Q$ orders;

\item[2.] the differential distributions in any $1/m_Q$ order are convergent 
for almost all allowed $q^2$, except for the region close to zero recoil; 

\item[3.]  before comparing the calculated differential distributions 
based on the expansion (\ref{exp2}) with the true distributions 
in the resonance region a proper smearing over some duality interval is required;

\item[4.] the $1/m_Q$ correction to the LO term is nonvanishing.

\end{itemize}

Clearly, the properties 1-3 are completely equivalent to the standard OPE,  
while the property 4 makes the quark-model expansion much less convenient than the 
standard OPE, at least for the calculation of the semileptonic decay rate. However:
\begin{itemize} 

\item[5.] the expansion (\ref{exp2}) turns out to be more suitable for the calculation of 
the differential distributions, e.g. for the calculation of $d\Gamma/dM_X$: 
in this case the LO result (the box diagram of Fig. \ref{fig:inclfig1}a) 
is well-defined in the whole kinematical region as well as the higher 
order corrections to it. Thus already the box diagram is appropriate for comparison 
with the experimental $d\Gamma/dM_X$ at all $M_X$ apart from the 
resonance region. 
Beyond the resonance region no additional smearing of the calculated 
$d\Gamma/dM_X$ is required.
\end{itemize}

In full QCD the situation is of course much more complicated. Hadrons are 
coherent states of infinite number of quarks and gluons. Nevertheless, many 
applications of the constituent quark model have proved the treatment of mesons 
as bound states of two constituent quarks to provide a reasonable description 
of their properties. From this viewpoint the arguments given above remain valid. 
Namely, the box diagram represents the main contribution to the hadronic tensor
which reproduces the free-quark decay in the infinite quark mass limit. 
However the hadronic tensor calculated from the box diagram contains $1/m_Q$ term 
compared with the free-quark decay tensor. This linear $1/m_Q$ term is known to be cancelled 
by the $1/m_Q$ contributions of higher order diagrams. In practice, however the 
$1/m_Q$ term of the box diagram is not so dangerous: namely,   
the hadronic tensor calculated from the box diagram, as well as all corrections given by the 
other graphs, are regular in the whole kinematical region. Thus {\it the box diagram 
should provide a reasonable description appropriate for comparison 
with experiment}. Moreover, the box-diagram result 
can be further improved by effectively taking 
into account the higher order term which kills the $1/m_Q$ 
correction contained in the box diagram. We follow this strategy in our analysis and perform a
relativistic treatment of quark-binding effects within a constituent quark picture. 

Our consideration of the quark binding effects in inclusive semileptonic decays is 
based on the relativistic dispersion formulation of the quark model 
discussed in detail in the previous chapters. 
 
The inclusive decay rates as well as the exclusive hadron transition form 
factors are given by double spectral representations in terms of the 
soft meson wave functions. Recall, that the double spectral densities of these  
spectral representations are obtained from the corresponding Feynman graphs, 
whereas subtraction terms should be fixed independently. Such subtraction terms 
in the spectral representations for exclusive 
transition form factors were fixed by requiring the structure of the 
heavy quark expansion in our approach to match the structure of the heavy quark expansion in QCD. 
In this Chapter we proceed along the same lines in inclusive processes. 

\begin{itemize}

\item We construct the double spectral representation of the hadronic tensor 
within the constituent quark model starting with $q^2<0$. The hadronic tensor 
is represented in terms of the soft wave function of the $B$ meson and the 
double spectral density of the box diagram. The hadronic tensor at $q^2>0$ 
is obtained by the analytical continuation. 
Then the $1/m_Q$ expansion of the spectral representation of the 
decay rate is performed and the LO term is shown to reproduce the free 
quark decay rate. The subtraction is defined in such a way  
that the $1/m_Q$ correction to the semileptonic decay rate is absent. 
This corresponds to effectively taking into account other terms  
beyond the box-diagram approximation which contribute in subleading $1/m_Q$ orders. 
Moreover an account of the $1/m_Q^2$ effects of the whole series 
within the box-diagram expression is possible. This is done by introducing a phenomenological cut in the double
spectral representation of the box-diagram which affects only the differential
distributions at large $q^2$. This cut brings the size of the $1/m_Q^2$ corrections in the 
$\Gamma(B \to X_c \ell \bar{\nu}_{\ell})$ in full agreement with the OPE result and keeps the LO and $1/m_Q$ 
correction unchanged. The cut yields differential distributions which are finite also 
in the endpoint $q^2$-region where the HQ expansion series is not properly convergent;

\item We calculate various differential distributions in terms of the $B$-meson soft 
wave function. These distributions are regular in the whole kinematically accessible region 
and, apart from the 
resonance region (where the exact distributions are dominated by single resonances and 
a proper smearing over the duality intervals is necessary), can be directly compared with the observable values. 
The main effect of quark-binding upon these distributions 
is determined unambiguously through the soft wave function, while the $1/m_Q^2$ corrections 
depend on the particular details of an account of the higher-order terms in the 
series (\ref{exp2}).  
However, in practice these 
details are not essential due to the following two reasons: 
first they are numerically small, and 
second, the size of the $1/m_Q^2$ corrections in the integrated semileptonic rate is close to the OPE result. 
So we expect that the size of the $1/m_Q^2$ corrections is 
reasonably reproduced also in other quantities;

\item We perform numerical estimates of various differential 
distributions in inclusive decays 
in terms of the $B$-meson wave function determined from the 
description of the exclusive processes.  
\end{itemize}
In the next section we present necessary formulas
for the free-quark decay and also the OPE prediction for the total integrated rate  
to $1/m_Q^2$ accuracy. In Section \ref{v.b} we construct the dispersion representation 
for the box diagram at $q^2<0$ and discuss its analytical continuation to
$q^2>0$. Section \ref{v.c} gives the $1/m_Q$ expansion of the hadronic tensor in the quark model, 
and Sections \ref{v.d} and \ref{v.d} present numerical results for 
the differential distributions. 

\subsection{\label{v.a}Free quark decay and OPE}
The semileptonic weak decay $Q_2\to Q_1$ is governed by the effective Hamiltonian 
\begin{eqnarray}
\label{heff}
H_{\rm eff}=\frac{G_F}{\sqrt{2}}V_{21}\cdot\bar Q_1(x)\gamma_\mu(1-\gamma_5)Q_2\cdot
\bar l_1\gamma_\mu(1-\gamma_5)l_2,   
\end{eqnarray}
where $V_{21}$ is the corresponding matrix element of the Cabibbo-Kabayashi-Maskawa 
quark mixing matrix. In what follows we assume both leptons to be massless. 
We imply $Q_2=b$ and $Q_1=c$, and therefore use both notations 
throughout this section. 
\begin{figure}[htb]
\vspace{0.25cm}
\begin{center}
\epsfig{file=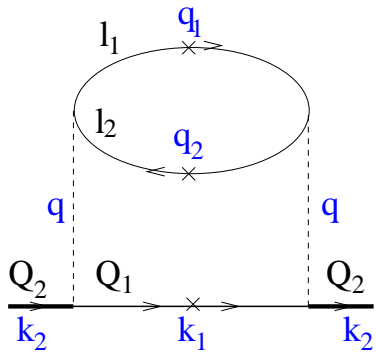,width=5cm}
\vspace{0.25cm}
\caption{The cut Feynman diagram corresponding to the free quark semileptonic decay  
$Q_2(k_2)\to Q_1(k_1)l_1(q_1)\bar l_2(q_2)$. 
Crossed lines mean that the corresponding particles are on the mass shell. 
\label{fig:freequarkdecay}}
\end{center}
\end{figure}
A tree-level decay rate of the free-quark semileptonic decay averaged over the 
polarizations of the initial quark $Q_2$ and summed over the polarizations of 
the final quark $Q_1$ is given the cut Feynman diagram of Fig. \ref{fig:freequarkdecay} 
and has the form 
\begin{eqnarray}
\label{gamma0ini}
\Gamma_0(Q_2\to Q_1 l_1\bar l_2)&=&\frac{G_F^2}{2}|V_{21}|^2\frac{(2\pi)^4}{2m_2}
\int\frac{dk_1 dq_1 dq_2}{(2\pi)^9}
\delta(k_2-k_1-q_1-q_2)
\nonumber\\
&&\times\frac{1}{2}{\rm Sp}\left(\gamma_\mu(1-\gamma_5)
(m_1+\hat k_1)\gamma_\nu(1-\gamma_5)(m_2+\hat k_2)\right)
\theta(k_1^0)\delta(k_1^2-m_1^2)
\nonumber\\
&&\times{\rm Sp}\left(\gamma_\mu(1-\gamma_5)\hat q_2
\gamma_\nu(1-\gamma_5)\hat q_1\right)\theta(q_1^0)\delta(q_1^2)
\theta(q_2^0)\delta(q_2^2)
\end{eqnarray}
It is convenient to introduce additional integration by inserting 
the unity factor 
\begin{eqnarray}
\int dq \delta(q-q_1-q_2) d\mu^2 \delta (q^2-\mu^2)=1 
\end{eqnarray}
in the integrand. Then we can rewrite the decay rate as follows
\begin{eqnarray}
\label{gamma0}
\Gamma_0(Q_2\to Q_1 l_1\bar l_2)&=&\frac{G_F^2|V_{21}|^2}{2}\frac{(2\pi)^4}{2m_2}
\int\frac{dk_1 dq}{(2\pi)^9}d\mu^2\theta(q^0)\delta(q^2-\mu^2)\theta(k_1^0)
\delta(k_1^2-m_1^2)
\nonumber\\
&&\times
\delta(k_2-k_1-q)w^0_{\mu\nu}(k_1,k_2)l_{\mu\nu}(q).
\end{eqnarray}
Here we have introduced the following quantites: 

\noindent The leptonic tensor $l_{\mu\nu}(q)$
\begin{eqnarray}
\label{leptonictensor}
l_{\mu\nu}(q)&=&\int{dq_1 dq_2}
\delta(q-q_1-q_2)\theta(q_1^0)\delta(q_1^2)\theta(q_2^0)\delta(q_2^2)
\;{\rm Sp}\left(\gamma_\mu(1-\gamma_5)\hat q_1 
\gamma_\nu(1-\gamma_5)\hat q_2\right)
\nonumber\\
&=&\frac{4\pi}{3}(q_\mu q_\nu-g_{\mu\nu}q^2). 
\end{eqnarray}
The free-quark tensor 
\begin{eqnarray}
w^0_{\mu\nu}(k_1,k_2)&=&\frac{1}{2}{\rm Sp}\left(\gamma_\mu(1-\gamma_5)
(m_1+\hat k_1)\gamma_\nu(1-\gamma_5)(m_2+\hat k_2)\right)
\nonumber\\
&=&4\left(k_{1\mu}k_{2\nu}+k_{2\mu}k_{1\nu}-g_{\mu\nu}k_1k_2
+i\epsilon_{\mu\nu\alpha\beta}k_{2\alpha}k_{1\beta}\right). 
\end{eqnarray} 
We also denote 
\begin{eqnarray}
\label{c}
C_{SL}(m_2^2,m_1^2,q^2)&=&\frac{1}{2}w^0_{\mu\nu}(k_1,k_2)(q_\mu q_\nu-g_{\mu\nu}q^2)
\nonumber\\
&&=2k_1q\cdot 2k_2q+q^2\cdot 2k_1k_2
\nonumber\\
&&=(m_2^2-m_1^2)^2+q^2(m_2^2+m_1^2)-2q^4. 
\end{eqnarray}
By virtue of these formulas we come to the representation 
\begin{equation}
\label{freerate}
\Gamma_0(Q_2\to Q_1l_1\bar l_2)=\frac{G_F^2 |V_{bc}|^2}{96 \pi^3}\frac{1}{m_2^3} 
\int d\mu^2\lambda^{1/2}(m_2^2,m_1^2,\mu^2) C_{SL}(m_2^2,m_1^2,\mu^2),
\end{equation}
Switching to $m_b=m_2$ and $m_c=m_1$, the differential decay rate takes the familiar form   
\begin{equation}
\label{freeq}
\frac{d\Gamma_0(b\to cl\nu)}{dq^2} = \frac{G_F^2 |V_{bc}|^2}{96 \pi^3}\frac{1}{m_b^3} 
\lambda^{1/2}(m_b^2,m_c^2,q^2) C_{SL}(m_b^2,m_c^2,q^2).
\end{equation}
The integrated semileptonic decay rate is then given by the expression 
\begin{equation}
\label{freeq_total}
\Gamma_0(b\to cl\nu) = \int\limits_0^{(m_b-m_c)^2} dq^2 {d\Gamma_0 \over dq^2} = 
{G_F^2 |V_{bc}|^2 \over 192 \pi^3} \cdot m_b^5 \cdot I_0(r)
\end{equation}
with 
\begin{equation}
I_0(r) = 1 - 8 r + 8 r^3 - r^4 - 12 r^2 {\rm ln}(r), \qquad r \equiv (m_c / m_b)^2. 
\end{equation}
The operator product expansion in QCD predicts the absence of the $1/m_b$ corrections to 
the ratio of the bound to free quark decay rates \cite{cgg,bsuv93}.  
With the $1/m_b^2$ accuracy the integrated rate of the inclusive semileptonic $B\to X_c$ 
decay is given by the expression
\begin{equation}
\label{OPE_total}
\Gamma(B\to X_cl\nu) = \Gamma_0(b\to cl\nu) \cdot 
\left( 1 + {\lambda_1 + 3 \lambda_2 \over 2 m_b^2} 
- 6 {\lambda_2 \over m_b^2} {(1 - r)^4 \over I_0(r)} \right).
\end{equation}
In this expression $\lambda_1$ and $\lambda_2$ are the hadronic matrix elements of the operators of 
dimension 5 appearing in the OPE of the product of the two weak currents. 
The value $\lambda_2 = 0.12 ~ GeV^2$ is well known from the $B-B^*$ mass splitting,
whereas the knowledge of $\lambda_1$ is loose and present estimates range from $-0.6 ~ GeV^2$ to $0$. 
Note that in the nonrelativistic quark potential model one has 
$\lambda_1 = -<\vec{k}^2>$, where $\vec{k}$ is the relative momentum of the constituent 
$Q \bar{q}$ pair. Typically, the nonrelativistic quark model estimates of $\lambda_1$ range 
from $-0.6$ to $-0.4 ~ GeV^2$.

\subsection{\label{v.b}Inclusive meson decay in the dispersion approach}

We now proceed to the inclusive rate for the decay of a pseudoscalar
meson with mass $M_1$ containing a heavy quark $Q_2$, which we
refer to as $P_{Q_2}$, induced by the quark semileptonic transition 
$Q_2\to Q_1 l_1 \bar l_2$. 

As discussed above, within the quark model the dominant
contribution to the inclusive decay rate of any process for $m_Q\to\infty$ 
is given by the box diagram of Fig. \ref{fig:inclfig1}a. 
For semileptonic inclusive decay the corresponding diagram is shown 
in Fig. \ref{fig:boundquarkdecay}. We start with this diagram and discuss the 
relevant modifications due to the other terms presented in Fig. \ref{fig:inclfig1}b,c later. 
\begin{figure}[htb]
\vspace{0.25cm}
\begin{center}
\epsfig{file=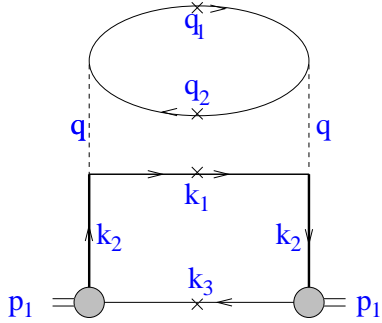,width=5cm}
\vspace{0.5cm}
\caption{The cut Feynman diagram corresponding to the decay
$P_{Q_2}\to X_{Q_1}l_1(q_1)\bar l_2(q_2)$. 
Crossed lines mean that the corresponding particles are on the mass shell. 
We denote $p_2=k_1+k_3$. Our notations follow those of Chapter 3 
where transition form factors have been considered. 
\label{fig:boundquarkdecay}}
\end{center}
\end{figure}
Recall that the quark structure of the pseudoscalar meson transition
into two real quarks is described by the vertex (\ref{pseudoscalarvertex})
\begin{eqnarray}
\frac{G(s_1)}{\sqrt{N_c}} \bar Q_2(k_2)i\gamma_5 q_3(k_3)\delta(\tilde p_1-k_2-k_3),
\end{eqnarray} 
where $\tilde p_1^2=s_1, k_2^2=m_2^2, k_3^2=m_3^2$. 

Notice that the upper leptonic block of the diagram in Fig. \ref{fig:boundquarkdecay}
is exactly the same block $l_{\mu\nu}$ as for the free quark decay. 
The leptonic tensor $l_{\mu\nu}$ is given by Eq. (\ref{leptonictensor}). 

Within the dispersion approach the differential decay rate 
reads
\footnote{This representation is appropriate for the calculation of the
$q^2$-distribution. A more detailed formula is necessary for the lepton energy spectrum. 
It will be considered later in this Chapter.}
\begin{eqnarray}
\label{gammadisp}
\frac{d\Gamma(P_{Q_2}\to X_{Q_1}l_1\bar l_2)}{dq^2}&=&
\frac{G_F^2\,|V_{21}|^2}{2}\frac{(2\pi)^4}{2M_1}\frac{1}{(2\pi)^{12}}
\\
&\times&\int \frac{ds_1\,G(s_1)}{s_1-M_1^2}\frac{ds'_1\,G(s'_1)}{s'_1-M_1^2}
\frac{\pi\,\lambda^{1/2}(s_1,s_2,q^2)}{2s_1}
l_{\mu\nu}(q)\tilde w_{\mu\nu}(\tilde p_1, \tilde p_2, \kappa)|_{\kappa^2\to 0}.
\nonumber
\end{eqnarray}
The hadronic tensor $\tilde w_{\mu\nu}(\tilde p_1, \tilde p_2,\kappa)$ 
represents the full double spectral density of the 
Feynman box diagram $w_{\mu\nu}(p_1, p_2,\kappa)$ of Fig. \ref{fig:boxoffforward} a. 
This quantity is constructed from the double discontinuity 
$\tilde w^D_{\mu\nu}(\tilde p_1, \tilde p_2,\kappa)$ 
\begin{eqnarray}
\tilde w^D_{\mu\nu}(\tilde p_1, \tilde p_2,\kappa)=
{\rm disc_{s_1}}{\rm disc_{s'_1}}w_{\mu\nu}(\tilde p_1, \tilde p_2,\kappa), 
\end{eqnarray}
by applying the appropriate subtraction prescription.  
Fig. \ref{fig:boxoffforward}b shows the momenta in the integrand of Eq. (\ref{gammadisp}). 
They satisfy the following relations 
\begin{eqnarray}
\tilde p'_1&=&\tilde p_1+\kappa, 
\nonumber\\
\tilde q'&=&\tilde q+\kappa,  
\nonumber\\
\tilde p_2&=&\tilde p_1-\tilde q, 
\nonumber\\
\tilde p_1^2&=&s_1, 
\nonumber\\
\tilde p'^2_1&=&s'_1, 
\nonumber\\
\tilde p^2_2&=&s_2. 
\nonumber\\
\tilde q^2&=&\tilde q'^2=q^2, 
\end{eqnarray} 
\begin{center}
\begin{figure}[htb]
\begin{tabular}{cc}
\epsfig{file=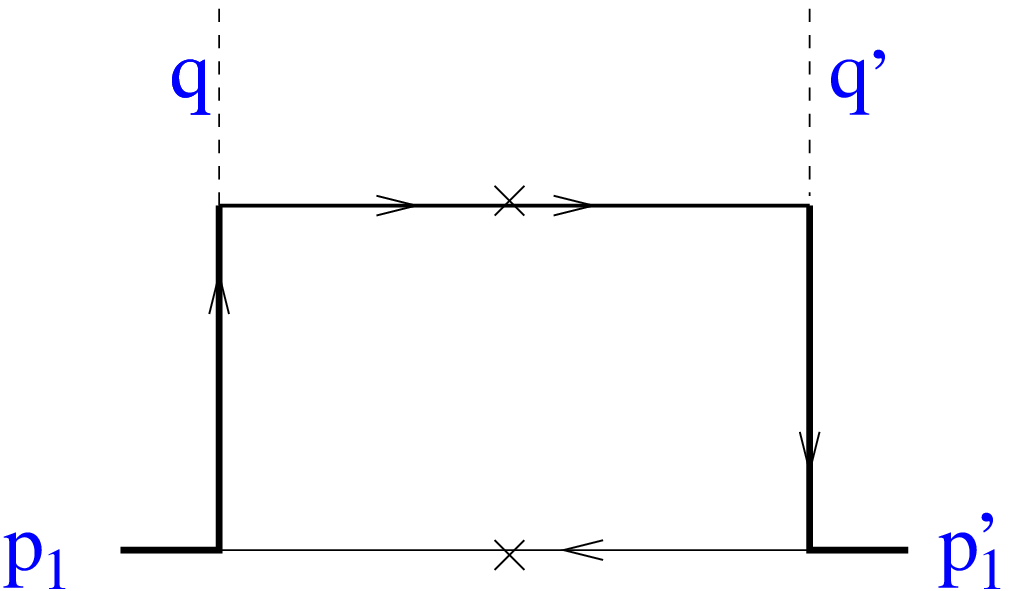,width=5cm}
\qquad&\qquad
\epsfig{file=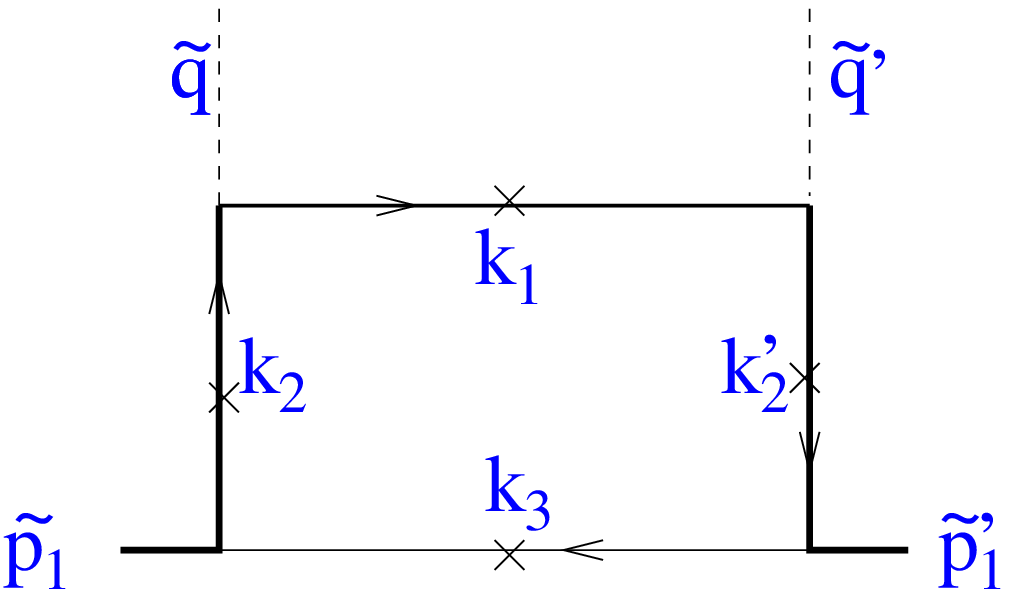,width=5cm}
\end{tabular}
\caption{a: The box diagram for $w_{\mu\nu}(p_1, p_2, q-q')$;  
b: its double discontinuity in the variables $s_1=\tilde p_1^2$ and $s_1'=\tilde p_1^{'2}$,  
$\tilde w_{\mu\nu}(\tilde p_1, \tilde p_2,\kappa=\tilde q-\tilde q')$. 
Crossed lines correspond to particles on the mass shell.  
\label{fig:boxoffforward}}
\vspace{.5cm}
\end{figure}
\end{center}

\subsubsection{The spacelike region}
The double discontinuity of the box diagram can be 
calculated in the region $q^2\le 0$ and $q'^2\le 0$ by placing all
intermediate particles in the loop on their mass shell as shown in 
Fig. \ref{fig:boxoffforward}b.  
To obtain the corresponding expression for positive values 
$q^2=q'^2>0$ corresponding to the real decay kinematics 
we shall perform the analytical continuation in $q^2$. 

At $q^2, ~ q'^2\le 0$ the explicit expression for 
$\tilde w^D_{\mu\nu}$ reads 
\begin{eqnarray}
\label{tildew}
&&\tilde w^D_{\mu\nu}(\tilde p_1, \tilde p_2,\kappa)=
\int dk_1 dk_2 dk'_2 dk_3 
\delta(k_1^2-m_1^2) 
\delta(k_2^2-m_2^2)
\delta(k'^2_2-m_2^2)
\delta(k_3^2-m_3^2)
\nonumber\\
&&\qquad\times
\delta(\tilde p_1-k_2-k_3)
\delta(\tilde p'_1-k'_2-k_3)
\delta(\tilde p_2-k_1-k_3) 
\nonumber\\
&&\qquad\times 
(-1)\,{\rm Sp}\left(i\gamma_5\,(m_3-\hat k_3)
i\gamma_5\,(m_2+\hat k'_2)\gamma_\mu(1-\gamma_5)
(m_1+\hat k_1)\gamma_\nu(1-\gamma_5)(m_2+\hat k_2)\right).    
\end{eqnarray} 
The function $\tilde w_{\mu\nu}(\tilde p_1, \tilde p_2,\kappa)$ 
contains several Lorentz structures multipled by the invariant amplitudes. The latter depend 
on the six independent scalar variables 
\begin{eqnarray}
s_1, s_1', s_2, q^2, q'^2, \kappa^2. 
\end{eqnarray} 
We are interested in the case $q^2=q'^2$ and $\kappa^2\to 0$. 

Let us consider the trace in the integrand of $\tilde w^D_{\mu\nu}$. 
Hereafter we neglect 
terms proportional to $\kappa$. Since all particles are on the mass shell, we find 
\begin{eqnarray}
-(m_2+\hat k_2)i\gamma_5\,(m_3-\hat k_3)
i\gamma_5\,(m_2+\hat k_2)=(2k_2k_3+2m_2m_3)(m_2+\hat k_2).
\end{eqnarray}
Therefore, 
\begin{eqnarray}
\label{hadronictrace}
&&-{\rm Sp}\left(i\gamma_5\,(m_3-\hat k_3)
i\gamma_5\,(m_2+\hat k_2)\gamma_\mu(1-\gamma_5)
(m_1+\hat k_1)\gamma_\nu(1-\gamma_5)(m_2+\hat k_2)\right)
\nonumber\\
&&\qquad\qquad=2(s_1-(m_2-m_3)^2)w^0_{\mu\nu}(k_1,k_2).
\end{eqnarray}
From the free quark decay we know that 
\begin{eqnarray}
w^0_{\mu\nu}(k_1,k_2)l_{\mu\nu}(q)=\frac{8\pi}{3}C_{SL}(m_2^2,m_1^2,q^2). 
\end{eqnarray}
Clearly, the contribution of the traces in the spectral representation 
for $d\Gamma/dq^2$ 
turns out to be independent of the internal integration variables. 
Therefore, we only have to consider the scalar part of $\tilde w^D_{\mu\nu}$  
defined according to the relation 
\begin{eqnarray}
\label{tildea}
&&\tilde A_D(s_1,s'_1,p^2_2,q^2,q'^2,\kappa^2)=
\int dk_1 dk_2 dk'_2 dk_3 
\delta(k_1^2-m_1^2) 
\delta(k_2^2-m_2^2)
\delta(k'^2_2-m_2^2)
\delta(k_3^2-m_3^2)
\nonumber\\
&&\qquad\times
\delta(\tilde p_1-k_2-k_3)
\delta(\tilde p'_1-k'_2-k_3)
\delta(\tilde p_2-k_1-k_3). 
\end{eqnarray}
Explicit calculations give for $\tilde A_D$ the following expression
\begin{eqnarray}
\label{ad}
\bar A_D(s_1,s_2,q^2)
=\frac{\pi\theta(...)}{2\lambda^{1/2}(m_1^2,m_2^2,q^2)}. 
\end{eqnarray}
Notice that the argument of the $\theta$ function in eq (\ref{ad}) is just the 
same as for the spectral density of the triangle graph for the form 
factor given by Eq. (\ref{thetafunction}). 

Eq. (\ref{tildea}) provides the double discontinuity but, as usual, we 
need to define in addition a subtracion prescription.  
The form of this prescription cannot be determined within the dispersion approach 
and should be fixed from some other arguments. 
We shall determine the subtraction prescription by matching to the known QCD result and 
requiring the absence of the $1/m_Q$ corrections in the ratio of the bound 
and free quark decay rates. As we have discussed, in the quark model the absence of the 
$1/m_Q$ corrections is provided by the other diagrams shown in Fig. \ref{fig:inclfig1}b,c 
which cancel the $1/m_Q$ term in the box diagram. Therefore requiring the absence of the 
$1/m_Q$ corrections corresponds to an effective account for these terms.  

Our subtraction prescription explicitly reads 
\begin{eqnarray}  
\label{prescr}
\tilde A(s_1,s_2,q^2)=
\frac{M_1}{\sqrt{s_1}}\tilde A_D(s_1,s_2,q^2).   
\end{eqnarray}
As we shall demonstrate in the next section, this prescription indeed provides the 
required absence of the $1/m_Q$ corrections in agreement with the OPE. 

For the analysis of the limit $m_Q\to\infty$ it is convenient to have the 
'external' integration over $s_1$ and the 'internal' integration over $s_2$ 
in the spectral representation for the decay rate. The solution of the $\theta$ 
function then takes the form 
\begin{eqnarray}
\label{limitss2}
(m_2+m_3)^2<s_1&<&\infty
\nonumber\\
s_2^{-}(s_1,q^2)<s_2&<&s_2^{+}(s_1,q^2)
\nonumber
\end{eqnarray}
where the limits $s_2^{\pm}$ are obtained by setting $\eta=\pm 1$ 
in the relation
\begin{eqnarray}
\label{s22eta}
s_2(s_1,q^2)&=&(m_1+m_3)^2+
\frac{m_1}{m_2}\left(s_1-(m_2+m_3)^2    \right)
+\frac{m_1}{m_2}(\omega -1)(s_1-m_2^2-m_3^2) \nonumber \\
&+&\eta\frac{m_1}{m_2}\lambda^{1/2}(s_1,m_2^2,m_3^2)\sqrt{\omega^2-1}. 
\end{eqnarray}
The quark recoil $\omega$ is defined according to the relation
\begin{eqnarray}
\label{omega}
q^2=(m_2-m_1)^2-2m_1m_2(\omega-1).
\end{eqnarray}
Isolating the free-quark decay amplitude we come to the following dispersion 
representation for the differential inclusive semileptonic decay rate 
at $q^2\le 0$
\begin{eqnarray}
\label{inclrate}
{d\Gamma \over dq^2}&=& K_0(q^2)
\int ds_1 \varphi^2(s)\frac{s_1-(m_2-m_3)^2}{8\pi^2s_1}
\frac{m_2^3}{\sqrt{s_1}}
\int\limits_{s_2^{-}(s_1,q^2)}^{s_2^{+}(s_1,q^2)}ds_2
\frac{\lambda^{1/2}(s_1,s_2,q^2)}{\lambda^{1/2}(m_1^2,m_2^2,q^2)}, 
\end{eqnarray}
where
\begin{eqnarray}
\label{k0}
K_0(q^2)\equiv\frac{1}{\lambda^{1/2}(m_2^2,m_1^2,q^2)}\frac{d\Gamma_0}{dq^2}.  
\end{eqnarray}
Note that in Eq. (\ref{inclrate}) the free-quark differential rate $d\Gamma_0 / dq^2$ 
factorizes out, so that the differential rate for a bound quark is a product of the free-quark 
differential rate and a bound state factor, as already noted in \cite{termarti}.
As in the previous chapters, we use the notation $\varphi(s)=G(s)/(s-M_1^2)$;  
the normalization condition for $\varphi(s)$ is given by Eq. (\ref{norma}). 

It is convenient to rearrange Eq. (\ref{inclrate}) by isolating
under the integral the structure similar to the structure of the
normalization condition (\ref{norma})
\begin{eqnarray}
\label{inclrate2}
{d\Gamma \over dq^2} & = & \frac{d\Gamma_0}{dq^2}
\int ds_1 \varphi^2(s_1)
\left[s_1-(m_2-m_3)^2\right]\frac{\lambda^{1/2}(s_1,m^2_2,m_3^2)}{8\pi^2s_1}
r(s_1,q^2),
\nonumber  \\
&&r(s_1,q^2)=\int\limits_{-1}^{1}\frac{d\eta}{2}
\frac{\lambda^{1/2}(s_1,s_2,q^2)}{\lambda^{1/2}(m^2_2,m_1^2,q^2)}, 
\end{eqnarray}
and $s_2$ is related to $\eta$ by Eq. (\ref{s22eta}). 
As we shall see later, 
\begin{eqnarray}
r(s_1,q^2)\sim 1  
\end{eqnarray}
in the heavy-quark limit, such that thanks to the normalization condition 
of the soft wave function Eq. (\ref{norma}) one gets 
\begin{eqnarray}
\frac{d\Gamma}{dq^2} \to \frac{d\Gamma_0}{dq^2} 
\end{eqnarray} 
as $m_Q \to \infty$. 

\subsubsection{The timelike region and the anomalous contribution}

To obtain the spectral representation at $q^2>0$ we perform the
analytical continuation in $q^2$. This procedure 
is done along the same lines as in the case of the transition form factor 
which has been discussed in
detail in Chapter II. 
As a result of this procedure 
in addition to the normal part which is just the expression
(\ref{inclrate2}) taken at $q^2>0$, 
the anomalous part emerges due to the non-Landau type 
singularities of the Feynman graph. 
The location of the singularities in the complex $s_1$ and $s_2$ planes 
corresponding to the situation of the 'external' $s_1$ integration 
is shown in Fig. \ref{fig:locations1s2}. 
\begin{center}
\begin{figure}[h]
\begin{tabular}{lr}
\mbox{\epsfig{file=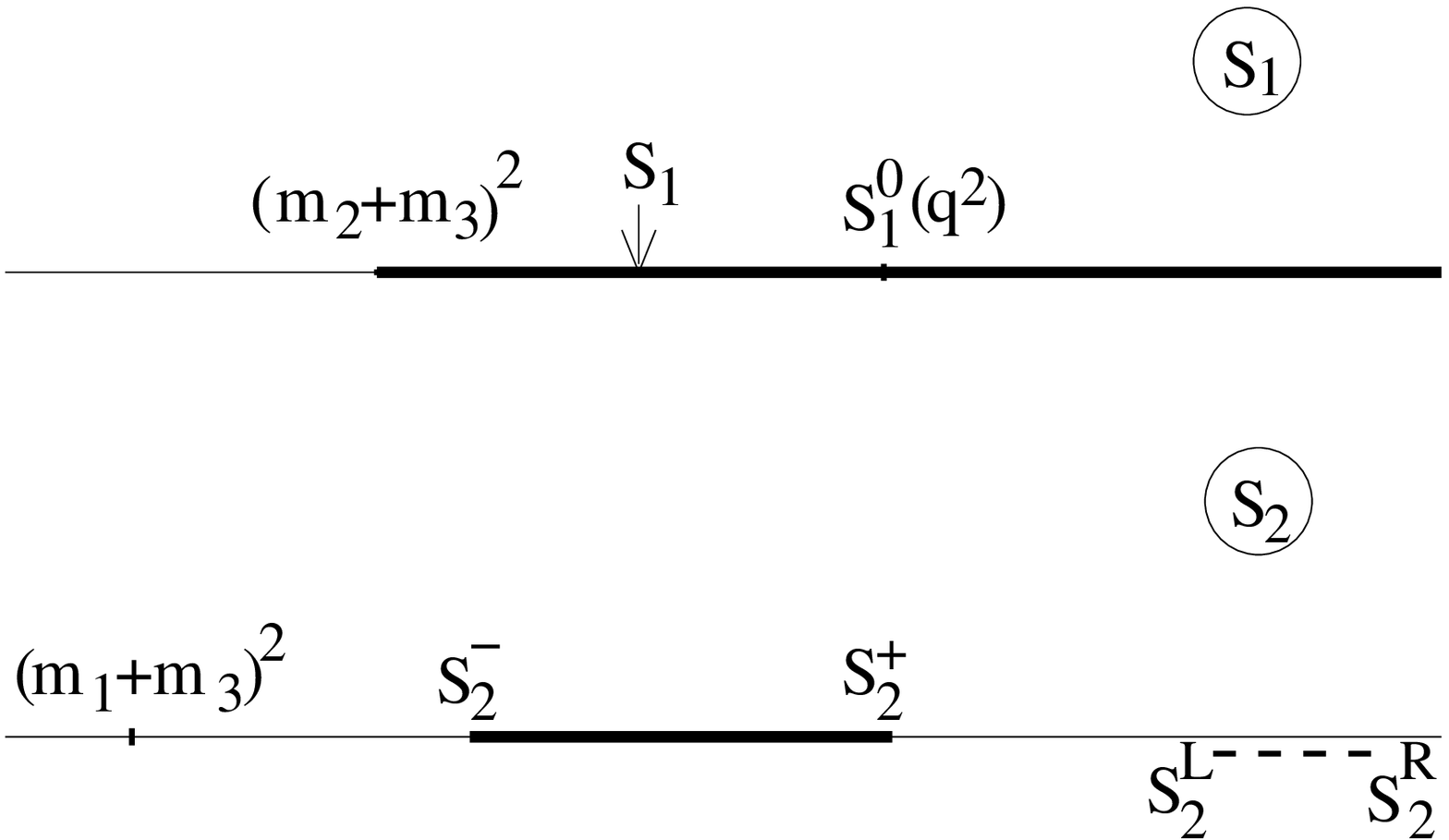,width=7.cm}}
\qquad
& 
\qquad
\mbox{\epsfig{file=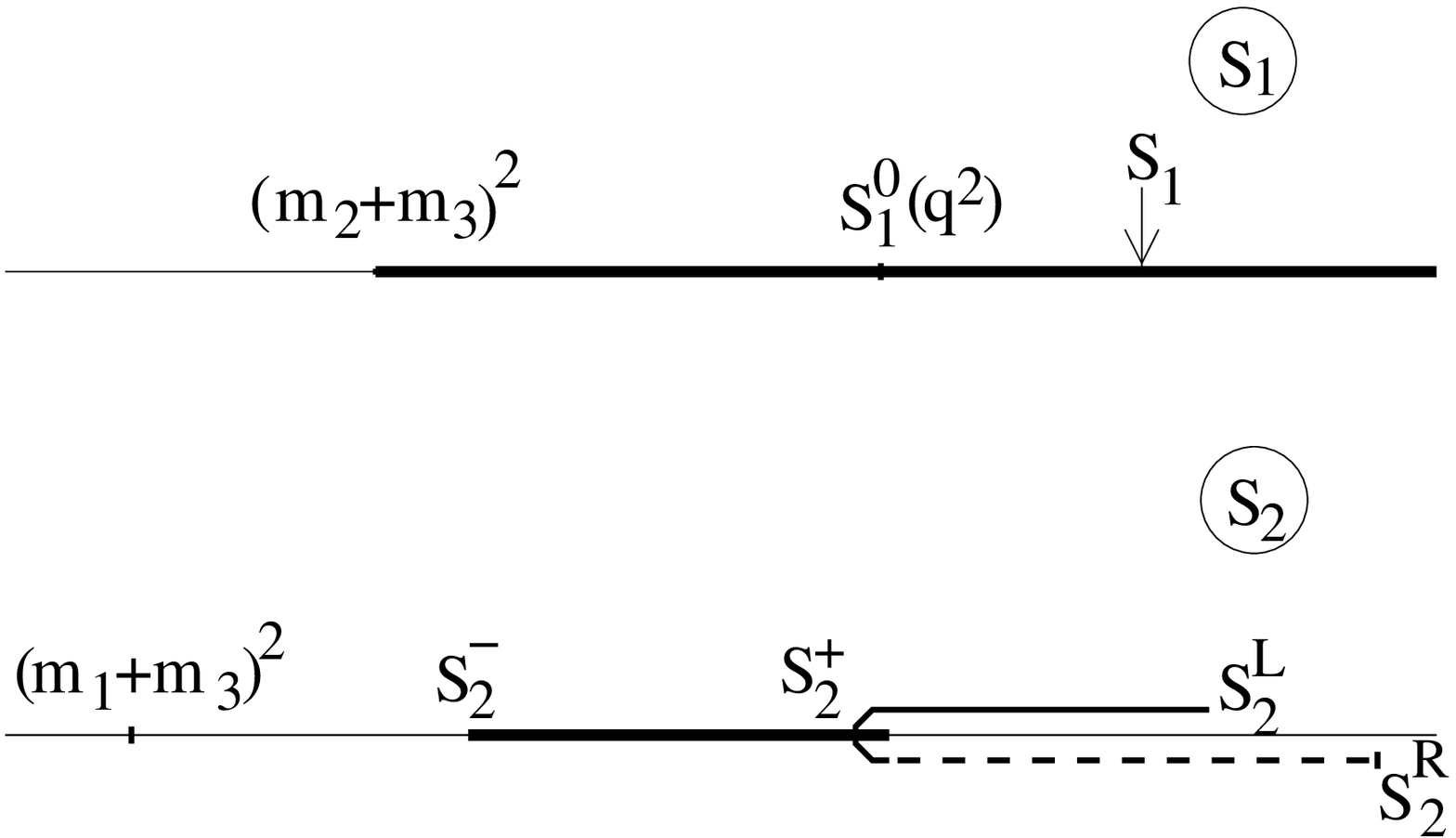,width=7.cm}}
\end{tabular}
\vspace{.8cm}
\caption{\label{fig:locations1s2}
The integration region for $m_2>m_1$ and $q^2>0$ for the order of 
integration $\int ds_1\int ds_2$: a. $s_1<s_1^0$, b. $s_1>s_1^0$.  
Solid lines - cuts on the physical sheet, 
dashed lines - cuts on the second sheet.}
\vspace{.5cm}
\end{figure}
\end{center}
The definition of 
$s_2^{\pm}$ is given by Eq. (\ref{limitss2}) and 
\begin{eqnarray}
s_2^R(s_1,q^2)&=&(\sqrt{s_1}+\sqrt{q^2})^2, \nonumber \\
s_2^L(s_1,q^2)&=&(\sqrt{s_1}-\sqrt{q^2})^2, \nonumber \\
\sqrt{s_1^0(q^2)}&=&\frac{q^2+m_2^2-m_1^2}{2\sqrt{q^2}}+\sqrt{
\left({\frac{q^2+m_2^2-m_1^2}{2\sqrt{q^2}}}\right)^2+m_3^2-m_2^2}.
\end{eqnarray}
As one can see, the anomalous cut appears on the physical $s_2$ sheet 
between $s_2^+(s_1,q^2)$ and $s_2^L(s_1,q^2)$ 
in the region $q^2\le (m_2-m_1)^2$ for the values $s_1>s^0_1(q^2)$. 

Respectively, the spectral density 
$r(s_1,q^2)$ acquires the anomalous piece and takes at $q^2>0$ the 
following form 
\begin{eqnarray}
&&r(s_1,q^2)=\frac{m_2}{\sqrt{s_1}}
\frac{m_2^2}{\lambda^{1/2}(s_1,m^2_2,m_3^2)\lambda(m^2_2,m_1^2,q^2)}\nonumber 
\\
&&\qquad\times
\left[{
\int_{s_2^-(s_1,q^2)}^{s_2^+(s_1,q^2)} ds_2\lambda^{1/2}(s_1,s_2,q^2)
+2\theta(q^2)\theta(s_1>s_1^0)\int_{s_2^+(s_1,q^2)}^{s_2^L(s_1,q^2)} ds_2\lambda^{1/2}(s_1,s_2,q^2)
}\right].  
\end{eqnarray}

The $q^2$-behavior of the anomalous
term is determined by the lower limit of the $s_1$ integration, $s^0_1(q^2)$. 
Namely, its contribution to the semileptonic rate reads 
\begin{eqnarray}
\frac{1}{\Gamma}\frac{d\Gamma^{anom}}{d\omega}\simeq 
\frac{\Lambda^3}{m_Q^3\sqrt{\omega-1}}R^{anom}(\omega),
\end{eqnarray}
where 
\begin{eqnarray}
R^{anom}(\omega)=\int\limits_{{m_Q}\sqrt{\omega-1}}^{\infty}
dk k^2 w^2(k).
\end{eqnarray}
Here $k= \lambda^{1/2}(s_1,m_2^2,m_3^2) / 2\sqrt{s_1}$ 
is the relative momentum of the quarks inside the $B$-meson. The 
wave function $w(k)$ is normalized according to the relation (\ref{normaw}) 
leading to 
\begin{eqnarray}
R^{anom}(1)=1.
\end{eqnarray}
Since the soft wave function is steeply falling beyond the 
confinement region where $\vec k^2\lesssim \Lambda^2$, the anomalous contribution 
becomes inessential already at $\omega-1\simeq \Lambda/m_b$. 
Only in the endpoint region $\omega-1\lesssim \frac{\Lambda^2}{m_Q^2}$ 
the anomalous contribution to the differential distribution becomes strong 
and diverges like 
\begin{eqnarray}
\frac{1}{\Gamma}\frac{d\Gamma^{anom}}{d\omega}\simeq \frac{\Lambda^3}{m_Q^3}
\frac{1}{\sqrt{\omega-1}}. 
\end{eqnarray}
The contribution of the anomalous term to the integrated semileptonic rate 
of the order 
\begin{eqnarray}  
\Gamma^{anom}/\Gamma\simeq \Lambda^2/m_Q^2 
\end{eqnarray}
comes from the endpoint region, whereas the rest of the decay phase space 
provides only the relative $\Lambda^3/m_Q^3$ 
anomalous contribution to the semileptonic decay rate. 

Therefore, the anomalous contribution is negligible at all $\omega$ 
except for the endpoint region $\omega-1\lesssim \frac{\Lambda^2}{m_Q^2}$, which 
is in fact a very narrow region near zero recoil. As we have discussed, the HQ expansion for the differential distributions 
is anyway ill-defined in this kinematical region. Contributions of 
the same order of magnitude come also from other terms in the expansion 
(\ref{exp2}), and keeping this anomalous contribution is beyond  
the accuracy of our considerations. 
Thus we shall systematically omit the anomalous contribution in numerical calculations.

\newpage
\subsection{\label{v.c}The $1/m_Q$ expansion of the semileptonic decay rate and $d\Gamma/dq^2$}

In this section we perform the $1/m_Q$ expansion of the meson inclusive decay rate. We show that:
\begin{itemize}

\item[a.] in the leading $1/m_Q$ order the heavy meson inclusive decay rate is equal to the free quark decay rate;

\item[b.] our subtraction prescription  (\ref{prescr}) leads to the 
differential distribution $d\Gamma/dq^2$ given by eq (\ref{inclrate2}) which satisfies the relation 
\begin{eqnarray}
\frac{d\Gamma(B\to X_c \ell \bar{\nu}_{\ell})/dq^2}
{d\Gamma_0(b\to c \ell \bar{\nu}_{\ell})/dq^2}=
1+O(1/m_Q^2)
\end{eqnarray}
in most of the $q^2$ region except for 
a close vicinity of zero recoil point. This property guarantees the absence of the $1/m_Q$ 
corrections in the ratio of the integrated rates, i.e. 
\begin{eqnarray}
\frac{\Gamma(B\to X_c \ell \bar{\nu}_{\ell})}{\Gamma_0(b\to c \ell \bar{\nu}_{\ell})}=1+O(1/m_Q^2);
\end{eqnarray}
\item[c.] the size of the $1/m_Q^2$ corrections can be tuned such that they become numerically 
close to the OPE prediction. This is done by introducing the cut in the spectral 
representation of the decay rate of the $B$ meson. This cut affects only the 
differential distribution $d\Gamma(B \to X_c \ell \bar{\nu}_{\ell})/dq^2$ at large $q^2$ 
near zero recoil, i.e. in the region 
$\omega\le 1+O(1/m_Q)$. 
\end{itemize}

An important feature of the whole approach is that already the zero order 
expression provides a realistic $M_X$-distribution. As we shall see later, the 
modifications b) and c) while affecting the 
total rate and the $q^2$-distributions at large $q^2$, only moderately affect 
the $M_X$-distribution, so that the latter is completely determined by the soft 
$B$-meson wave function.

Let us consider the $1/m_Q$ expansion of the spectral density $r_D$. 
The normal part of $r_D(s_1,q^2)$ has a simple form 
\begin{eqnarray}
\label{r2}
r_D(s_1,q^2)=\frac{m_2}{M_1}
\int_{-1}^{1}\frac{d\eta}{2}\frac{\lambda^{1/2}(s_1,s_2,q^2)}{\lambda^{1/2}(m_2^2,m_1^2,q^2)}.
\end{eqnarray}
This representation is a convenient starting point for performing the $1/m_Q$  
expansion. Let us point out that although the integration in $\eta$ can be
easily performed, it is more convenient to work out the HQ expansion
before the integration. 

Assuming that $m_2$ is large and that the meson wave function is localized in the
region $z_1\simeq \Lambda$ we 
obtain the following expression for $\lambda(s_1,s_2,q^2)$ valid at all $q^2$ 
\begin{eqnarray}
\label{lambda}
\lambda(s_1,s_2,q^2)&\to& m_2^4\left({1+\frac{z_1+m_3}{m_2}}\right)^2
\nonumber\\
&&\times\left[{
\lambda(1,\hat q^2,\hat r^2)
+\frac{2\eta}{m_2}\chi(z_1)\sqrt{z_1(z_1+2m_3)}(1+\hat q^2-\hat r^2)
\lambda^{1/2}(1,\hat q^2,\hat r^2)
}\right.
\nonumber\\
&&\left.{+\frac{z_1(z_1+2m_3)}{m_2^2}(1+\hat q^2-\hat r^2)^2
+\frac{\eta^2}{m_2^2}z_1(z_1+2m_3)\lambda(1,\hat q^2,\hat r^2)}\right]
\end{eqnarray}
where $\hat q^2=q^2/m_b^2$, $\hat r=m_1/m_2$ and $\chi(z_1)=1-(z_1+m_3)/2m_2$. 
In the limit $m_2\to\infty$ we can expand the $\lambda(s_1,s_2,q^2)$ in powers
of $1/m_2$. Notice however that an actual expansion parameter is not $1/m_2$ but
rather 
\begin{eqnarray}
\frac{\sqrt{z_1(z_1+2m_3)}}{m_2\lambda^{1/2}(1,\hat q^2,\hat r^2)}, 
\end{eqnarray}
and the averaging over the $B$ meson state implies 
$\sqrt{z_1(z_1+2m_3)}\simeq \Lambda$. 
Hence the region where the expansion is fastly converging is 
\begin{eqnarray}
m_2\lambda^{1/2}(1,\hat q^2,\hat r^2)\gg \Lambda. 
\end{eqnarray}
This relation can be written as 
\begin{eqnarray}
|\vec k_1|=\lambda^{1/2}(m_2^2,m_1^2,q^2)/2m_2\gg \Lambda,
\end{eqnarray}
which means that in the rest frame of the $b$ quark the 
daughter quark has a 3-momentum much bigger than $\Lambda$. 

The final expression reads 
\begin{eqnarray}
\label{r2b}
r_D(s_1,q^2)&\to&\frac{m_2}{M_1}\left({1+\frac{z_1+m_3}{m_2}}\right)
\left[{
1+\frac{z_1(z_1+2m_3)}{2m_2^2}\left({1+\frac{8\hat q^2}{3\lambda(1,\hat q^2,\hat r^2)}
}\right)}\right]. 
\end{eqnarray}
The $1/m_Q$ term in the ratio of the bound to free quark distributions is
generated by the $(m_2+z_1+m_3)/M_1$ term in $r_D$.

As we have discussed this linear $1/m_Q$ term contained in 
the box diagram cancels against the $1/m_Q$ terms coming from other terms 
in the expansion (\ref{exp2}). Thus the main contribution of these  
terms can be taken into account by performing the 
subtraction in the spectral representation for the box diagram which kills the 
$1/m_Q$ term. This is implemented by defining the subtraction prescription 
as follows: 
\begin{eqnarray}
\label{r3}
r(s_1,q^2)=\frac{M_1}{\sqrt{s_1}}r_D(s_1,q^2).  
\end{eqnarray}
After performing the subtraction and taking into account that in the heavy
quark limit $z(z+2m_3)=\vec k^2$ we come to the following 
relation  
\begin{eqnarray}
\label{r4}
R(q^2)\equiv \frac{d\Gamma(B\to X_c \ell \bar{\nu}_{\ell}) / dq^2}{d\Gamma_0(b\to c \ell \bar{\nu}_{\ell}) / dq^2} \to
1+\frac{<\vec k^2>}{2m_2^2}\left({1+\frac{8\hat q^2}{3\lambda(1,\hat q^2,\hat r^2)}
}\right).  
\end{eqnarray}
This expansion is valid in the region of $q^2$ such that $\lambda(1,\hat
q^2,\hat r^2)=O(1)$, i.e. in most 
of the $q^2$ phase space except for the region near zero recoil where $\lambda(1,\hat q^2,\hat r^2) \simeq 0$.  

The expression (\ref{r4}) has the following features:
\begin{itemize}

\item[1.] In the leading $1/m_Q$ order the ratio $R(q^2)$ is equal to one. 
Thus the decay rate of the free and the bound quark 
coincide in the heavy quark limit for all $q^2$. 
Beyond the leading order, the differential distribution coincide also 
within the $1/m_Q$ accuracy in most of the $q^2$ phase space, except for the region near zero recoil. 
This guarantees the absence of the $1/m_Q$ corrections in the ratio of the integrated rates as well. 
Thus, our description is in full agreement with the OPE results within the $1/m_Q$ order; 

\item[2.] Since the box diagram represents only a part of the $1/m_Q^2$ corrections, 
we cannot expect the box diagram alone to reproduced correctly the $1/m_Q^2$ term 
in the ratio 
of the integrated rates $\Gamma / \Gamma_0$. 
In fact, the sign of the 
$1/m_Q^2$ correction in eq. (\ref{r4}) turns out to be opposite to the OPE result 
(cf., e.g., with the results of refs. \cite{bsuvd94,mw} at $q^2 = 0$). 
Moreover, the $1/m_Q^2$ term generated by the box diagram is expressed 
merely in terms of $\langle B|\vec k^2|B\rangle$ , 
whereas the $1/m_Q^2$ corrections of the OPE series include also $\langle B|\hat{V}_1|B\rangle $,  
where $\hat{V}_1$ is the $1/m_Q$ term appearing in the expansion (\ref{Vpot}) of the effective 
potential (e.g., the chromomagnetic operator in QCD).

\end{itemize}

We argue however that it is possible to further modify the spectral 
representation of the box diagram to bring the size of the $1/m_Q^2$ 
term developed by this modified representation in agreement with the OPE result. 
This procedure corresponds to phenomenologically taking into account 
the contribution of other $1/m_Q^2$ terms of the expansion (\ref{exp2}). 

Omitting the anomalous contribution for the reasons explained above, the differential decay rate 
takes the form 
\begin{eqnarray}
\label{uncut}
\frac{d\Gamma}{dq^2}&=&K_0(q^2) 
\int_{(m_1+m_3)^2}^{\infty} ds_1 
\varphi^2(s_1)\frac{s_1-(m_2-m_3)^2}{8\pi^2s_1}\lambda^{1/2}(s_1,m_2^2,m_3^2)
\frac{m_2}{\sqrt{s_1}}\int_{-1}^{1}\frac{d\eta}{2}\lambda^{1/2}(s_1,s_2,q^2).
\end{eqnarray}
where $s_2$ depends on $\eta$ through Eq. (\ref{limitss2}). 

Now, our goal is to modify the differential distribution of Eq. (\ref{uncut}) 
in such a way that the leading order result and the $1/m_Q$ correction in the integrated rate remain
intact whereas the $1/m_Q^2$ term numerically reproduces the OPE-based estimate. 

Obviously, we may allow a strong deformation of the 
differential $q^2$-distribution at large $q^2$ near zero recoil, where the 
heavy-quark expansion is anyway ill-defined; we may also require $d\Gamma/dq^2(q^2=0)$ 
to exactly reproduce the OPE result within the $1/m_Q^2$ accuracy. 

Most easily this program may be implemented through the following two steps: 

First, by introducing the factor $F(s_1)=1/(1+\vec k^2/m_Q^2)$ 
which sets the $1/m_Q^2$ term in the {\it differential} rate at $q^2=0$;  

Second, by cutting the $s_2$-integration in (\ref{uncut}) at some upper limit 
$s_2^{max}(q^2)$ which tunes the size of the $1/m_Q^2$ effects in the {\it integrated} rate. 

The resulting differential $q^2$ differential distribution takes the form 
\begin{eqnarray}
\label{cut}
\frac{d\Gamma}{dq^2}&=&K_0(q^2) 
\int
\limits_{(m_1+m_3)^2}^{\infty} 
ds_1 
\varphi^2(s_1)\frac{s_1-(m_2-m_3)^2}{8\pi^2s_1}\lambda^{1/2}(s_1,m_2^2,m_3^2)
\nonumber\\
&&\times
\frac{m_2}{\sqrt{s_1}}F(s_1)\int_{-1}^{1} \frac{d\eta}{2} 
\lambda^{1/2}(s_1,s_2,q^2)\theta\left(s_2<s_2^{max}(q^2)\right). 
\end{eqnarray}
In order not to affect the integrated rate in the leading and the subleading $1/m_Q$ orders, 
the function $s_2^{max}(q^2)$ should satisfy certain properties. Let us look at them more closely. 

Recall, that the soft wave function $\varphi(s_1)$ is localized in the region
\begin{eqnarray}
s_1\le s_1^{max} \simeq (m_2+m_3+\gamma)^2
\nonumber
\end{eqnarray} 
where $\gamma$ is a constant of order $\Lambda$ 
which does not scale with $m_Q$. Let us determine $q_0^2$ through the equation 
\begin{eqnarray}
s_2^+(s_1^{max},q_0^2)=s_2^{max}(q_0^2)
\end{eqnarray}
where $s_2^+$ is the maximal value of $s_2$ corresponding to $\eta=1$ in (\ref{limitss2}). Furthermore, assume that $s_2^{max}(q^2)$ decreases with $q^2$, and take into account that $s_2^+(s_1,q^2)$ is a monotonous rising function of both $s_1$ and $q^2$. 
Then, at $q^2<q_0^2$ for all $s_1<s_1^{max}$ one finds the relation 
$s_2^+(s_1,q^2)<s_2^{max}(q^2)$, and thus
the $q^2$-distribution does not feel the presence of the cut at all. 
For $q^2>q_0^2$ the cut becomes really effective and strongly
influences the $q^2$-distribution. In order these changes in the cut $q^2$-differential 
distribution not to change the integrated rate in the LO and $1/m_Q$ order, we need the 
$q_0^2$ to be not far from zero recoil such that the corresponding $\omega_0=1+O(1/m_Q)$. 
Choosing the cut in the form 
\begin{eqnarray}
\label{rcut}
\sqrt{s_2^{max}(q^2)}=m_1+m_3+a\left(m_2-m_1-\epsilon-\sqrt{q^2}\right), 
\end{eqnarray}
where $\epsilon\simeq \Lambda$ and $a$ is a rising function of $m_Q$, satisfies 
these requirements. 

The parameter $\epsilon$ accounts for a mismatch between the quark and the hadron threshold, and 
the form of  
$a(m_Q)$ can be found from fitting the size of the $1/m_Q^2$ corrections in the 
integrated rate to the OPE prediction.    
Notice also that the $q^2$ distributions obtained through the cut expression are even 
more realistic than those obtained from the uncut spectral representations.

We could of course choose a more sophisticated parameterization of $s_2^{max}(q^2)$ 
to reproduce a correct $q^2$-behaviour of $d\Gamma/dq^2$ near zero recoil point. For instance, 
taking into account that the lightest final meson is pseudoscalar, we can write  
$\sqrt{s_2^{max}(q^2)}=m_1+m_3+a_P(m_Q) (M_{P_Q}-M_{P_{Q'}}-\sqrt{q^2})^{3/2}$ yielding 
the correct behavior near $\sqrt{q^2}=M_{P_Q}-M_{P_{Q'}}$ where only one $P-$wave decay channel  
$P_Q\to P_{Q'} \ell \bar{\nu}_{\ell}$ is opened. In addition, in the heavy quark limit the $S-$wave transition 
$P_Q\to V_{Q'} \ell \bar{\nu}_{\ell}$ requiring another functional dependence 
$\sqrt{s_2^{max}(q^2)}=m_1+m_3+a_V(m_Q)(M_{P_Q}-M_{P_{Q'}}-\sqrt{q^2})^{1/2}$
is opened at $\sqrt{q^2}=M_{P_Q}-M_{V_{Q'}}$ with only small delay in $q^2$ of order $\Lambda^2$. 
So the effects of opening this channel are even more important and should be also taken into account. 
However in the region of large $q^2$ with few opened channels the inclusive consideration 
is anyway not working properly, and taking into account such subtle effects is beyond the 
accuracy of the method. So in numerical calculations we proceed with the phenomenological 
cut provided by eq (\ref{rcut}).

The parameters of the model, such as the quark masses and the wave functions 
were already fixed in Chapter III and we now use them for inclusive decays. 

We choose the parameters of the cut according to the criteria above in the following form 
\begin{eqnarray}
\epsilon&=&(m_Q-m_{Q'})-(M_Q-M_{Q'})
\nonumber\\
a&=& 1.82 + 0.029 m_Q, 
\nonumber
\end{eqnarray}
where $m_{Q'}$ is the mass of the parent heavy quark and 
$M_{Q'}$ is the final meson lowest mass. Effectively, this means that 
$q^2_{max}=(M_Q-M_{Q'})^2$.  

A remark is in order here: To compare the calcuated integrated rate $\Gamma$ as a 
function of $m_Q$ with the OPE result 
within the $1/m_Q^2$ order, it is reasonable to adopt the expansions of 
the hadron masses $M_Q$ and $M_{Q'}$ also up to the second order in 
$1/m_Q$. Then one finds 
$\epsilon = -\frac{1}{2}(\lambda_1 + 3 \lambda_2) \cdot (1/m_{Q'} - 1/m_{Q})$. 
Setting $\lambda_1 = -0.44 ~ GeV^2$ and $\lambda_2 = 0.12 ~ GeV^2$, 
yields $\epsilon = 0.10 / m_Q$ at $m_{Q'} / m_Q = m_c / m_b \simeq 0.28$.
We point out that in calculations for the real $B$ decays we use experimental values of 
hadronic masses (involving all orders in $1/m_Q$). 
 
The corresponding integrated semileptonic decay rate $\Gamma$ 
is plotted in Fig. \ref{fig:inclrate} as a function of $1/m_Q$ and compared with 
the OPE predictions (\ref{OPE_total}). 
\begin{figure}[htb]
\vspace{0.25cm}
\begin{center}
\epsfig{file=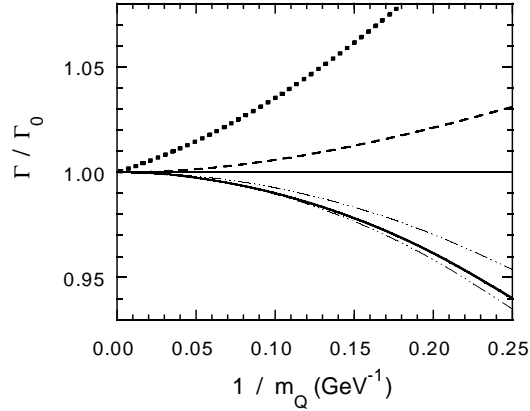,width=7cm}
\vspace{0.25cm}
\caption{ The ratio of the integrated rates of the bound-to-free quark semileptonic  
decay $Q \to Q' \ell \bar{\nu}_{\ell}$ vs the inverse heavy-quark mass  
$1/m_Q$ at fixed value of $m_{Q'} / m_Q = m_c / m_b = 0.28$. Dotted line - the rate 
calculated from the initial spectral representation of the box diagram which 
contains $1/m_Q$ correction, dashed - with the proper subtraction killing 
the $1/m_Q$ term but without tuning the size of the $1/m_Q^2$ effects. Solid 
- final result which also includes the cut bringing the size of the 
$1/m_Q^2$ effects in agreement with the OPE prediction.  Upper and lower 
dot-dashed lines are the OPE results (\ref{OPE_total}) corresponding to 
$\lambda_1 = 0$ and $\lambda_1 = -0.6 ~ GeV^2$ (with $\lambda_2 = 0.12 ~ GeV^2$), respectively.
\label{fig:inclrate}}
\end{center}
\end{figure}
One can clearly see that the calculations based on both the spectral density $r_D$ of Eq. (\ref{r2}) 
and the spectral density $r$ of Eq. (\ref{r3}) which includes a subtraction 
predict a larger rate for a bound heavy quark compared to the free one, that contradicts OPE. 

The introduction of the cut $s_2^{max}(q^2)$ (\ref{rcut}) brings our dispersion approach 
results in perfect agreement with the standard OPE framework for the whole range of the 
considered values of $m_Q$.

Figure \ref{fig:inclrq2} shows the influence of the cut upon the 
$q^2$-distribution for the $B \to X_c \ell \bar{\nu}_{\ell}$ decay. 
\begin{figure}[htb]
\vspace{0.25cm}
\begin{center}
\epsfig{file=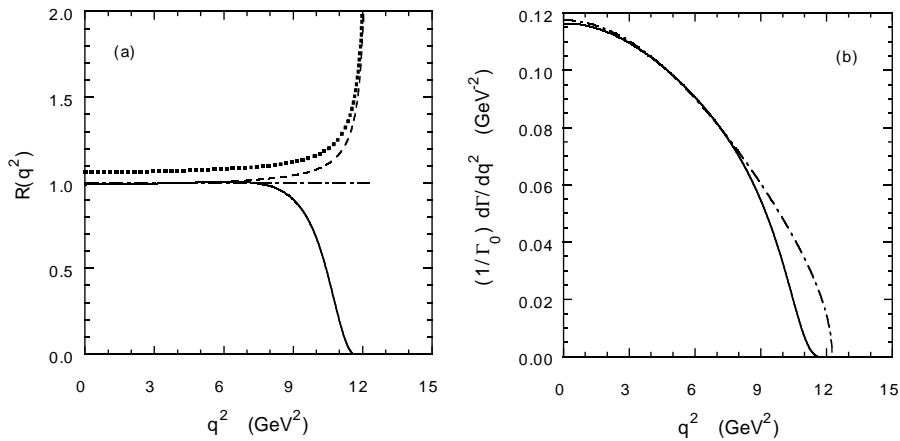,width=12cm}
\vspace{0.25cm}
\caption{Distribution $d\Gamma/dq^2$ in $B \to X_c \ell \bar{\nu}_{\ell}$ decays vs the 
squared four-momentum transfer $q^2$: (a) ratio of the bound-to-free quark decay 
$R(q^2) \equiv (d\Gamma / dq^2) / (d\Gamma_0 / dq^2)$, 
notation of lines same as in Fig. \protect\ref{fig:inclrate}. (b) Differential distribution 
in the bound 
(solid) and free (dot-dashed) semileptonic quark decay. Parameters of the cut (\ref{rcut}) are 
$a=1.96$ and $\epsilon=0.091\;GeV$. 
\label{fig:inclrq2}}
\end{center}
\end{figure}
As already discussed, the introduction of the $q^2$-dependent cut in the spectral 
representation does not change the differential $q^2$ distributions at small $q^2$ but it strongly 
affects the region of large $q^2$. In particular, the cut provides the vanishing of 
$d\Gamma/dq^2$ at the {\it physical} threshold $q^2=(M_B-M_D)^2=11.6\;GeV^2$. Let us point out again that 
this cutting procedure does not affect the {\it integrated} rate at the leading and subleading 
$1/m_Q$ orders. Note also that the differential distributions $d\Gamma/dq^2$ given by the dashed and solid lines in Fig \ref{fig:inclrq2}(a), 
are equal to each other at $q^2=0$ and match the OPE result for $d\Gamma/dq^2(q^2=0)$.

 
As we shall see later, the improvements on the $q^2$-differential
distributions by approximate account of higher order graphs 
affect only moderately the $M_X$-distribution in $B\to X_c \ell \bar{\nu}_{\ell}$ 
(as well as the photon lineshape in the rare $B\to X_s\gamma$ decay). The latter are thus 
determined by the $B$-meson wave function. 

This property allows us to obtain a realistic energy distribution and other observables 
through the soft wave function of the heavy meson. Thus we do not need to introduce 
any unknown 'smearing function' describing the motion of the $b$ quark inside the $B$ 
meson, but rather directly calculate the effects of the $b$ quark motion with the 
soft meson wave function.

\subsection{\label{v.d}$d\Gamma/dM_X$}
We now proceed to the calculation of the $M_X$-distribution. 

Recall that in the free-quark decay, 
which is the leading-order process within the OPE framework, 
one finds the $M_X$ distribution in the form $\delta(M_X-m_c)$ (neglecting the 
radiative corrections). 
Inclusion of the $1/m_Q$ corrections yields a singular series 
containing derivatives of the $\delta$ function. For the interpretation of these
results one needs an introduction of a smearing function. In the quark model 
the $M_X$-spectrum is obtained already smeared because of the Fermi-motion of the 
$b$ quark in the $B$ meson. Therefore the shape of the $M_X$-spectrum is calculable 
through the $B$ meson wave function determined in the Chapter II. 

Let us calculate $d\Gamma/dM_X$ from our spectral representation for the decay rate. 
To this end it is necessary to change the order of integration 
in Eq. (\ref{inclrate2}) as follows ($M_X^2=s_2$ in our notations)
\begin{eqnarray}
\frac{d\Gamma}{dq^2} & = & K_0(q^2)
\int\limits_{(m_1+m_3)^2}^{s_2^{max}(q^2)} ds_2 
\int\limits_{s_1^-(s_2,q^2)}^{s_1^+(s_2,q^2)}
ds_1\varphi^2(s_1)\frac{s_1-(m_2-m_3)^2}{8\pi^2}
\left(\frac{m_2}{\sqrt{s_1}}\right)^3
\frac{\lambda^{1/2}(s_1,s_2,q^2)}{\lambda^{1/2}(m^2_2,m_1^2,q^2)}
\nonumber
\end{eqnarray}
This representation immediately gives $d^2\Gamma/dq^2 dM_X^2$. 
Integrating the latter over $q^2$ one obtains $d\Gamma/dM_X^2$.  
The calculated distribution is reported in Fig.\ref{fig:inclrmx}. 
Our result should be compared with the leading order OPE result $\delta(m_X-m_c)$. 
One can see that already the box diagram of the quark model provides a smooth and 
reasonable distribution (beyond the resonance region), which is only moderately 
affected by a proper account of the subleading $1/m_Q$ effects. 
Most important is that {\it at large $M_X$ the calculated 
distribution does not require any additional smearing and can be directly compared with the 
experimental results}. 
\begin{figure}[htb]
\vspace{0.25cm}
\begin{center}
\epsfig{file=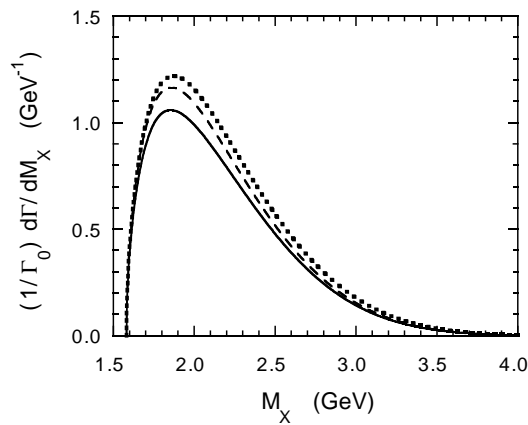,width=7cm}
\vspace{0.25cm}
\caption{Distribution $d\Gamma / dM_X$ in $B \to X_c \ell \bar{\nu}_{\ell}$ decays vs the invariant mass of the produced hadronic system $M_X$. Lines same as in Fig. \protect\ref{fig:inclrate}.  
\label{fig:inclrmx}}
\end{center}
\end{figure}

\newpage
\subsection{\label{v.e}The electron energy spectrum}
In distinction to the Lorentz-invariant distributions $d\Gamma/dq^2$ and
$d\Gamma/dM_X$ considered above, the lepton energy spectrum should be referred 
to a specific reference frame. 

The starting point of the electron energy spectrum calculation 
is the double spectral representation (\ref{gammadisp}) for the 
inclusive semileptonic decay rate. 
We define the variable $E$ to be the energy of the lepton $l_1$ in the 
reference frame specified by the condition 
\begin{eqnarray}
\label{inclrefframe}
\tilde p_1=(\sqrt{s_1}, 0). 
\end{eqnarray}
This is the rest frame of the $Q_2\bar q_3$ quark pair with the invariant mass
$\sqrt{s_1}$. 

As the first step, let us consider the hadronic tensor $\tilde w^D_{\mu\nu}$
given by Eq. (\ref{tildew}). Using the expression (\ref{hadronictrace}) for the trace
and performing the integration over $dk_1 dk_2$ leads to the following result 
\begin{eqnarray}
\label{a4}
\tilde w_{\mu\nu}(\tilde p_1,\tilde q)=
\frac{\pi\theta(\dots)\left(s_1-(m_2-m_3)^2\right)}
{\lambda^{1/2}(m_2^2,m_1^2,q^2)}w_{\mu\nu}(\tilde p_1,\tilde q)
\end{eqnarray}
where $w_{\mu\nu}$ is represented in terms of the 'dispersion momenta'  
$\tilde p_1$ and $\tilde q=\tilde p_1-\tilde p_2$ 
($\tilde p_1^2=s_1$, $\tilde p_2^2=s_2$, $\tilde q^2=q^2$) as follows 
\begin{eqnarray}
\label{a5}
w_{\mu\nu}(\tilde p_1,\tilde q)=-g_{\mu\nu}w_1
+\frac{\tilde p_{1\mu}\tilde p_{1\nu}}{s_1}w_2
+i\epsilon_{\mu\nu\alpha\beta}\frac{\tilde p_{1\mu}}{\sqrt{s_1}}\tilde q_\beta w_3
+\frac{\tilde p_{1\mu}\tilde q_\nu+\tilde p_{1\nu}\tilde q_\mu}{\sqrt{s_1}}w_4
+\tilde q_\mu \tilde q_\nu w_5. 
\end{eqnarray}
Lorentz structures containing the momentum $\kappa$ are omitted in this expression. 

The functions $w_{1-3}$ necessary for the energy distribution have the form 
\begin{eqnarray}
&w_1&=4(m_1^2+m_2^2-q^2)-16\beta,\qquad 
\nonumber\\
q^2 w_1+\frac{\vec q^2}{3} w_2&=&\frac{4}{3}C_{SL}(m^2_1,m^2_2,q^2),\qquad 
\nonumber\\
&w_3&=8\sqrt{s_1}(1-\alpha_1-\alpha_2),
\end{eqnarray}
with $\alpha_1$, $\alpha_2$, and $\beta$ given by Eqs. (\ref{alpha1}-\ref{beta}).  
We have introduced 
\begin{eqnarray}
\label{a61}
|\vec q|=\frac{\lambda^{1/2}(s_1,s_2,q^2)}{2\sqrt{s_1}}, \qquad
q^0=\frac{s_1+q^2-s_2}{2\sqrt{s_1}},
\end{eqnarray}
the components of the momentum $q$ in the reference frame (\ref{inclrefframe})

It might be interesting to note that at $q^2<0$, and using the reference frame 
$q_+=0$, $p_{1\perp}=0$ ($q^2=-q_\perp^2$) one obtains 
\begin{eqnarray}
\beta=-\left(k_\perp^2-\frac{(k_\perp q_\perp)^2}{q_\perp^2}\right),\qquad
1-\alpha_1-\alpha_2=1-x_3, 
\end{eqnarray}
where $x_3$ and $k_\perp$ are the $(+)$ and $(\perp)$ components of the spectator
quark momentum, respectively. In the heavy meson one finds 
\begin{eqnarray}
\label{a11}
\beta\simeq \Lambda^2,\qquad x_3\simeq \Lambda/m_Q.
\end{eqnarray}
Although at $q^2>0$ the 
interpretation of $\beta$ and $\alpha_1+\alpha_2$ in terms of 
$k_\perp$ and $x_3$ is not straightforward, the estimates 
(\ref{a11}) remain valid also at $q^2>0$. 
This means that in $B \to X_c \ell \bar{\nu}_{\ell}$ decay our 
$w_i$'s differ only slightly from the corresponding free-quark expressions.

Now everything is prepared for the calculation of the lepton energy spectrum. 
The following steps are standard: 
We must take the convolution of the trace over the lepton loop with the hadronic
tensor $\tilde w_{\mu\nu}$, and integrate the result over the part of the 
lepton pair phase space leaving out the integration over $dE$. 

As a result, we obtain the following double differential distribution 
\begin{eqnarray}
\label{double}
\frac{d^2\Gamma}{dEdq^2}&=& \frac{G_F^2|V_{21}|^2}{128\pi^3}
\int
\limits_{(m_2+m_3)^2}^{\infty}
ds_1\varphi^2(s_1)
\frac{s_1-(m_2-m_3)^2}{8\pi^2s_1}\lambda^{1/2}(s_1,m_2^2,m_3^2)\frac{F(s_1)}{m_2^2}
\nonumber\\
&&
\int\limits_{-1}^{1}\frac{d\eta}{2} \theta\left(q_0>E+\frac{q^2}{4E}\right)
\theta\left(s_2<s_2^{max}(q^2)\right)
\nonumber\\
&&\times
\left\{2q^2 w_1(s_1,s_2,q^2)
+[4E(q^0-E)-q^2] w_2(s_1,s_2,q^2)
+2q^2(2E-q^0) w_3(s_1,s_2,q^2) \right\}. 
\end{eqnarray}
In this formula $s_2$ is connected with $\eta$ through Eq. (\ref{limitss2}), 
and $q_0$ is given in terms of the integration variables by Eq. (\ref{a61}).

Eq. (\ref{double}) includes modifications 
which tune the size of the $1/m_Q^2$ corrections as explained above.  

Let us check that the integration of the double differential distribution 
(\ref{double}) over $E$ brings us to the expression (\ref{cut})
for $d\Gamma/dq^2$ obtained earlier. 

By virtue of the relations  
\begin{eqnarray}
&&\int dE\theta(q^0>E+q^2/4E)=|\vec q|,\\
&&\int E dE\theta(q^0>E+q^2/4E)=q^0|\vec q|/2,\\
&&\int E^2 dE\theta(q^0>E+q^2/4E)=\frac{1}{12}|\vec q|(3q_0^2+\vec q^2).
\end{eqnarray}
one finds 
\begin{eqnarray}
\label{check}
\int dE 
\{2q^2 w_1+[4E(q^0-E)-q^2]w_2+2q^2(2E-q^0)w_3 \}\theta(q^0>E+q^2/4E)
=2|\vec q|(q^2w_1+\frac{1}{3}\vec q^2w_2). 
\end{eqnarray}
Thus, integrating the double differential distribution (\ref{double}) over $E$ 
and using (\ref{check}) gives $d\Gamma/dq^2$ in the form (\ref{cut}). 
Clearly, for the calculation of $d\Gamma/dq^2$ 
we do not need to know all $w_i$, but only their linear combination 
(\ref{check}). On the other hand, for calculating the electron energy spectrum
$d\Gamma/dE$ all functions $w_{1-3}$ are needed. 

The electron spectrum $d\Gamma/dE$ is obtained by integrating (\ref{double}) over $q^2$. 
Fig. \ref{fig:inclre} plots the results of the calculations. 
Here the effects of the subleading $1/m_Q$ orders are more pronounced but 
nevertheless lead only to a moderate change of the quark model box-diagram 
result. We also compare the dispersion approach prediction with the electron spectrum 
in the free-quark decay process. At low values of the electron energy the 
spectrum in free quark decay and in the inclusive $B$ decay are practically
indistinguishable. As desired the spectrum is nonzero at large $E$ above the 
free-quark threshold. At the intermediate $E$ the spectrum in $B$ decay is 
substantially than the spectrum in the free-quark decay, providing the known 
suppression of the integrated $\Gamma(B\to X_cl\nu)$ compared to 
$\Gamma_0(b\to cl\nu)$. 

\begin{figure}[htb]
\vspace{0.25cm}
\begin{center}
\epsfig{file=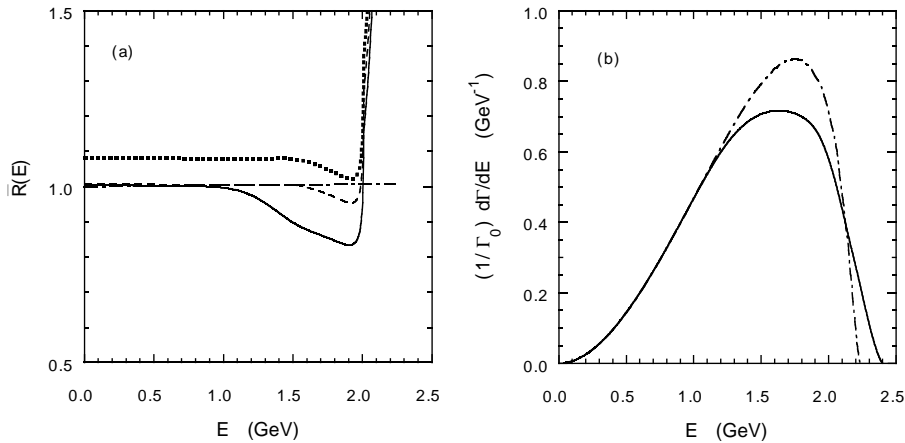,width=12cm}
\vspace{0.25cm}
\caption{Distribution $d\Gamma / dE$ in $B \to X_c \ell \bar{\nu}_{\ell}$ decays vs the lepton energy $E$:
(a) ratio of the bound-to-free quark decay $\bar{R}(E) \equiv (d\Gamma/dE)/(d\Gamma_0 / dE)$, lines same as in Fig. \protect\ref{fig:inclrate}, 
(b) Our quark-model result (solid) vs free-quark result (dot-dashed).
\label{fig:inclre}}
\end{center}
\end{figure}

\newpage
\subsection{\label{v.f}Discussion}

In this Chapter we discussed the dispersion approach to the description of quark-binding 
effects in the inclusive decays of heavy mesons. Our approach allows to express 
kinematical distributions in inclusive decays of heavy mesons in terms of the heavy meson 
soft wave function. This soft wave function describes long-distance effects both in 
exclusive and inclusive processes. 

Summarising the main results of this Chapter:
\begin{itemize}

\item[1.] We analysed the hadronic tensor in the quark model. We have shown that 
the diagrammatic representation of the hadronic tensor in the quark model yields
an expansion in inverse powers of the heavy quark mass as well as the standard OPE. However 
in distinction to the standard OPE which is based on the free quark decay as the LO process, the 
LO process in the quark model is described by the box diagram 
with a free $Q\bar q$ in the final state. This yields some specific features of the hadronic 
tensor calculated within the quark model, both negative and positive.

On the one hand, a consideration based on the box diagram alone reproduces the correct 
LO result but contains also $1/m_Q$ corrections. These $1/m_Q$ terms are cancelled 
against contributions of other graphs thus leading to the agreement with the standard 
OPE result. Hence the effects of the subleading diagrams should be taken into account 
for obtaining a consistent approach.

On the other hand, the hadronic tensor calculated from the box diagram is a regular function in 
the kinematically allowed region of all variables, as well as all the subleading terms of 
order $1/m_Q$.  
This feature makes the quark model calculation of the hadronic tensor 
very suitable for describing the differential distributions; 

\item[2.] We have constructed the double spectral representation of the hadronic tensor for the $B \to X_c \ell \bar{\nu}_{\ell}$ decay
in terms of the $B$ meson soft wave function and the double spectral density of the box diagram, 
and analysed its $1/m_Q$ expansion in the case of the heavy-to-heavy inclusive transition. 
The spectral representation is further modified in order to take into account essential 
effects of the other diagrams contributing in subleading orders. Namely,  
the subtraction term in this dispersion representation is determined such that the $1/m_Q$ correction 
in the integrated semileptonic rate is absent in agreement with OPE. Furthermore, a phenomenological 
cut function is introduced into the spectral representation to bring the size of the $1/m_Q^2$ terms in the 
differential $q^2$-distribution at $q^2=0$ and in the integrated rate into full 
agreement with OPE. Thus our representation of the hadronic tensor obeys the OPE predictions in the regions 
where the latter are expected to be valid; 

\item[3.] We have obtained numerical results on differential distributions in inclusive $B \to X_c \ell \bar{\nu}_{\ell}$ decays using the 
$B$-meson wave function and other quark model parameters previously determined from the 
description of exclusive meson transition form factors within the dispersion approach. So, basically our predictions are parameter-free.
We notice that 
modifications of the spectral representations which take into account the subleading $1/m_Q$ 
effects 
within the box-diagram representation, introduce some uncertainties in our results. However, they  
do not affect our predictions strongly, and the main features of the 
inclusive distributions are determined by the soft meson wave function. Moreover, the size of 
the 
subleading corrections is in perfect agreement with the OPE result for the integrated rate, 
and we expect 
to describe also these subleading effects in differential distributions in a proper 
quantitative way.

\end{itemize} 

The proposed approach can be applied to the inclusive $B\to X_{u,s}$ transitions.  
In particular it is especially suitable for the description of the photon line shape in $B \to X_s \gamma$ decays.  
However, certain subtleties in heavy-to-light transitions 
compared with the heavy-to-heavy decays emerge. They  
are mostly connected with the fact that in heavy-to-light transition the kinematically allowed 
$q^2$-interval of the hadron semilaptonic decay is larger than that of the quark decay, while in case of the 
heavy-to-heavy transitions the situation is just opposite and the $q^2$-region of the quark decay is larger. 
This feature requires a detailed analysis of the $q^2$-region near zero recoil in heavy-to-light 
inclusive decays. 

It is also worth noting that our approach takes into account only 
non-perturbative effects in inclusive decays of heavy mesons. 
Perturbative corrections have been ignored. So, for 
comparing our results with the experimental differential distributions perturbative corrections 
should be also included into consideration. 

\newpage
\section{Conclusion}
We presented a review of the relativistic dispersion approach to meson transitions
induced by the weak currents. This approach is based on the constituent quark
picture and takes into account the leading two-particle singularities of the
Feynman diagrams making use of spectral representations over the mass variables. 

We reported the following main results: 

\vspace{.5cm}
1. Relativistic meson $Q\bar q$ wave function has been defined which 
describes in a universal way strong-QCD effects related to the meson structure 
in various meson interactions. Due to confinement, the wave function 
in terms of the relative quark momentum is localized in the region 
of the order of the confinement scale. The normalization condition for the wave 
function corresponds to the electric charge conservation and is thus related to 
the meson elastic electromagnetic form factor at zero momentum transfer. 

\vspace{.5cm}
2. We considered the elastic electromagnetic form factor of a pseudoscalar meson and 
form factors describing weak transition of a pseudoscalar meson to 
pseudoscalar and vector mesons at spacelike momentum transfers.  
Relativistic double spectral representations for these form factors have been  
constructed in terms of the double spectral densities of the corresponding 
triangle Feynman graphs and the wave functions of the participating mesons. 

The subtraction prescription has been fixed by requiring the dispersion approach 
form factors to respect all properties of the transition form factors known from  
long-distance QCD in the limit of heavy quark mass. 

\vspace{.5cm}
3. The dispersion approach form factors 
were studied in the two limiting cases of the heavy-to-heavy and heavy-to-light 
weak transitions.
\begin{itemize}
\item
{\it Heavy-to-heavy} transition  

In this case the masses of the initial 
quark $m_Q$ and the final quark $m_{Q'}$ participating in the weak transition 
$Q\to Q'$ are taken to be much larger than the confinement scale $\Lambda$
\begin{eqnarray}
m_Q\simeq m_{Q'}\gg\Lambda.
\nonumber  
\end{eqnarray} 
The form factors from the dispersion approach have been shown to fully respect the  
structure of the $1/m_Q$ expansion known from QCD in the leading and 
subleading $1/m_Q$ orders. 
The Isgur-Wise function and the subleading-order universal form factors 
have been calculated in terms of the wave function of a meson containing 
an infinitely heavy quark.  
\item
{\it Heavy-to-light} transition 

In this case the quark masses satisfy the relation    
\begin{eqnarray} 
m_Q\gg m_{Q'}\simeq\Lambda.
\nonumber  
\end{eqnarray}
The dispersion approach form factors have been shown to 
obey the QCD-based relations between the form factors of the vector, axial vector 
and tensor currents valid within the $1/m_Q$ accuracy in the region of large $q^2$. 
\end{itemize}

\vspace{.3cm}
4. We performed a detailed comparison of the dispersion approach with the 
light-front quark model for the description of weak transitions. 
Both approaches describe the dominance of the $q\bar q$ components of the meson wave 
function in meson interactions and therefore should lead to the same results. 

Indeed, we have shown that for the transition between pseudoscalar mesons 
the results from both approaches in the region of spacelike momentum transfers $q^2<0$ 
are identical. On the other hand, for some of the form factors describing the pseudoscalar to vector meson
transition the results are different. This is related to a different treatment of
a vector meson, a composite system with spin-1, within both approaches. 
In the language of spectral representations, the
difference between the two approaches can be traced back to different subtraction 
prescriptions.  
We have shown that the subtraction prescription implemented in the light-cone quark
model does not fully respect the next-to-leading order constraints from QCD for the 
heavy-to-heavy transition. 

Another advantage of the dispersion approach is the possibility for a consistent 
description of the region of timelike momentum transfers $q^2>0$ relevant for the decay
kinematics. In the dispersion approach this is achieved by performing the analytical 
continuation in the variable $q^2$. The light-front quark model in general cannot 
consistently treat the form factor at $q^2>0$: 
it describes only a reference-frame dependent part of the form factor, 
the partonic contribution, leaving out the nonpartonic contribution. 

\vspace{.5cm}
5. We obtained the form factors in the region $q^2>0$ by the analytical continuation. 
We observed the appearance of the anomalous cut in the complex $s$-plane, where 
$s$ is the invariant mass of the $q\bar q$ pair. Accordingly, the spectral 
representation for the transition form factor acquires the anomalous contribution 
in the decay region. The anomalous contribution is small for small $q^2$ but 
becomes increasingly important in the region of large $q^2$ near zero recoil. 

\vspace{.5cm}
6. A model for pseudoscalar mesons was considered, which allowed us to study 
the transition regime to the heavy quark limit. Meson decay constants and transition form 
factors as functions of the quark masses have been analysed. 
It was found that the corrections to the leading order $1/m_Q$ relations can be as big 
as 10-20\% for the quark masses in the region of the realistic $c$ and $b$ quarks. 

\vspace{.5cm}
7. We presented the results  for the form factors for weak decays of $B_{(s)}$ 
and $D_{(s)}$ mesons to light pseudoscalar and vector mesons. 
\begin{itemize}
\item
The effective quark masses and meson wave functions were determined 
by fitting the quark model parameters to the available lattice QCD results for the 
$B\to \rho$ transition form factors at large momentum transfers and to 
the measured $D\to (K,K^*)l\nu$ decay rates. 
The knowledge of the quark model parameters allowed us to predict numerous form 
factors for many decay channels and for all kinematically accessible $q^2$ values. 

In spite of the rather different masses and properties of mesons
involved in weak transitions, all existing data on the form factors can be 
understood in the constituent quark picture, i.e. all 
form factors can be described by the few degrees of
freedom of constituent quarks. 
Details of the soft wave functions are not crucial; only the
spatial extention  of these wave functions of order of the confinement scale
is important. In other words, only the meson radii are essential.
\item
The calculated transition form factors are in good agreement
with the results from lattice QCD and from sum rules
in their regions of validity. 

\item
Our predictions agree well with all available experimental data. 
An important example is the ratio of the semileptonic branching 
fractions of the $B$ meson to light $\pi$ and $\rho$ mesons, 
${\cal B}(B\to\rho l\nu)/{\cal B}(B\to\pi l\nu)$. A good agreement of the predicted 
ratio with the results of the experiment leads to close values of the 
matrix element $V_{ub}$ extrtacted from the $B\to\rho$ and $B\to\pi$ channels. 

\item 
We estimated the products of the meson weak and strong coupling constants by 
extrapolating the form factors to the meson pole. The value of each coupling constant 
can be obtained independently from the residues of several form factors. 
In all cases the values extracted from the different form factors 
agree with each other within the 5-10\% accuracy. They also agree with the results of the
independent direct calculations of these decay constants within the dispersion approach. 
First direct measurement of the coupling constant $g_{D^*D\pi}$ which has become 
available recently is in perfect agreement with our estimate. 
This gives additional argument in support of the reliability of our results 
for the form factors. 
\end{itemize}

8. Weak annihilation in a rare radiative decay was analysed. The relevant 
$B\to\gamma l\nu$ form factors were calculated in terms of the $B$ 
meson wave function. Parameter-free estimates were obtained and found 
to agree well with estimated from perturbatibe QCD and QCD sum rules. 
The situation with contact terms was clarified. 

A new contribution to the weak annihilation amplitude was reported,   
related to the photon emission from the light-quark loop. 
This contribution  
was believed to vanish in the limit of zero quark masses, and therefore 
neglected in previous analyses. A detailed study shows that this 
contribution stays finite and should be taken into account. 

\vspace{.5cm}
9. Non-factorizable effects in the $B^0-\bar B^0$ mixing amplitude 
due to the soft gluon exchanges were studied 
assuming the local gluon condensate dominance. 
The corresponding correction to the factorizable amplitude was demonstrated 
to be negative. This correction contains specific $B$ meson transition form
factors which were calculated within the dispersion approach. Numerical 
estimate for the nonfactorizable effect was obtained.  

\vspace{.5cm}
10. The dispersion approach was applied to inclusive semileptonic decays. 
Spectral representations for the differential and integrated semileptonic decay rate 
$\Gamma(B\to X_c l\nu)$ were obtained.  
The decay rates calculated within the dispersion approach satisfy an important property 
known from the operator product expansion in QCD in the heavy quark limit: 
namely, the order $1/m_Q$ correction in the ratio of the bound to free quark decay rate 
$\Gamma(B\to X_c l\nu)/\Gamma_0(b\to cl\nu)$ is absent. 

Different kinematical distributions
such as the dilepton $q^2$-spectrum $d\Gamma/dq^2$, $d\Gamma/dM_X$ and the electron energy
spectrum $d\Gamma/dE_l$ were calculated in terms of the $B$ meson wave function.  
This wave function determines the $B$ meson form factors and has been 
tested in exclusive $B$ decays.


\vspace{1cm}
\acknowledgments
I take pleasure to express my gratitude to 
Vladimir Anisovich, Michael Beyer, Michail Kobrinsky, Alain Le Yaouanc, Bernard Metsch, 
Nikolai Nikitin, Victor Nikonov, Luis Oliver, Olivier P\'ene, Herbert Petry, 
Jean-Claude Raynal, Silvano Simula, and Berthold Stech for the most pleasant 
collaboration on many of the issues discussed in this paper. 

I am grateful to Yasha Azimov, Damir Becirevic, Gregory Korchemsky, 
Otto Nachtmann, and Lidia Smirnova for useful and stimulating discussions. 

I would like to thank the Alexander von Humboldt-Stiftung for financial 
support.


\begin{thebibliography}{150}
\bibitem{expd2pi}The rates for the $D\to \pi$ and $D\to\rho$ are obtained by
combining the values for the $D\to K$ and $D\to K^*$ with the measurements of the
ratios of the semileptonic branching fractions $D\to \pi/D\to K$ by CLEO 
\cite{cleopik} and
$D\to \rho/D\to K^*$ by E653 \cite{e653}.
\end{thebibliography}
\end{document}